\documentclass[onecolumn,noshowpacs,nofootinbib,11pt]{revtex4-1}

\usepackage{float,xcolor,upgreek,yfonts,tikz}
\usepackage{color}
\newcommand{\bes}{\begin{subequations}}
\newcommand{\ees}{\end{subequations}}
\usepackage{graphicx,bigints}
\usepackage{graphicx,float}\usepackage[all]{xy}
\usepackage{amsmath,upgreek}
 
  \newcommand{\blt}{\textcolor{black}}
   \usepackage{caption}
      \newcommand{\cclt}{\textcolor{cyan}}
   \newcommand{\clt}{\textcolor{black}}
 \usepackage{booktabs,makecell}
\usepackage{subcaption}
\usepackage{amssymb}

\usepackage{alphabeta}
 \usepackage{cancel}
\usepackage{dcolumn}
\usepackage{textgreek}
\usepackage{multirow} 
\usepackage{units}
\usepackage{overpic}
\usepackage{physics}
\usepackage{bm}

\newcommand{\beq}{\begin{eqnarray}}
\newcommand{\eeq}{\end{eqnarray}}
\newcommand{\be}{\begin{equation}}
\newcommand{\ee}{\end{equation}}

\newcommand{\bea}{\begin{eqnarray}}
\newcommand{\eea}{\end{eqnarray}}

\newcommand{\ba}{\begin{eqnarray}}
\newcommand{\ea}{\end{eqnarray}}
\usepackage[colorlinks,hyperindex,unicode]{hyperref}
\definecolor{green1}{RGB}{0,128,0} 
\hypersetup{hidelinks,backref=true,pagebackref=true,hyperindex=true,colorlinks=true,breaklinks=true,urlcolor= blue}
\hypersetup{%
  colorlinks = true,
  linkcolor  = blue,
  citecolor = cyan,
}
\usepackage{bookmark,textgreek}
\usepackage{hyperref,color,xcolor}
\hypersetup{hidelinks,hyperindex=true,colorlinks=true,breaklinks=true,urlcolor= blue}
\hypersetup{%
  colorlinks = true,
  linkcolor  = blue
}
\newcommand\orcidcasadio{{\href{https://orcid.org/0000-0002-1330-7787}{\orcidicon}}}

\newcommand\orcidchristian{{\href{https://orcid.org/0009-0009-5593-0479}{\orcidicon}}}

\newcommand\orcidroldao{{\href{https://orcid.org/0000-0003-3978-532X}{\orcidicon}}}
\newcommand{\orcidicon}{%
	\begin{tikzpicture}
	\draw[lime, fill=lime] (0,0)
		circle [radius=0.16]
		node[white] {{\fontfamily{qag}\selectfont \tiny ID}};
	\draw[white, fill=white] (-0.0625,0.095)
		circle [radius=0.007];
	\end{tikzpicture}	\hspace{-2mm}
}

\begin{document}
\title{Quantum gravitational corrections at third-order curvature, acoustic analog black holes and their quasinormal modes}

\author{R.~Casadio\orcidcasadio}
\affiliation{Dipartimento di Fisica e Astronomia, Universit\`a di Bologna, via Irnerio~46, 40126 Bologna, Italy}
\affiliation{I.N.F.N., Sezione di Bologna, I.S.~FLAG, viale B.~Pichat~6/2, 40127 Bologna, Italy}
\email{casadio@bo.infn.it}

\author{C. Noberto Souza\orcidchristian}
\affiliation{Center for Natural and Human Sciences, Federal University of ABC, 09210-580, Santo Andr\'e, Brazil}
\email{christian.noberto@ufabc.edu.br}

\author{R. da Rocha\orcidroldao}
\affiliation{Center of Mathematics, Federal University of ABC, 09210-580, Santo Andr\'e, Brazil}
\email{roldao.rocha@ufabc.edu.br}

\medbreak
\begin{abstract} 
{\color{black} Quasinormal modes for bosonic  (scalar, electromagnetic, and axial gravitational) and fermionic field perturbations, radiated from black holes that carry quantum gravitational corrections at third order in the curvature to the Schwarzschild solution, are scrutinized from the propagation of analog transonic sound waves across a de Laval nozzle.  The thermodynamic variables, the nozzle geometry, the Mach number, and the thrust coefficient are computed as functions of the parameter driving the effective action for quantum gravity containing a dimension-six local operator beyond general relativity. The quasinormal modes for quantum gravitational corrected analog black holes are also determined for higher overtones, yielding a more precise description of the quantum-corrected ringdown process and the gravitational waveform way before the fundamental mode sets in.}  
\end{abstract}



\maketitle
\section{Introduction}
\label{sec1} 
Black holes (BHs) are expected to emit quantum radiation, which makes them remarkable tools for exploring the connection
between quantum mechanics, thermodynamics, and gravity.
The interplay among these theoretical aspects is encompassed by the Bekenstein--Hawking entropy.
In particular, the entropy of a Schwarzschild BH can be accounted for by using effective field theoretical
methods to calculate quantum gravitational corrections to the Bekenstein--Hawking entropy,
which induce quantum corrections to the black body factors of Schwarzschild BHs~\cite{Calmet:2021lny,Calmet:2019eof}. 

Quantum gravity corrections are expected to emerge in gravitational backgrounds beyond the Einstein--Hilbert action
of General Relativity (GR).
Invariance under diffeomorphism is a fundamental symmetry of gravity, and the Einstein--Hilbert action can therefore
be viewed as the deep-infrared and lowest-energy approximation of a more complete action involving higher-order curvature
operators that induce quantum corrections. 
The sum of operators constituting the effective action for quantum gravity has to be truncated somehow, for the task of
viably deriving the equations of motion that correspond to physically sound
solutions~\cite{Casadio:2021rwj,Kuntz:2019lzq,Asorey:2024mkb,Bonora:2014dfa,Barvinsky:1994hw,Odintsov:1991nd}.
Higher-order curvature operators are commonly used in effective approaches to quantum gravity.
Relevant phenomenological aspects of quantum gravity,  quantum gravitational corrected BHs, and some of their
physical signatures were addressed in Refs.~\cite{Wang:2025fmz,Nastase:2016sji,Perrucci:2024qrr,Calmet:2024pve,Donoghue:2018izj,Kuntz:2019omq,Casadio:2017cdv,Wang:2019rcf,deBrito:2020xhy,Casadio:2017sze}
also encompassing effective field theories (EFTs)~\cite{Moreira:2024ewr,Lessa:2023dbd,Ovalle:2018ans,Bazeia:2014xxa,Meert:2020sqv}. 
Quantum gravity corrections were explored in the context of hydrodynamics in Refs.~\cite{daRocha:2020gee,daRocha:2021xwq,daRocha:2009gb}, with several applications.

Corrections to the Einstein--Hilbert gravity at 1-loop were computed by ’t Hooft and Veltman~\cite{tHooft:1974toh},
whereas Goroff and Sagnotti  addressed quantum gravitational corrections at third-order in curvature as 2-loop quantum
gravitational corrections to gravity~\cite{Goroff:1985th}.
In this context, Refs.~\cite{Lessa:2023thi,Bueno:2016lrh,DeFelice:2023vmj,Rachwal:2021bgb} studied Einstein cubic gravity
and effective theories of gravity, while quantum gravity with a third-order curvature term was studied in Refs.~\cite{Lambiase:2024vkz,Calmet:2023gbw,Matyjasek:2020bzc,Marciu:2023hdb,Bukhari:2025fhd}.
The effect of third-order curvature corrections to gravity and the consequence for thermal and hydrodynamical properties
of a dual gauge theory describing quantum-corrected transport coefficients of the quark-gluon plasma were analyzed in Ref.~\cite{daRocha:2024lev}. 
 
One can explore bosonic and fermionic perturbations of BHs with quantum corrections for the Schwarzschild metric,
as analytical solutions to the equations of motion associated with a more general action involving higher-order curvature
operators. 
Quasinormal (QN) modes are a fundamental feature of the gravitational signal emitted by compact objects in several
astrophysical processes.
Their eigenfrequencies manifest relevant information about the nature of the emitting source and its inner structure as well. 
The BH stability under perturbations is intricately associated with the inner features of the BH itself.
Typically, the BH stability can be investigated by examining the evolution of field perturbations on
the BH background or during BH mergers. 
The end state of astrophysical binary BH mergers is a perturbed single BH, characterized by the final
remnant mass and angular momentum. 
When a BH is perturbed, it can emit gravitational waves (GWs), primarily characterized by QN modes.
The term quasinormal contrasts with the usual normal modes in Newtonian gravity, since they damp after emitting
GWs~\cite{Kokkotas:1999bd,Nollert:1999ji,Cardoso:2019rvt,Konoplya:2011qq}.  

Collisions between BHs go through three stages: the inspiral, the merger, and the ringdown phases.
With the LIGO/Virgo unprecedented discovery of GWs from merging BHs~\cite{LIGOScientific:2016aoc,VIRGO:2014yos},
the ringdown phase was detected in the GW signal, consisting of fast decaying oscillations characterised by eigenfrequencies
over characteristic timescales.
The QN spectra of more regular compact objects differ drastically from those that originate from perturbed BHs,
although they still exhibit a comparable ringdown phase~\cite{Cavalcante:2024kmy}. 
QN modes could also appear as echoes~\cite{Cardoso:2016rao,Maggio:2019zyv,2022zym,Abedi:2016hgu,Cavalcanti:2022cga,Kuntz:2019gup}.
Several studies have robustly established that the QN frequencies of a BH are solely determined by the characteristics
of the BH itself and the fields present in the perturbation process~\cite{Ferrari:2007dd,Berti:2009kk,Gong:2023ghh,Oliveira:2018oha}. 
QN modes of spherically symmetric sources can be split into spin-weighted spherical harmonics of order $\ell$ and degree $m$.
For each pair $(\ell,m)$, there exists a discrete set of complex frequencies denoted $\omega_{\ell m n}$, where $n$ indexes the overtone.
The oscillatory behaviour is described by the real part of $\omega_{\ell m n}$, whereas the imaginary part 
 is related to the damping timescale, or equivalently, the inverse of the decay rate.  
 The overtone index orders the QN modes for decreasing damping timescales so that the fundamental mode $n=0$ 
corresponds to the least-damped mode and is the longest-lived~\cite{Konoplya:2011qq}.

The main goal of this work is to study BHs that carry quantum gravitational corrections at third order in the curvature
expansion through the QN modes of analog models.
Beyond providing insights into the stability of the BH spacetime, QN frequencies are crucial for determining the parameters
that characterise these quantum-corrected BHs. 
Acoustic waves travelling through inviscid and inhomogeneous fluid flows have been shown to emulate waves
on BH backgrounds.
In any transonic fluid flow, while sound waves can move from subsonic to supersonic regions, they are prevented
from propagating in the opposite direction.
Thus, the critical sonic point, where the sound velocity matches the fluid velocity, behaves as an acoustic horizon,
similar to an event horizon for sound waves.
For a fluid flow in a nozzle, this horizon may emerge at the nozzle throat, which is the narrowest part of the tube~\cite{Visser:1997ux}.
Numerous analog gravity models have been formulated, and a wide range of experiments have been conducted
and designed to observe the analog of QN ringing.
Apart from the importance of experimental validation, analog gravity models are essential in theoretical contexts
to enhance our comprehension of BH physics. 
Refs.~\cite{Okuzumi:2007hf,Furuhashi:2006dh,Abdalla:2007dz} indicated that the QN ringing in sound waves
is generated by acoustic BHs, analogous to the way BHs emit QN ringing in GWs.
It offers the potential to observe BH QN ringing in laboratory conditions involving de Laval nozzles~\cite{Sarkar:2017puh,Casadio:2024uwj,Anacleto:2019rfn,Yang:2024fql}.
Since some gravitational excitations in BH scenarios can be described similarly to QN modes of sound waves
in a nozzle, one can indirectly investigate quantum gravity corrections to BHs within the context of aerodynamics. 

In this work, we analyse QN modes of BHs that incorporate quantum gravitational corrections
at third order in the curvature expansion through transonic waves in a de Laval nozzle. 
We will show how BHs carrying quantum gravitational corrections can be mapped into analog gravity models,
to test some of their features in a laboratory.
Given the absence of observational support for quantum corrections to BHs, such experiments in aerodynamics
can improve our understanding of the physical signatures of quantum gravity. \blt{The closer a BH mass is to the Planck mass, the larger the quantum gravitational corrections are expected to set in. We also investigate a large range of BH masses, from the Planck scale to stellar and astrophysical scales, showing that quantum gravitational corrections to the nozzle geometry, thermodynamic variables, Mach number, and thrust coefficients are more significant for smaller masses. This could be relevant for primordial BHs and their analogues. }
Sec.~\ref{sec2} aims to present quantum gravitational corrections in third order in the curvature expansion concerning BHs.
In Sec.~\ref{sec3}, we explore the relationship between perturbations on BH geometries and sound waves within
a de Laval nozzle, outlining the conditions and constraint equations under which the analogy holds.
Spinor, scalar, vector, and tensor perturbations of fluid flows in analog aerodynamics are proposed to probe quantum
gravity-corrected BHs experimentally.
Sec.~\ref{sec4} demonstrates how the parameters affecting quantum gravitational corrections at third order
in the curvature expansion concerning BH parameters correspond to the nozzle geometry, thermodynamic variables,
Mach number, and thrust coefficient. For it, the QN mode frequencies are calculated using the Mashhoon method,
followed by a computation of the quality factor for the analog de Laval nozzle. Higher overtones are also computed and
discussed, yielding a precise description of the GW form way before the fundamental mode dominates.
The ringdown of the quantum gravitational-corrected BH is therefore  
addressed, improving the extraction of information from quantum gravity-corrected BH sources.
Finally, Sec.~\ref{sec5} reviews the primary findings and presents concluding remarks. \blt{Appendix \ref{ap10} analyzes the de Laval nozzle geometry, thermodynamic variables, Mach number, and thrust coefficient for fixed representative values of the quantum gravitational correction parameter and BH masses typically varying in the range from the Planck mass to astrophysical BHs}. 
\section{Quantum gravitational corrections at third-order curvature}
\label{sec2}
Quantum gravitational corrections to the entropy of Schwarzschild BHs can be obtained
using effective field theoretical methods to calculate the Bekenstein--Hawking entropy~\cite{Calmet:2021lny,Calmet:2019eof}. 
Starting from the Wald entropy formula 
\begin{eqnarray}
S_{{\scalebox{.6}{\textsc{Wald}}}}
=
-2\pi \lim_{{r\to r_{{\scalebox{.7}{\textsc{h}}}}}}
\int \dd\Upsigma \  \upepsilon_{\mu\nu} \upepsilon_{\rho\sigma}
 \frac{\partial \mathcal{L}}{\partial R_{\mu\nu\rho\sigma}}
\ ,
\label{wald11}
\end{eqnarray}
where $\dd\Upsigma=r^2 \sin \theta\, d\theta\, d\varphi$ is the area element on spheres, 
$\mathcal{L}$ is the Lagrangian of the model, $R_{\mu\nu\rho\sigma}$ denotes the Riemann tensor, and
$r_{{\scalebox{.7}{\textsc{h}}}}= 2 G_{{\scalebox{.55}{\textsc{N}}}} M$ is the horizon radius. Quantum corrections in the metric modify the position of the event horizon and, hence, Eq. (\ref{wald11}). The effective action of quantum gravity, at second order in the curvature, with  cosmological constant set to
zero, has a local sector ~\cite{Barvinsky:1990up,Donoghue:1994dn,Delhom:2022xfo}
\begin{align}\label{EFTaction}
S_{{\scalebox{.55}{\textsc{EFT}}}}
=
\int \sqrt{|g|}\dd^4x  \left[ \frac{R}{16\pi G_{{\scalebox{.55}{\textsc{N}}}}} + c_1(\upmu) R^2 + c_2(\upmu) R_{\mu\nu} R^{\mu\nu} + c_3(\upmu) R_{\mu\nu\rho\sigma} R^{\mu\nu\rho\sigma} + \mathcal{L}_{{\scalebox{.65}{\textsc{m}}}} \right] \ ,
\end{align}
where $\upmu$ denotes the 
renormalization scale, and $\mathcal{L}_{{\scalebox{.65}{\textsc{m}}}}$ is the matter Lagrangian, whereas the nonlocal sector is given by 
\begin{align}\label{nonlocalaction}
	\!\!\!\! \Upgamma_{{\scalebox{.55}{\textsc{NL}}}}^{\scriptstyle{(2)}} \!=\int\!\!  \sqrt{|g|}\dd^4x\! \left[ \upalpha R \log\!\left(\frac{\Box}{\upmu^2}\right)R \!+\! \upbeta R_{\mu\nu} \log\left(\frac{\Box}{\upmu^2}\right) R^{\mu\nu} \!+\! \upgamma R_{\mu\nu\rho\sigma} \log\left(\frac{\Box}{\upmu^2}\right)\!R^{\mu\nu\rho\sigma}\!+\!\mathcal{O}(M_{{\scalebox{.6}{\textsc{Pl}}}}^{-2}) \right],
	\end{align}
	where $M_{{\scalebox{.6}{\textsc{Pl}}}}$ is the reduced Planck mass and $c_i(\upmu), \upalpha,\upbeta,\upgamma$ denote the Wilson coefficients.
Ref. \cite{Calmet:2021lny} showed the absence of quantum corrections to the Schwarzschild metric, up to second order in curvature~\cite{Calmet:2017qqa,Calmet:2018elv}. It implies that the horizon radius remains unchanged and the quantum-corrected Wald entropy can be computed at second order, by Eqs.~\eqref{EFTaction} and \eqref{nonlocalaction}, as~\cite{Calmet:2021lny}
\begin{eqnarray} \label{quad_entropy}
S_{{\scalebox{.6}{\textsc{Wald}}}}^{(2)}  &=&\frac{A}{4 G_{{\scalebox{.55}{\textsc{N}}}}} +  64 \pi^2\left\{c_3(\upmu) + \upgamma \left[ \log \left (4 G_{{\scalebox{.55}{\textsc{N}}}}^2 M^2 \upmu^2\right) -2 +2 \upgamma_{{\scalebox{.55}{\textsc{E}}}} \right]\right\},
\end{eqnarray}
where $A=16 \pi (G_{{\scalebox{.55}{\textsc{N}}}} M)^2$ is the BH area and Euler's constant $\gamma_{{\scalebox{.55}{\textsc{E}}}}
\approx 0.5772156$. As there are no corrections to the metric, the temperature remains unchanged, and nonlocal quantum corrections yield a pressure $P$  for the BH. The first law of thermodynamics therefore reads 
\begin{eqnarray}\label{eq50}
T\dd S-P\dd V=\left(1+\frac{16 \pi\,\upgamma }{G_{{\scalebox{.55}{\textsc{N}}}} M^2}\right) \dd M.
\end{eqnarray}
One identifies $T\dd S=\dd M$ and $ 16 \pi\,\upgamma/(G_{{\scalebox{.55}{\textsc{N}}}} M^2) \dd M= - P \dd V$  with $\dd V=32 \pi G_{{\scalebox{.55}{\textsc{N}}}}^3 M^2 \dd M$, yielding the BH pressure 
\begin{eqnarray} \label{pressure}
P=- \frac{\upgamma}{2 G_{{\scalebox{.55}{\textsc{N}}}}^4 M^4},
\end{eqnarray}
which can take negative values, as for spin-0, spin-1/2, and spin-2 fields, one has $\upgamma>0$; or positive values, as $\upgamma<0$ for spin-1 fields. 
\clt{The variation of the nonlocal action \eqref{nonlocalaction} yields an effective energy-momentum tensor which has an effective radial pressure component \eqref{pressure},  
 as the nonlocal terms modify the energy equilibrium appearing in the first law of black hole thermodynamics. 
The identification $T\dd S = \dd M$ therefore continues to hold for the leading-order 
Bekenstein--Hawking term $S = \frac{A}{4 G_{{\scalebox{.55}{\textsc{N}}}}}$, representing the dominant entropy contribution, since quantum corrections to the entropy do not alter the thermodynamic differential structure and only shift the internal energy by a small amount already encoded in the work term $P\dd V$. 
Within the EFT framework, nonlocal corrections act as small backreaction effects encoded in the thermodynamic identity 
rather than in the spacetime geometry itself~\cite{Calmet:2021lny}. 
}

At third order in curvature, the effective action contains a dimension six local operator
\begin{eqnarray}\label{sixlocal}
{\cal L}^{(3)}=c_6 G_{{\scalebox{.55}{\textsc{N}}}}  R^{\mu\nu}_{\ \ \alpha\sigma} R^{\alpha\sigma}_{\ \ \beta\rho} R^{\beta\rho}_{\ \ \mu\nu}  \ ,
\end{eqnarray}
where $c_6$ is a dimensionless parameter \clt{corresponding to a Wilson coefficient controlling the first cubic curvature correction in vacuum}. There is only one invariant involving only Riemann tensors in vacuum\footnote{\clt{In the sense that while other cubic curvature invariants exist, in vacuum these reduce to linear combinations of $ R^{\mu\nu}_{\ \ \alpha\sigma} R^{\alpha\sigma}_{\ \ \beta\rho} R^{\beta\rho}_{\ \ \mu\nu}$, making $c_6$ the unique 2-loop vacuum coefficient in pure gravity.}}, with a corresponding nonlocal operator $R^{\mu\nu}_{\ \ \alpha\sigma} \log{\Box} R^{\alpha\sigma}_{\ \ \beta\rho}R^{\beta\rho}_{\ \ \mu\nu}$ that can be neglected \cite{Goroff:1985th}, \clt{and cannot be removed by lower-order field redefinitions.}  \clt{In fact, in EFT of gravity, nonlocal terms arise as the low-energy manifestation of 1-loop effects of massless fields and carry additional suppression by inverse powers of the Planck mass. 
Their contribution to the gravitational action is therefore subleading,  compared to the local curvature invariants at the same order in the derivative expansion. 
Second, for backgrounds of slowly varying curvature, such as the Schwarzschild geometry of a macroscopic black hole, the nonlocal logarithmic operator produces corrections that are proportional to 
$\log(\Box/\mu^2)$ acting on curvature tensors. 
In this case, the relevant curvature scales as $R \sim G_{{\scalebox{.55}{\textsc{N}}}} M/r^3$, and the $\Box$ operator acts on quantities that vary only over distances of order $r \gg \ell_{{\scalebox{.6}{\textsc{Pl}}}}$, yielding $\Box R / R \ll M_{{\scalebox{.6}{\textsc{Pl}}}}^2$. 
Consequently, the logarithmic kernel produces at most mild, subleading corrections to the local term $R^3$. 
Moreover, the nonlocal effects become significant only near or above the cutoff of the EFT, when curvatures approach the Planck scale, or for geometries with rapidly varying fields. 
Since the EFT description of gravity is valid only for$R/M_{{\scalebox{.6}{\textsc{Pl}}}}^2 \ll 1$ and for horizon radii much larger than the Planck length ($r_{\text{h}} \gg \ell_{{\scalebox{.6}{\textsc{Pl}}}}$), neglecting the nonlocal $R\log\Box R^2$-type terms at cubic order is self-consistent. 
This approximation ensures that only the dominant, local dimension-six operator contributes to the leading quantum correction to the Schwarzschild metric \eqref{ck}, with metric coefficients \eqref{eq:calmet-kuipers-metric}.}

The dimension six local operator leads to a metric
\begin{eqnarray}
\dd s^2 = g_{\mu\nu}\dd x^\mu \dd x^\nu= -B(r) \dd t^2 + A(r) \dd r^2 + r^2 \dd \Omega^2,\label{ck}
\end{eqnarray}
with components~\cite{Calmet:2021lny}
\bes
\label{eq:calmet-kuipers-metric}
\begin{eqnarray}
    B(r) &=& 1-\frac{2G_{{\scalebox{.55}{\textsc{N}}}} M}{r} + 640\pi\,c_6 \frac{G^5_{{\scalebox{.55}{\textsc{N}}}} M^3}{r^7}\label{eq:calmet-kuipers-metric1} \\
    A(r) &=& \qty[1-\frac{2G_{{\scalebox{.55}{\textsc{N}}}} M}{r} + 3456\pi\,c_6\frac{G^4_{\scalebox{.55}{\textsc{N}}} M^2}{r^6}\,\qty(1 - \frac{49 G_{{\scalebox{.55}{\textsc{N}}}} M}{27r})]^{-1}.\label{eq:calmet-kuipers-metric2}
\end{eqnarray}
\ees
\clt{Although the terms that are beyond the usual Schwarzschild solution in Eqs. (\ref{eq:calmet-kuipers-metric1}, \ref{eq:calmet-kuipers-metric2}) are highly suppressed for astrophysical black holes ($r \gg \ell_{{\scalebox{.6}{\textsc{Pl}}}}$), they are crucial in providing a well-defined EFT parameterization of short-distance quantum gravity effects. The relation $
c_6 G_{\scalebox{.55}{N}} \sim {1}/{M_{{\scalebox{.6}{\textsc{Pl}}}}^2}$ 
exhibits the canonical EFT suppression by two powers of the reduced Planck mass, reflecting the hierarchy of higher-derivative corrections in low-energy gravity.
The coefficient $c_6$ arises from 2-loop divergences in the Goroff--Sagnotti calculation~\cite{Goroff:1985th}, being a genuine quantum gravitational coupling, rather than a classical higher-curvature ambiguity, and sets the leading nontrivial correction to the vacuum gravitational action in the EFT expansion.}

\clt{In pure Einstein gravity, the 1-loop effective action is finite on shell, but Goroff and Sagnotti~\cite{Goroff:1985th} showed that at two loops the effective action develops a divergence proportional to the same cubic invariant
\begin{equation}
\Gamma_{\rm div}^{(2)} = \frac{209}{2880 (4\pi)^4 \epsilon} 
\int \dd^4x \sqrt{|g|} \,
R^{\mu\nu}{}_{\alpha\sigma} 
R^{\alpha\sigma}{}_{\beta\rho} 
R^{\beta\rho}{}_{\mu\nu}.\label{gorsagn}
\end{equation}
The numerical factor in Eq. (\ref{gorsagn}) is the canonical Goroff--Sagnotti coefficient that multiplies the 2-loop counterterm in pure Einstein gravity~\cite{Goroff:1985th}. 
The term $\epsilon = 4-d$ is the standard parameter employed in the  dimensional regularization procedure, where calculations are performed in $d$ spacetime dimensions and the result, as usual, is expanded around $d=4$, to isolate the divergent parts. 
This divergence is absorbed by the counterterm $c_6$, leaving a finite, renormalized coefficient 
\begin{equation}
c_6(\mu) = c_6^{{\scalebox{.6}{\textsc{bare}}}} + \frac{209}{2880 (4\pi)^4} \frac{1}{\epsilon} + c_6^{{\scalebox{.6}{\textsc{finite}}}}(\mu),
\end{equation}
where $c_6^{{\scalebox{.6}{\textsc{finite}}}}(\mu)$ depends on the renormalization scale $\upmu$ and the ultraviolet (UV) completion of gravity.
Therefore, the coefficient $c_6$ can be thought of as the finite part of a loop counterterm absorbing the Goroff--Sagnotti divergence. 
The coefficient $c_6$ runs logarithmically with $\upmu$,  according to  the equation $\upmu \frac{\dd c_6}{\dd \upmu} = \beta_{c_6}$, where  $\beta_{c_6}$ is the renormalization group beta function computable from loop diagrams, indicating how the coefficient $c_6$ runs as one probes gravity at different energies. The dependence on the renormalization scale encodes how the UV description of gravity feeds into low-energy predictions for higher-curvature effects.}

\clt{
It is worth noting that gravitational theories containing higher-derivative operators, such as the cubic curvature term \eqref{sixlocal}, are generically subject to the Ostrogradsky instability, which arises from the presence of higher-order time derivatives in the action and typically indicates the existence of ghostlike degrees of freedom~\cite{Woodard:2015zca,Simon:1990ic}, including the case with unsuppressed cubic curvature terms \cite{DeFelice:2023vmj}.  
Within the EFT framework adopted here, however, these operators are treated perturbatively as higher-order corrections suppressed by powers of the Planck scale, ensuring that no additional propagating degrees of freedom appear below the cutoff. 
The would-be ghost poles associated with the higher-derivative terms lie far above the EFT validity range and therefore do not correspond to physical excitations in the low-energy regime. 
Consequently, the EFT remains consistent with GR as the deep-infrared and lowest-energy approximation, with the higher-curvature terms encoding virtual quantum gravitational effects rather than introducing new dynamical fields~\cite{Donoghue:1994dn,Burgess:2003jk,Donoghue:2017ovt}. 
This interpretation is consistent with the standard treatment of quadratic and cubic curvature corrections in the EFT of gravity~\cite{Barvinsky:1985an,Barvinsky:1990up,Calmet:2018elv,Calmet:2021lny,Holdom:2022zzo}, where higher-order operators parameterize nonlocal quantum effects and remain under perturbative control. }

Other quantum gravitational corrections from the dimension-six operator describing stellar distributions have been studied in
Ref.~\cite{Solodukhin:2019xwx}.

\section{Quantum gravitational corrected black holes and the de~Laval nozzle}
\label{sec3}

This section explores the conditions under which sound waves propagating through a fluid in a de~Laval nozzle can emulate bosonic (massless scalar fields, electromagnetic fields, and axial gravitational perturbations) and fermionic (Dirac) field perturbations on the quantum-corrected metric~\eqref{ck}. Specifically, the focus is on the QN ringing modes. Perturbations in an actual BH often lead to the emission of GWs, characterized by an initial blast of strong-field radiation, followed by a phase of damped oscillations dominated by QN modes. The QN modes are fingerprints of the BH geometry and are pivotal for understanding its underlying stability and dynamics.

The relativistic equations for massless scalar fields $\Phi$, electromagnetic fields $A_\mu$, and Dirac fields $\Uppsi$  in a background described by the quantum gravitational metric $g_{\mu\nu}$ in Eq. (\ref{ck}), with coefficients \eqref{eq:calmet-kuipers-metric}, can be respectively expressed as
\bes
\begin{eqnarray}
\label{eq:Klein-Gordon}
\frac{1}{\sqrt{-g}} \partial_\mu \qty(\sqrt{-g} g^{\mu \nu} \partial_\nu) \Phi \qty(x^\mu) &=& 0,
\\
\label{eq:Electromagnetic Field}
\clt{\frac{1}{\sqrt{-g}}\partial_\mu \qty(F_{\rho\sigma}g^{\mu\rho}g^{\tau\sigma}\sqrt{-g})} &=& 0,
\\
\label{eq:Dirac Field}
\upgamma^\alpha e_{\alpha}^{\ \mu}\qty(\partial_\mu - \Upgamma_\mu)\Uppsi\qty(x^\mu) &=& 0,
\end{eqnarray}
\ees
where $g=\text{det} \qty(g_{\mu \nu})$, $F_{\mu\nu} = \partial_\mu A_\nu - \partial_\nu A_\mu$, $\upgamma^\alpha$ are the Dirac matrices, $\Upgamma_\mu = -\frac{1}{8} \qty[\upgamma^{\rho},\upgamma^{\sigma}] g_{\alpha\tau}e_{\rho}^{\ \alpha}\nabla_{\mu}e_{\sigma}^{\ \tau}$ denotes the spin connection, and $e_{\mu}^{\ \nu}$ stands for the tetrad field which expressed the metric $g_{\mu\nu}$ in terms of the Minkowski metric $\eta_{\mu\nu}$ by $g_{\mu\nu}=\eta_{\lambda\sigma}e_\mu^{\ \lambda}e_\nu^{\ \sigma}$.

The first step to compute QN modes is to introduce a tortoise coordinate for the quantum-corrected BH metric~\eqref{ck}, 
\begin{equation}
    \label{tortoise coordinate}
    \dv{r_{{\scalebox{.65}{$\star$}}}}{r} = \sqrt{\frac{A(r)}{B(r)}}
    \ .
\end{equation}
The perturbations can be decomposed in modes of frequency $\omega$, to wit
\begin{equation}\label{ztr}
    \Phi\qty(t,r_{{\scalebox{.65}{$\star$}}},\theta,\varphi)=e^{-i\omega t} \frac{R(r_{{\scalebox{.65}{$\star$}}})}{r_{{\scalebox{.65}{$\star$}}}} Y_{\ell}^{m}\qty(\theta,\varphi),
\end{equation}
where spherical harmonics of degree $\ell$ and order $m$ are given by
\begin{equation}
    Y_{\ell}^{m}\qty(\theta,\varphi) = \frac{\qty(\sin \theta)^{|m|}}{2^\ell \ell!} \qty(\dv{}{\cos \theta})^{|m|+\ell} \qty(1-\cos^2 \theta)^{-\ell} \ e^{im\varphi}
    \ .
\end{equation}
Eqs.~\eqref{eq:Klein-Gordon}-\eqref{eq:Dirac Field} then reduce to a Schrödinger-like differential equation~\cite{Kokkotas:1999bd,Berti:2009kk,Konoplya:2011qq}
\begin{equation}
    \label{quasinormal modes equation}
    \qty(\dv[2]{}{r_{{\scalebox{.65}{$\star$}}}} + \omega^2 -  V_{{\scalebox{.65}{{\textsc{eff}}}}}(r_{{\scalebox{.65}{$\star$}}})) R(r_{{\scalebox{.65}{$\star$}}}) = 0,
\end{equation}
for the radial part $R(r_{{\scalebox{.65}{$\star$}}})$. 
 The effective potential, respectively for integer and semi-integer values of the spin, reads:
\begin{eqnarray}
 V_{{\scalebox{.65}{{\textsc{eff}}}}}(r) &=&\begin{cases}
 \displaystyle B(r) \frac{\ell\qty(\ell+1)}{r^2} + \frac{1-
 \clt{s^2}
 }{2r} \dv{}{r}\left(\frac{B\qty(r)}{A\qty(r)}\right),\qquad     \text{for} \;\;s\in\mathbb{Z},\label{final effective potential int}
\\
 \displaystyle \frac{B(r)}{r^2} \qty(\ell+\frac{1}{2})^2 \pm \qty(\ell+\frac{1}{2}) \sqrt{\frac{B(r)}{A(r)}} \dv{}{r}\left(\frac{\sqrt{B\qty(r)}}{r}\right),\qquad \text{for}\;\;s\in\mathbb{Z}+\frac12.\label{final effective potential half}\end{cases}
\end{eqnarray}
The Dirac field is governed by two isospectral wave functions, which can be transformed into one another using the Darboux transformation \cite{Dubinsky:2024fvi}. Hence, choosing only $V_{{\scalebox{.65}{{\textsc{eff}}}}}^{+}(r)$, corresponding to the positive sign on the right-hand side of the second equation in \eqref{final effective potential half} is sufficient for the analysis of QN modes~\cite{Li:2013fka,Castello-Branco:2004rzk}. 

\clt{
The effective potential for integer spins, calculated with the quantum corrected Schwarzschild Metric, Eq.~\eqref{eq:calmet-kuipers-metric}, can be written in terms of the parameter $c_6$ as follows:
\begin{align} \label{eq:veff_ck_integer}
    V_{{\scalebox{.65}{{\textsc{eff}}}}}^\text{bosonic}(r) = V_0(r) + V_1(r) c_6 + V_2(r)  c_6^2,
\end{align}
where $V_0(r)$ is the effective potential for the Schwarzschild metric, whereas $V_1(r)$ and $V_2(r)$ are corrections coupled with $c_6$ and $c_6^2$, respectively:
\bes
\begin{align}
   V_0(r) &=  \qty(1-\frac{2G_N M}{r})\qty(
   \frac{\ell\qty(\ell+1)}{r^{2}}  -
    \frac{ 2G_N M }{r^3} \qty(s^2-1) )
\\
    V_1(r) &= \qty[
    640\frac{ G_N M }{r}\ell\qty(\ell+1) +
    \qty(10368- 
    43904\frac{ G_N M }{r} + 
    45056\frac{ G_N^{2} M^{2} }{r^{2}}
    ) \qty(s^2-1)
    ]\frac{G_N^{4} M^{2}}{r^{8}} \pi 
\\
    V_2(r) &= \qty(
    14376960-28098560\frac{  G_N M}{r})
   \qty(s^2-1) \frac{G_N^{9} M^{5}}{r^{15}} \pi^{2} .
\end{align}
\ees
The quadratic structure of Eq.~\eqref{eq:veff_ck_integer} shows that integer-spin particles sense the quantum correction up to $c_6^2$.
It is now easy to see that the effective potential saturates with the parameter
\begin{equation}\label{eq:c6 saturation integer}
    c_6^\text{saturation} = -\frac{1}{2}\frac{V_1(r)}{V_2(r)}.
\end{equation}
}
\clt{
The effective potential for particles with spin-1/2 is given by
\begin{multline}
    \label{eq:veff_ck_half}
    V_{{\scalebox{.65}{{\textsc{eff}}}}}^\text{fermionic}(r) = \frac{1}{r^2}\qty(\ell+\frac{1}{2})^2 \qty(1-\frac{2G_\text{N} M}{r} + 640 \pi c_6 \frac{G_\text{N}^5 M^3}{r^7})  \\ 
    - \frac{1}{r^2}\qty(\ell+\frac{1}{2})\sqrt{1-\frac{2G_{{\scalebox{.55}{\textsc{N}}}} M}{r} + 3456\pi\,c_6\frac{G^4_{\scalebox{.55}{\textsc{N}}} M^2}{r^6}\,\qty(1 - \frac{49 G_{{\scalebox{.55}{\textsc{N}}}} M}{27r})} 
    \\ \times \qty(1-\frac{3G_\text{N}M}{r} + 2880 \pi c_6 \frac{G_\text{N}^5 M^3}{r^7}).
\end{multline}
Different from the integer case, $V_{{\scalebox{.65}{{\textsc{eff}}}}}^\text{fermionic}(r)$ \cclt{is not a polynomial in $c_6$, and thus} this case should be more sensitive to quantum corrections.
To compare the two cases, the saturation values of $c_6$ and the maximum value of $V_{{\scalebox{.65}{{\textsc{eff}}}}}(r)$ can be analysed. Fig.~\ref{fig:saturation_c6_vs_Veff} shows the value of $c_6$ that saturates $V_{{\scalebox{.65}{{\textsc{eff}}}}}(r)$, and its value. Fig.~\ref{fig:saturation_c6_vs_radius} shows the orbit radius $r_\text{orbit}$, varying $c_6$ from -0.05 to 0.05. 
The fermionic trajectory in Fig.~\ref{fig:saturation_c6_vs_Veff} exhibits 
non-monotonic behaviour: as $c_6\lesssim0$, $V_\text{eff}^\text{max}$ increases with $c_6$, and when $c_6\gtrsim0$, $V_\text{eff}^\text{max}$ decreases with $c_6$, faster than the bosonic case. This behaviour will be present in the results in Sec.~\ref{sec4}.
It is important to point out that $V_{{\scalebox{.65}{{\textsc{eff}}}}}^\text{fermionic}(r)$ reaches its global maximum value near Schwarzschild, so spin-1/2 particles experience the maximum attraction when $c_6\approx 0$ and rapidly change with a small quantum perturbation, with bigger slopes near Schwarzschild (minimum global slope value $\approx -0.722$ at $c_6 \approx0.01$ and maximum global slope value $\approx 0.264$ at $c_6 \approx-0.01$).
}

\clt{
The integer spin case is less sensitive to variations of $c_6$. The curve is monotonic, so as $c_6$ increases, $V_\text{eff}^\text{max}$ decreases. The behaviour only changes with higher values for $c_6$, for which no maximum $V_\text{eff}^\text{max}$ can be defined. For reasonable values of $c_6$, the bosonic curve shown in Fig.~\ref{fig:saturation_c6_vs_Veff} does not have a well-defined global maximum or minimum. Similar to the fermionic case, the minimum global slope is located at $c_6 \approx 0.01$ with value $\approx -0.592$. 
}

\clt{
Fig.~\ref{fig:saturation_c6_vs_radius} shows that fermionic orbits are closer to the event horizon than the bosonic case, in every $c_6$ value. The minimum stable orbit, for either the fermionic or the bosonic case, is reached with $c_6 \neq 0$, so one can expect a setup different from the Schwarzschild metric that minimises the total energy of the system, and thus preferred by nature.
}
\clt{
\begin{figure}[H]
    \centering
    \includegraphics[width=0.8\linewidth]{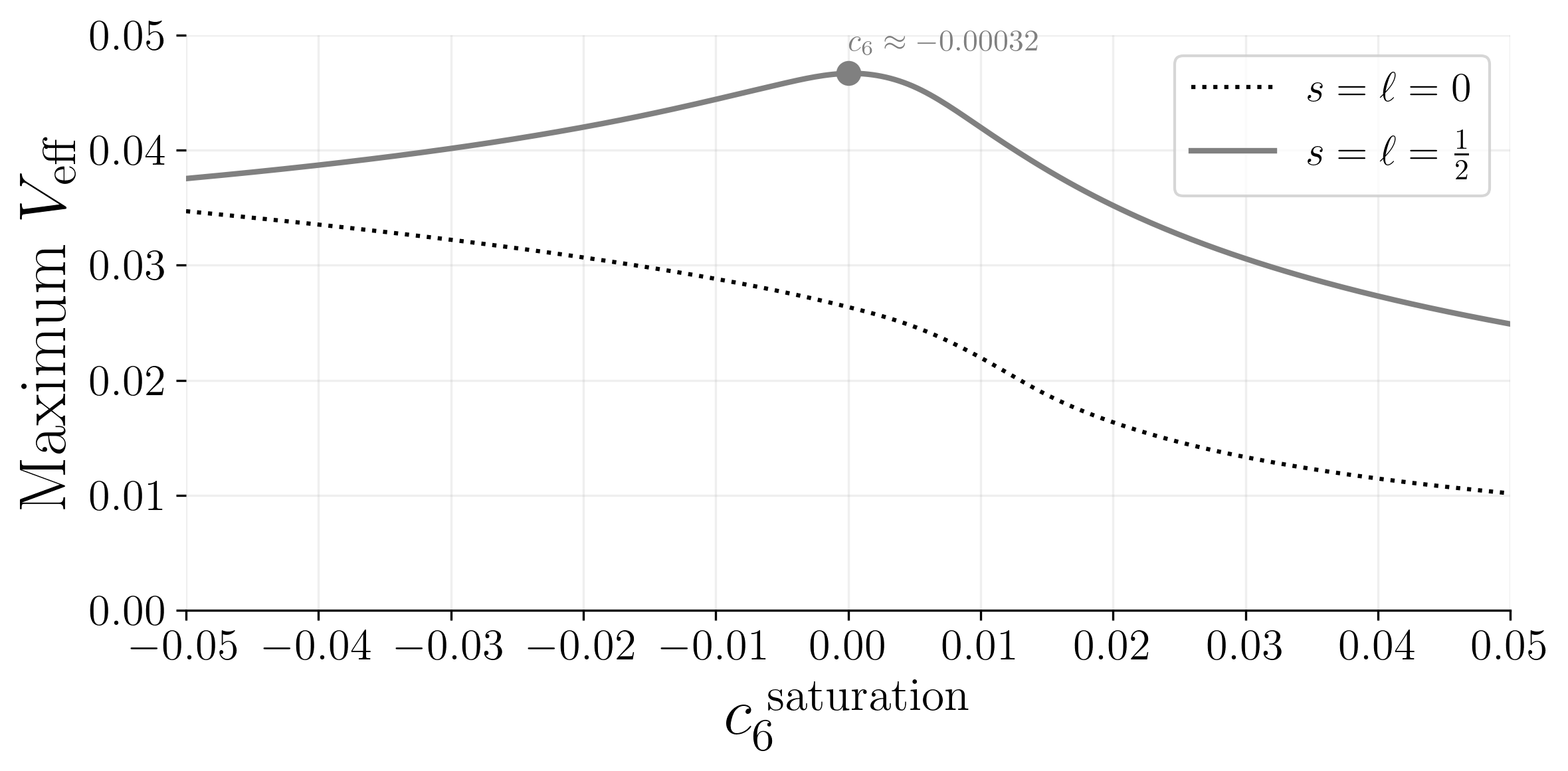}
    \caption{
    \clt{
    Saturation value for $c_6$ that maximises $V_\text{eff}$, versus its maximum effective potential value. The circle represents the global maximum value. The dotted black line considers $s=\ell = 0$, whereas the solid grey line considers $s=\ell=1/2$. Values calculated with natural units and $M=1$.
    }
    }
    \label{fig:saturation_c6_vs_Veff}
\end{figure}
\begin{figure}[H]
    \centering
    \includegraphics[width=0.8\linewidth]{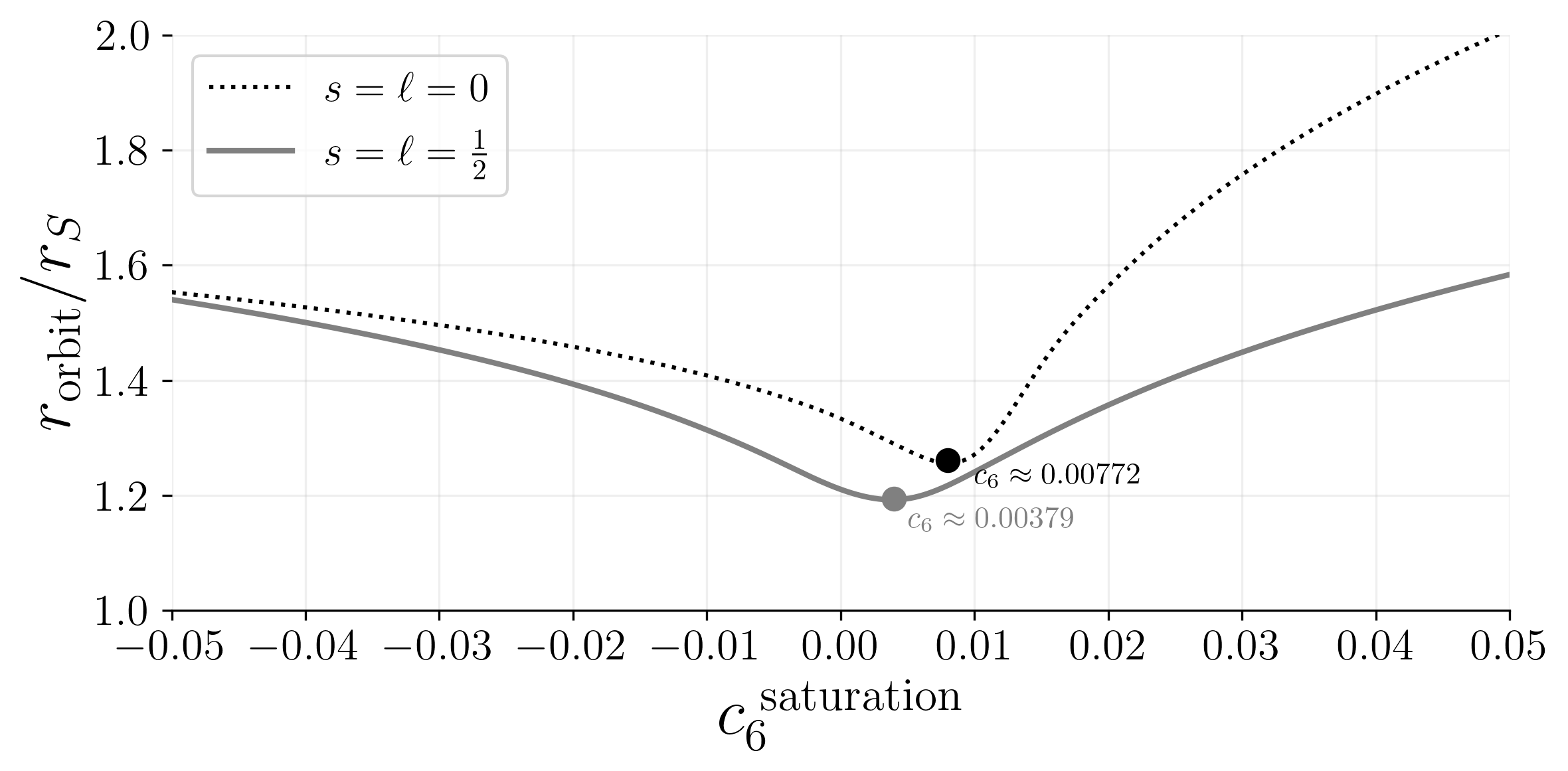}
    \caption{
    \clt{
    Saturation value for $c_6$ that maximises $V_\text{eff}$, versus the orbital radius value relative to the Schwarzschild Radius $r_S = 2G_\text{N}M$. The circles represent the global minimum value for each combination of $s$ and $\ell$. The dotted black line considers $s=\ell = 0$, whereas the solid grey line considers $s=\ell=1/2$.
    }
    }
    \label{fig:saturation_c6_vs_radius}
\end{figure}
}

To establish the connection with BH perturbations and fluid dynamics, one can express the Navier--Stokes equations for a longitudinal horizontal de Laval nozzle
as~\cite{Casadio:2024uwj}
\begin{equation}\label{eq:nse1}
         \rho\qty[\pdv{\vb{u}}{t} + \qty(\vb{u}\cdot \vb{\nabla})\vb{u}] = - \nabla p - \rho \nabla \Upphi + \eta \nabla^2 \vb{u} + \qty(\xi + \frac{1}{3}\eta) \nabla\qty(\nabla\cdot\vb{u}).
\end{equation}
Eq.~\eqref{eq:nse1} describes the dynamics of a viscous heat-conducting fluid flow, with flow velocity profile ${\bf u}={\bf u}(x^\mu)$, where $\eta$ is the dynamic shear viscosity, $\xi$ represents the bulk viscosity, $\rho=\rho(x^\mu)$ denotes the fluid density, and $\Upphi=\Upphi(x^\mu)$ is a scalar potential. These equations are reduced to the Euler equations in the absence of viscosity. Isentropic flows are both adiabatic and reversible, meaning that no heat is added to the fluid flow, even those due to friction or dissipative effects. For isentropic flows,  the flow velocity field can be expressed as $\vb{u} = - \nabla\upvarphi$, where $\upvarphi$ is the velocity potential. Under these assumptions, the equations ruling the fluid flow are simplified to:
\bes
\begin{eqnarray}
\label{eq:cont}
\pdv{(\rho {\scalebox{.87}{\textsc{A}}})}{t} + \nabla\cdot{(\rho {\scalebox{.87}{\textsc{A}}}\vb{u})} &=& 0,
\\
\label{eq:bern}
-\pdv{\upvarphi}{t} + h + \frac{1}{2} (\nabla\upvarphi)^2 + \Upphi &=& 0,
\end{eqnarray}
\ees
where ${\scalebox{.87}{\textsc{A}}}$ denotes the cross-sectional area of a de~Laval nozzle, and $h$ represents the specific enthalpy,
encoding the internal energy of the fluid flow and the product of its pressure and volume per unit mass.
To investigate wave dynamics within this setup, applying perturbations to Eqs.~\eqref{eq:cont} and \eqref{eq:bern} yields
\begin{equation}
\label{eq:soundpropagation}
-\pdv{}{t}\qty{\frac{\rho {\scalebox{.87}{\textsc{A}}}}{c_{\scalebox{.57}{\textsc{s}}}^2}\qty[\pdv{(\delta\upvarphi)}{t} +\vb{u}\cdot \qty(\nabla(\delta\upvarphi))]}
+ \nabla\cdot{
\qty{
\rho {\scalebox{.87}{\textsc{A}}} \nabla (\delta\upvarphi)
- \qty{\frac{\rho {\scalebox{.87}{\textsc{A}}}}{c_{\scalebox{.57}{\textsc{s}}}^2}\qty[\pdv{(\delta\upvarphi)}{t} +\vb{u}\cdot \qty(\nabla(\delta\upvarphi))]} \vb{u}
}} = 0,
\end{equation}
where $c_{\scalebox{.57}{\textsc{s}}}$ is the speed of sound in the medium.

The Venturi effect governs the behavior of fluid flows within de~Laval nozzles. 
In inviscid fluid dynamics, the velocity of a fluid flow must rise as it flows through the nozzle constriction, according to mass conservation. In contrast, the fluid flow static pressure must decrease by the Bernoulli principle and the Euler equations. Therefore, any increase in kinetic energy that the fluid experiences due to its increased velocity through a constriction is counterbalanced by a reduction in pressure resulting from a decrease in potential energy. When a fluid passes through a constricted region of a tube with varying cross-sectional area ${\scalebox{.87}{\textsc{A}}}(x)$, the pressure decreases while the velocity increases. The mass flow rate for a compressible fluid increases as the upstream pressure increases, yielding the increment of the fluid density through the nozzle constriction, although the flow velocity remains unaltered. This is the principle governing de Laval nozzles. The higher the source temperature, the higher the local sonic velocity is, permitting the mass flow rate to increase. However, it occurs only if the de Laval nozzle area also increases, to compensate for the resulting decrement in density.
For an ideal gas under non-viscous, adiabatic, and isentropic conditions, the equation of state $p = \rho R T$ applies, where $T$ represents the fluid temperature and $R=8.3144$	JK$^{-1}$ mol$^{-1}$ denotes the ideal gas constant. Key thermodynamic properties of the gas include its heat capacities at constant pressure ($c_{\scalebox{.6}{\textsc{P}}}$) and constant volume ($c_{\scalebox{.6}{\textsc{V}}}$), related by $R = c_{\scalebox{.6}{\textsc{P}}} - c_{\scalebox{.6}{\textsc{V}}}$. The adiabatic index $\gamma = c_{\scalebox{.6}{\textsc{P}}} /c_{\scalebox{.6}{\textsc{V}}}$ characterizes the fluid flow, as the speed of sound depends on this factor. Isentropic flows satisfy the relation 
\begin{equation}\label{isen}
p = \rho^\gamma = T^{\frac{\gamma}{\gamma-1}},
\end{equation}
ensuring shock-free and continuous flow properties.

The Mach number,
${\scalebox{.85}{\textsc{M}}}(x)=  \qty|\vb{u}(x)|/{c_{\scalebox{.57}{\textsc{s}}}(x)}$,
is a dimensionless parameter that quantifies the flow velocity ratio to the local speed of sound, with $c_{\scalebox{.57}{\textsc{s}}}^2 = {\dd p}/{\dd \rho} = \gamma RT$ representing the speed of sound. Here, $x$ denotes the longitudinal coordinate along the convergent-divergent nozzle. The mass flux $\dd m/\dd t \equiv \dot{m}$ measures the rate at which mass flows through the nozzle cross-section per unit of time and must be constant to ensure the continuity equation derived from mass conservation. Assuming that the nozzle radius $r = r(x)$ varies gradually along $x$, perturbations in the fluid flow can be approximated as quasi-one-dimensional ones, and we can write ${\bf u} = u\hat\imath$. Under these assumptions, Eq.~\eqref{eq:soundpropagation} simplifies to describe scalar field perturbations along a single propagation direction:
\begin{align}
\label{eq:one-dimensional wave equation}
\qty[\qty(\pdv{}{t} + \pdv{u}{x}) \frac{\rho {\scalebox{.87}{\textsc{A}}}}{c^2_s} \qty(\pdv{}{t} + u\pdv{}{x}) - \pdv{}{x} \qty(\rho {\scalebox{.87}{\textsc{A}}} \pdv{}{x})] \delta\upvarphi = 0.
\end{align}

Analogous to the QN modes described in Eq.~\eqref{quasinormal modes equation}, stationary solutions can be expressed through a Fourier transform
\be
\delta\upvarphi(x,t) = \frac{1}{2\pi} \int \dd \omega e^{-i\omega t} \delta\upvarphi_\omega(x)
\ .
\ee 
Substituting this into Eq.~\eqref{eq:one-dimensional wave equation} results in a time-independent differential equation for $\delta\upvarphi_\omega$, expressed as
\begin{multline}
\label{eq: time-independent differential equation}
\frac{1}{2\pi} \int \dd \omega  e^{-i \omega t}\left\{ \rho {\scalebox{.87}{\textsc{A}}}\qty(1-\frac{u^{2}}{c_{\scalebox{.57}{\textsc{s}}}^{2}}) \dv[2]{}{x} + \qty[\dv{(\rho {\scalebox{.87}{\textsc{A}}})}{x} + 2 i \omega \frac{\rho {\scalebox{.87}{\textsc{A}}} u}{c_{\scalebox{.57}{\textsc{s}}}^{2}} - \dv{}{x} \qty(\frac{\rho {\scalebox{.87}{\textsc{A}}} u^{2}}{c_{\scalebox{.57}{\textsc{s}}}^{2}})] \dv{}{x} \right. \\ \left.
+ \qty[\omega^{2} \frac{\rho {\scalebox{.87}{\textsc{A}}}}{c_{\scalebox{.57}{\textsc{s}}}^{2}} + i \omega \dv{}{x}\qty(\frac{\rho {\scalebox{.87}{\textsc{A}}} u}{c_{\scalebox{.57}{\textsc{s}}}^{2}})] \right\}\delta\upvarphi_{\omega} = 0.
\end{multline}
To simplify the analysis, auxiliary quantities can be introduced, such as the transfer function \cite{Okuzumi:2007hf,Furuhashi:2006dh,Abdalla:2007dz}
\begin{equation}
 F_\omega (x) = \sqrt{g_{\scalebox{.63}{\textsc{c}}}} \int \dd t \exp\qty{i\omega \qty[t - \int \dd x \frac{u}{(c^2_s - u^2)}]} \delta\upvarphi(t,x),   
\end{equation}
where $g_{\scalebox{.63}{\textsc{c}}} = {\rho {\scalebox{.87}{\textsc{A}}}}/{c_{\scalebox{.57}{\textsc{s}}}}$.
Additionally, a coordinate transformation $x = x(x_{{\scalebox{.65}{$\star$}}})$ based on the tortoise coordinate for the canonical acoustic BH can be applied:
\begin{equation}
\label{eq: acoustic BN tortoise coordinate}
\dv{x_{{\scalebox{.65}{$\star$}}}}{x} = \frac{{c_{\scalebox{.57}{\textsc{s0}}}}}{c_{\scalebox{.57}{\textsc{s}}}(1 - {\scalebox{.85}{\textsc{M}}}^2)},
\end{equation}
where ${c_{\scalebox{.57}{\textsc{s0}}}}$ represents the stagnation sound speed, and $x_{{\scalebox{.65}{$\star$}}}$ serves as the acoustic analog of the tortoise coordinate. Under these transformations, Eq.~\eqref{eq: time-independent differential equation} assumes the form of a Schrödinger-wavelike differential equation:
\begin{equation}
\label{eq:schrodinger acoustic BN}
\qty(\dv[2]{}{x_{{\scalebox{.65}{$\star$}}}} + \frac{\omega^2}{{c_{\scalebox{.57}{\textsc{s0}}}}^2} - V_{{\scalebox{.65}{{\textsc{eff}}}}}(x_{{\scalebox{.65}{$\star$}}})) F_\omega(x_{{\scalebox{.65}{$\star$}}}) = 0,    
\end{equation}
with the effective potential given by \cite{Abdalla:2007dz}
\begin{equation}
\label{eq: effective potential acoustic BN}
V_{{\scalebox{.65}{{\textsc{eff}}}}}(x_{{\scalebox{.65}{$\star$}}}) = \frac{1}{2g_{\scalebox{.63}{\textsc{c}}}} \dv[2]{g_{\scalebox{.63}{\textsc{c}}}}{x_{{\scalebox{.65}{$\star$}}}} - \qty(\frac{1}{2g_{\scalebox{.63}{\textsc{c}}}} \dv{g_{\scalebox{.63}{\textsc{c}}}}{x_{{\scalebox{.65}{$\star$}}}})^2.
\end{equation}

With the equations and effective potentials for both gravitational and aerodynamic systems now established, the next step involves their application in experimental contexts. First, a precise relationship between the fluid density $\rho$ and the parameter $g_{\scalebox{.63}{\textsc{c}}}$ must be determined. Subsequently, the constraint equations linking the Schrödinger-wavelike equations -- Eq.~\eqref{quasinormal modes equation} for gravitational QN modes and the dual analog  Eq.~\eqref{eq:schrodinger acoustic BN} for acoustic BHs -- must be analyzed to fully understand the QN mode dynamics in both analog systems.

To ensure consistency between the wave dynamics described by Eq.~\eqref{eq:schrodinger acoustic BN} and the effective potential in Eq.~\eqref{eq: effective potential acoustic BN}, it is convenient to express the de Laval nozzle cross-sectional area ${\scalebox{.87}{\textsc{A}}}$ in terms of $g_{\scalebox{.63}{\textsc{c}}}$. For a perfect fluid under isentropic flow conditions, the nozzle longitudinal area ${\scalebox{.87}{\textsc{A}}}$ relative to the area at the nozzle throat ${\scalebox{.87}{\textsc{A}}}_{{\scalebox{.65}{$\star$}}}$ can be described by
\begin{equation}
    \frac{{\scalebox{.87}{\textsc{A}}}}{{\scalebox{.87}{\textsc{A}}}_{{\scalebox{.65}{$\star$}}}} = \frac{1}{{\scalebox{.85}{\textsc{M}}}} \qty[\frac{2}{\gamma+1}\qty(1+\frac{\gamma-1}{2}{\scalebox{.85}{\textsc{M}}}^2)]^{\frac{\qty(\gamma+1)}{2(\gamma-1)}},
\end{equation}
Remembering that ${\scalebox{.85}{\textsc{M}}}$ represents the Mach number and $\gamma$ is the adiabatic index.

To simplify the analysis, hereon both ${\scalebox{.87}{\textsc{A}}}$ and $\rho$ are normalized by their respective throat values, ${\scalebox{.87}{\textsc{A}}}_{{\scalebox{.65}{$\star$}}}$ and $\rho_0$. Using these normalized quantities, one can  express:
\bes
\begin{align}
\label{eq:g em funcao de rho e A}
    g_{\scalebox{.63}{\textsc{c}}} &= \frac{1}{2}\rho^{(3-\gamma)/2} {\scalebox{.87}{\textsc{A}}}, \\
    {\scalebox{.87}{\textsc{A}}}^{-1} &= \rho \sqrt{1-\rho^{(\gamma-1)}}.
\end{align}
\ees
The density $\rho$ can be related to $g_{\scalebox{.63}{\textsc{c}}}$ as:
\begin{align}
\label{eq: rho em funcao de g}
\rho^{1-\gamma} = 2g_{\scalebox{.63}{\textsc{c}}}^2\qty(1-\frac{\sqrt{g_{\scalebox{.63}{\textsc{c}}}^2-1}}{g_{\scalebox{.63}{\textsc{c}}}}).
\end{align}
Substituting this into the expression for ${\scalebox{.87}{\textsc{A}}}$ yields:
\begin{align}
{\scalebox{.87}{\textsc{A}}} = \sqrt{2 g_{\scalebox{.63}{\textsc{c}}} \qty(\sqrt{g_{\scalebox{.63}{\textsc{c}}}^2 - 1} + g_{\scalebox{.63}{\textsc{c}}})} \qty[2g_{\scalebox{.63}{\textsc{c}}}^2 \qty(1-\sqrt{1-g_{\scalebox{.63}{\textsc{c}}}^{-2}})]^{\frac{1}{\gamma-1}}.
\end{align}
Using the relationships between local and total quantities for isentropic flows with Eq. \eqref{eq:g em funcao de rho e A} yields
\begin{eqnarray}
\label{eq: mach em funcao de g}
    {\rho}^{-1} = \qty(1 + \frac{\gamma - 1 }{2} {\scalebox{.85}{\textsc{M}}}^2)^{\frac{1}{\gamma - 1}}, 
    \end{eqnarray}
        implying that 
    \begin{eqnarray}
    g_{\scalebox{.63}{\textsc{c}}}   = \frac{1}{{\scalebox{.85}{\textsc{M}}} \sqrt{2(\gamma-1)}}\qty(1+\frac{\gamma-1}{2}{\scalebox{.85}{\textsc{M}}}^2),
\end{eqnarray}
where ${\scalebox{.85}{\textsc{M}}} = 1$ at the event horizon $r_{{\scalebox{.7}{\textsc{h}}}}$, corresponding to the nozzle throat.
For air flow ($\gamma \approx 1.40$), $g_{\scalebox{.63}{\textsc{c}}}$ satisfies:
\begin{align}
\label{eq: cond contorno g 1}
    g_{\scalebox{.63}{\textsc {c}}}(r_{{\scalebox{.7}{\textsc{h}}}}) = \frac{\gamma+1}{2 \sqrt {2 (\gamma - 1)}} = \frac {3\sqrt {5}} {{5}}> 1.
\end{align}
This condition provides a boundary constraint for numerical integrations. The second boundary constraint is 
\begin{align}
\label{eq: cond contorno g 2}
    \lim_{r\to r_{{\scalebox{.7}{\textsc{h}}}}} \dv{\scalebox{.87}{\textsc{A}}}{x} = 0,
\end{align}
or, equivalently, 
\begin{align}  \lim_{r\to r_{{\scalebox{.7}{\textsc{h}}}}} \dv{g_{\scalebox{.63}{\textsc {c}}}}{x} = 0,
\end{align}
indicating that the nozzle geometry is smooth at the throat.




The Schrödinger-type equations~\eqref{quasinormal modes equation} and \eqref{eq:schrodinger acoustic BN}
are derived from an effective potential and the tortoise coordinate framework.
Establishing the equivalence between the effective potentials requires the tortoise coordinates from both systems to be equal, leading to the condition
\begin{equation}
\label{eq: coordenadas iguais}
    \dd r_{{\scalebox{.65}{$\star$}}} = \dd x_{{\scalebox{.65}{$\star$}}} = \sqrt{\frac{A(r)}{B(r)}} \dd r =
    \frac{\qty(\gamma-1)\sqrt{2g_{\scalebox{.63}{\textsc{c}}}^2\qty(1-\sqrt{1-g_{\scalebox{.63}{\textsc{c}}}^{-2}})}}{\gamma+1 - 4g_{\scalebox{.63}{\textsc{c}}}^2\qty(1-\sqrt{1-g_{\scalebox{.63}{\textsc{c}}}^{-2}})} \dd x.
\end{equation}
By defining $\mathtt{F}(r) \equiv \sqrt{B(r)/A(r)}$, this relation simplifies to \cite{Abdalla:2007dz}
\begin{equation}
\label{eq:xemfuncaoder123}
    \dv{x}{r} = \frac{\gamma+1-4g_{\scalebox{.63}{\textsc{c}}}(r)^2\qty(1-\sqrt{1-g_{\scalebox{.63}{\textsc{c}}}(r)^{-2}})}
    {\mathtt{F}(r)\qty(\gamma-1)\sqrt{2g_{\scalebox{.63}{\textsc{c}}}(r)^2 \qty(1-\sqrt{1-g_{\scalebox{.63}{\textsc{c}}}(r)^{-2}})}}.
\end{equation}

The function $g_{\scalebox{.63}{\textsc{c}}}(r)$ can be  determined by rewriting the effective potential, Eq.~\eqref{eq: effective potential acoustic BN}, as
\begin{equation}
    \label{eq: pot em funcao de r}
      \frac{g_{\scalebox{.63}{\textsc{c}}}''(r)}{g_{\scalebox{.63}{\textsc{c}}}(r)} +  \frac{g_{\scalebox{.63}{\textsc{c}}}'(r)}{g_{\scalebox{.63}{\textsc{c}}}(r)}\qty(\frac{\mathtt{F}'(r)}{\mathtt{F}(r)}  - \frac{1}{2} \frac{g_{\scalebox{.63}{\textsc{c}}}'(r)}{g_{\scalebox{.63}{\textsc{c}}}(r)})
    = \frac{2 V_{{\scalebox{.65}{{\textsc{eff}}}}}(r)}{\mathtt{F}^2(r)}.
\end{equation}
Singularities in solving this equation are resolved via the substitution $g_{\scalebox{.63}{\textsc{c}}}\qty(r) \equiv \chi^2\qty(r)$, yielding:
\begin{align}\label{eq: lambda}
\chi''(r) + \frac{\mathtt{F}'(r)}{\mathtt{F}(r)}\chi'(r) - \frac{V_{{\scalebox{.65}{{\textsc{eff}}}}}(r)}{\mathtt{F}^2(r)} \chi(r) = 0.
\end{align}
Eq.~\eqref{eq: lambda} has the appropriate form to apply the Frobenius method.
For the Schwarzschild limit ($c_6\to0$), the solution reduces to:
\begin{equation}
\label{gc12}
g_{\scalebox{.63}{\textsc{c}}} 
=
\frac{\gamma+1}{2\sqrt{2}\sqrt{\gamma-1}}
\qty[\frac{\Gamma(1+\ell+{s}){}_2F_1\qty({s}-\ell,{s}+\ell+1,1+2{s},\dfrac{r}{2G_{{\scalebox{.55}{\textsc{N}}}}M})}{(2{s})!(\ell-{s})!}\qty(\frac{r}{2G_{{\scalebox{.55}{\textsc{N}}}}M})^{{s}+1}]^2,
\end{equation}
consistent with prior results \cite{Abdalla:2007dz}. Near the horizon $r_{{\scalebox{.7}{\textsc{h}}}}$, the Frobenius method is applied using the series expansion
\begin{equation}
    \chi\qty(r) = \sum_{k=0}^\infty a_k \qty(r-r_{{\scalebox{.7}{\textsc{h}}}})^{k+p}, \label{eq:frob1}
\end{equation}
with indicial conditions
\begin{equation}
\label{eq:condinilamb}
a_0 = \sqrt{\frac{\gamma+1}{2\sqrt{2}\sqrt{\gamma-1}}}, \quad a_1 = 0.
\end{equation}
It generates the recurrence relation
\begin{equation}\label{eq: recurrence eq}
    \qty(k +2)\qty(k+1) a_{k+2} + \qty(k+1)a_{k+1} \frac{\mathtt{F}'(r)}{\mathtt{F}(r)} - a_k\frac{V_{{\scalebox{.65}{{\textsc{eff}}}}}(r)}{\mathtt{F}^2(r)} =0,
\end{equation}
which leads to
\begin{eqnarray}
    g_{\scalebox{.63}{\textsc{c}}}\qty(r) &=& {\frac{\gamma+1}{2\sqrt{2}\sqrt{\gamma-1}}} \qty[1 +\frac{1}{2} \frac{V_{{\scalebox{.65}{{\textsc{eff}}}}}(r)}{\mathtt{F}^2(r)} \qty(r-r_{{\scalebox{.7}{\textsc{h}}}})^2 - \frac{1}{6} \frac{V_{{\scalebox{.65}{{\textsc{eff}}}}}(r)}{\mathtt{F}^2(r)} \frac{\mathtt{F}'(r)}{\mathtt{F}(r)} \qty(r-r_{{\scalebox{.7}{\textsc{h}}}})^3 + \cdots]^2.
\end{eqnarray}
Numerical integration via Runge–Kutta methods then determines $g_{\scalebox{.63}{\textsc{c}}}(r)$, which is subsequently mapped to the nozzle longitudinal coordinate $x$ using Eq.~\eqref{eq:xemfuncaoder123}. This setup applies to the quantum-corrected metric~\eqref{ck}, for analyzing analog quantum-corrected nozzle profiles and their QN modes.

\section{de Laval nozzle analog of quantum gravitational corrected  black holes, QN modes and their overtones}

\label{sec4}

The analogy between sound waves in a de~Laval nozzle and quantum-corrected BHs~\eqref{ck} goes beyond qualitative comparisons, allowing for quantitative analysis within experimental precision. This enables the examination of fluid flow propagation features in de~Laval nozzles with numerical accuracy. QN modes, described by the wave equation~\eqref{quasinormal modes equation} in gravitational systems and by Eq.~\eqref{eq:schrodinger acoustic BN} for acoustic BHs, demonstrate a numerical correspondence. This equivalence implies that the effective potential governing perturbations in a specific aerodynamic configuration matches that of quantum gravitational corrected BHs.
Consequently, acoustic waves in de~Laval nozzles create an analog physical system replicating the effective potential generating QN modes by perturbations of quantum-gravitational corrected BHs.

\subsection{The quantum gravitational corrected  analog de Laval nozzle}

To solve numerically Eq.~\eqref{eq: lambda} requires fixing the value of the parameter $ c_6 $ in the components~\eqref{eq:calmet-kuipers-metric}, alongside the spin $ s $ and multipole quantum number $ \ell $ in Eq.~(\ref{final effective potential int}). The parameter $ c_6 $, corresponding to third-order curvature corrections in the effective field theory, must satisfy $ |c_6| \lesssim \mathcal{O}(1) $ to preserve perturbative validity~\cite{Calmet:2021lny}. For astrophysical BHs ($ M \gg M_{{\scalebox{.65}{\textsc{Pl}}}} $), values beyond this limit violate the consistency of the effective field theory, while Planck-scale systems ($ M \sim M_{{\scalebox{.65}{\textsc{Pl}}}} $) require renormalization group analysis to relate $ c_6 $ to the the renormalization scale $\upmu$~\cite{Calmet:2021lny}.  
The wave modes are classified by the multipole quantum number $ \ell $. We have $ \ell = 0 $ ($ s $-wave, spherically symmetric perturbations), $ \ell = 1 $ ($ p $-wave, dipole axial oscillations), $ \ell = 2 $ ($ d $-wave, quadrupole gravitational modes), $ \ell = {1}/{2} $ (fermionic $s$-wave, spin-${1}/{2}$ field excitations), and $ \ell = {3}/{2} $ (fermionic $ p $-wave, spin-${1}/{2}$ field excitations). 

\begin{figure}[H]
    \centering
    \begin{subfigure}{0.49\textwidth}
        \centering
        \includegraphics[width=\linewidth]{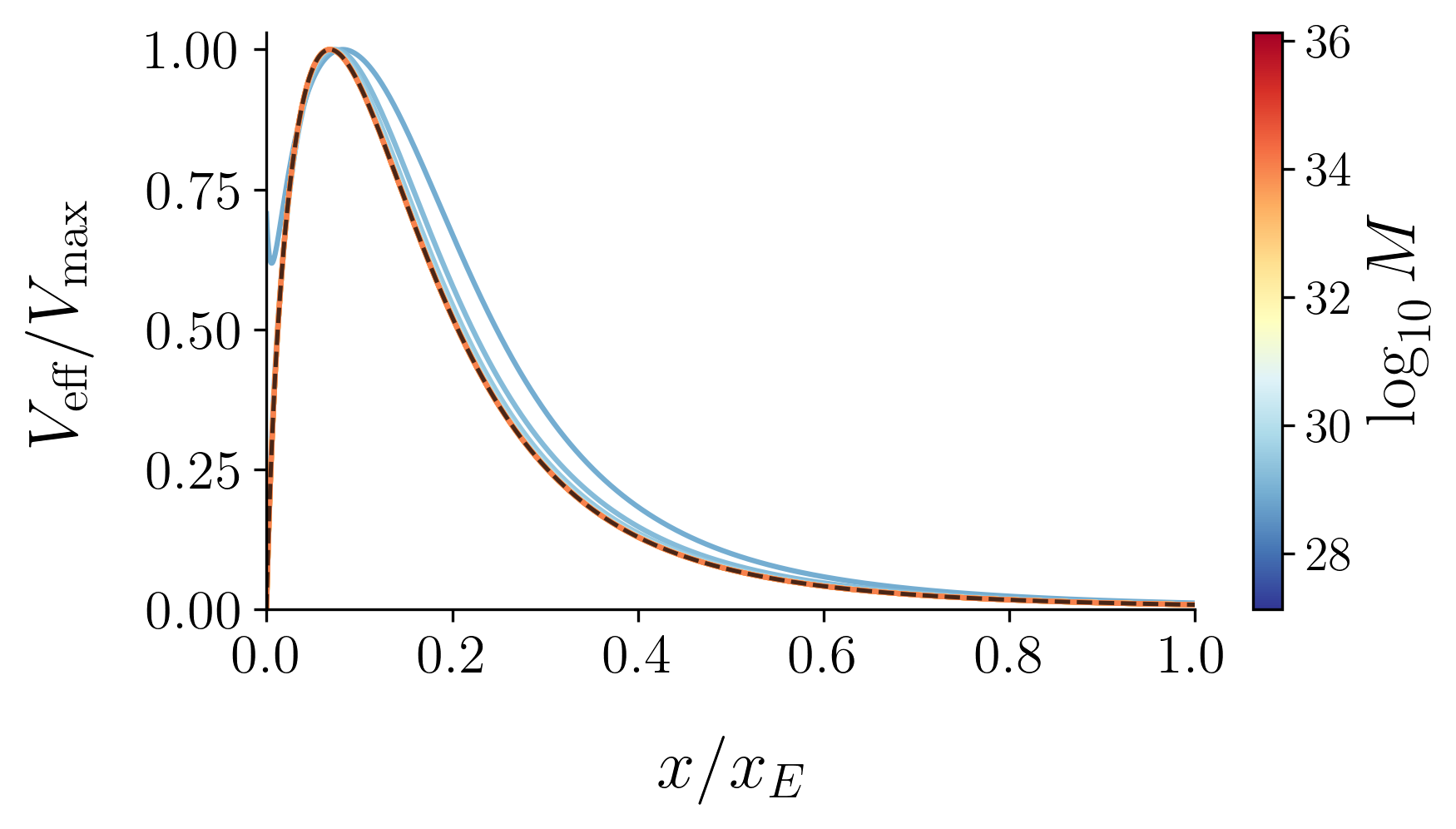}
         \caption{\footnotesize $s = \ell = 0$, $c_6 = 0.005$.}
        \label{fig:veffa123}
    \end{subfigure}
    \begin{subfigure}{0.49\textwidth}
        \centering
        \includegraphics[width=\linewidth]{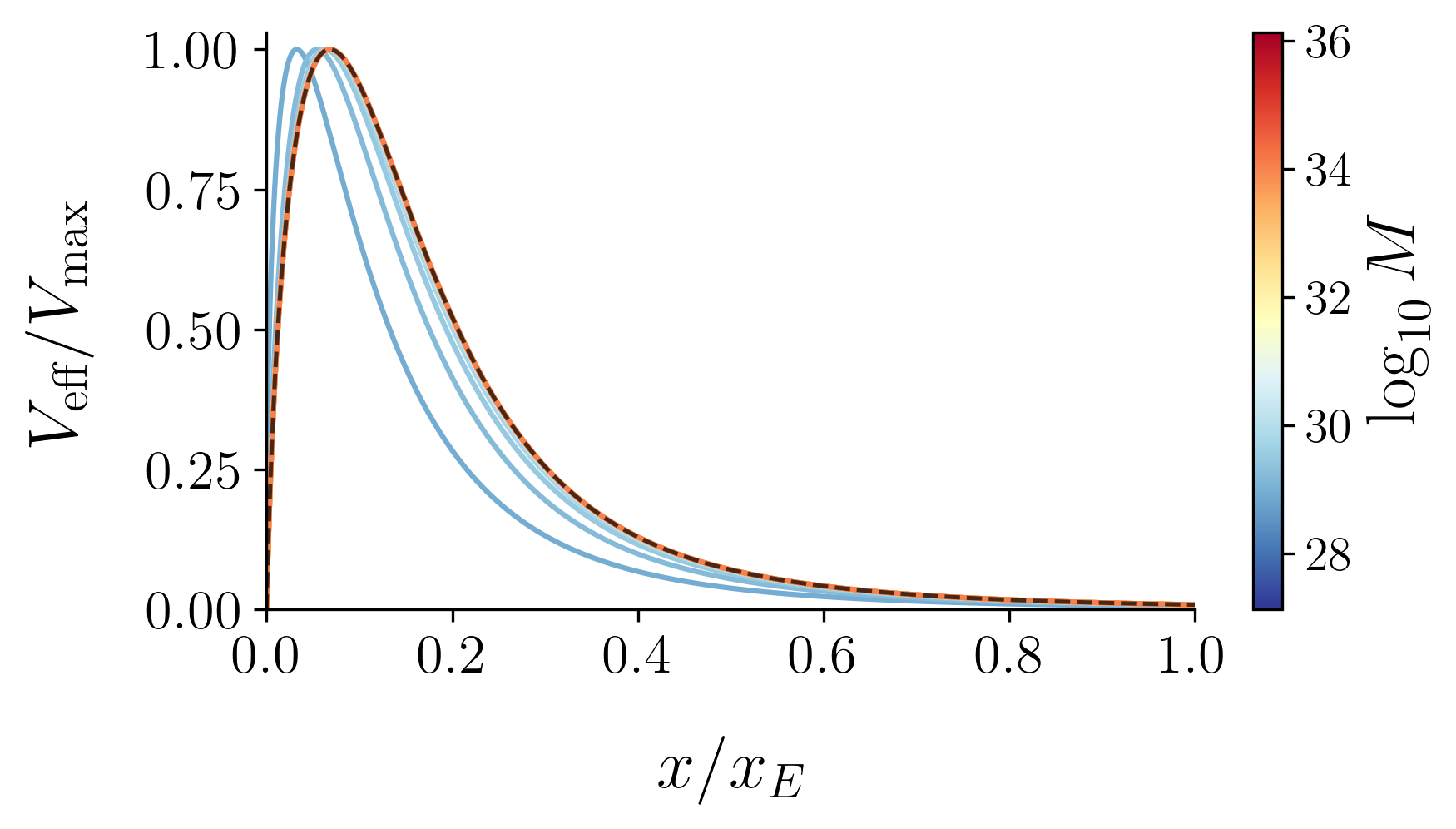}
         \caption{\footnotesize $s = 0$ and $\ell=0$, $c_6 = -0.01$.}
        \label{fig:veffb11123}  \end{subfigure}
                  \caption{\footnotesize Effective potential as a function of the longitudinal direction of the de~Laval nozzle, varying the BH mass in the range $10^{27}$ kg $< M < 10^{36}$ kg and keeping $c_6$ fixed.}
\end{figure}

The metric~\eqref{ck} imposes constraints on $ c_6 $ through the corrected event horizon radius:  
\begin{equation}\label{eq:eh-ck}
r_{{\scalebox{.7}{\textsc{h}}}} = 2G_{{\scalebox{.55}{\textsc{N}}}}M\left(1 - c_6 \frac{5\pi}{G_{{\scalebox{.55}{\textsc{N}}}}^2M^4}\right),      
\end{equation}
which requires $ c_6 < \mathcal{O}(G_{{\scalebox{.55}{\textsc{N}}}}^2M^4/\pi) $ to ensure $ r_{{\scalebox{.7}{\textsc{h}}}} \in \mathbb{R}^+ $. For $ M \sim M_{{\scalebox{.65}{\textsc{Pl}}}} $, this simplifies to the reasonable $ |c_6| \leq 1 $. Numerical solutions of Eqs.~\eqref{eq:xemfuncaoder123} and~\eqref{eq: lambda} are therefore restricted to $ |c_6| \leq 1 $, avoiding unphysical horizons or divergent potentials.  
\blt{More precisely, Eq. \eqref{eq:eh-ck}  implies that $c_6 \ll G_{{\scalebox{.55}{\textsc{N}}}}^2M^4/5\pi$, for all possible values of the BH mass $M$ in units with $G_{{\scalebox{.55}{\textsc{N}}}} = 1/M_{{\scalebox{.6}{\textsc{Pl}}}}^2$. Since the minimum admissible BH mass has order $M \sim M_{{\scalebox{.6}{\textsc{Pl}}}}$, and as the parameter driving quantum gravitational corrections is assumed to be independent of $M$, the quantum gravitational correction parameter $c_6$ can be constrained using $M \sim M_{{\scalebox{.6}{\textsc{Pl}}}}$, yielding the bound  
\beq\label{boundc6}
|c_6| < \frac{1}{5\pi} \sim 0.0636.
\eeq
The de~Laval nozzle characteristics are analyzed as functions of the quantum-correction parameter $c_6$, with spin configurations $s = 0, 1, 2, {1}/{2}$ and multipole numbers $\ell = s +n$, for $n\in\mathbb{N}$.
More precisely, spinor, scalar, vector, and tensor perturbations of fluid flows in analog aerodynamics are proposed to experimentally probe QN modes of quantum gravity-corrected BHs, also for distinct overtones. }
\blt{ Figs.~\ref{fig:veffa123} and \ref{fig:veffb11123} illustrate the effective potential $ V_{{\scalebox{.65}{{\textsc{eff}}}}} $ as a function of the longitudinal coordinate in the de~Laval nozzle (normalized by the total length $x_{\scalebox{.67}{\textsc{e}}}$ of the nozzle), evaluated for distinct stellar and astrophysical masses varying systematically in the range $10^{27}$ kg $< M < 10^{36}$ kg. The parameter $c_6$, regulating quantum gravity corrections, is chosen to obey the constraint \eqref{boundc6}. Scalar $s$-wave perturbations, given by $s =\ell = 0$ are shown in Figs.~\ref{fig:veffa123} and \ref{fig:veffb11123}. The nozzle throat center at $ x = 0 $ corresponds to the analog event horizon in these configurations.}
\blt{Fig.~\ref{fig:veffa123} considers the  positive value $c_6 = 0.005$ and shows that the effective potential starts to decay, after the peak, at higher values of the longitudinal coordinate $x$, for astrophysical objects with masses around the  Solar mass, although their asymptotic values remain very similar. On the other hand, Fig. \ref{fig:veffb11123} regards the negative value $c_6 = -0.01$, illustrating the behaviour of the effective potential, this time decaying at lower values of the longitudinal coordinate $x$, for 
stellar objects with masses around the  Solar mass. Similarly to Fig.~\ref{fig:veffa123}, their asymptotic values, at the nozzle exit, are practically independent of the BH mass. Both 
Figs.~\ref{fig:veffa123} and \ref{fig:veffb11123} corroborate the fact that the effect of $c_6$ reduces for increasing masses. They also show that quantum gravitational effects essentially disappear for values of the BH mass higher than Solar masses. It means that quantum gravity corrections are potentially discernible in primordial BHs.}

\begin{figure}[H]
    \centering
    \begin{subfigure}{0.49\textwidth}
        \centering
        \includegraphics[width=\linewidth]{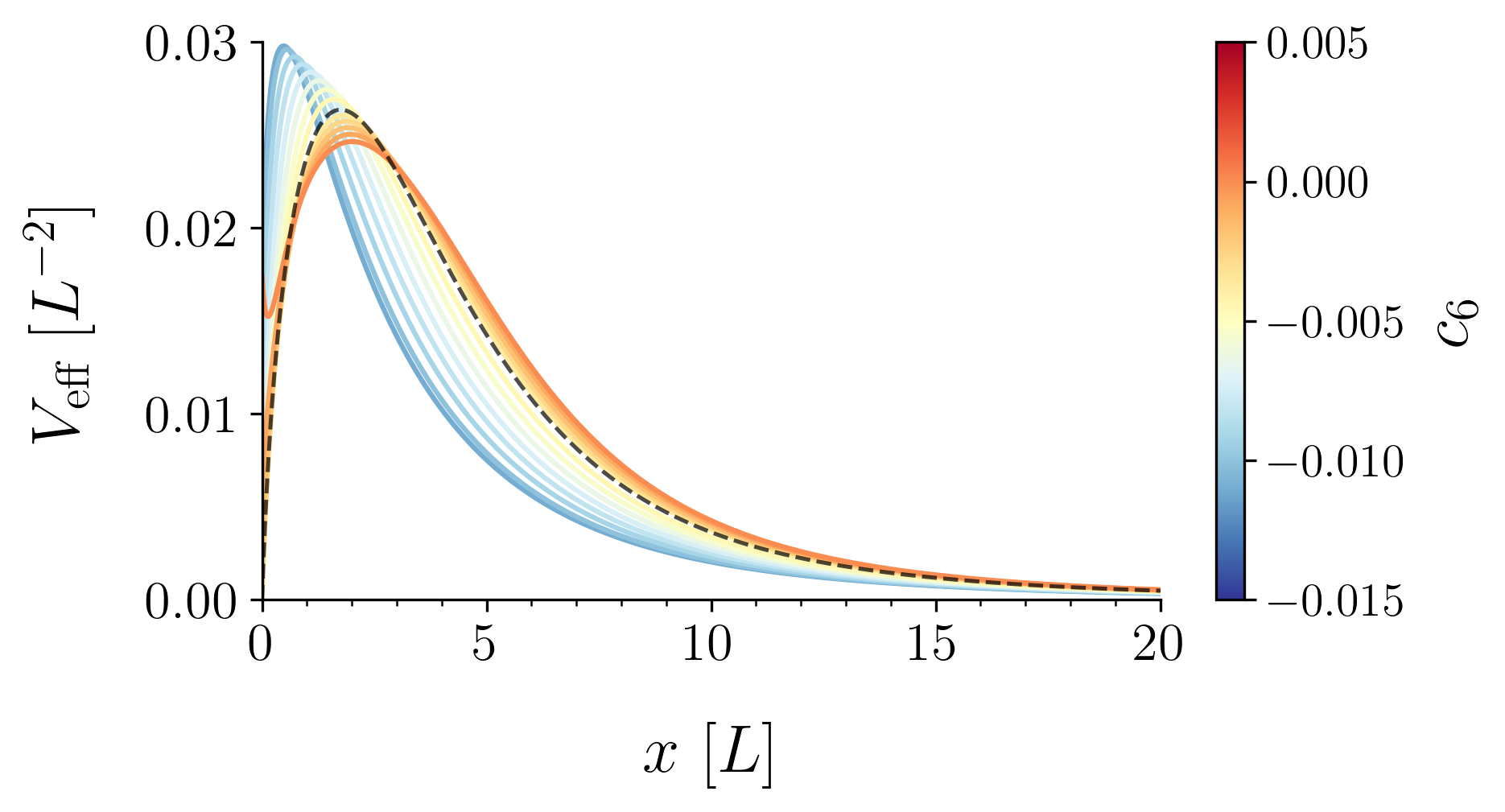}
         \caption{\footnotesize $s = \ell = 0$.}
        \label{fig:veffa}
    \end{subfigure}
    \begin{subfigure}{0.49\textwidth}
        \centering
        \includegraphics[width=\linewidth]{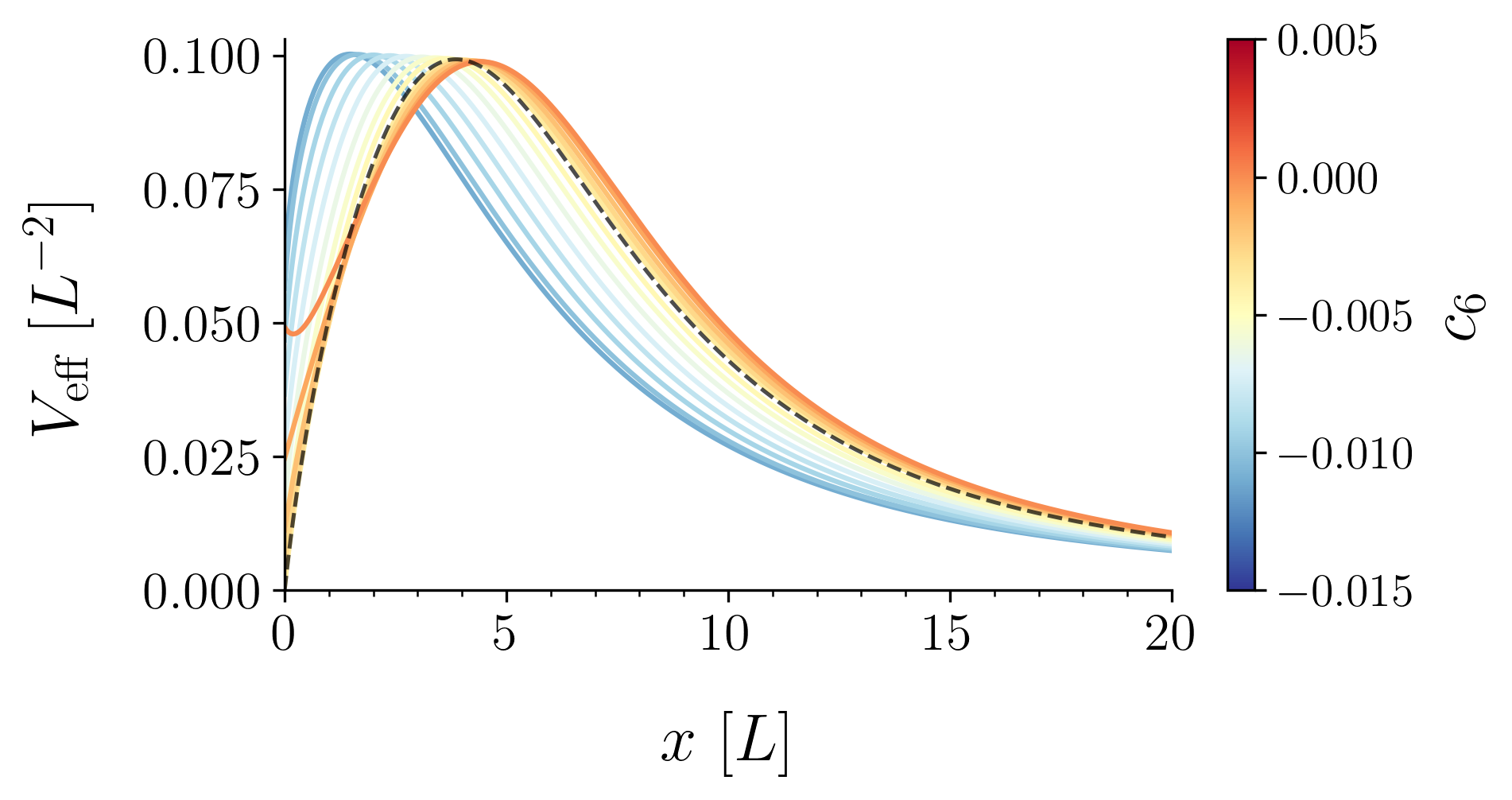}
         \caption{\footnotesize $s = 0$ and $\ell=1$.}
        \label{fig:veffb1}  \end{subfigure}
        
    \begin{subfigure}{0.49\textwidth}
        \centering
        \includegraphics[width=\linewidth]{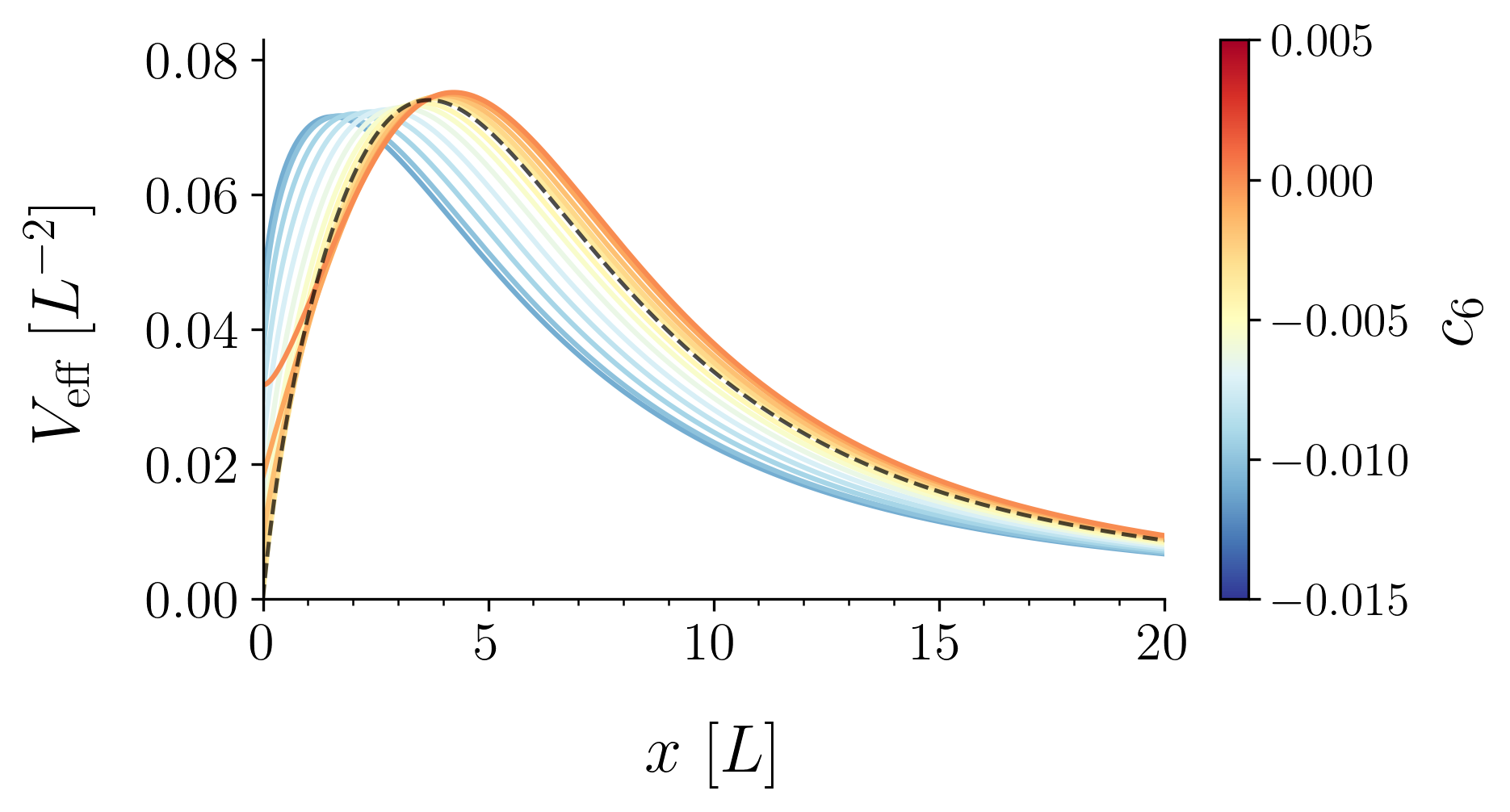}
         \caption{\footnotesize $s = \ell=1$.}
        \label{fig:veffb2}
    \end{subfigure}
    \begin{subfigure}{0.49\textwidth}
        \centering
        \includegraphics[width=\linewidth]{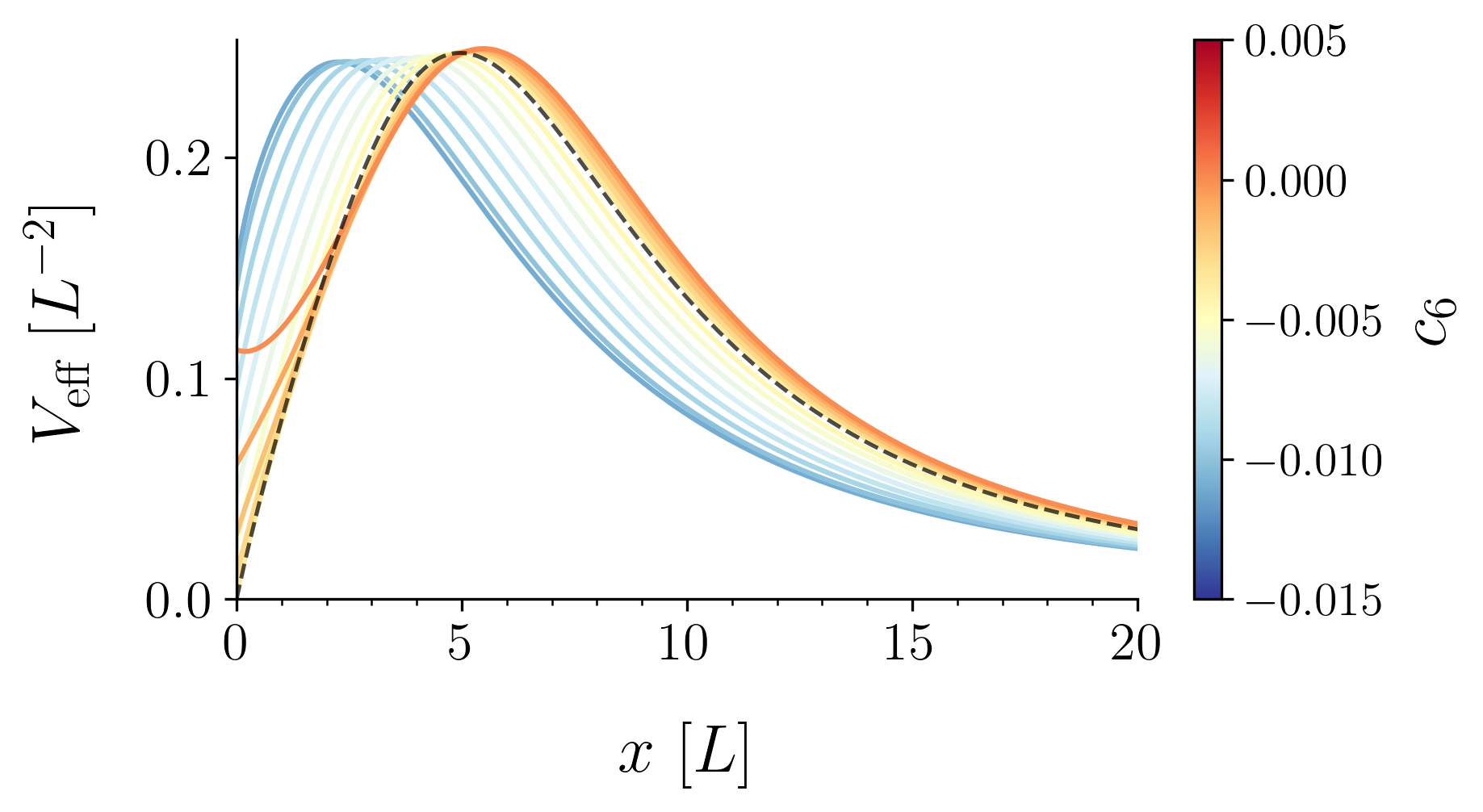}
         \caption{\footnotesize $s = 0$ and $\ell=2$.}
        \label{fig:veffc1} 
    \end{subfigure}
    
    \begin{subfigure}{0.49\textwidth}
        \centering\includegraphics[width=\linewidth]{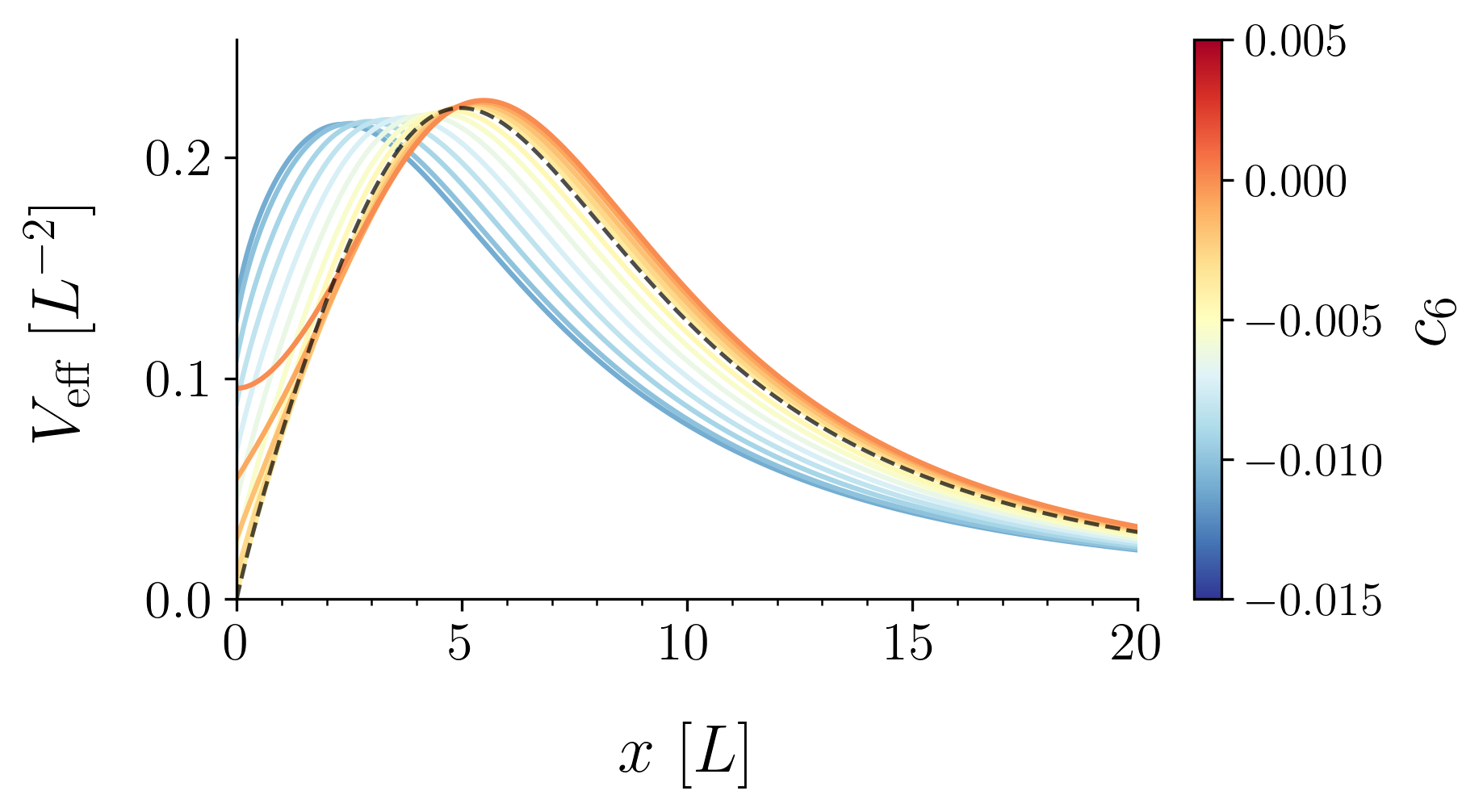}
         \caption{\footnotesize $s = 1$ and $\ell=2$.}
        \label{fig:veffc2}
    \end{subfigure}
    \begin{subfigure}{0.49\textwidth}
        \centering\includegraphics[width=\linewidth]{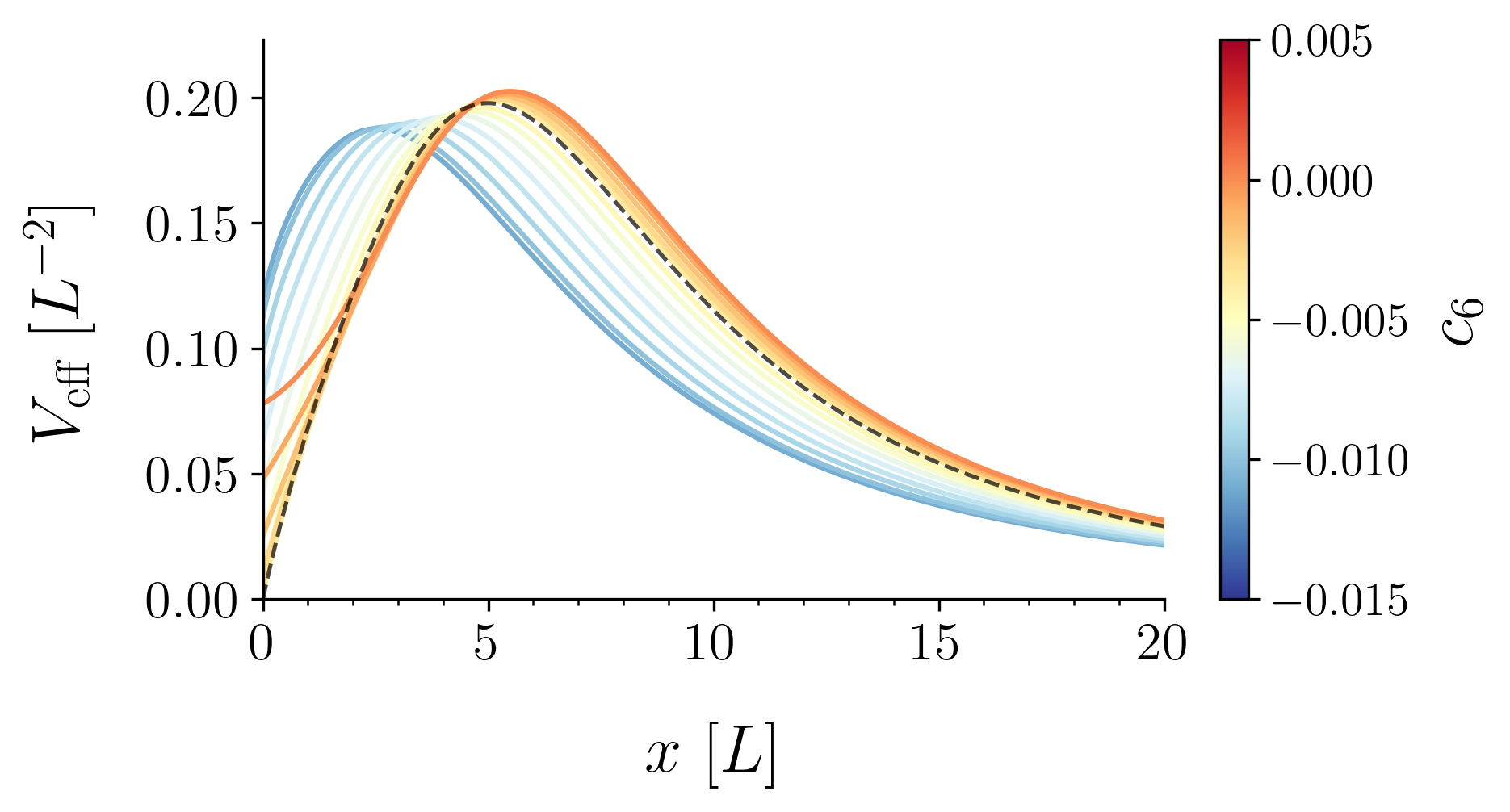}
         \caption{\footnotesize $s=\ell=2$.}
        \label{fig:veffc3}
    \end{subfigure} 
    \begin{subfigure}{0.49\textwidth}
        \centering
        \includegraphics[width=\linewidth]{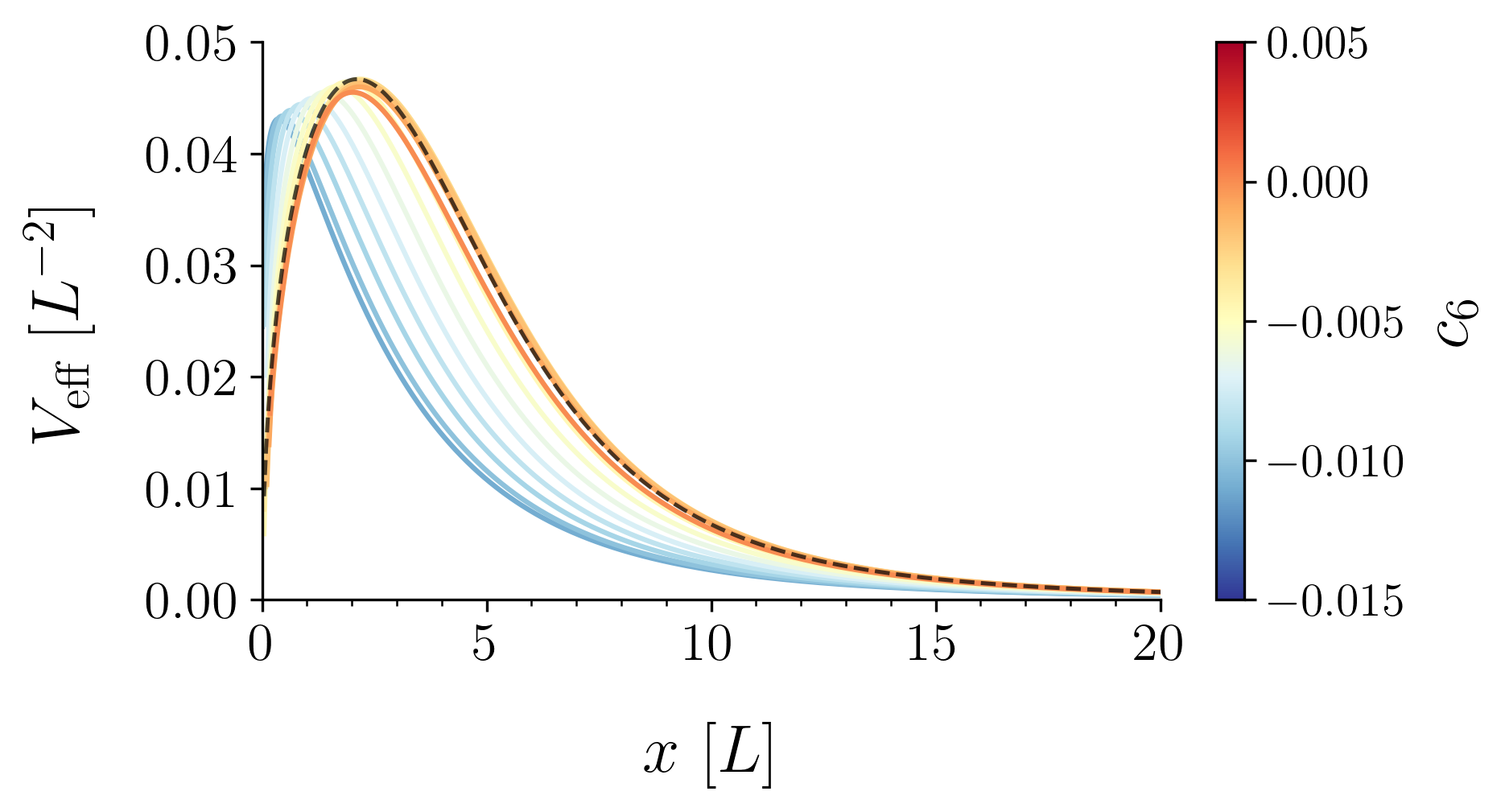}
         \caption{\footnotesize $s = \ell= {1}/{2}$.}
        \label{fig:veffe1}
    \end{subfigure}
    \begin{subfigure}{0.49\textwidth}
        \centering
        \includegraphics[width=\linewidth]{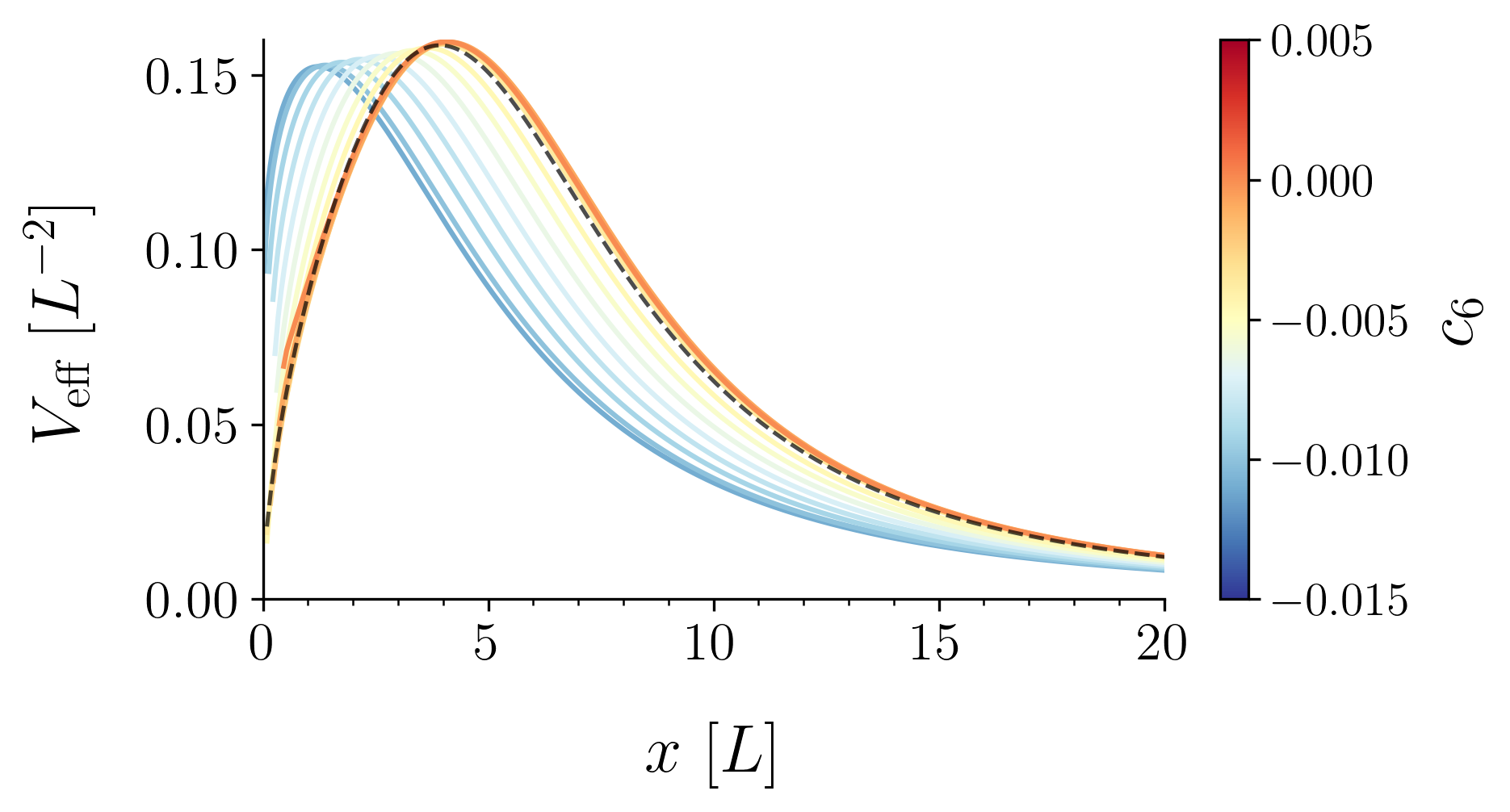}
         \caption{\footnotesize $s = {1}/{2}$ and $\ell= {3}/{2}$.}
        \label{fig:veffe2}  \end{subfigure}
     \caption{\footnotesize Effective potential as a function of the longitudinal direction of the de~Laval nozzle, varying $c_6$ for the quantum-corrected Schwarzschild metric \eqref{ck}. The dashed black line represents the  Schwarzschild solution. The values are calculated using $G_{{\scalebox{.55}{\textsc{N}}}} = M = 1$.}
    \label{fig:veff}
\end{figure}

\begin{figure}[H]
    \centering
    \begin{subfigure}{0.49\textwidth}
        \centering
        \includegraphics[width=\linewidth]{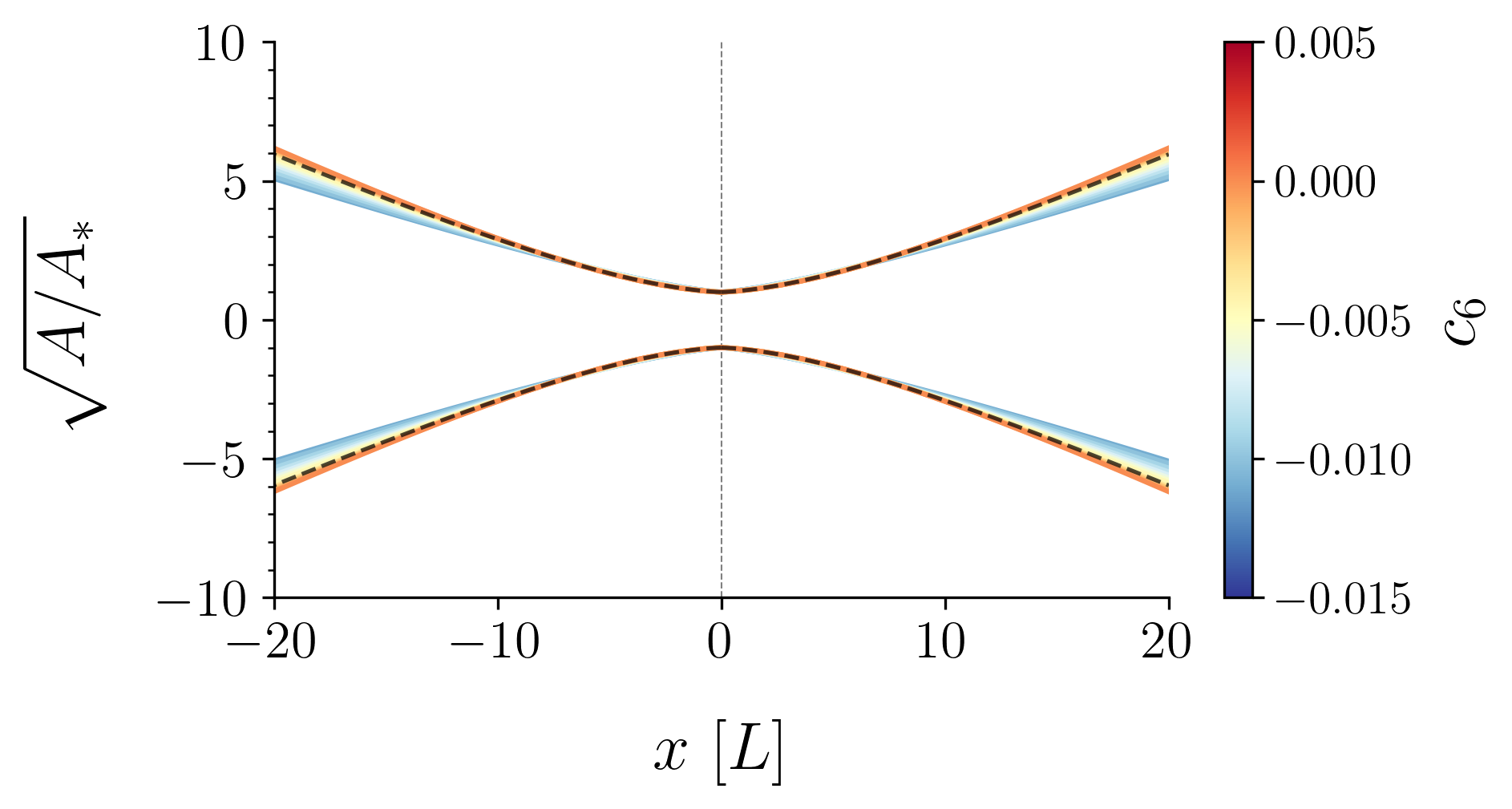}
         \caption{\footnotesize $s = \ell = 0$.}
        \label{fig:shapea}
    \end{subfigure}
    \begin{subfigure}{0.49\textwidth}
        \centering
        \includegraphics[width=\linewidth]{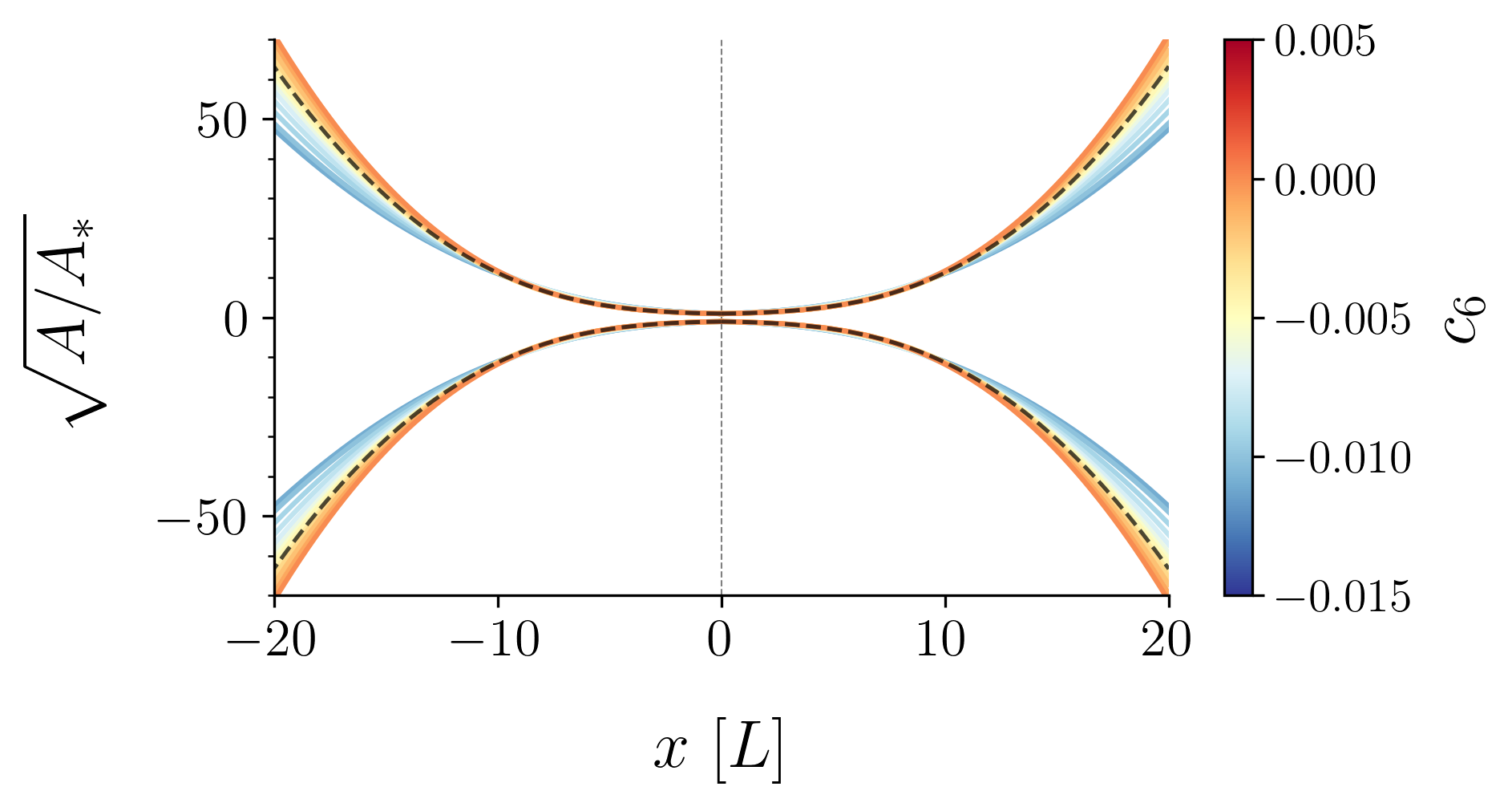}
         \caption{\footnotesize $s = 0$ and $\ell=1$.}
        \label{fig:shapeb1}  \end{subfigure}
        
    \begin{subfigure}{0.49\textwidth}
        \centering
        \includegraphics[width=\linewidth]{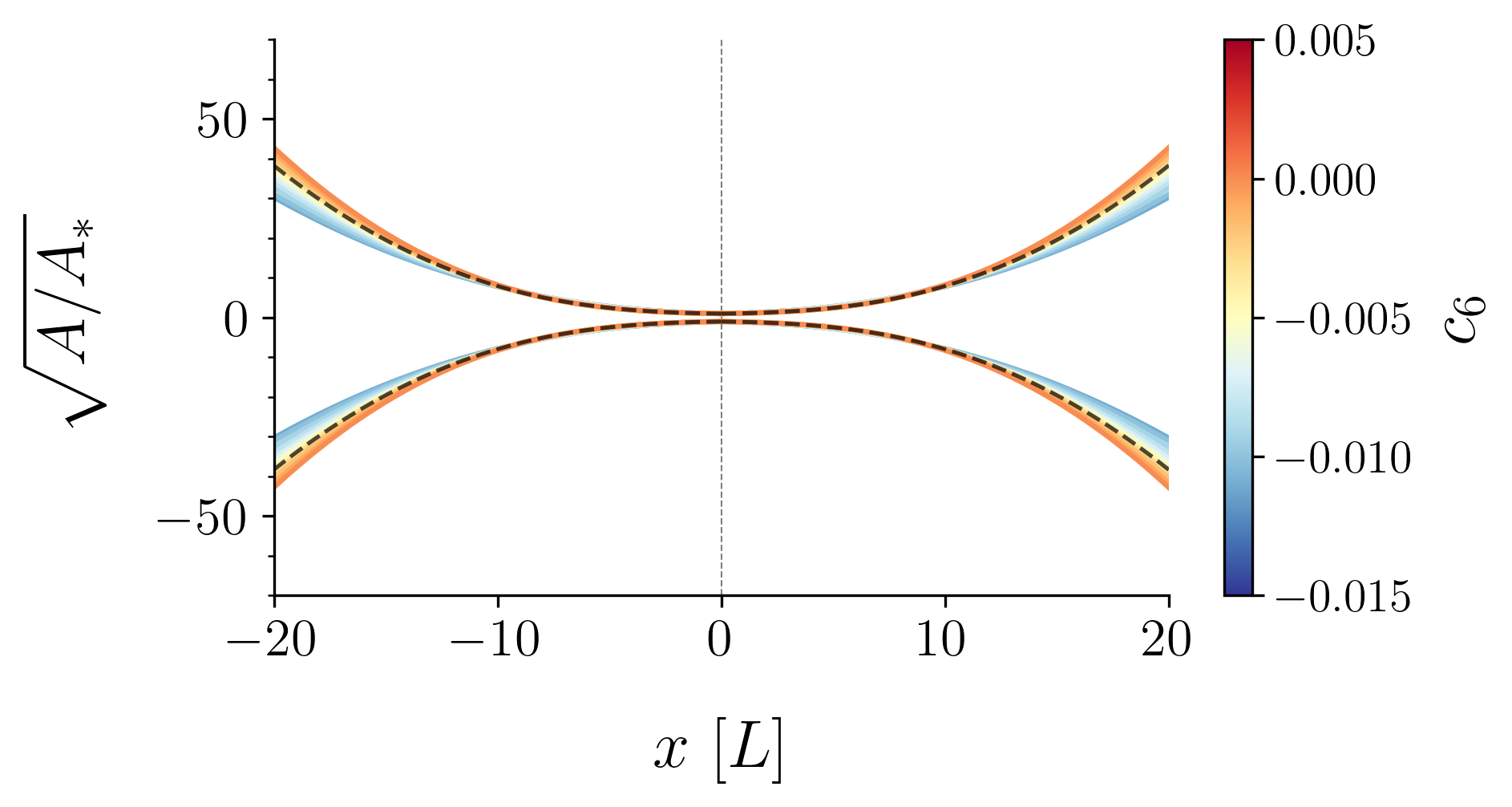}
         \caption{\footnotesize $s =\ell=1$.}
        \label{fig:shapeb2}
    \end{subfigure}
    \begin{subfigure}{0.49\textwidth}
        \centering
        \includegraphics[width=\linewidth]{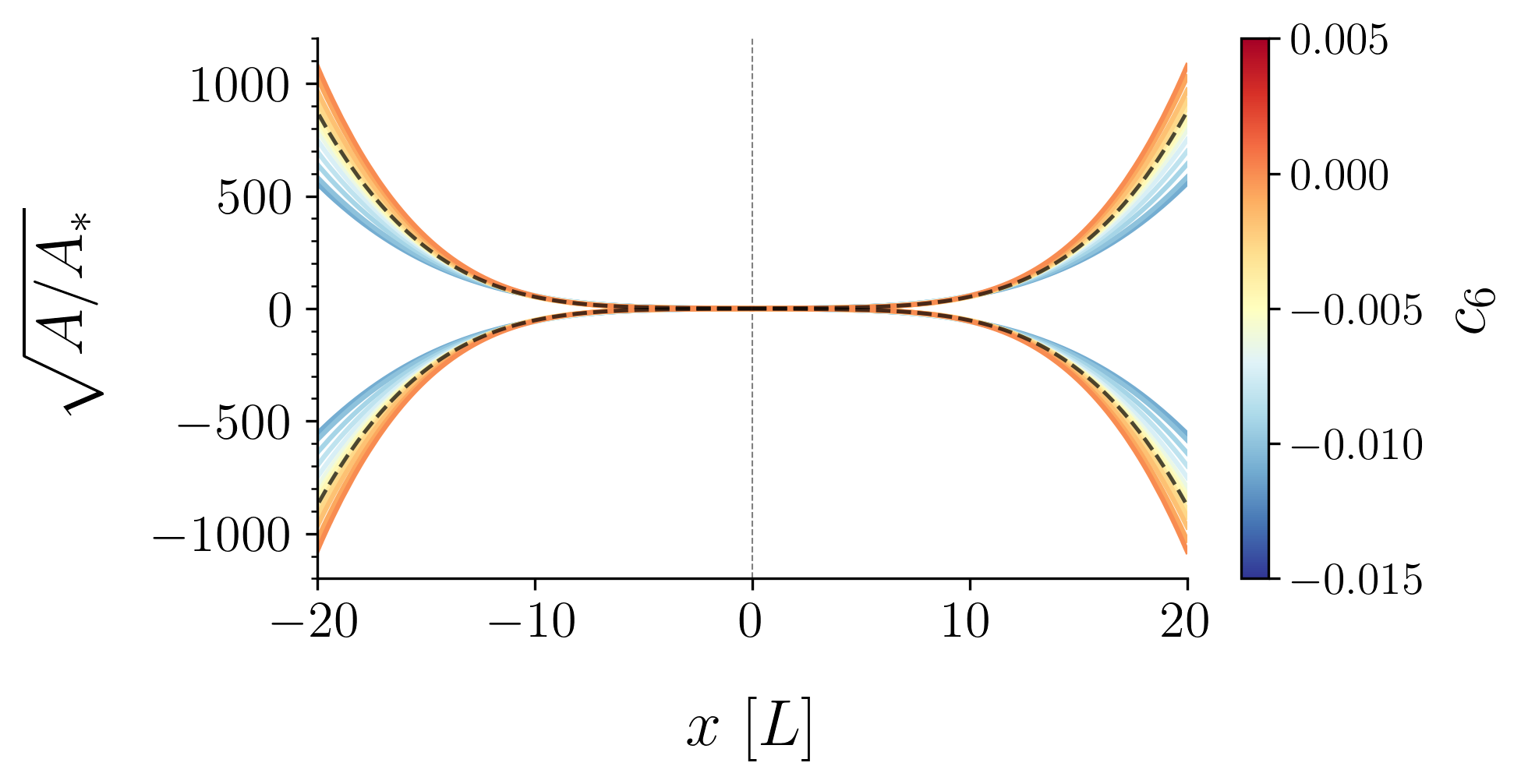}
         \caption{\footnotesize $s = 0$ and $\ell=2$.}
        \label{fig:shapec1} 
    \end{subfigure}
    
    \begin{subfigure}{0.49\textwidth}
        \centering\includegraphics[width=\linewidth]{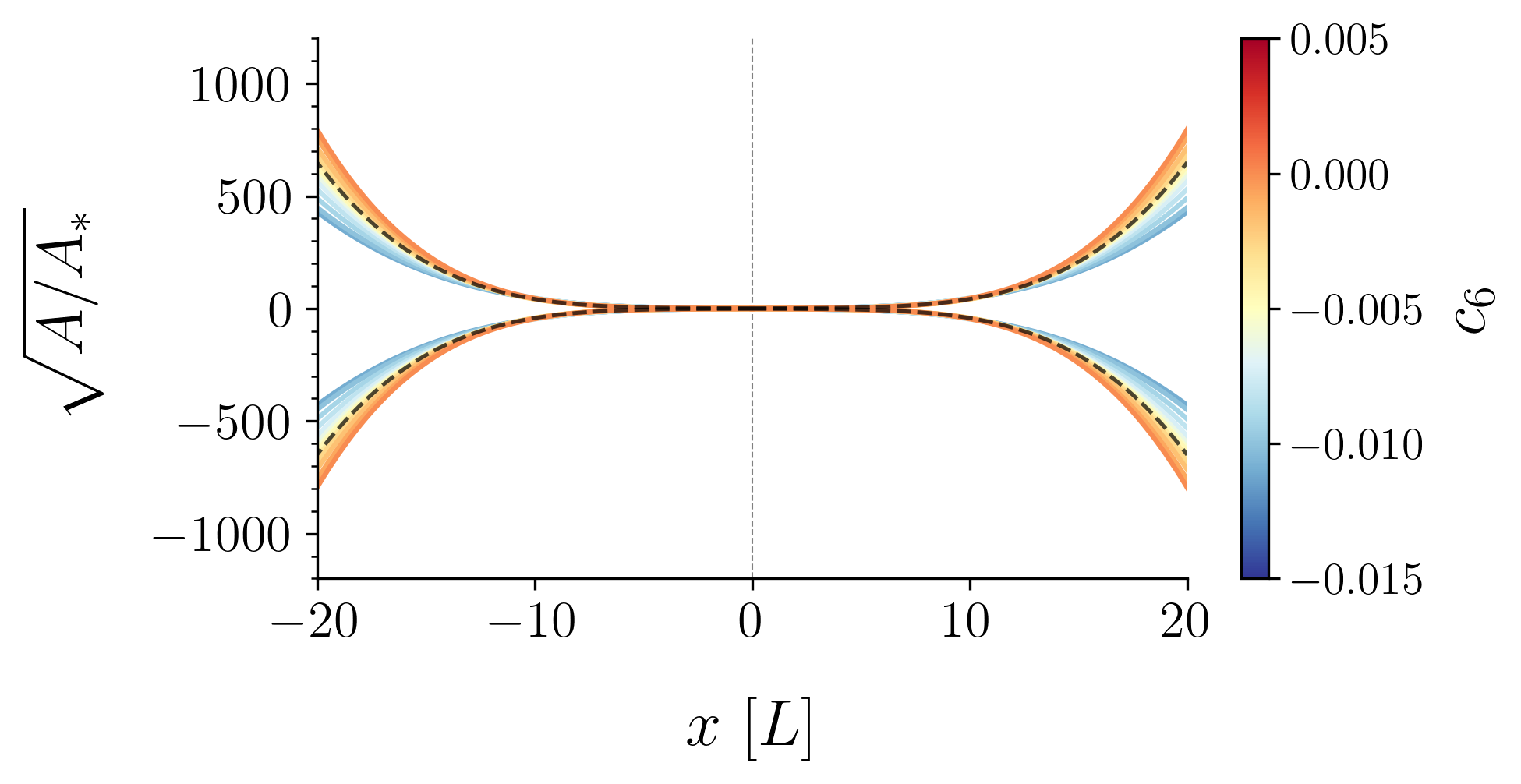}
         \caption{\footnotesize $s = 1$ and $\ell=2$.}
        \label{fig:shapec2}
    \end{subfigure}
    \begin{subfigure}{0.49\textwidth}
        \centering\includegraphics[width=\linewidth]{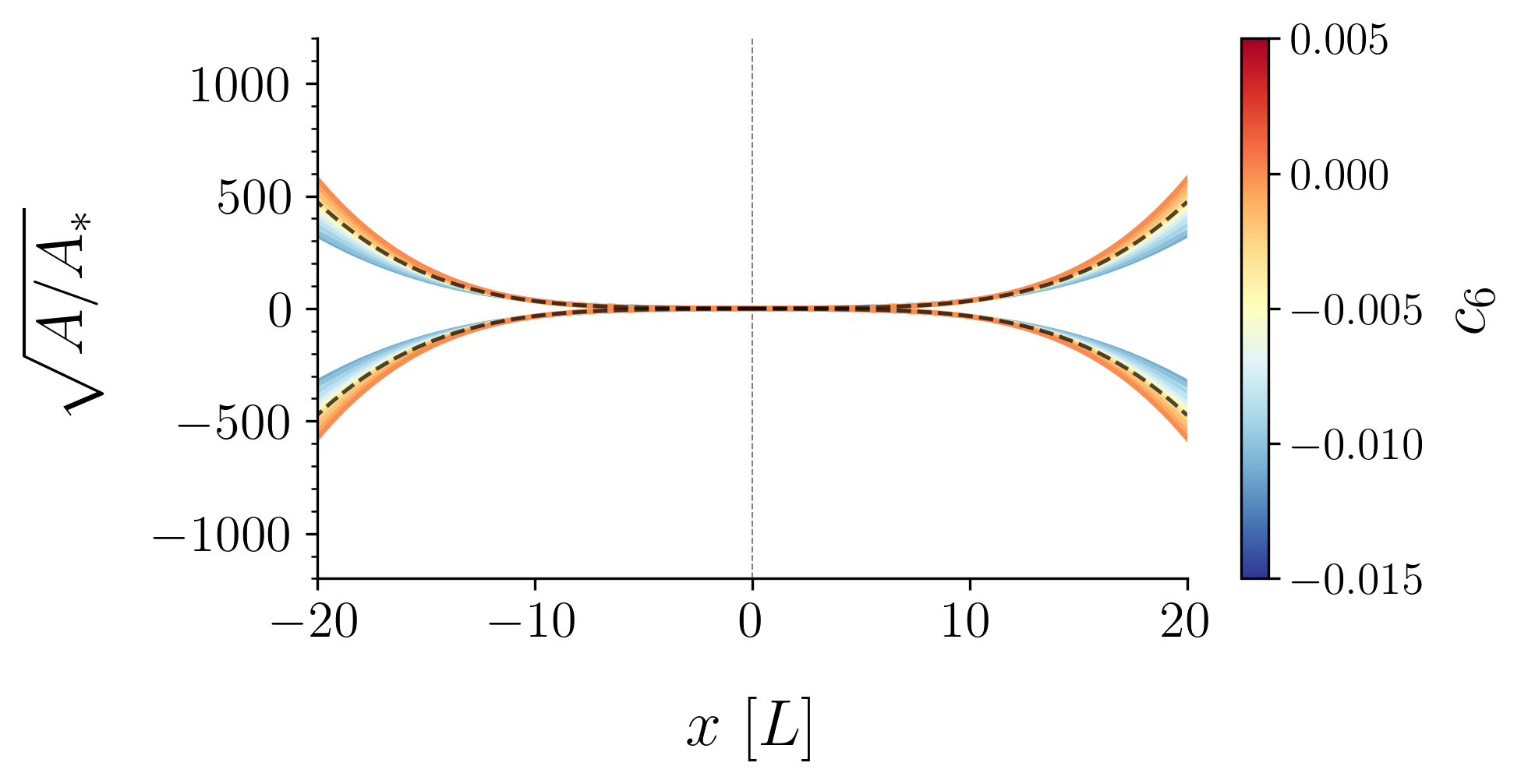}
         \caption{\footnotesize $s=\ell=2$.}
        \label{fig:shapec3}
    \end{subfigure} 
    \begin{subfigure}{0.49\textwidth}
        \centering
        \includegraphics[width=\linewidth]{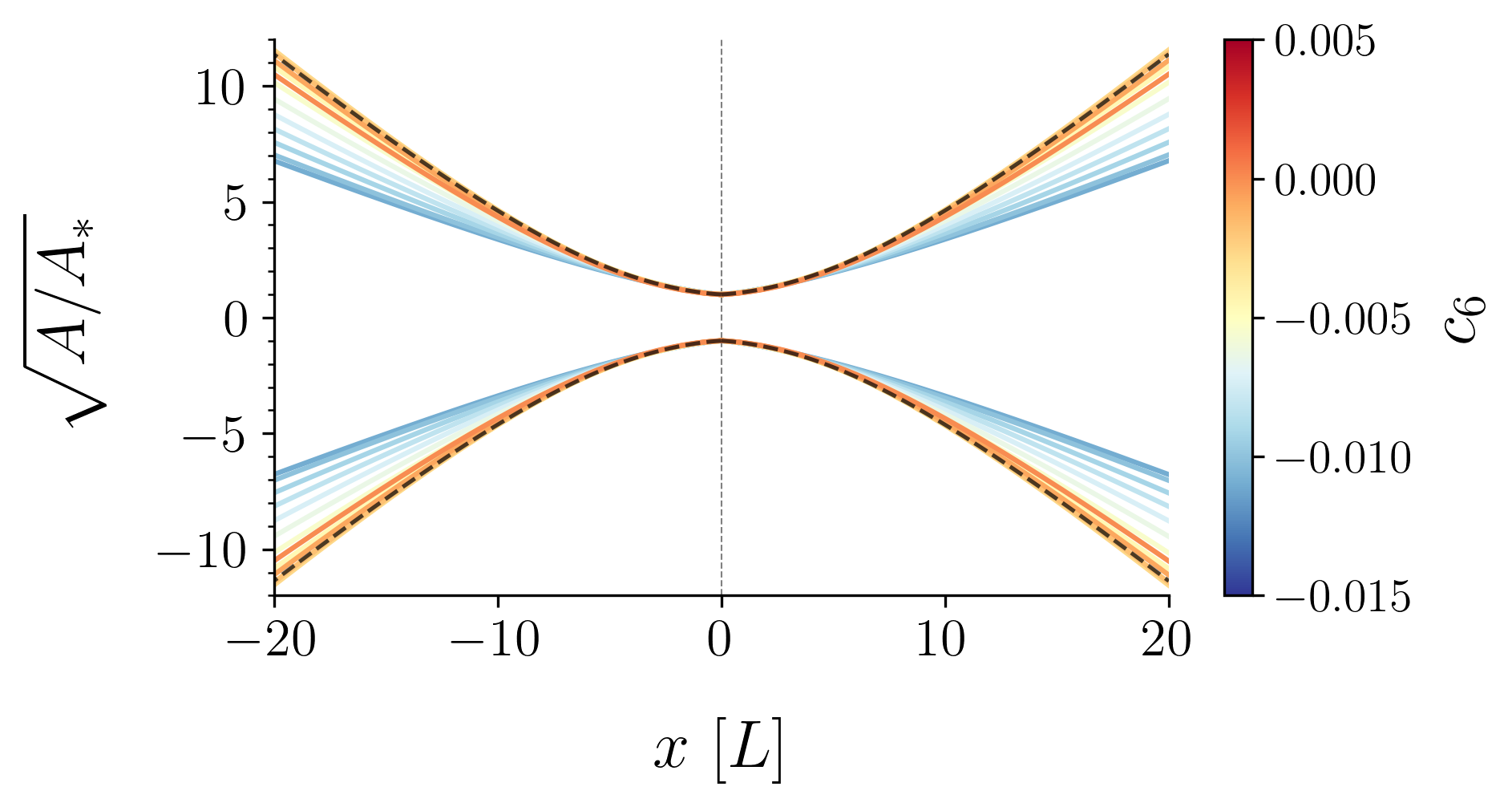}
         \caption{\footnotesize $s = \ell= {1}/{2}$.}
        \label{fig:shapee1}
    \end{subfigure}
    \begin{subfigure}{0.49\textwidth}
        \centering
        \includegraphics[width=\linewidth]{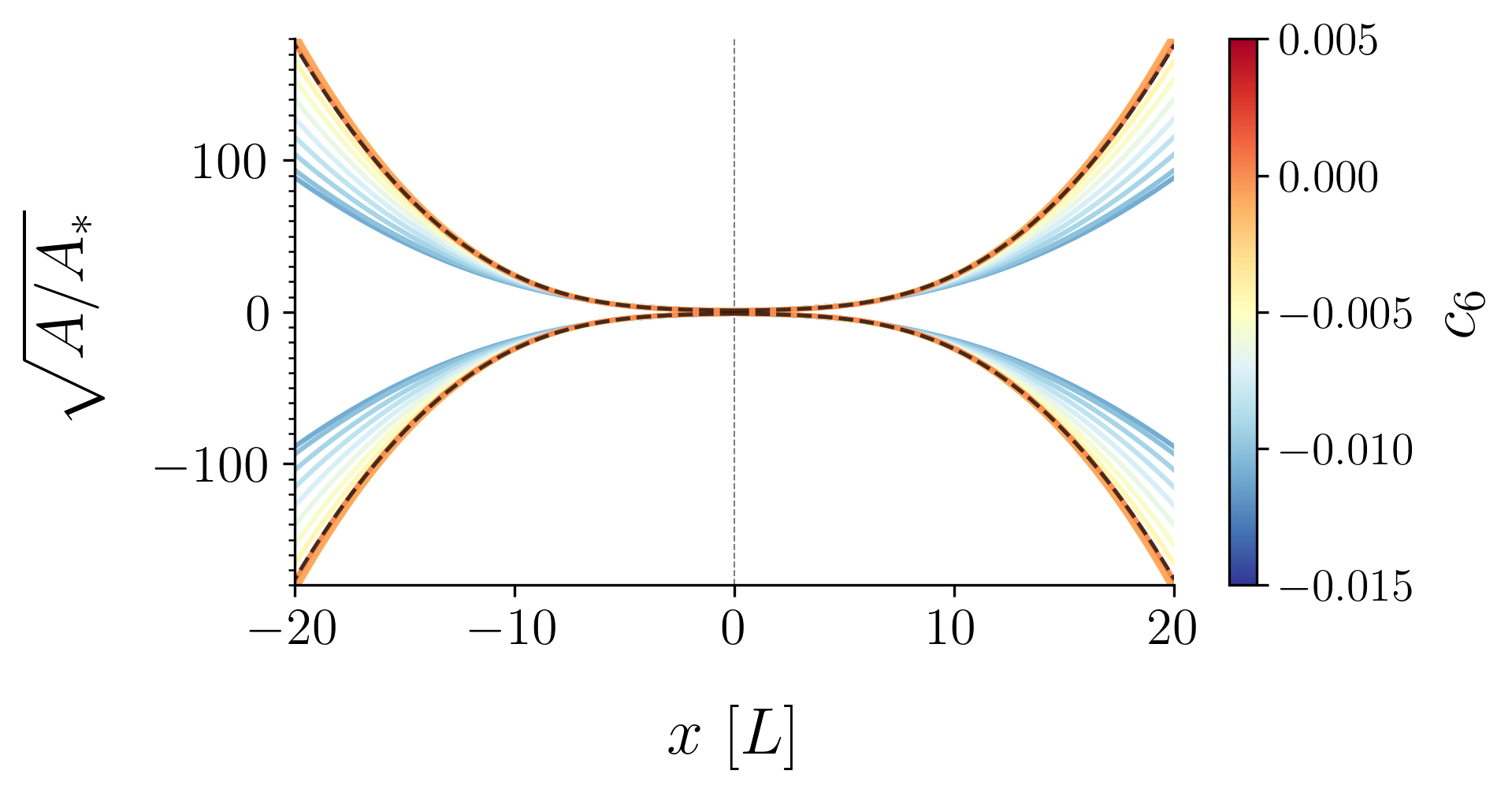}
         \caption{\footnotesize $s = {1}/{2}$ and $\ell= {3}/{2}$.}
        \label{fig:shapee2}  \end{subfigure}
     \caption{\footnotesize Nozzle shape geometry 
    as a function of the longitudinal direction of the de~Laval nozzle, varying $c_6$ for the quantum-corrected Schwarzschild metric~\eqref{ck}. The dashed black line represents the Schwarzschild solution. The values are calculated using $G_{{\scalebox{.55}{\textsc{N}}}} = M = 1$.}
    \label{fig:shape}
\end{figure}

\begin{figure}[H]
    \centering
    \begin{subfigure}{0.49\textwidth}
        \centering
        \includegraphics[width=\linewidth]{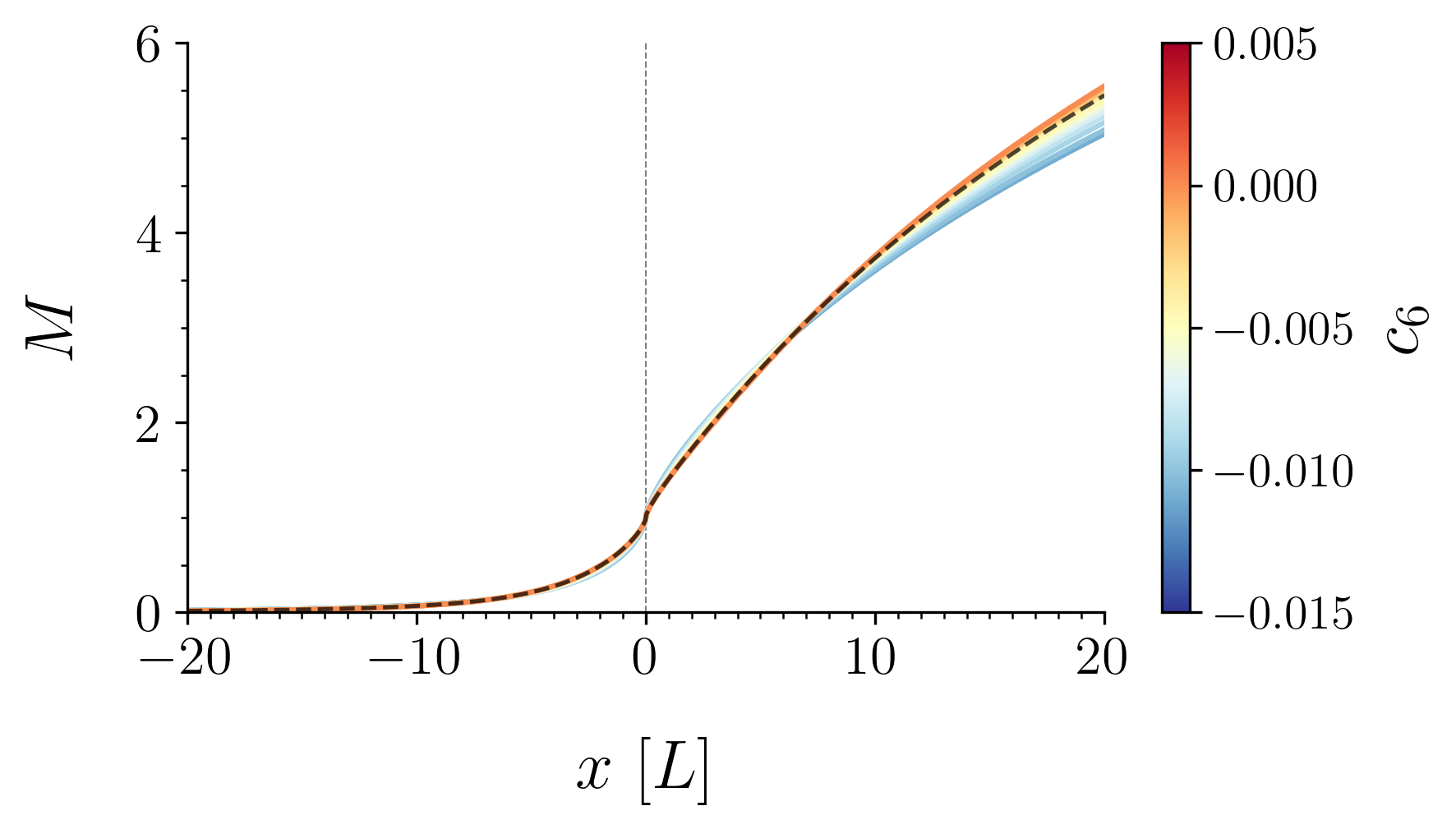}
         \caption{\footnotesize $s = \ell = 0$.}
        \label{fig:Macha}
    \end{subfigure}
    \begin{subfigure}{0.49\textwidth}
        \centering
        \includegraphics[width=\linewidth]{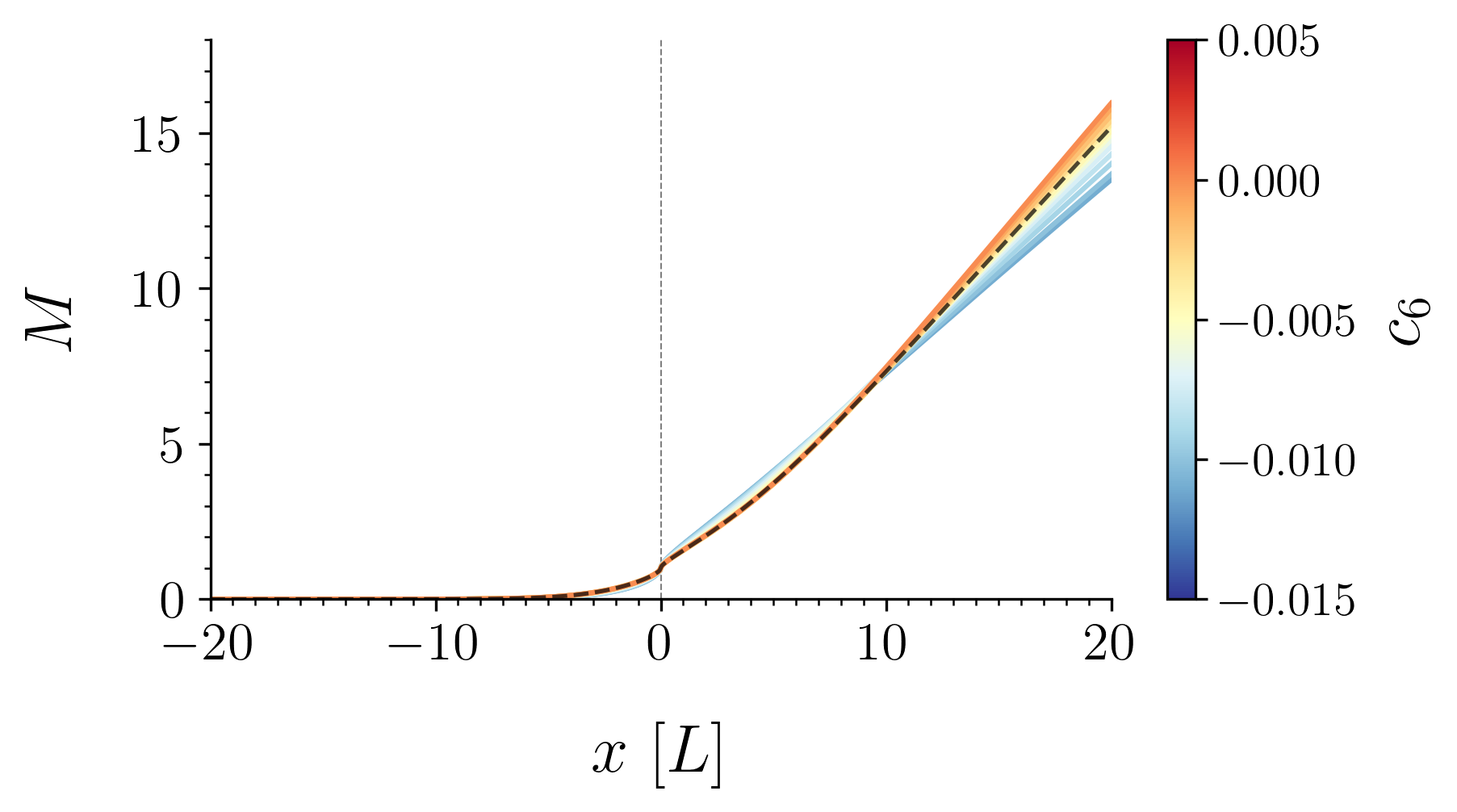}
         \caption{\footnotesize $s = 0$ and $\ell=1$.}
        \label{fig:Machb1}  \end{subfigure}
        
    \begin{subfigure}{0.49\textwidth}
        \centering
        \includegraphics[width=\linewidth]{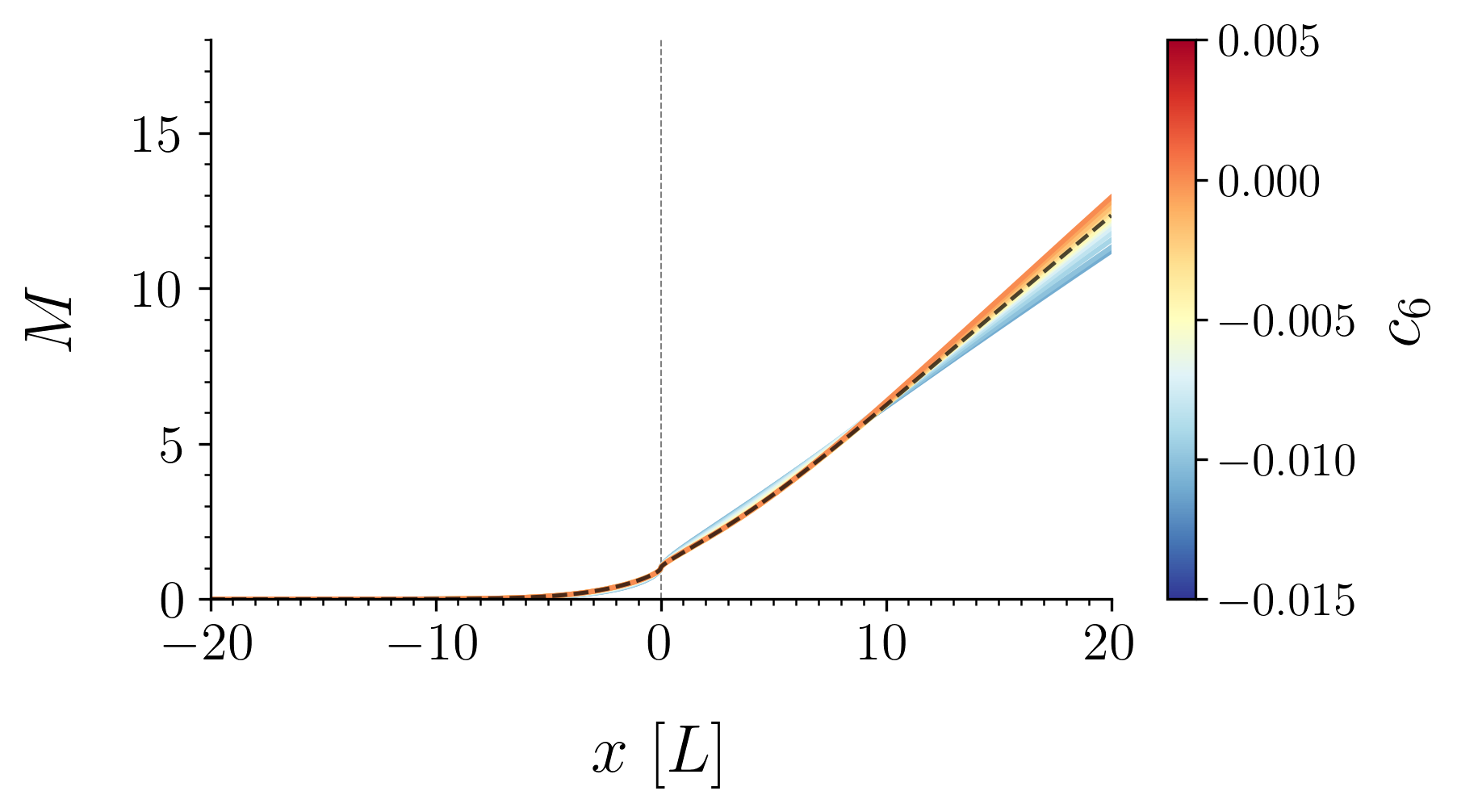}
         \caption{\footnotesize $s =\ell=1$.}
        \label{fig:Machb2}
    \end{subfigure}
    \begin{subfigure}{0.49\textwidth}
        \centering
        \includegraphics[width=\linewidth]{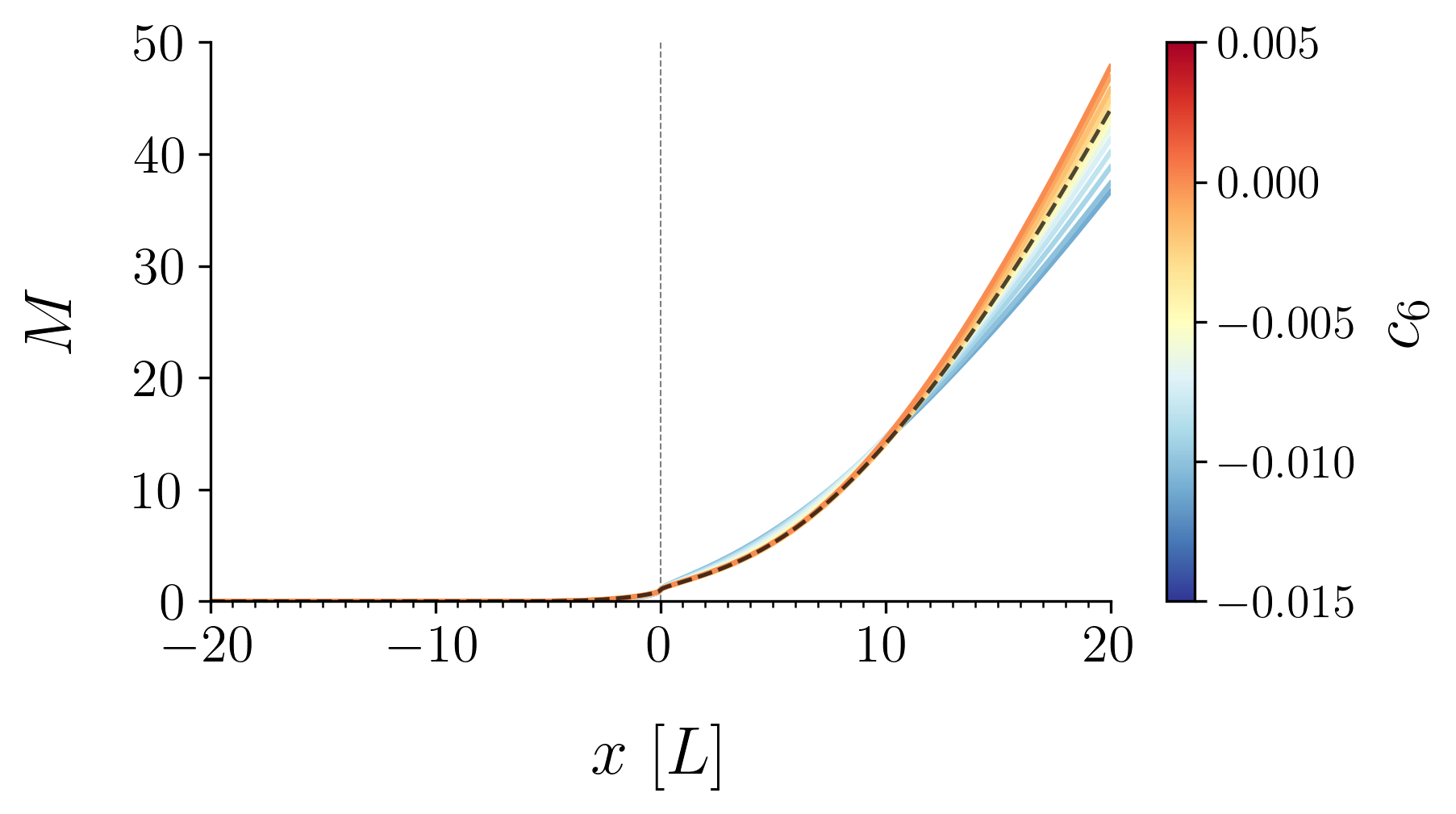}
         \caption{\footnotesize $s = 0$ and $\ell=2$.}
        \label{fig:Machc1} 
    \end{subfigure}
    
    \begin{subfigure}{0.49\textwidth}
        \centering\includegraphics[width=\linewidth]{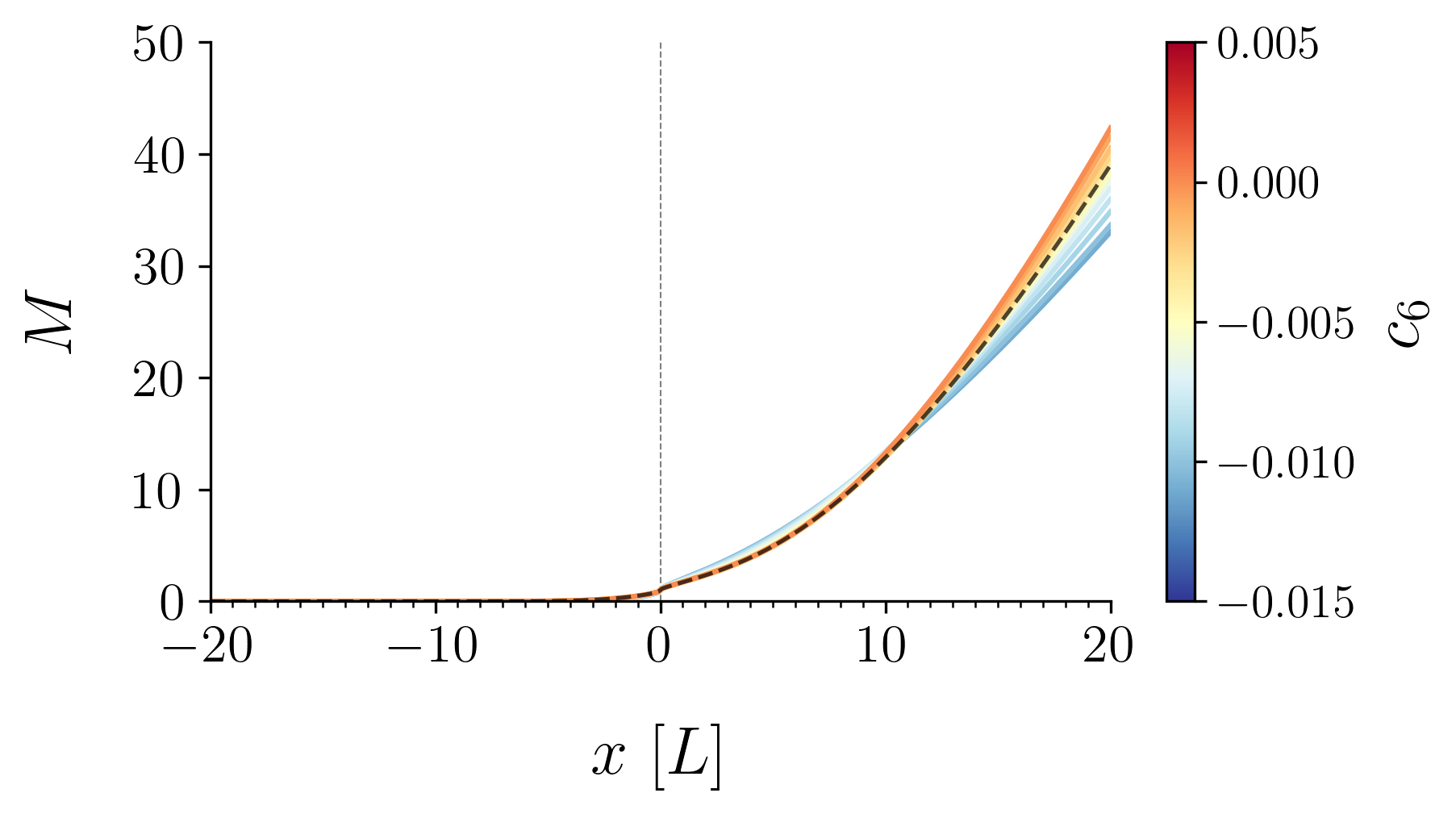}
         \caption{\footnotesize $s = 1$ and $\ell=2$.}
        \label{fig:Machc2}
    \end{subfigure}
    \begin{subfigure}{0.49\textwidth}
        \centering\includegraphics[width=\linewidth]{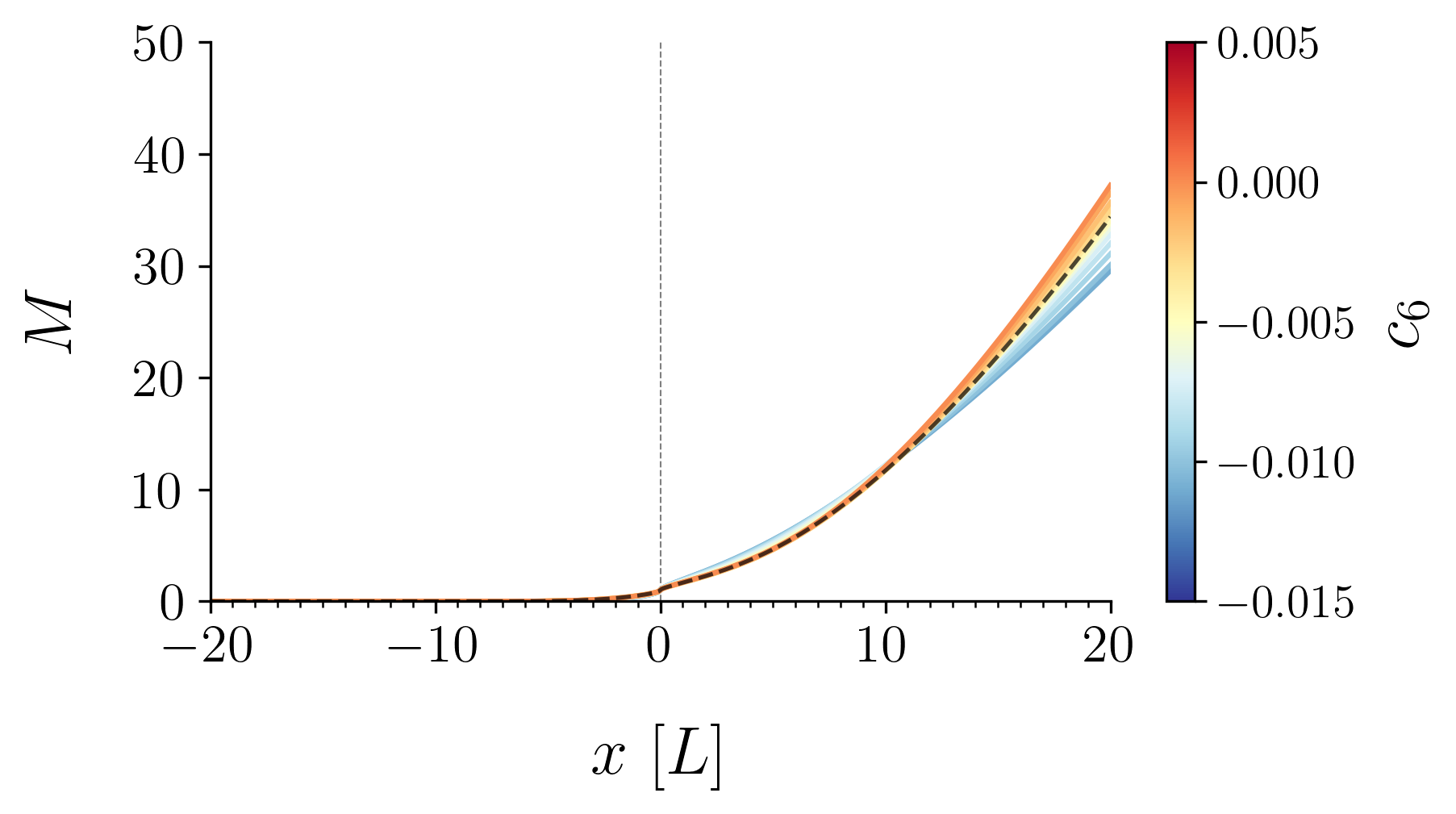}
         \caption{\footnotesize $s=\ell=2$.}
        \label{fig:Machc3}
    \end{subfigure} 
    \begin{subfigure}{0.49\textwidth}
        \centering
        \includegraphics[width=\linewidth]{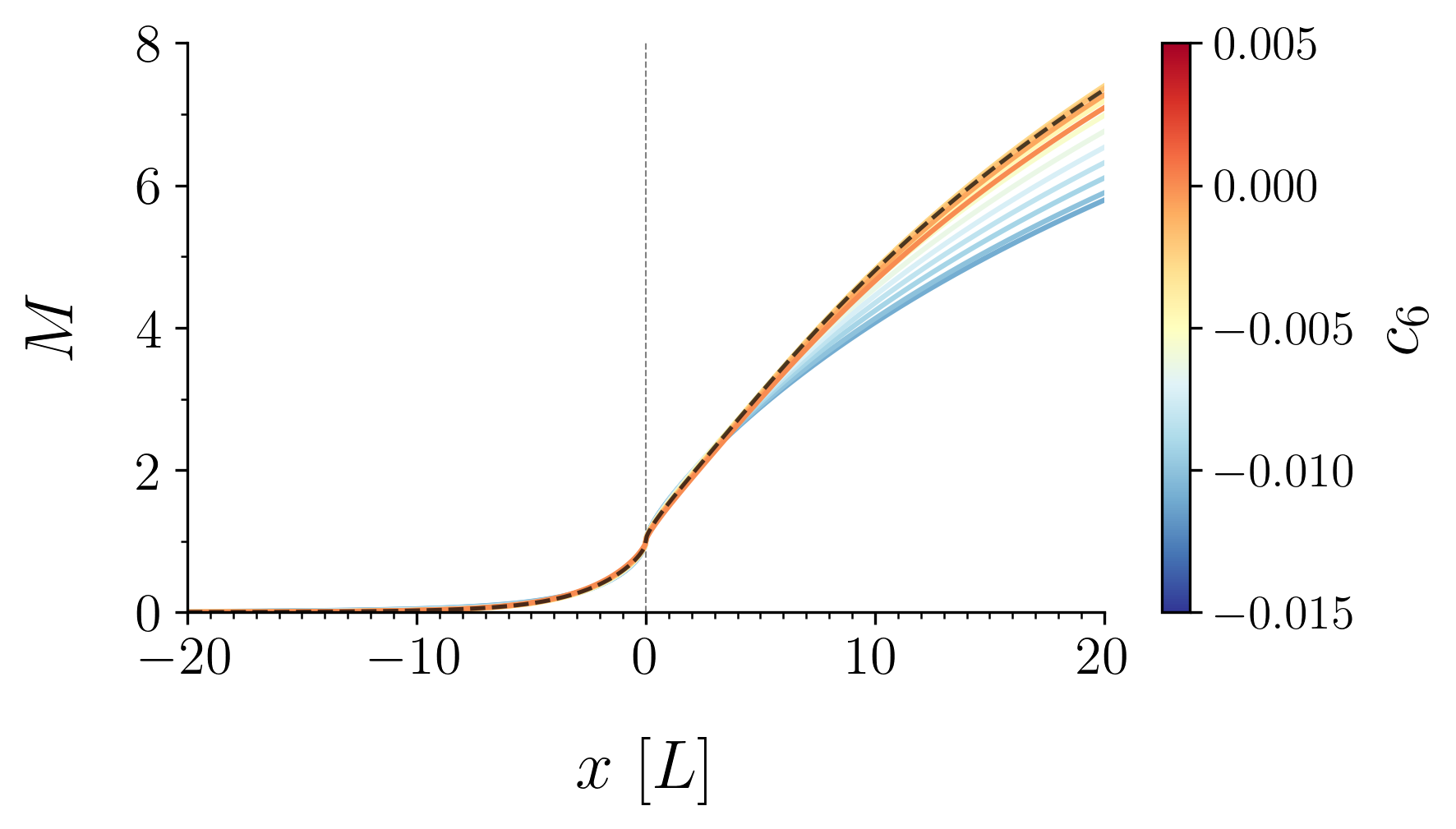}
         \caption{\footnotesize $s = \ell= {1}/{2}$.}
        \label{fig:Mache1}
    \end{subfigure}
    \begin{subfigure}{0.49\textwidth}
        \centering
        \includegraphics[width=\linewidth]{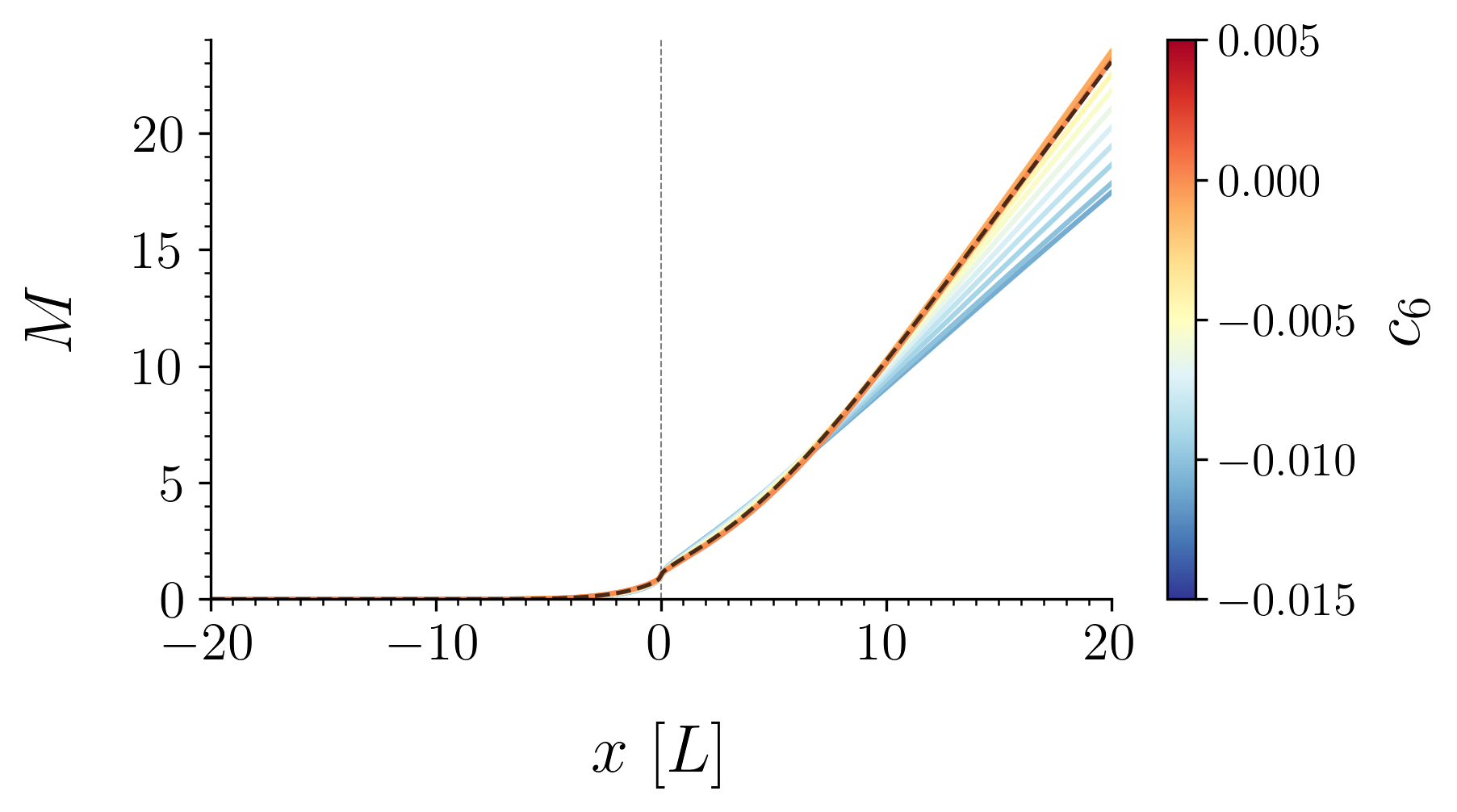}
         \caption{\footnotesize $s = {1}/{2}$ and $\ell= {3}/{2}$.}
        \label{fig:Mache2}  \end{subfigure}
     \caption{\footnotesize Mach number 
    as a function of the longitudinal direction of the de~Laval nozzle, varying $c_6$ for the quantum-corrected Schwarzschild metric~\eqref{ck}. The dashed black line represents the Schwarzschild solution. The values are calculated using $G_{{\scalebox{.55}{\textsc{N}}}} = M = 1$.}
    \label{fig:Mach}
\end{figure}

\begin{figure}[H]
    \centering
    \begin{subfigure}{0.49\textwidth}
        \centering
        \includegraphics[width=\linewidth]{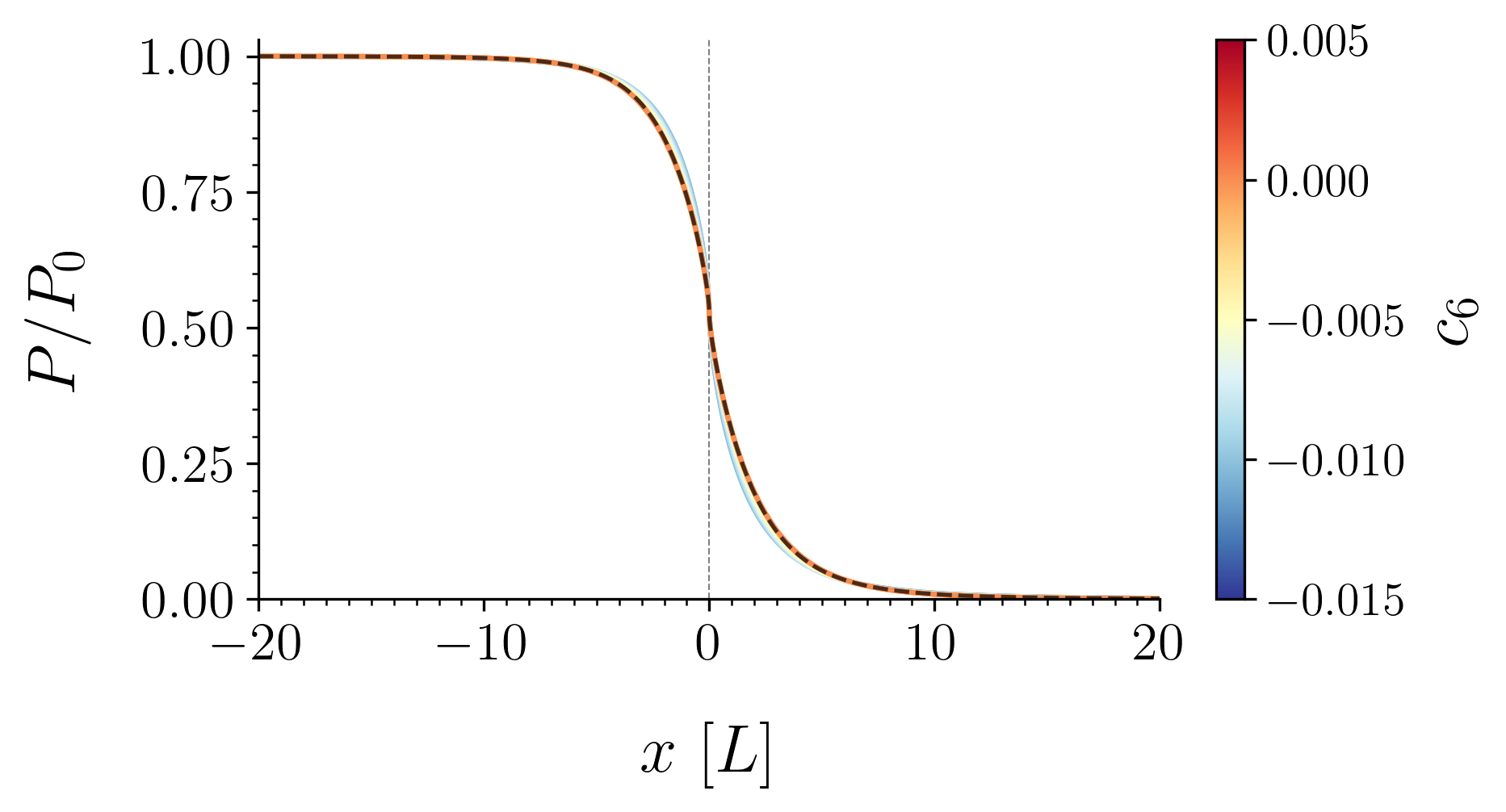}
         \caption{\footnotesize $s = \ell = 0$.}
        \label{fig:Pressurea}
    \end{subfigure}
    \begin{subfigure}{0.49\textwidth}
        \centering
        \includegraphics[width=\linewidth]{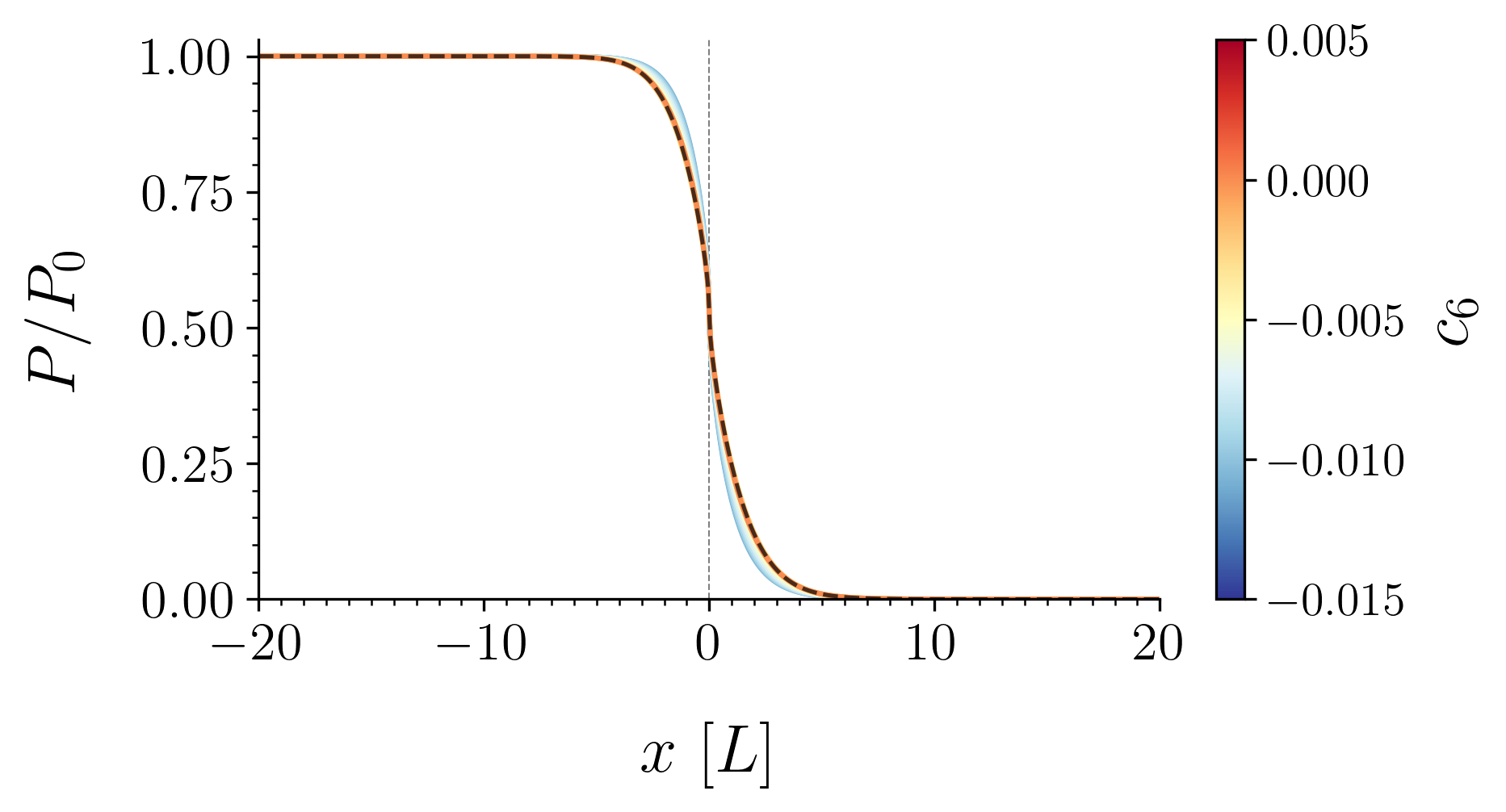}
         \caption{\footnotesize $s = 0$ and $\ell=1$.}
        \label{fig:Pressureb1}  \end{subfigure}
        
    \begin{subfigure}{0.49\textwidth}
        \centering
        \includegraphics[width=\linewidth]{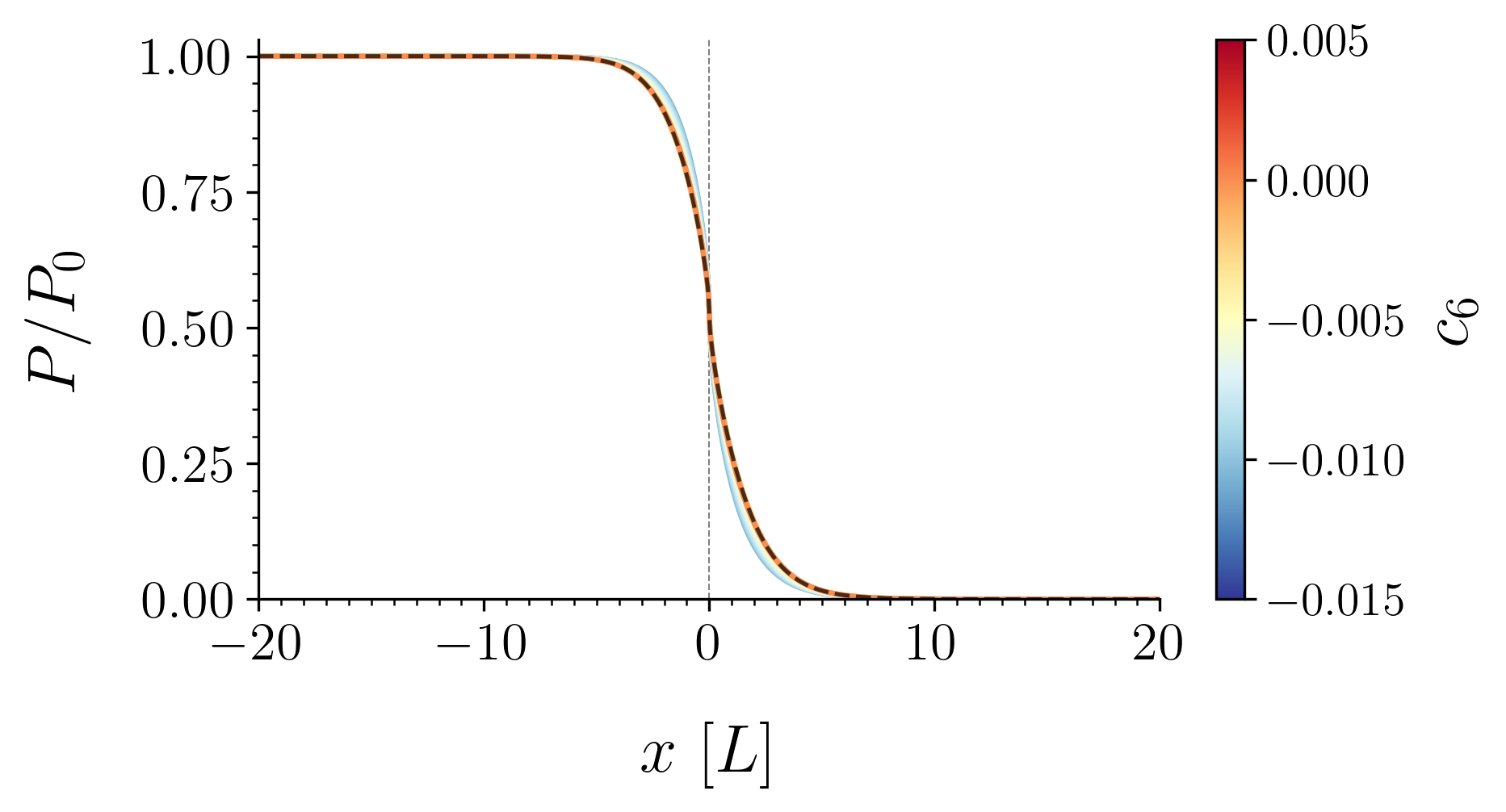}
         \caption{\footnotesize $s =\ell=1$.}
        \label{fig:Pressureb2}
    \end{subfigure}
    \begin{subfigure}{0.49\textwidth}
        \centering
        \includegraphics[width=\linewidth]{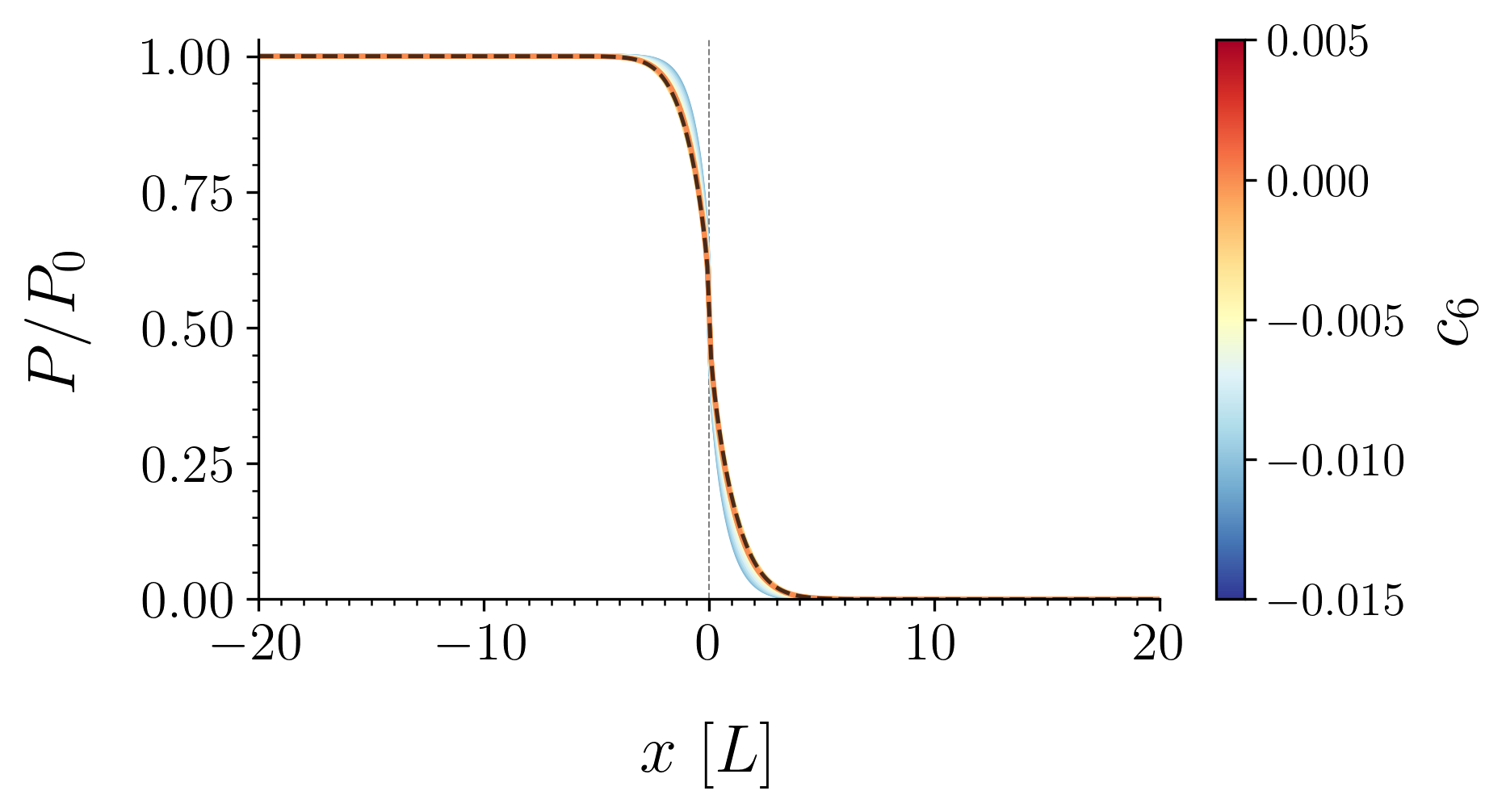}
         \caption{\footnotesize $s = 0$ and $\ell=2$.}
        \label{fig:Pressurec1} 
    \end{subfigure}
    
    \begin{subfigure}{0.49\textwidth}
        \centering\includegraphics[width=\linewidth]{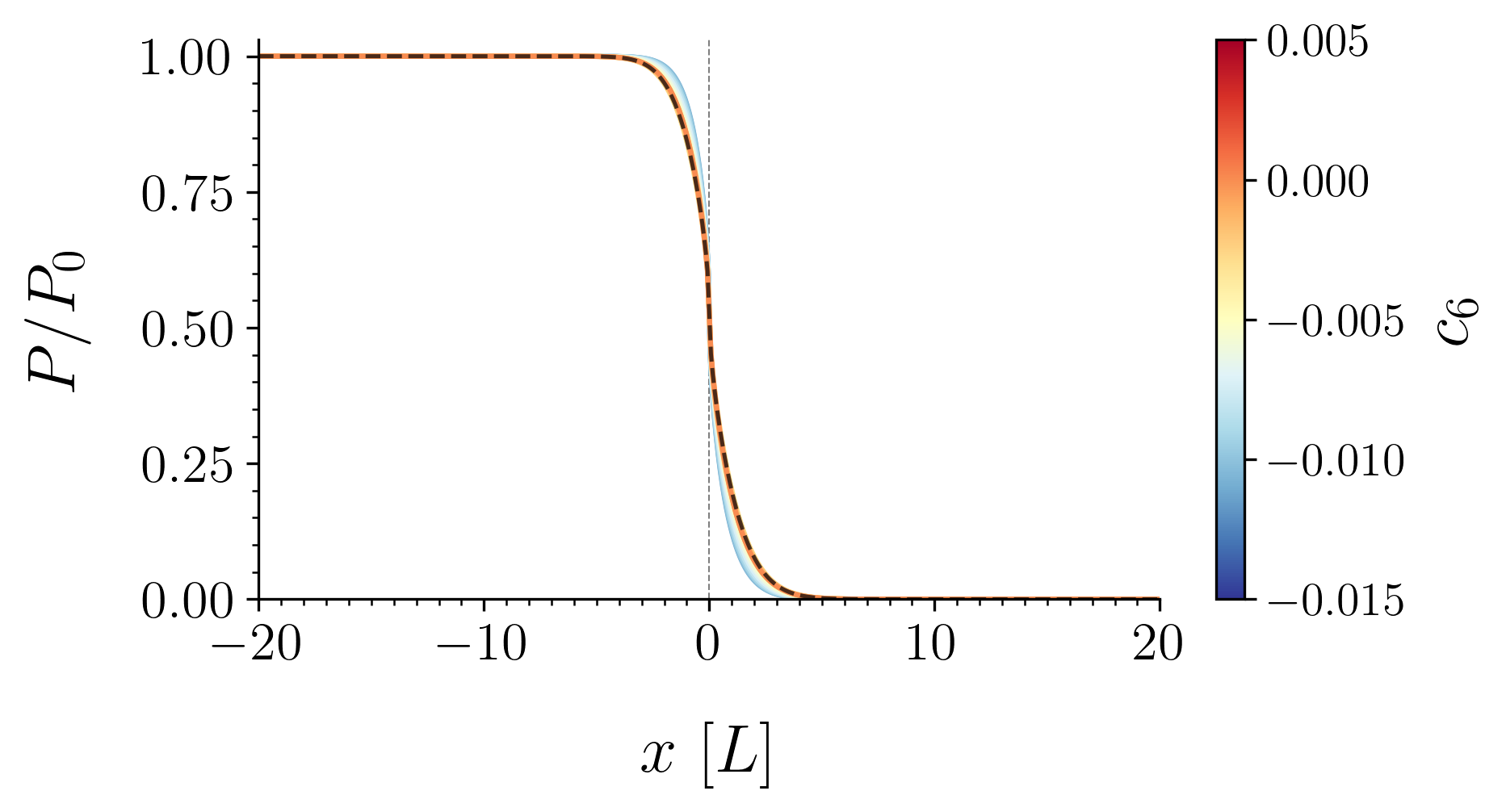}
         \caption{\footnotesize $s = 1$ and $\ell=2$.}
        \label{fig:Pressurec2}
    \end{subfigure}
    \begin{subfigure}{0.49\textwidth}
        \centering\includegraphics[width=\linewidth]{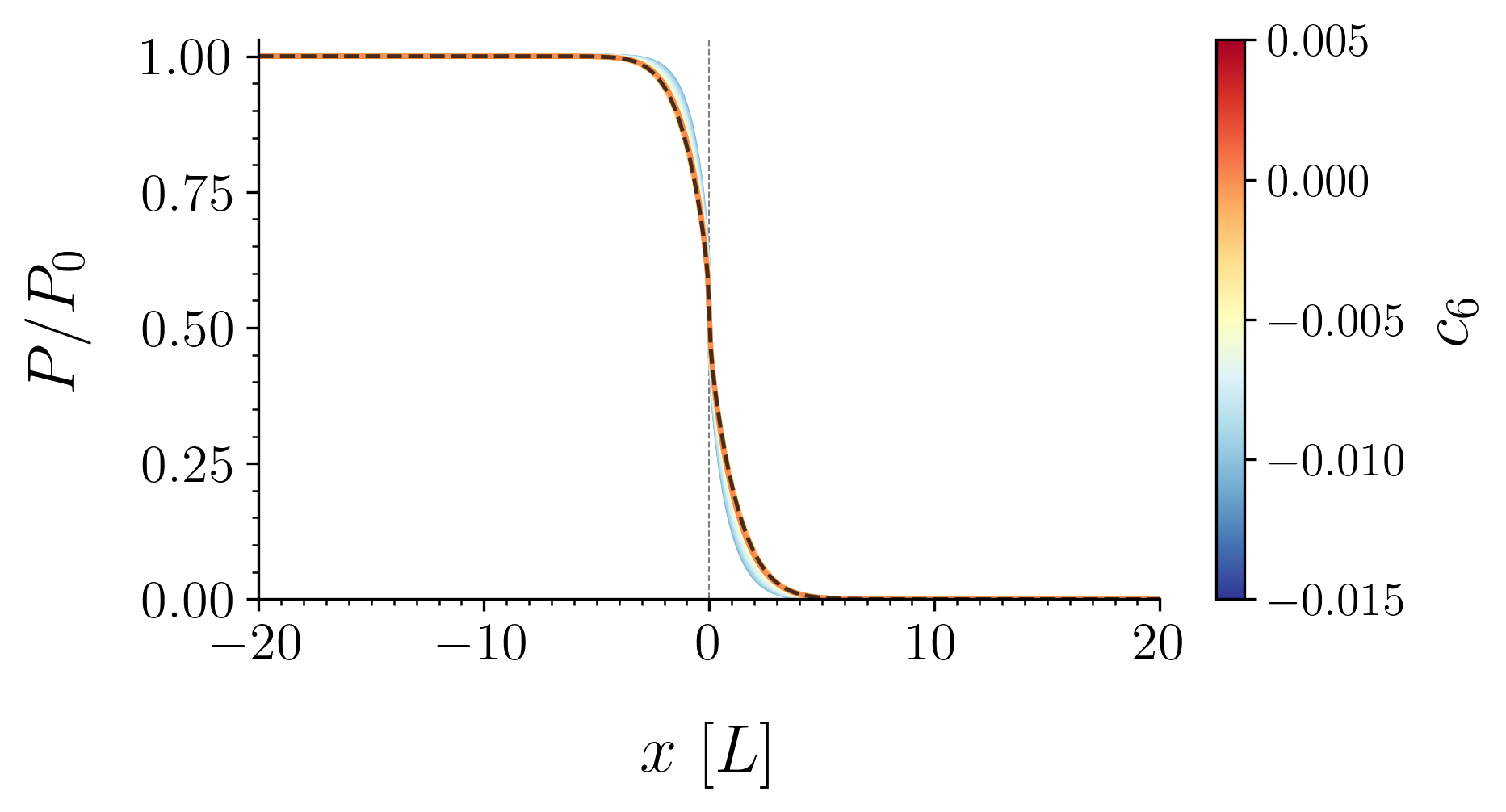}
         \caption{\footnotesize $s=\ell=2$.}
        \label{fig:Pressurec3}
    \end{subfigure} 
    \begin{subfigure}{0.49\textwidth}
        \centering
        \includegraphics[width=\linewidth]{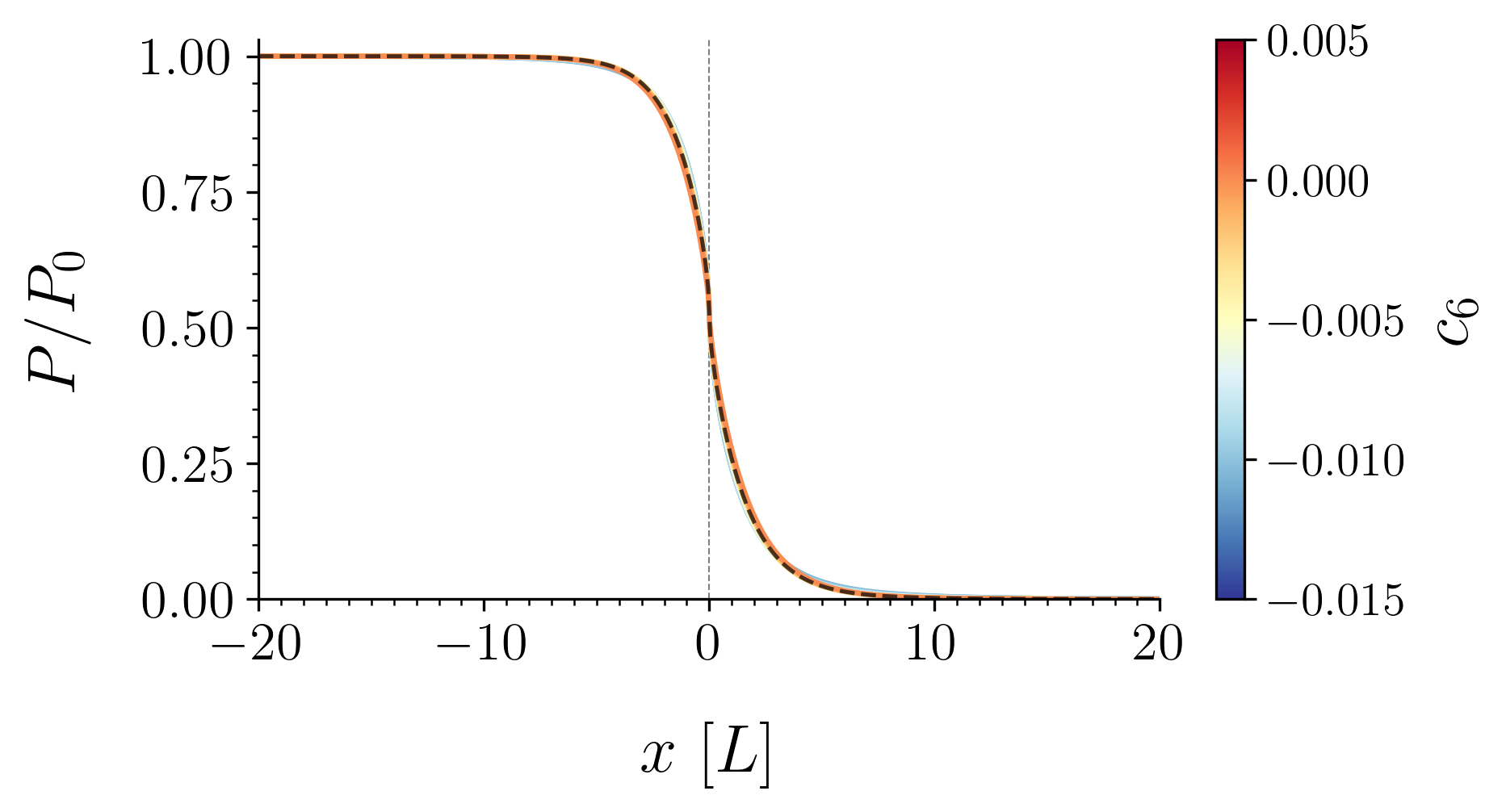}
         \caption{\footnotesize $s = \ell= {1}/{2}$.}
        \label{fig:Pressuree1}
    \end{subfigure}
    \begin{subfigure}{0.49\textwidth}
        \centering
        \includegraphics[width=\linewidth]{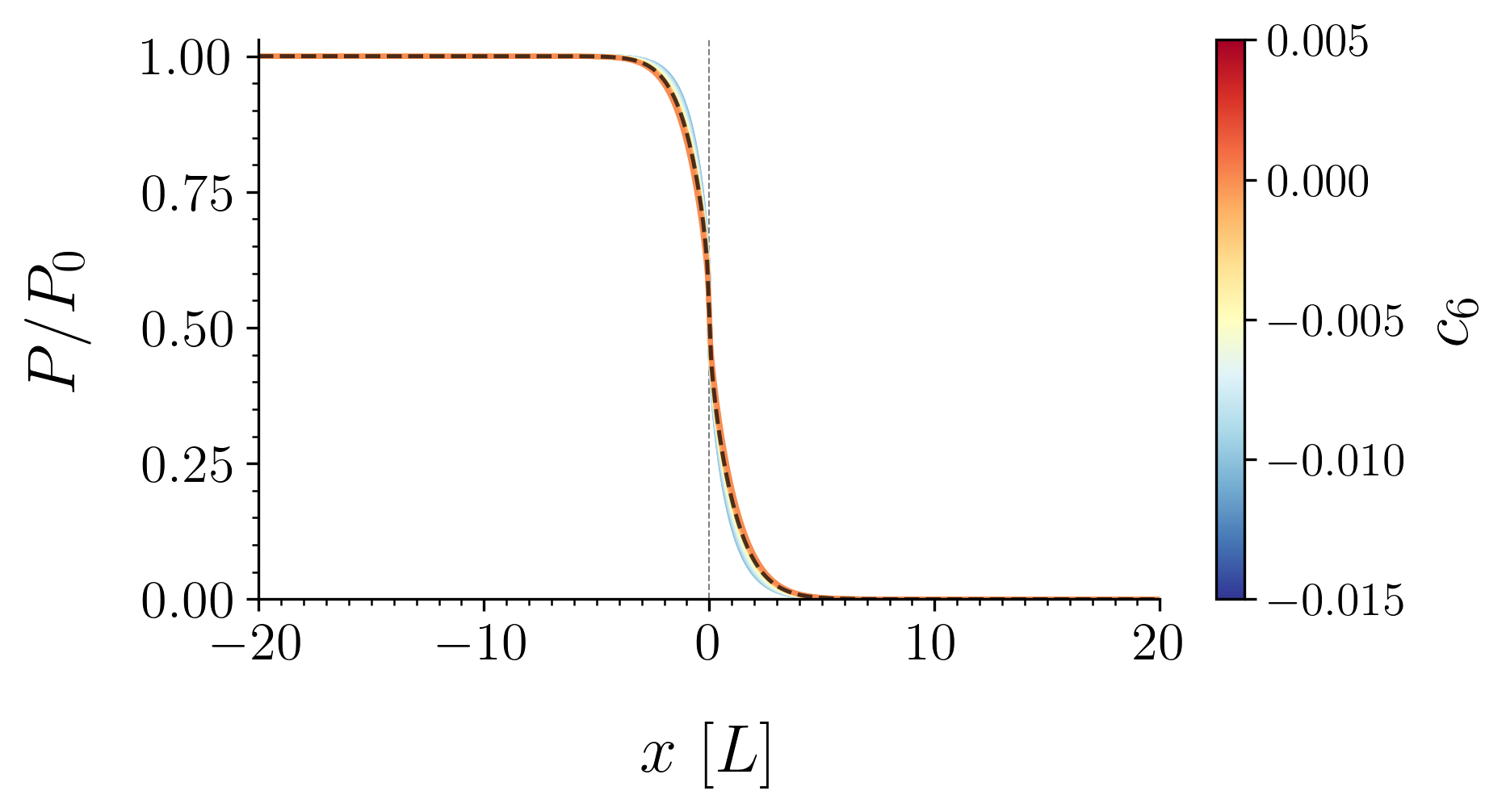}
         \caption{\footnotesize $s = {1}/{2}$ and $\ell= {3}/{2}$.}
        \label{fig:Pressuree2}  \end{subfigure}
     \caption{\footnotesize Relative pressure 
    as a function of the longitudinal direction of the de~Laval nozzle, varying $c_6$ for the quantum-corrected Schwarzschild metric~\eqref{ck}. The dashed black line represents the Schwarzschild solution. The values are calculated using $G_{{\scalebox{.55}{\textsc{N}}}} = M = 1$.}
    \label{fig:Pressure}
\end{figure}

\begin{figure}[H]
    \centering
    \begin{subfigure}{0.49\textwidth}
        \centering
        \includegraphics[width=\linewidth]{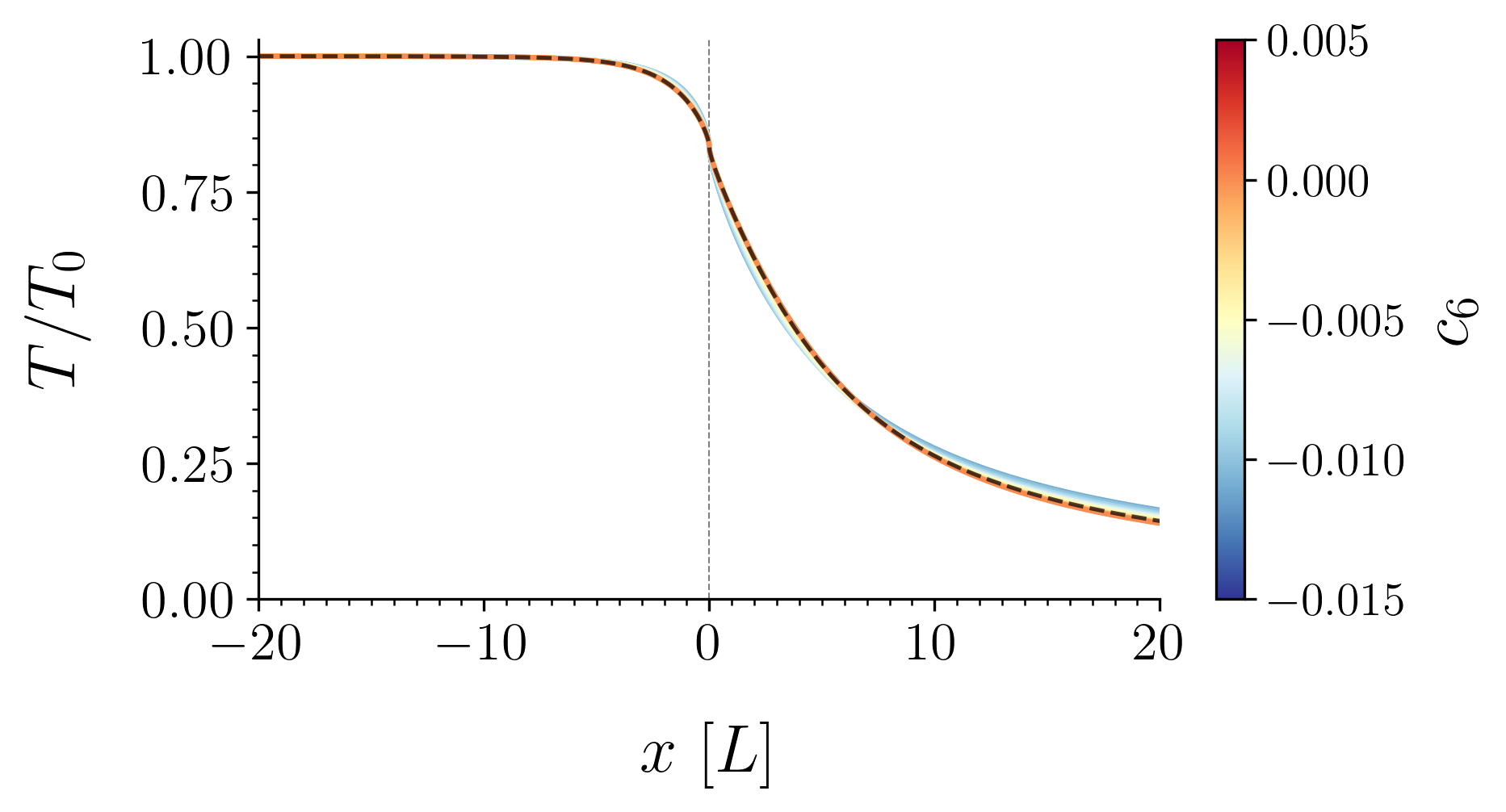}
         \caption{\footnotesize $s = \ell = 0$.}
        \label{fig:Temperaturea}
    \end{subfigure}
    \begin{subfigure}{0.49\textwidth}
        \centering
        \includegraphics[width=\linewidth]{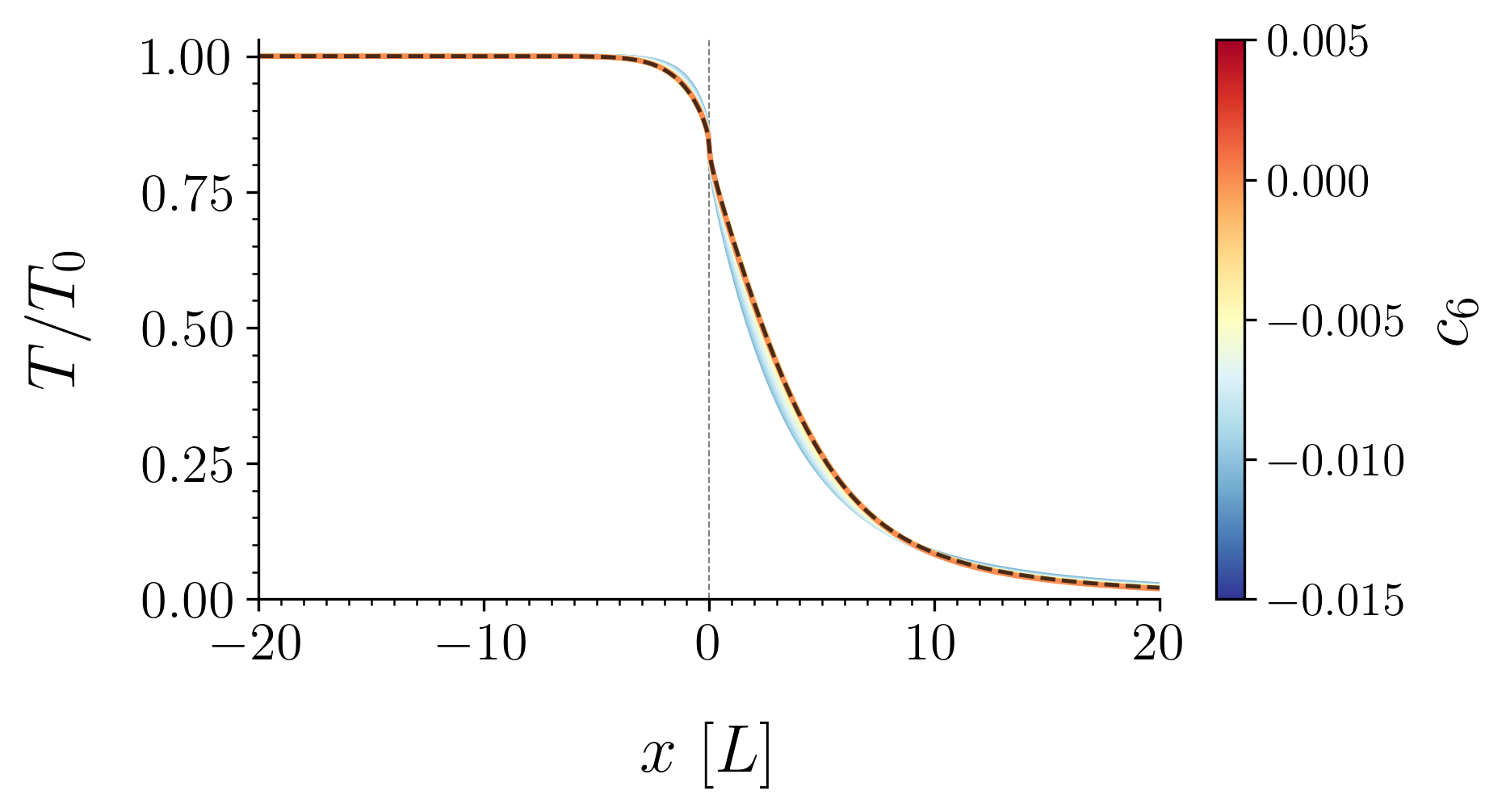}
         \caption{\footnotesize $s = 0$ and $\ell=1$.}
        \label{fig:Temperatureb1}  \end{subfigure}
        
    \begin{subfigure}{0.49\textwidth}
        \centering
        \includegraphics[width=\linewidth]{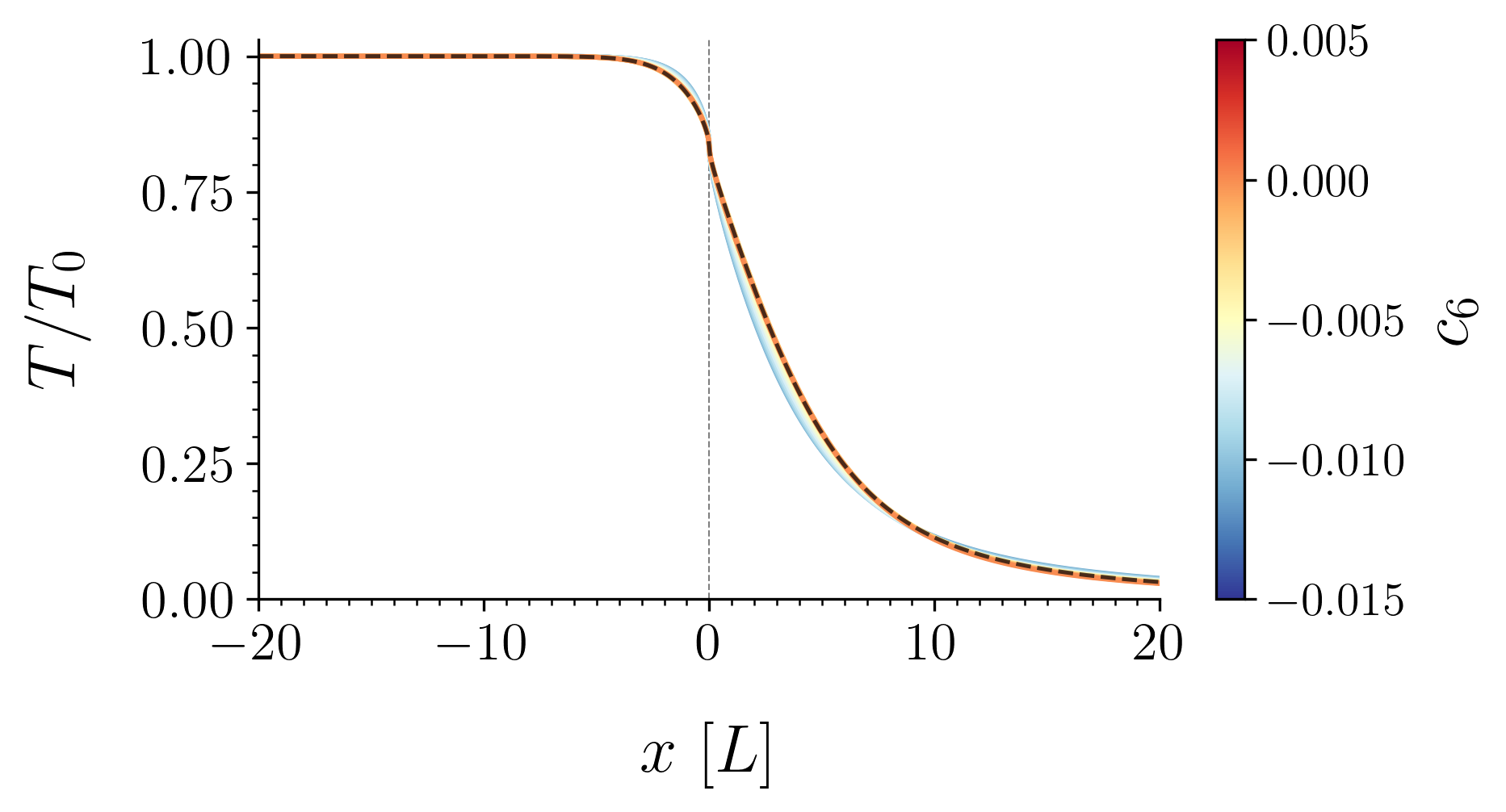}
         \caption{\footnotesize $s = \ell=1$.}
        \label{fig:Temperatureb2}
    \end{subfigure}
    \begin{subfigure}{0.49\textwidth}
        \centering
        \includegraphics[width=\linewidth]{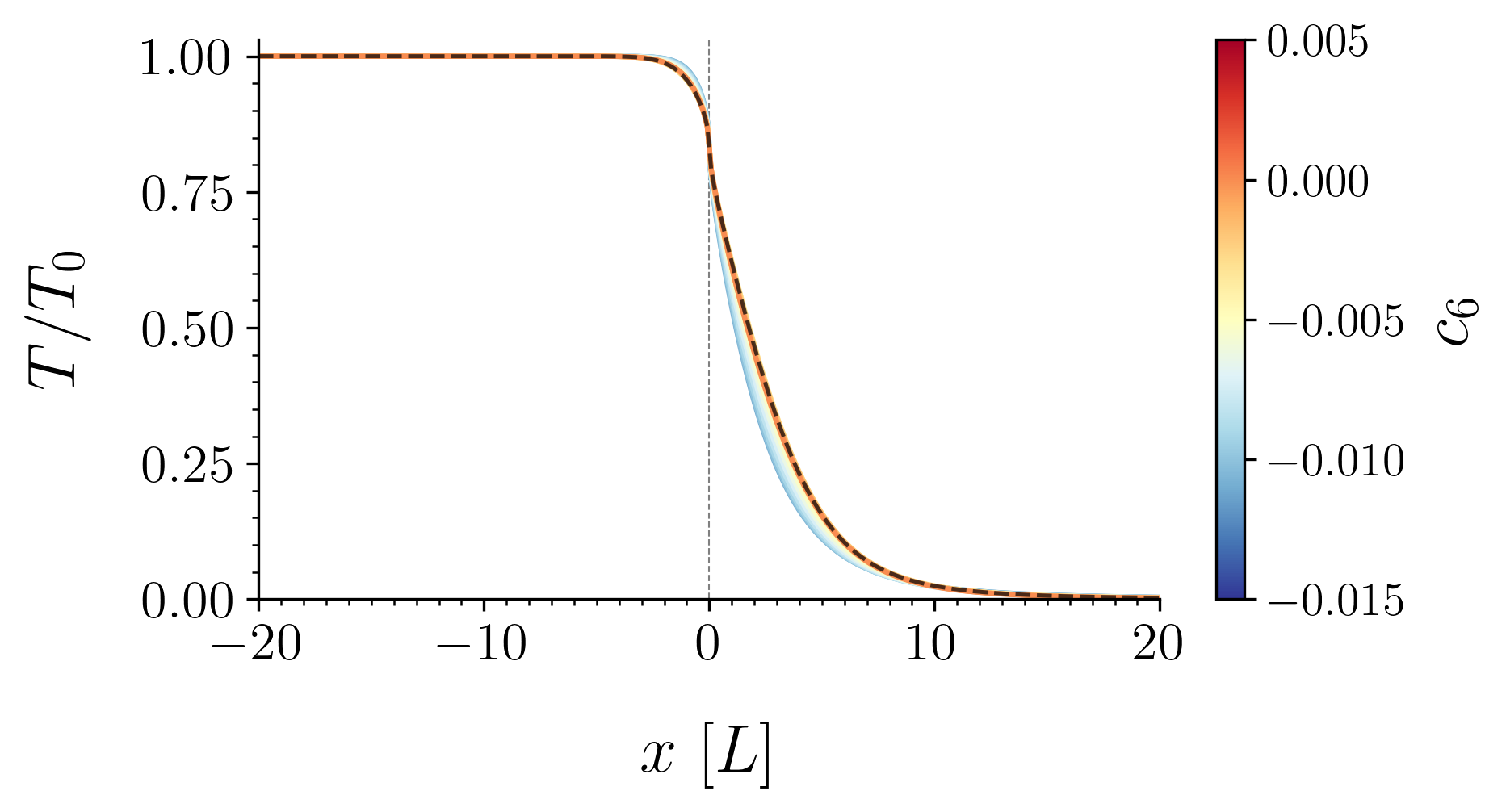}
         \caption{\footnotesize $s = 0$ and $\ell=2$.}
        \label{fig:Temperaturec1} 
    \end{subfigure}
    
    \begin{subfigure}{0.49\textwidth}
        \centering\includegraphics[width=\linewidth]{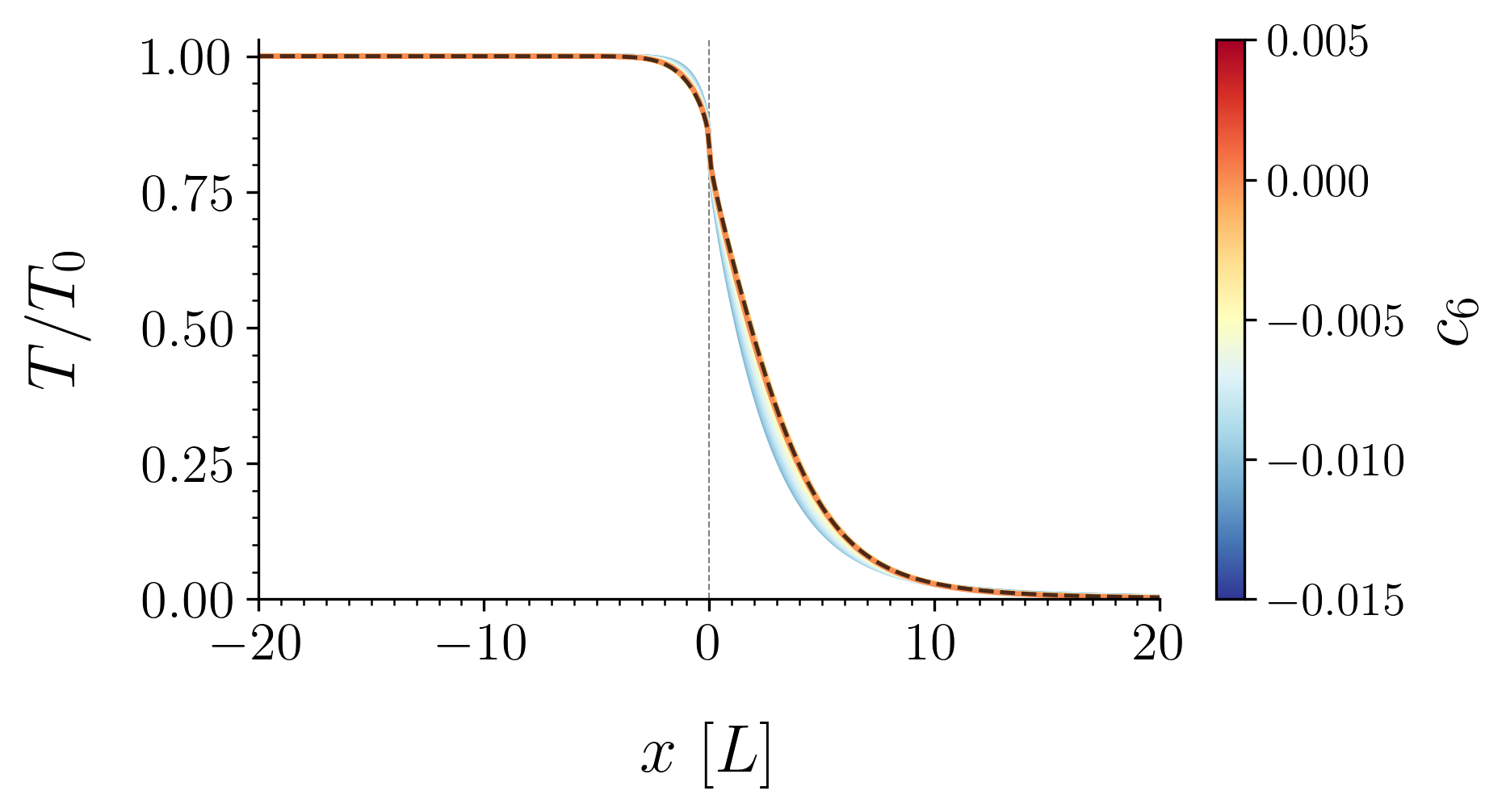}
         \caption{\footnotesize $s = 1$ and $\ell=2$.}
        \label{fig:Temperaturec2}
    \end{subfigure}
    \begin{subfigure}{0.49\textwidth}
        \centering\includegraphics[width=\linewidth]{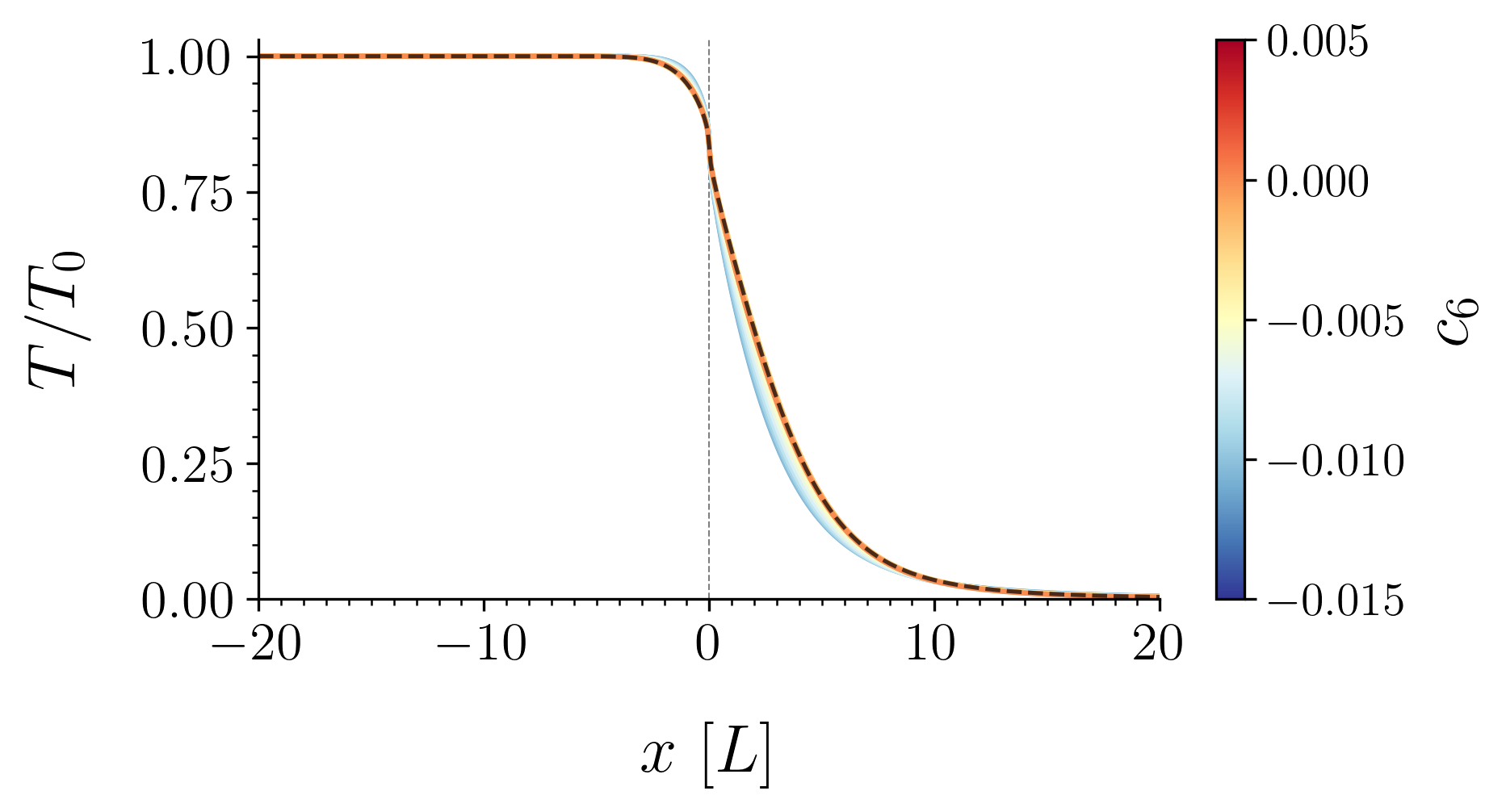}
         \caption{\footnotesize $s= \ell=2$.}
        \label{fig:Temperaturec3}
    \end{subfigure} 
    \begin{subfigure}{0.49\textwidth}
        \centering
        \includegraphics[width=\linewidth]{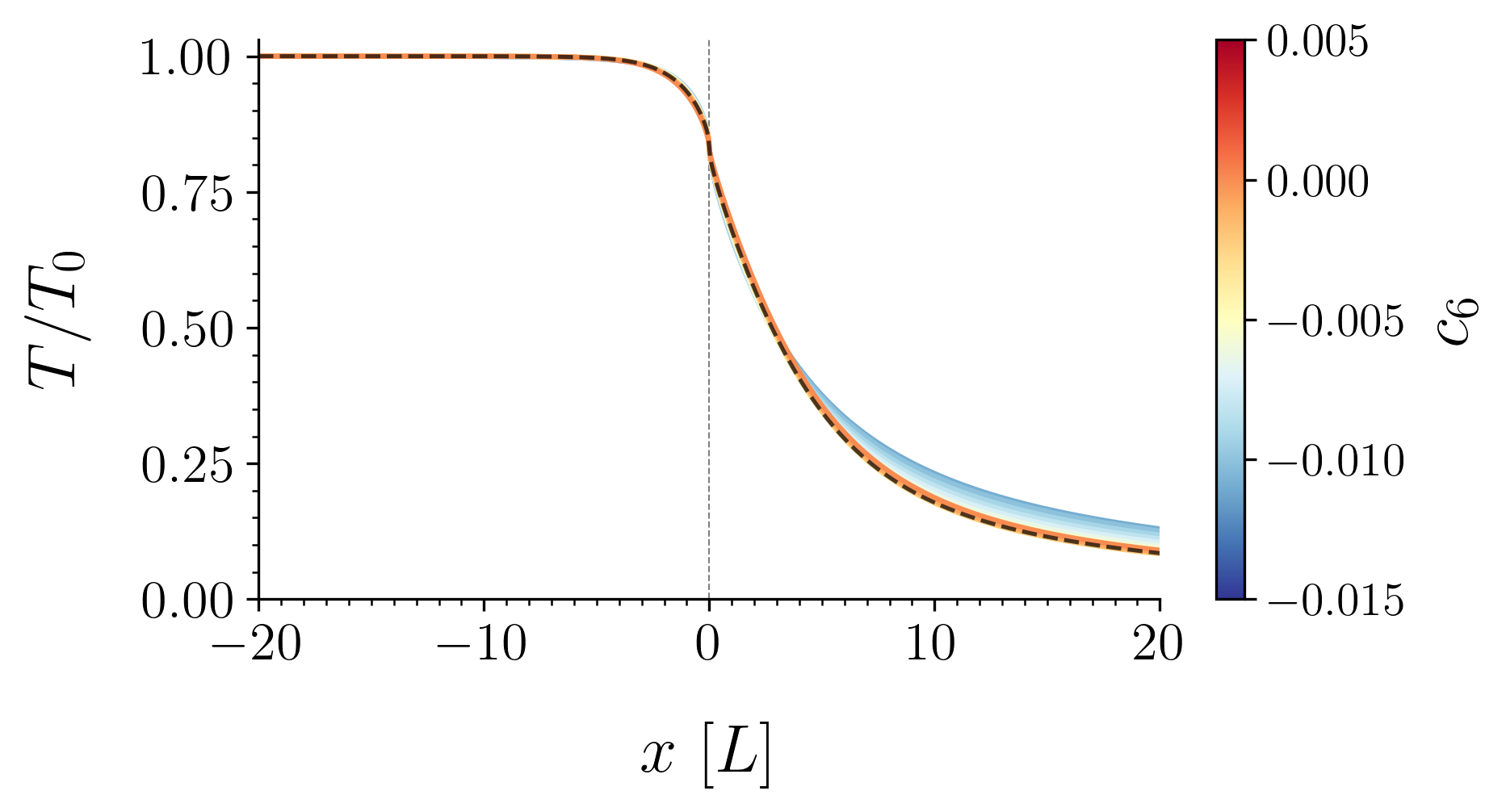}
         \caption{\footnotesize $s = \ell= {1}/{2}$.}
        \label{fig:Temperaturee1}
    \end{subfigure}
    \begin{subfigure}{0.49\textwidth}
        \centering
        \includegraphics[width=\linewidth]{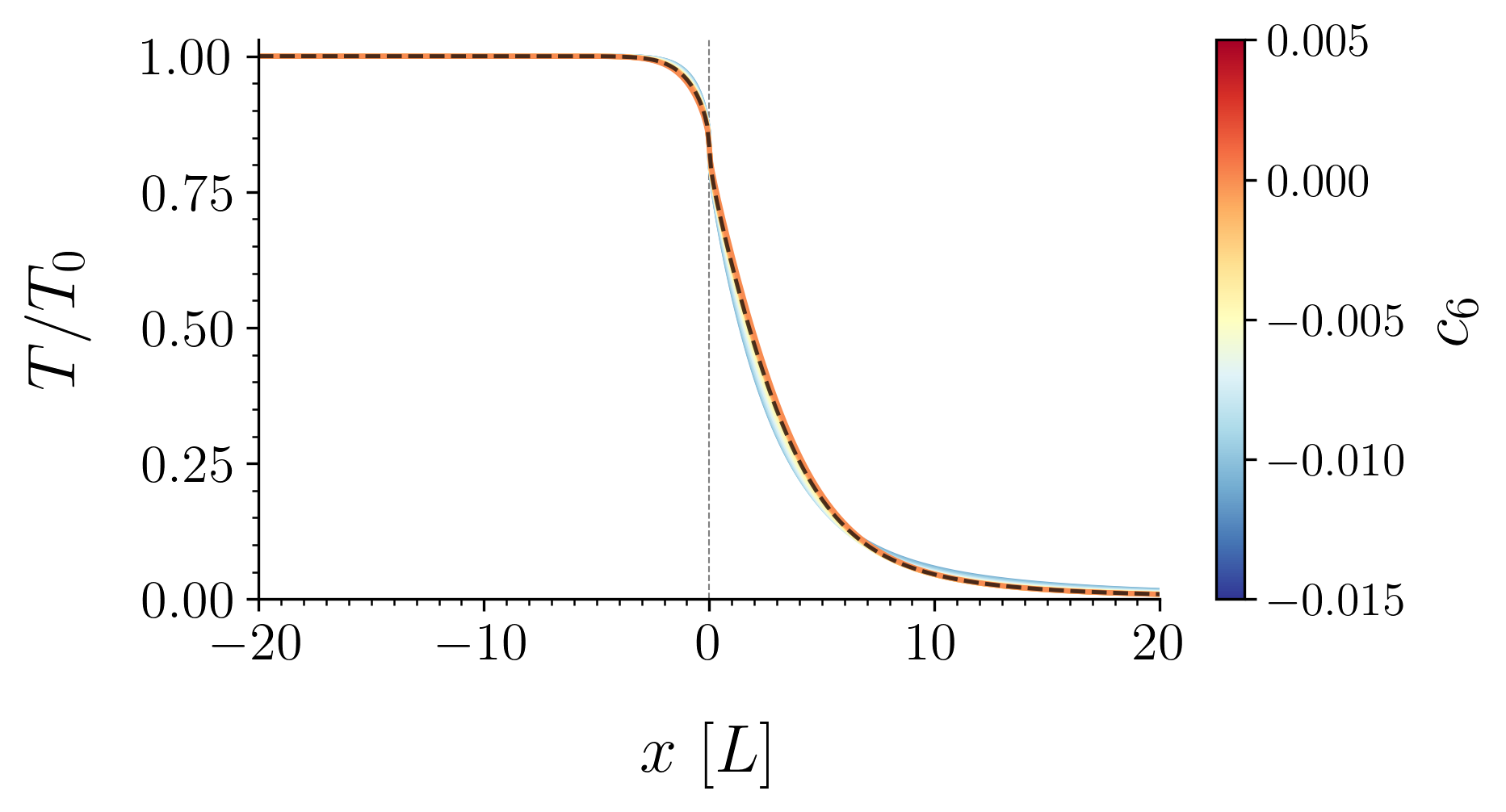}
         \caption{\footnotesize $s = {1}/{2}$ and $\ell= {3}/{2}$.}
        \label{fig:Temperaturee2}  \end{subfigure}
     \caption{\footnotesize Relative temperature 
    as a function of the longitudinal direction of the de~Laval nozzle, varying $c_6$ for the quantum-corrected Schwarzschild metric~\eqref{ck}. The dashed black line represents the Schwarzschild solution. The values are calculated using $G_{{\scalebox{.55}{\textsc{N}}}} = M = 1$.}
    \label{fig:Temperature}
\end{figure}

\begin{figure}[H]
    \centering
    \begin{subfigure}{0.49\textwidth}
        \centering
        \includegraphics[width=7.35cm]{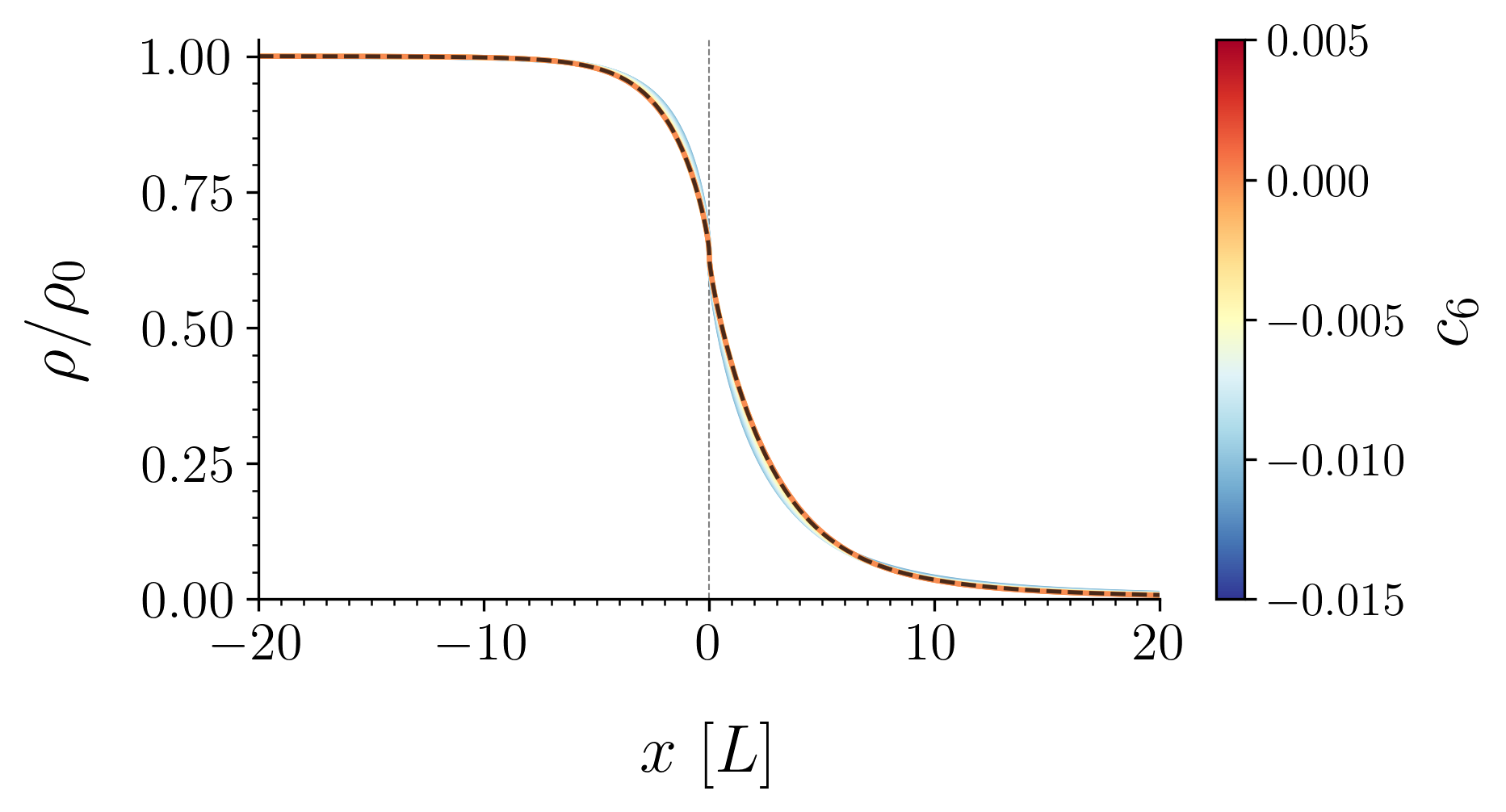}
         \caption{\footnotesize $s = \ell = 0$.}
        \label{fig:Densitya}
    \end{subfigure}
    \begin{subfigure}{0.49\textwidth}
        \centering
        \includegraphics[width=7.35cm]{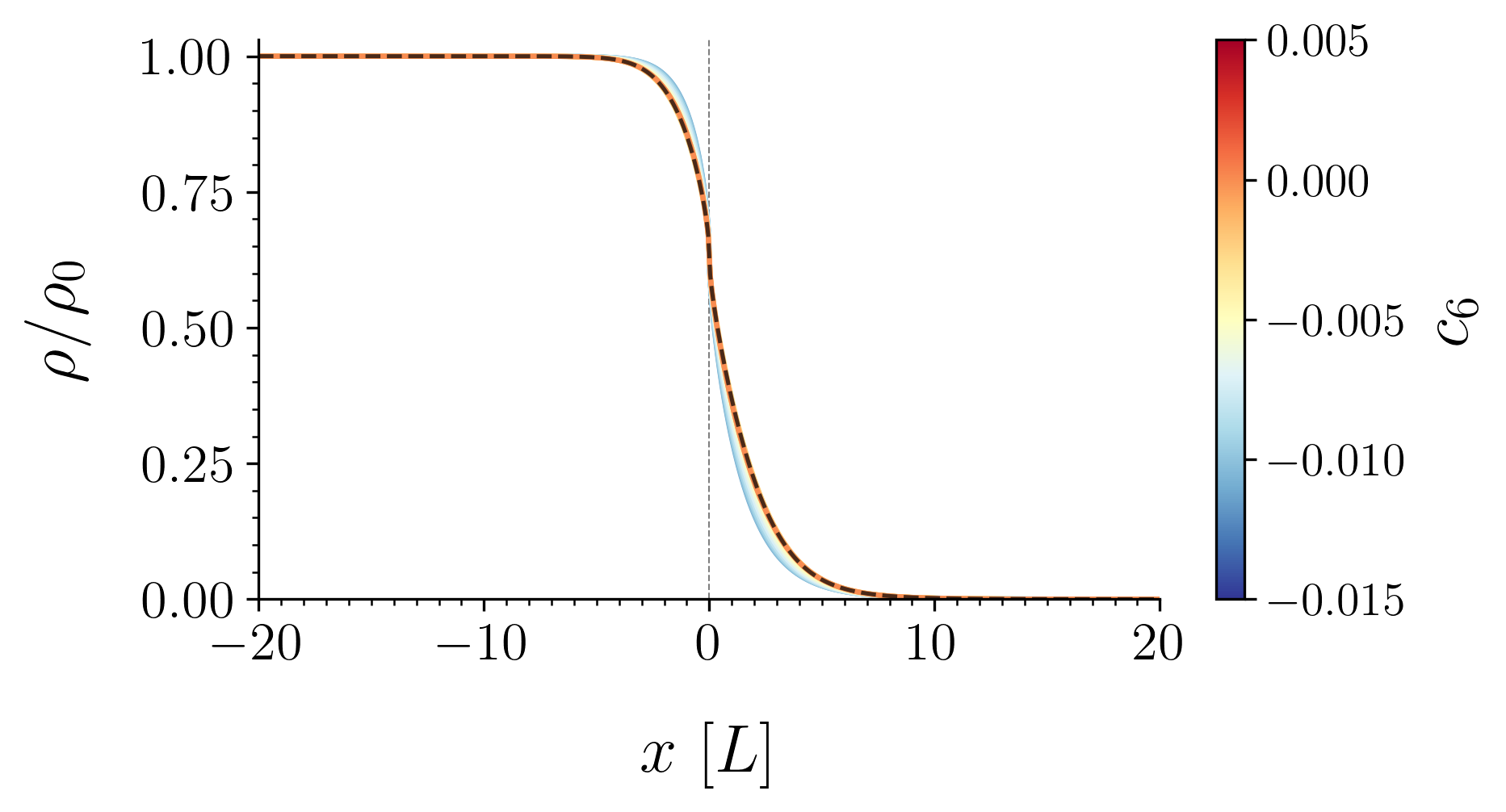}
         \caption{\footnotesize $s = 0$ and $\ell=1$.}
        \label{fig:Densityb1}  \end{subfigure}
        
    \begin{subfigure}{0.49\textwidth}
        \centering
        \includegraphics[width=7.35cm]{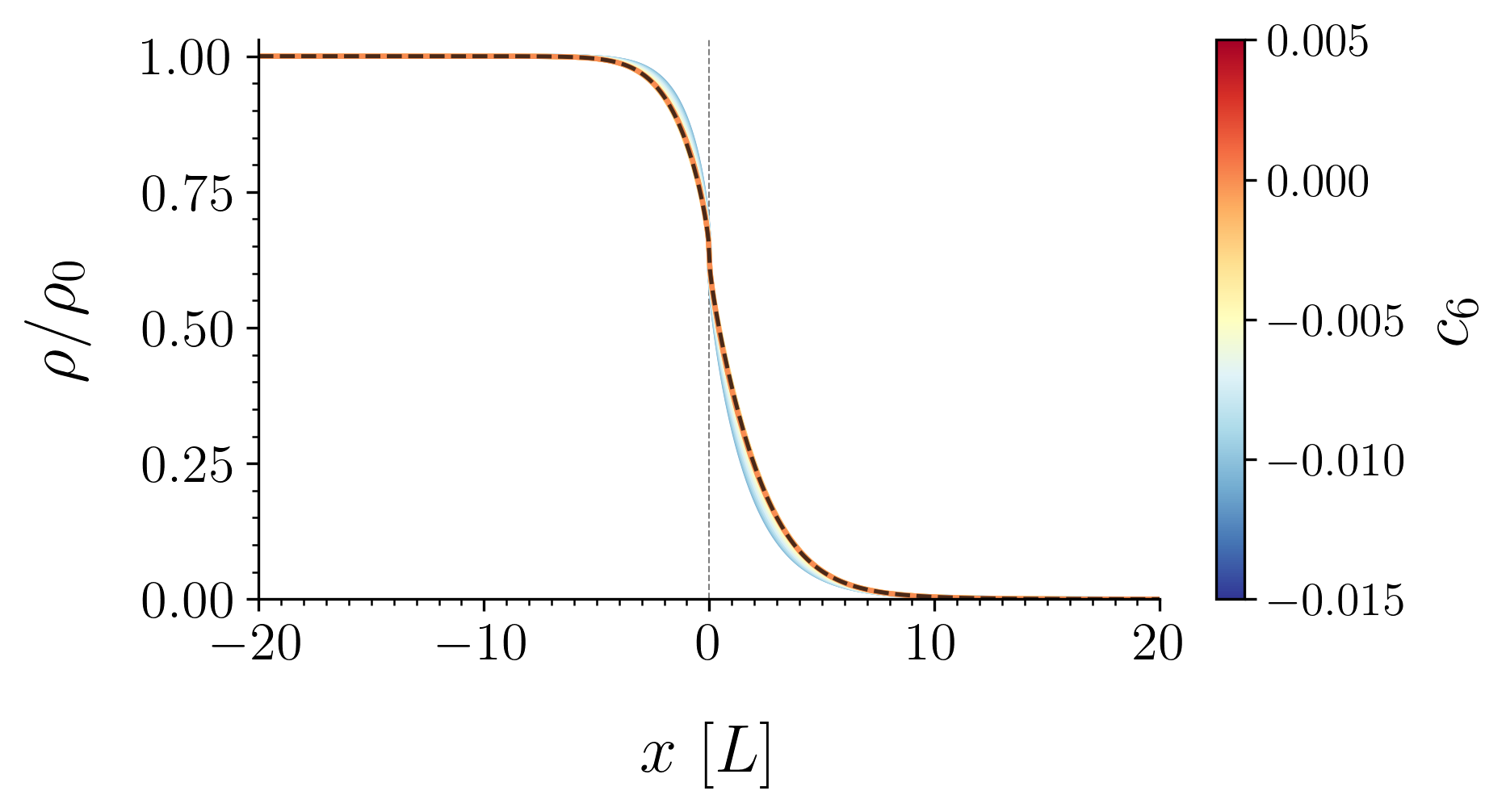}
         \caption{\footnotesize $s = \ell=1$.}
        \label{fig:Densityb2}
    \end{subfigure}
    \begin{subfigure}{0.49\textwidth}
        \centering
        \includegraphics[width=7.35cm]{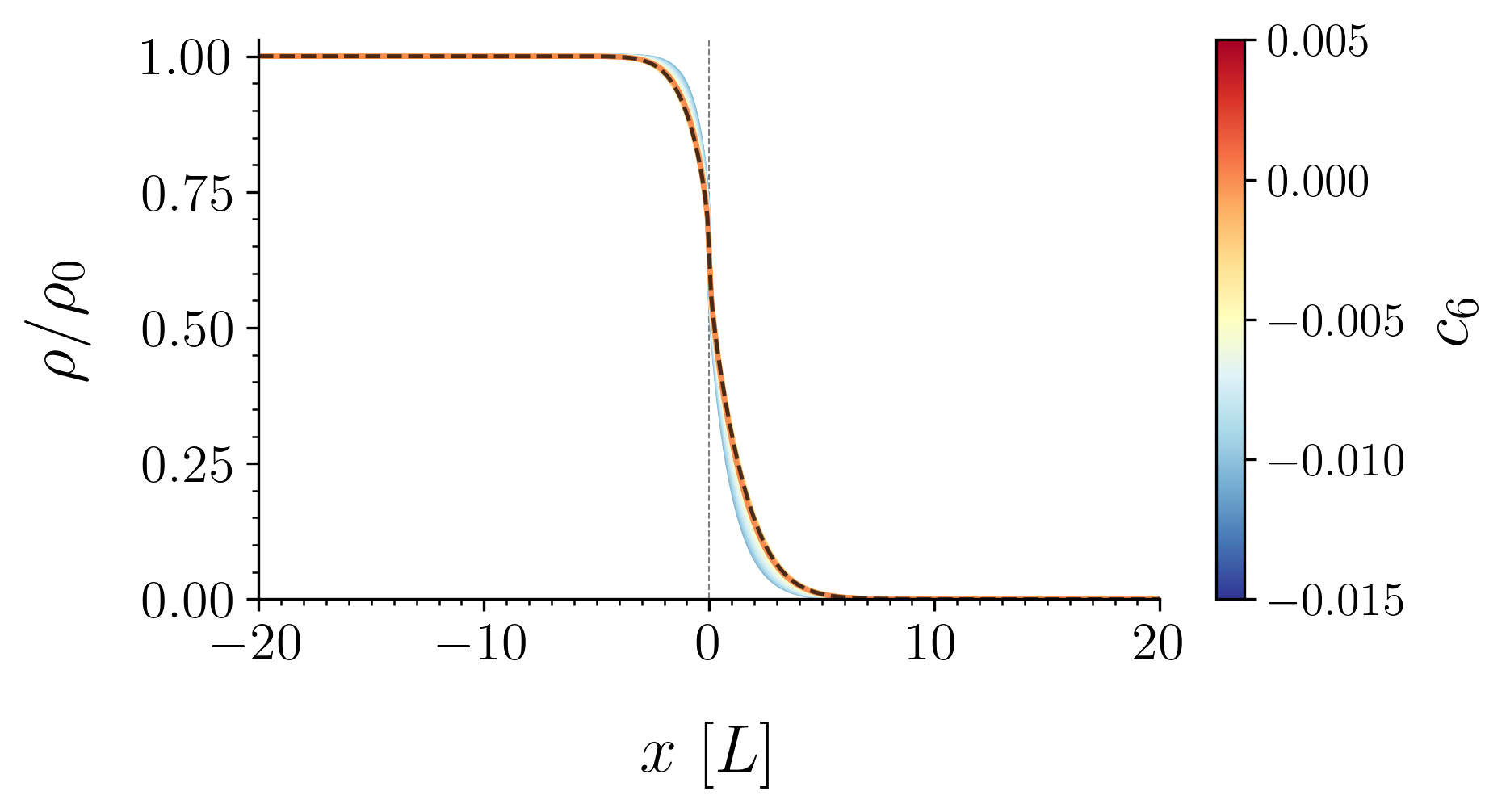}
         \caption{\footnotesize $s = 0$ and $\ell=2$.}
        \label{fig:Densityc1} 
    \end{subfigure}
    
    \begin{subfigure}{0.49\textwidth}
        \centering\includegraphics[width=7.35cm]{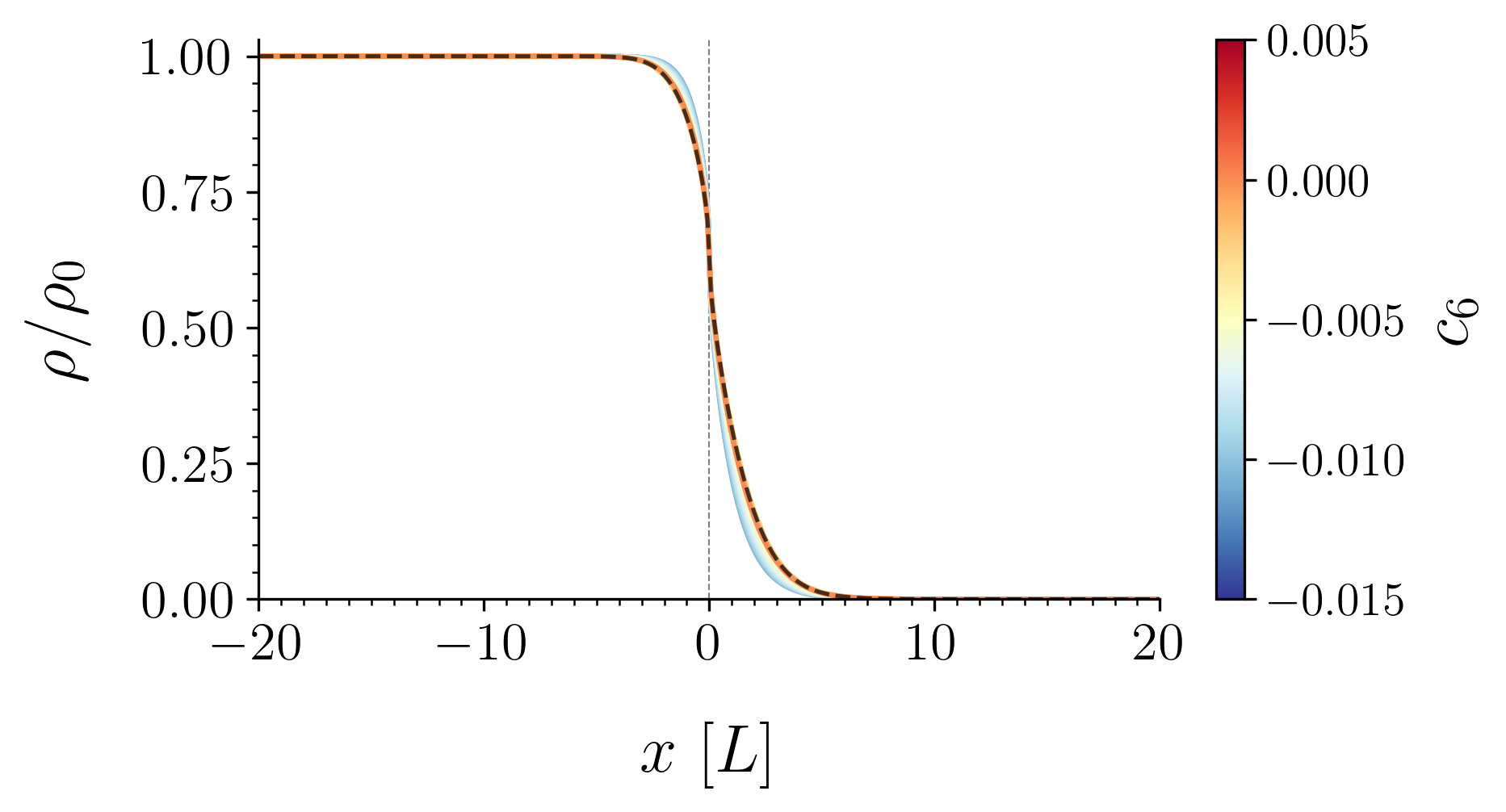}
         \caption{\footnotesize $s = 1$ and $\ell=2$.}
        \label{fig:Densityc2}
    \end{subfigure}
    \begin{subfigure}{0.49\textwidth}
        \centering\includegraphics[width=7.35cm]{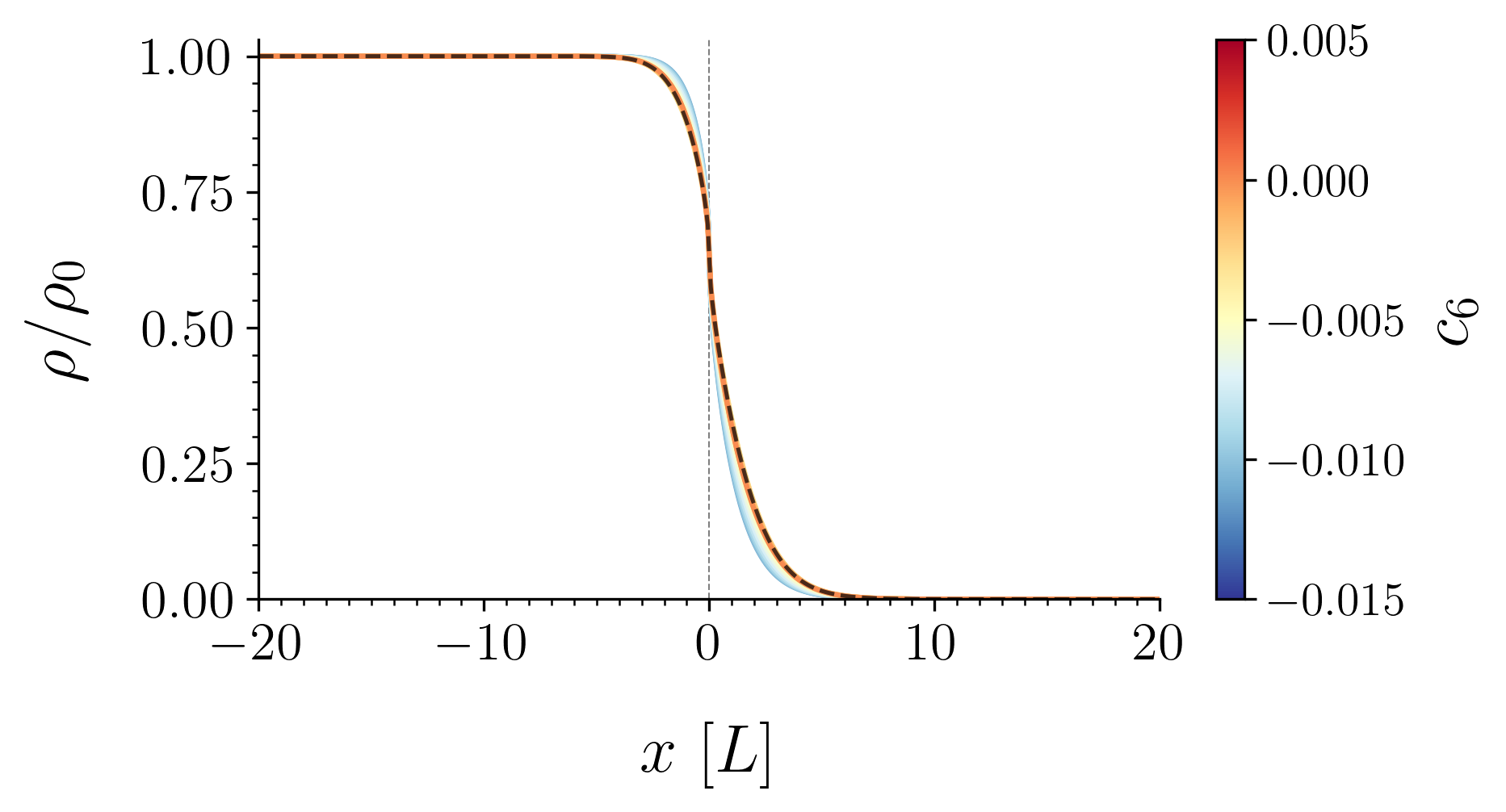}
         \caption{\footnotesize $s= \ell=2$.}
        \label{fig:Densityc3}
    \end{subfigure} 
    \begin{subfigure}{0.49\textwidth}
        \centering
        \includegraphics[width=7.35cm]{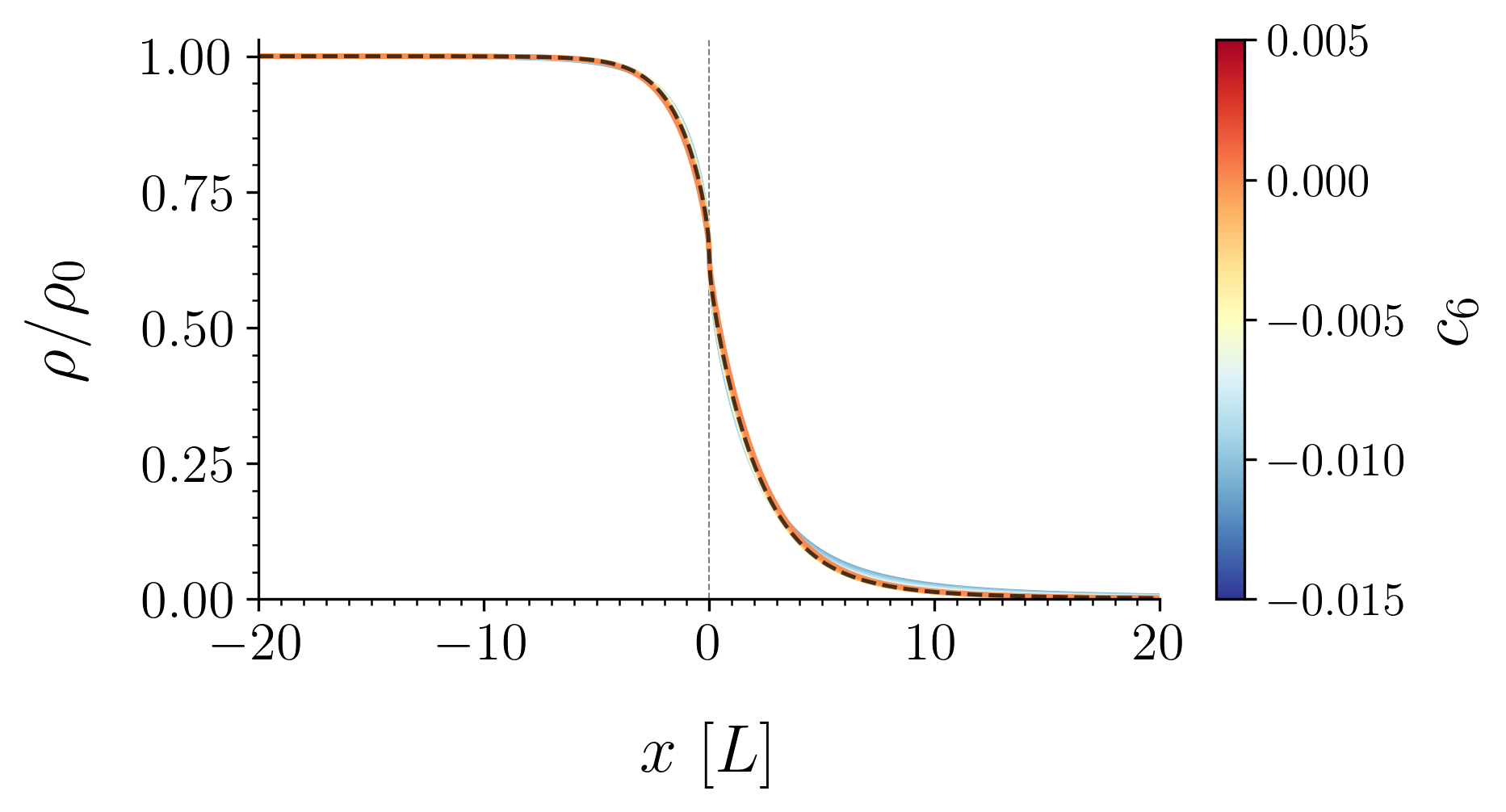}
         \caption{\footnotesize $s = \ell= {1}/{2}$.}
        \label{fig:Densitye1}
    \end{subfigure}
    \begin{subfigure}{0.49\textwidth}
        \centering
        \includegraphics[width=7.35cm]{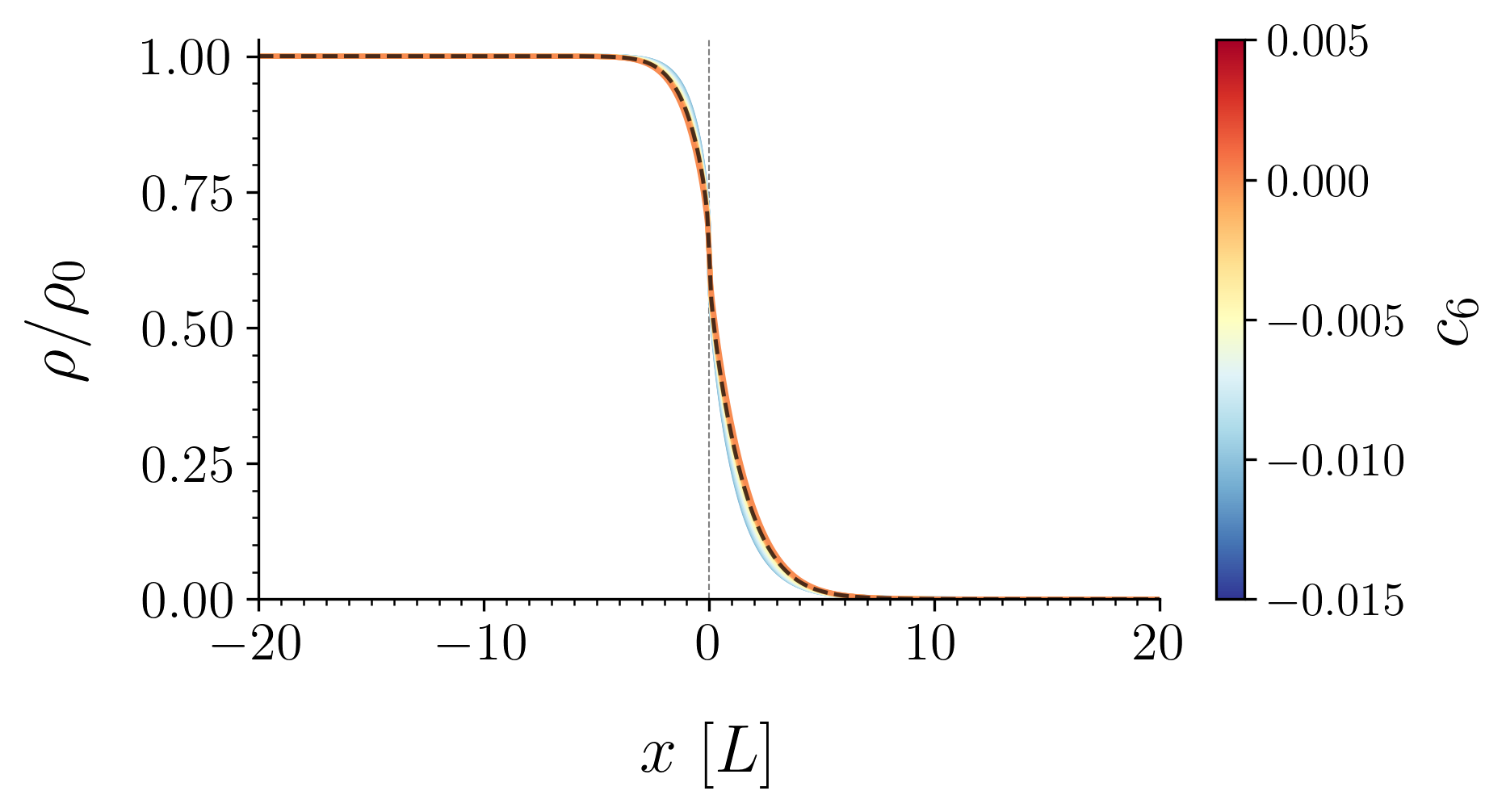}
         \caption{\footnotesize $s = {1}/{2}$ and $\ell= {3}/{2}$.}
        \label{fig:Densitye2}  \end{subfigure}
     \caption{\footnotesize Relative density 
    as a function of the longitudinal direction of the de~Laval nozzle, varying $c_6$ for the quantum-corrected Schwarzschild metric \eqref{eq:calmet-kuipers-metric}. The dashed black line represents the  Schwarzschild solution. The values are calculated using $G_{{\scalebox{.55}{\textsc{N}}}} = M = 1$.}
    \label{fig:Density}
\end{figure}

\begin{figure}[H]
    \centering
    \begin{subfigure}{0.49\textwidth}
        \centering
        \includegraphics[width=\linewidth]{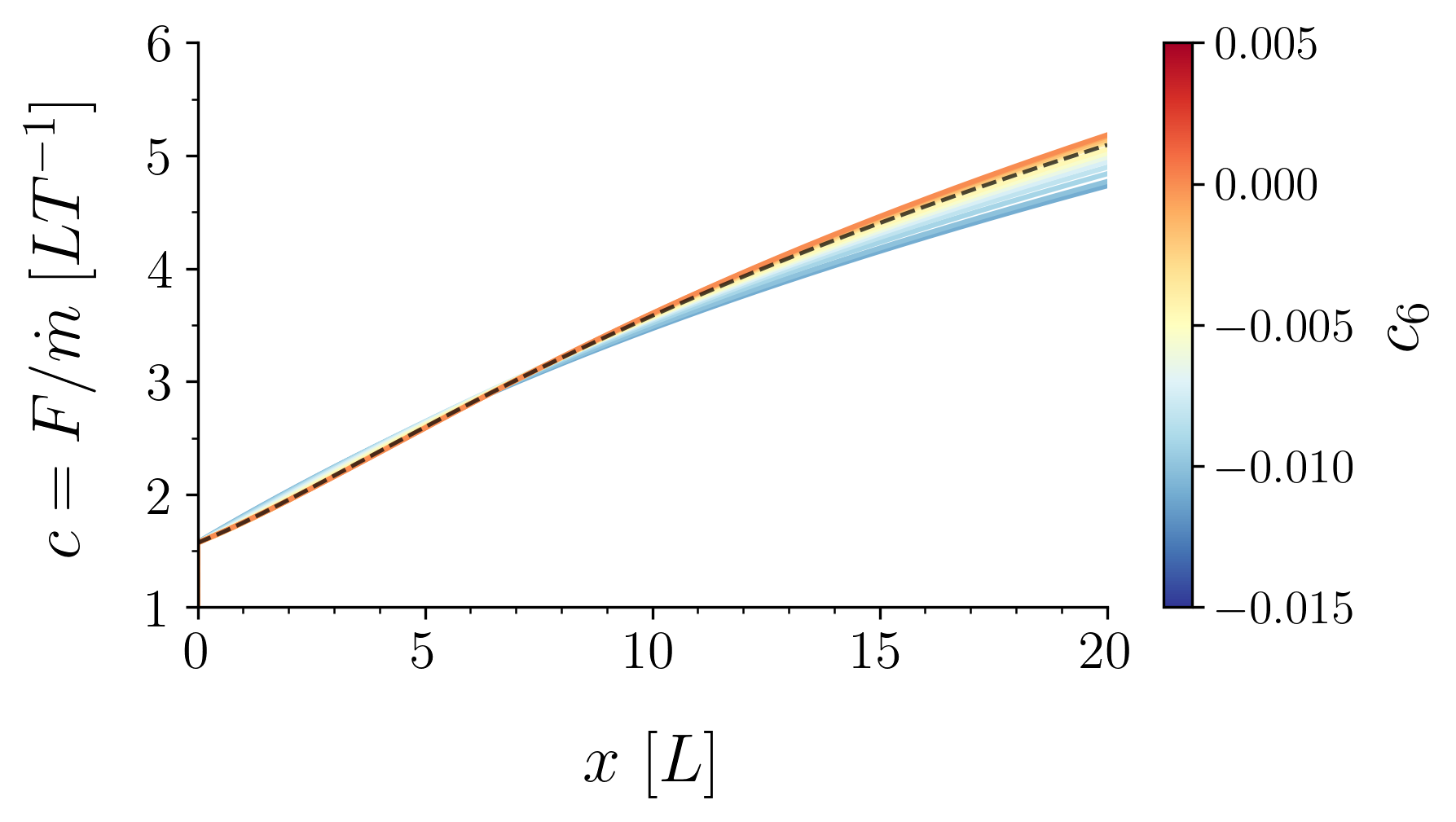}
         \caption{\footnotesize $s = \ell = 0$.}
        \label{fig:thrusta}
    \end{subfigure}
    \begin{subfigure}{0.49\textwidth}
        \centering
        \includegraphics[width=\linewidth]{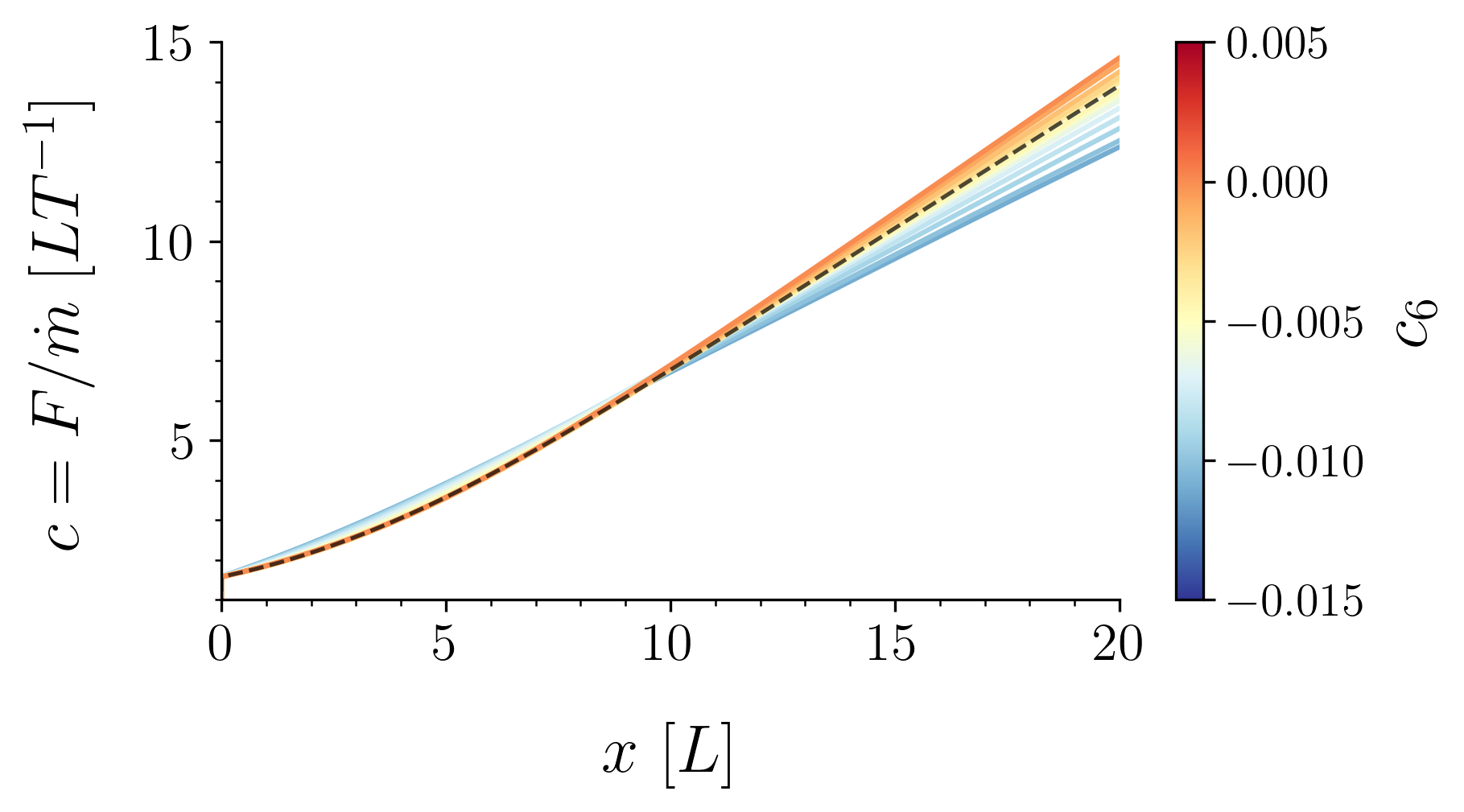}
         \caption{\footnotesize $s = 0$ and $\ell=1$.}
        \label{fig:thrustb1}  \end{subfigure}
        
    \begin{subfigure}{0.49\textwidth}
        \centering
        \includegraphics[width=\linewidth]{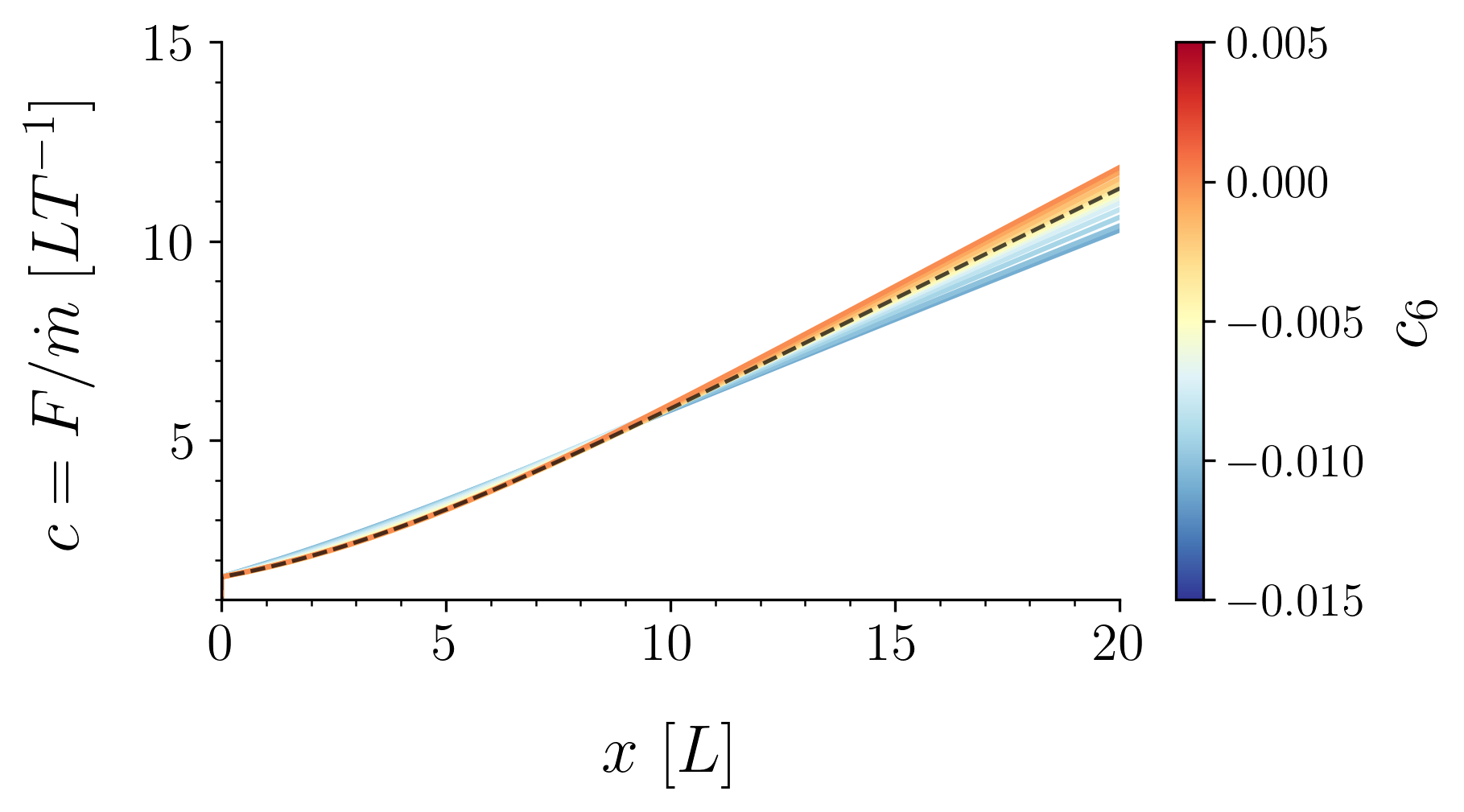}
         \caption{\footnotesize $s = \ell=1$.}
        \label{fig:thrustb2}
    \end{subfigure}
    \begin{subfigure}{0.49\textwidth}
        \centering
        \includegraphics[width=\linewidth]{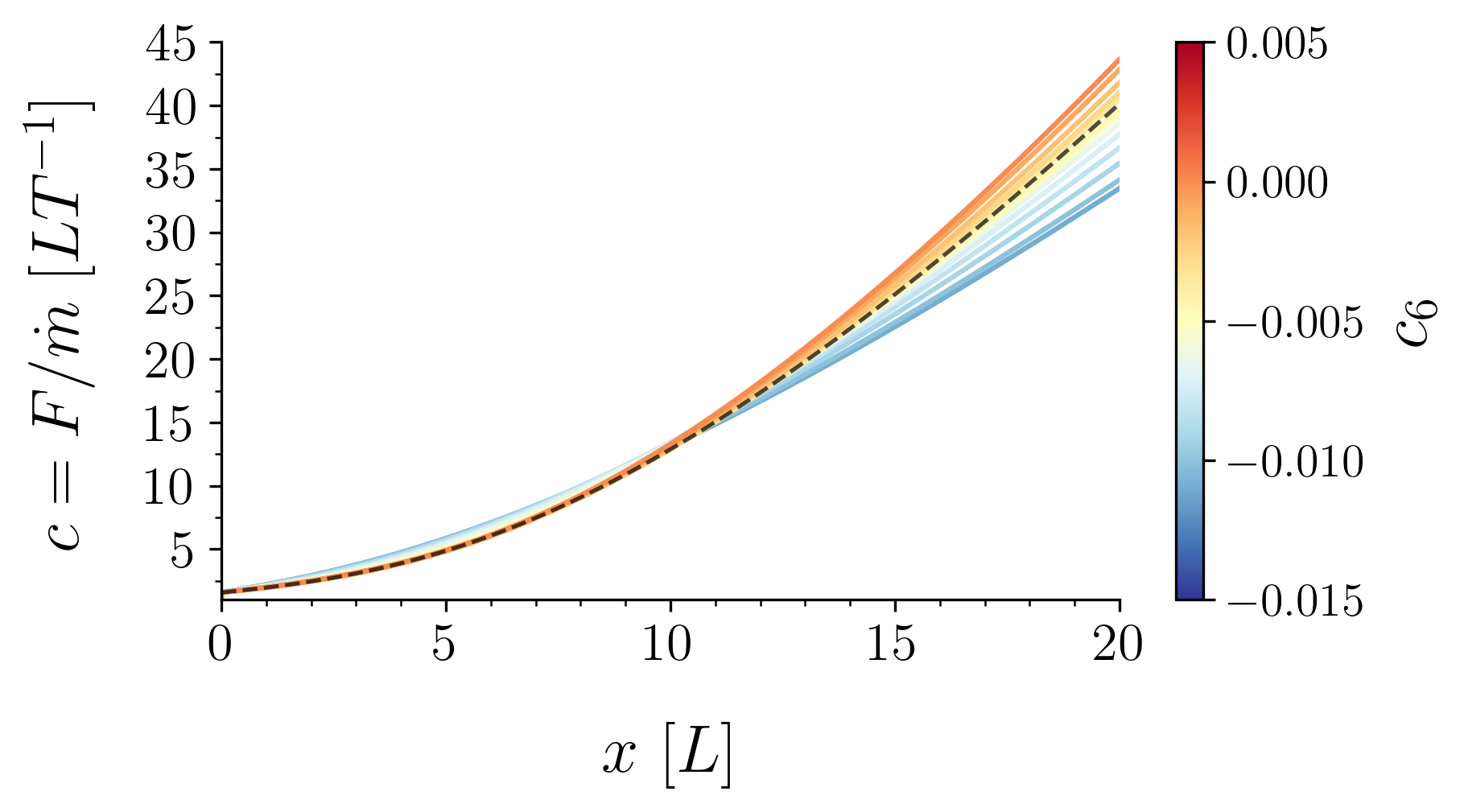}
         \caption{\footnotesize $s = 0$ and $\ell=2$.}
        \label{fig:thrustc1} 
    \end{subfigure}
    
    \begin{subfigure}{0.49\textwidth}
        \centering\includegraphics[width=\linewidth]{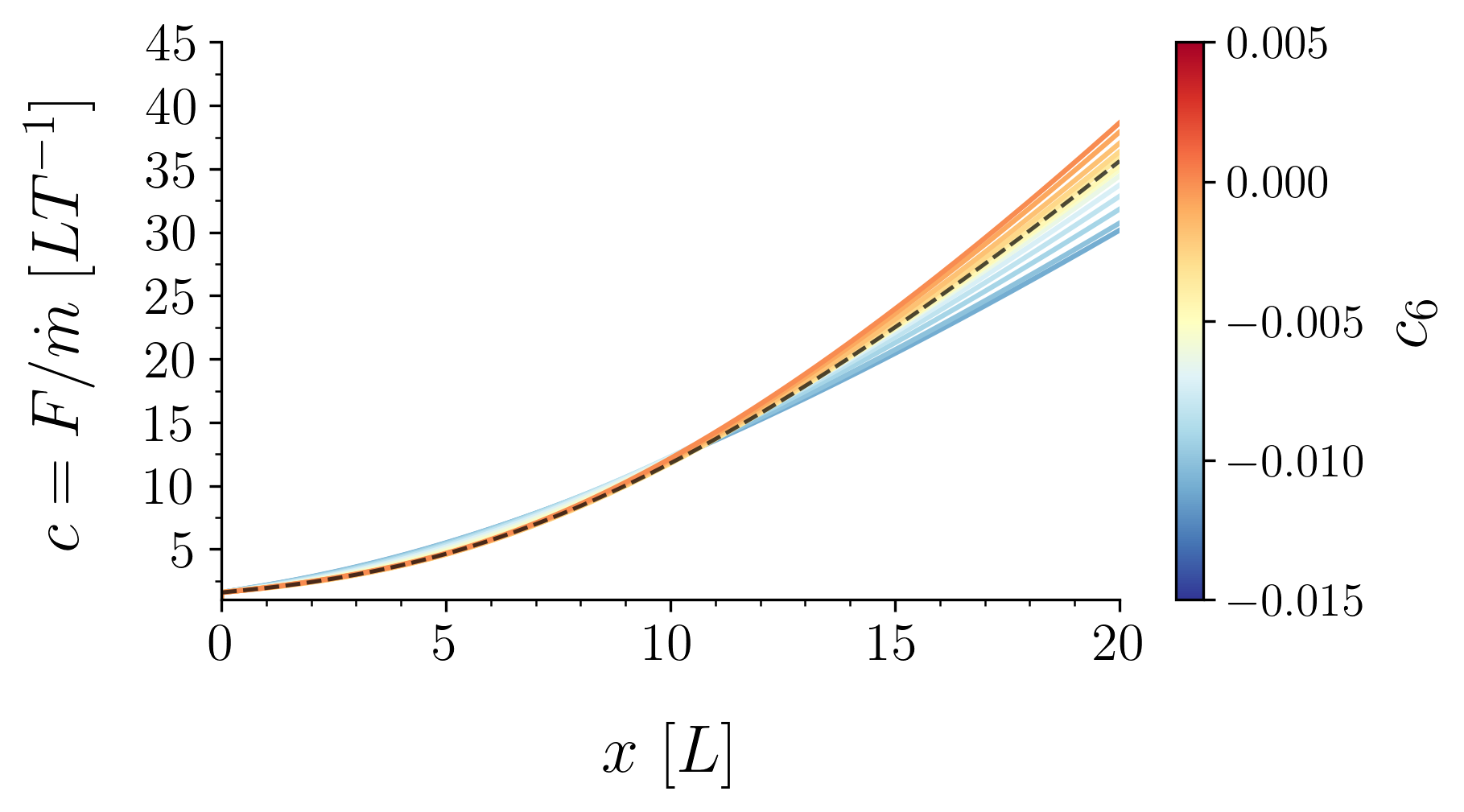}
         \caption{\footnotesize $s = 1$ and $\ell=2$.}
        \label{fig:thrustc2}
    \end{subfigure}
    \begin{subfigure}{0.49\textwidth}
        \centering\includegraphics[width=\linewidth]{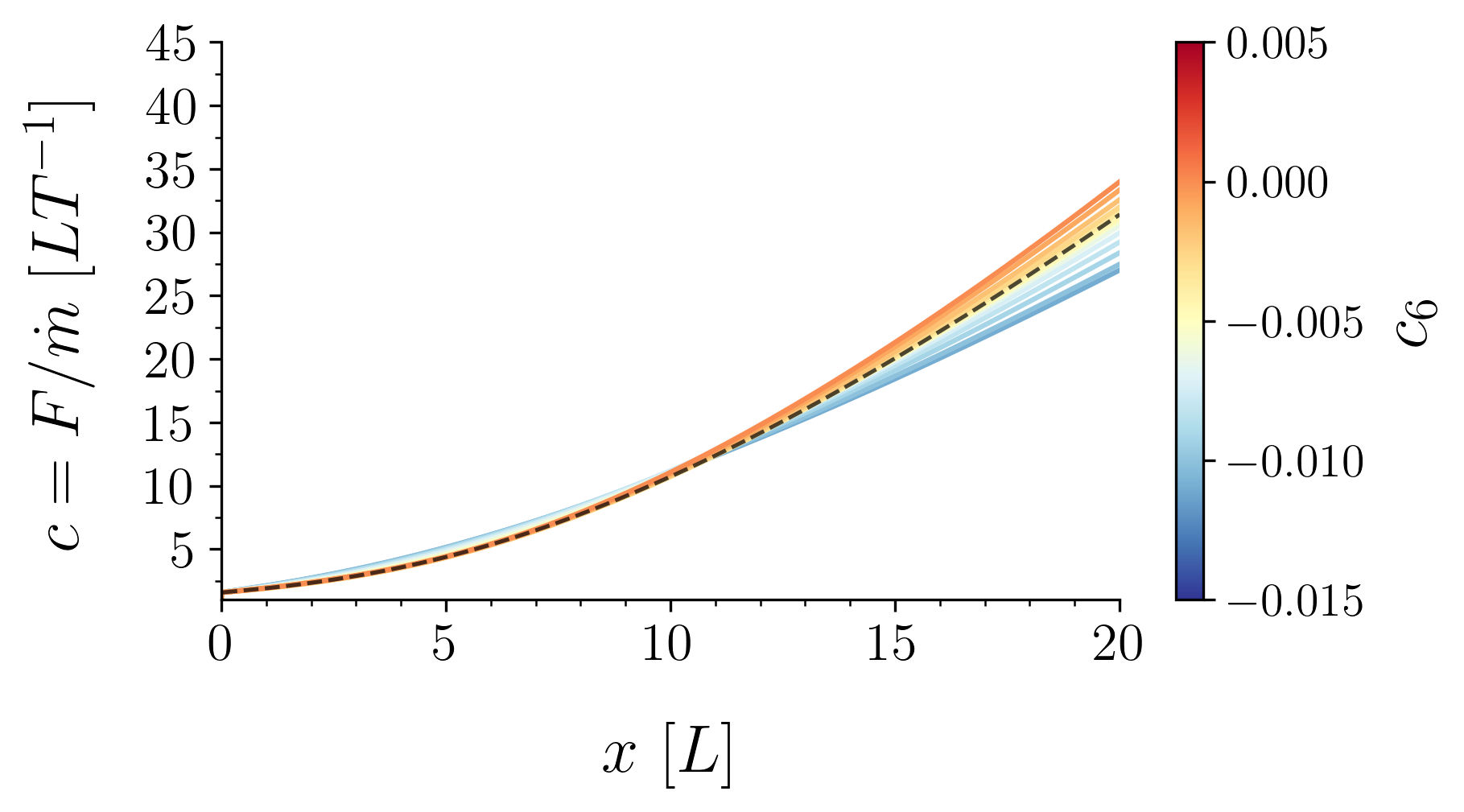}
         \caption{\footnotesize $s= \ell=2$.}
        \label{fig:thrustc3}
    \end{subfigure} 
    \begin{subfigure}{0.49\textwidth}
        \centering
        \includegraphics[width=\linewidth]{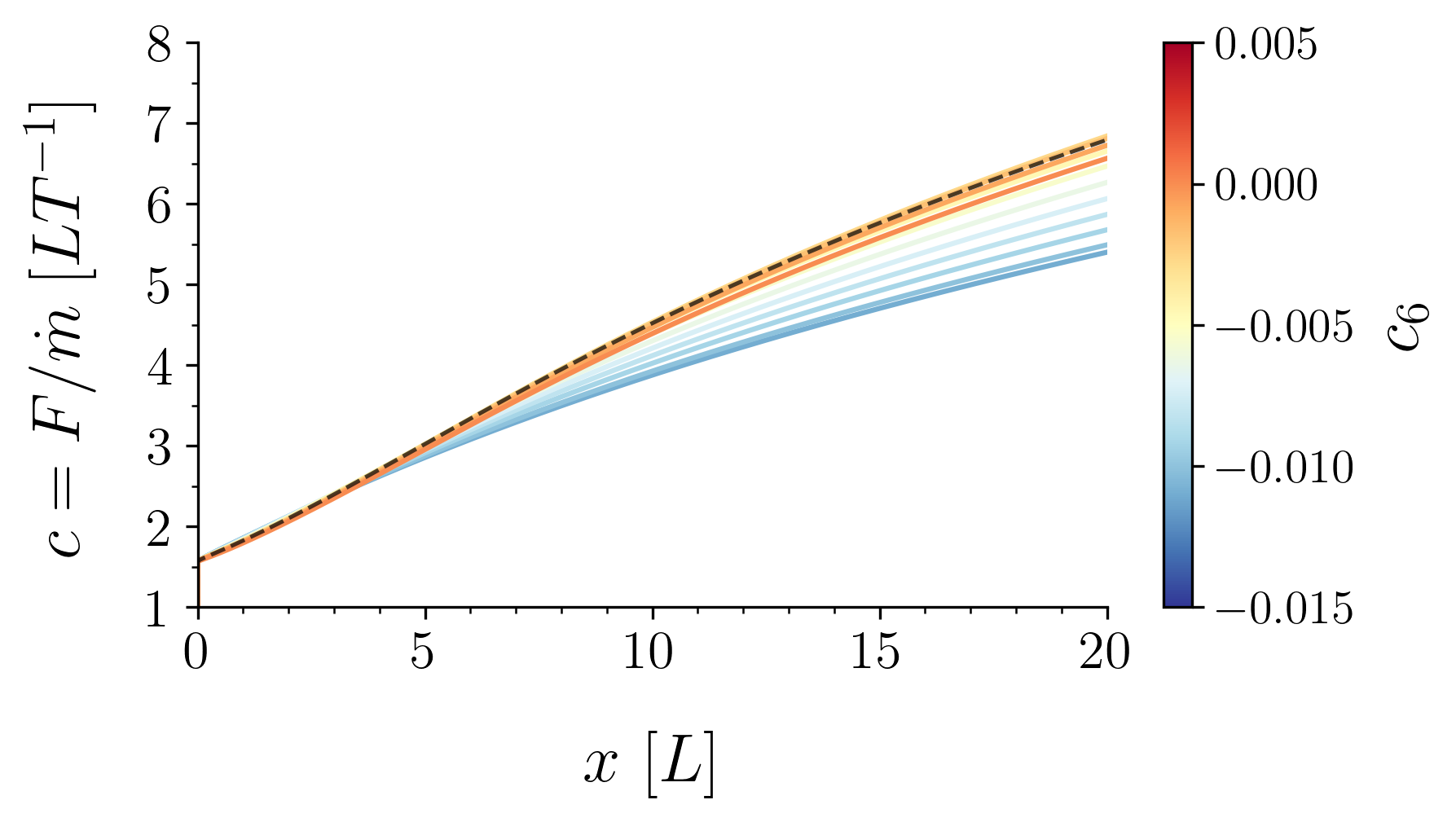}
         \caption{\footnotesize $s = \ell= {1}/{2}$.}
        \label{fig:thruste1}
    \end{subfigure}
    \begin{subfigure}{0.49\textwidth}
        \centering
        \includegraphics[width=\linewidth]{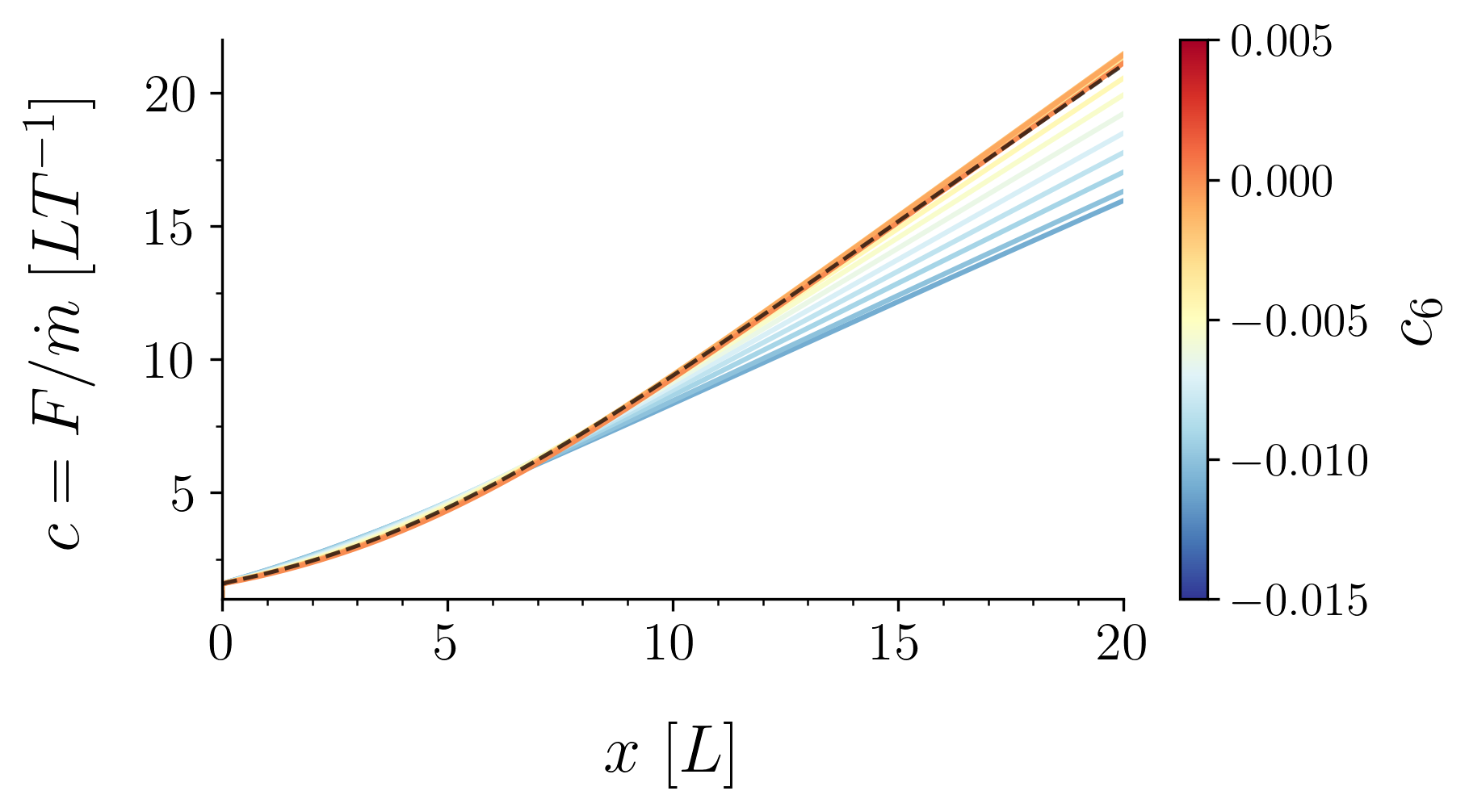}
         \caption{\footnotesize $s = {1}/{2}$ and $\ell= {3}/{2}$.}
        \label{fig:thruste2}  \end{subfigure}
     \caption{\footnotesize Effective exhaust velocity as a function of the longitudinal direction of the de~Laval nozzle, varying $c_6$ for the quantum-corrected Schwarzschild metric~\eqref{ck}. The dashed black line represents the Schwarzschild solution. The values are calculated using $G_{{\scalebox{.55}{\textsc{N}}}} = M = 1$.}
    \label{fig:thrust}
\end{figure}

The thermodynamic variables, the nozzle geometry, the Mach number, and the thrust coefficient are computed as functions of $c_6$.  Calculations of QN modes incorporate spin and multipole values, which are critical for overtone analysis and for the calculation of the quality factor $q_n \sim \Re\qty(\omega)/\Im\qty(\omega)$. The QN mode spectrum is computed for various values of $c_6$.
Eqs.~\eqref{eq:calmet-kuipers-metric}, \eqref{final effective potential int} and \eqref{eq:eh-ck} reveal that solutions with $c_6 \geq G_{{\scalebox{.55}{\textsc{N}}}}^2 M^4/5\pi$ yield non-physical solutions, as they either do not provide a real event horizon radius $ r_{{\scalebox{.7}{\textsc{h}}}} $ or the effective potentials $ V_{{\scalebox{.65}{{\textsc{eff}}}}} $ associated with them are not well behaved. Consequently, all subsequent results in Figs.~\ref{fig:veff}--\ref{fig:thrust} are generated for fixed values of $ s $ and $ \ell $, while $ c_6 $ is varied systematically. The nozzle throat center at $ x = 0 $ corresponds to the analog event horizon in these configurations.
To simplify the analysis, from now on we consider natural units with unitary BH mass, $G_{{\scalebox{.55}{\textsc{N}}}} = M = 1$. 

Fig.~\ref{fig:veff} displays the effective potential $ V_{{\scalebox{.65}{{\textsc{eff}}}}} $ as a function of the longitudinal coordinate in the de~Laval nozzle, evaluated for distinct values of the parameter $ c_6 $ driving quantum gravity corrections, spin $ s $, and multipole $ \ell $. Scalar $s$-wave perturbations, given by $s =\ell = 0$, are shown in Fig.~\ref{fig:veffa}, while dipole configurations with $ \ell = 1 $ are explored in Figs.~\ref{fig:veffb1} and~\ref{fig:veffb2}. Quadrupole modes with $ \ell = 2 $ are presented in Figs.~\ref{fig:veffc1} -- \ref{fig:veffc3}, and spin-${1}/{2}$ perturbations are detailed in Figs.~\ref{fig:veffe1} and~\ref{fig:veffe2}. 
Variations in $c_6$ modulate the potential depth and curvature, reflecting quantum gravity corrections to the nozzle acoustic geometry. 

The analysis of effective potential modulation by quantum gravity corrections shows trends across integer spin $s$ and multipole $\ell$ configurations. As the quantum gravity correction parameter $c_6$ increases, the peak of the effective potential shifts farther from the analog event horizon at $x = 0$. This trend has strong linear correlations, with the coefficient of determination $R^2 \geq 97.8\%$, between $c_6$ and the longitudinal coordinate $x$ at which the peak occurs. However, half-integer values of $s$ and $\ell$ combinations, such as $s=\ell= 1/2$ in Fig.~\ref{fig:veffe1}, exhibit a non-monotonic behaviour\clt{, the same discussed in Sec.~\ref{sec3}}, with a saturation limit at $c_6 \approx 0.002$, whereas for $\ell= 3/2$ in Fig.~\ref{fig:veffe2} the limit is $c_6 \approx 0.004$. For $s = \ell = 0$, Fig.~\ref{fig:veffa}, the potential peak shifts from $x = 0.277$ (corresponding to $c_6 = -0.015$, at the point $r_\text{peak} = 2.87$, with maximum value of the potential $V_\text{max} = 0.03$) to $x = 1.664$ (for $c_6 = 0.005$, at the peak longitudinal coordinate $r_\text{peak} = 2.556$, with maximum potential $V_\text{max} = 0.025$), with the Schwarzschild case in the case where $c_6 \to 0$ 
 at $x = 1.416$, yielding a linear fit with coefficient of determination $R^2 = 99.5\%$. Similarly, for the scalar case $s = 0$ and $\ell = 1$ illustrated in Fig.~\ref{fig:veffb1}, increasing the quantum gravity correction parameter $c_6$ from $-0.015$ to $0.005$ moves the peak from $x = 1.071$, corresponding to $r_\text{peak} = 3.01$ and  maximum value of the potential $V_\text{max} = 0.1$, to $x = 4.001$ (with $r_\text{peak} = 2.83$, and maximum value of the potential $V_\text{max} = 0.099$), within $R^2 = 99.1\%$. The case of $s =\ell = 1$ portrayed in Fig.~\ref{fig:veffb2} shows a comparable displacement from the longitudinal coordinate $x = 1.18$, (for $c_6 = -0.015$, $r_\text{peak} = 3.141$, $V_\text{max} = 0.072$) to the value $x = 3.826$ (corresponding to $c_6 = 0.005$, $r_\text{peak} = 2.928$, $V_\text{max} = 0.075$), within  $R^2 = 98.7\%$. For higher multipoles ($\ell = 2$), the peak displacement scales with the spin $s$. In fact, for the scalar case $s = 0$ in Fig.~\ref{fig:veffc1}, the longitudinal coordinate $x$ shifts from 1.797 to 5.058, within $R^2 = 98.4\%$; for the $s = 1$ case in Fig.~\ref{fig:veffc2},  the longitudinal coordinate $x$ shifts from 1.912 to 5.066, within $R^2 = 98.2\%$; and for the $s = 2$ case in Fig.~\ref{fig:veffc3}, the shift in $x$ goes from 2.097 to 5.073, within $R^2 = 97.8\%$, as $c_6$ varies from $-0.015$ to $0.005$. The half-integer spin case $s=\ell=1/2$ breaks the linear trend, with the longitudinal coordinate attaining the value $x = 0.649$, corresponding to $r_\text{peak} = 2.99$ and the maximum value of the potential $V_\text{max} = 0.041$ for $c_6 = -0.015$. 
  On the other hand, the longitudinal coordinate $x = 2.969$ regards $r_\text{peak} = 2.807$ and  $V_\text{max} = 0.04$, for $c_6 = 0.005$. This non-monotonic displacement suggests a critical threshold beyond which QN modes may exhibit enhanced/diminished quality factors $q_n$. Across all cases, the astrophysical distance $r_\text{peak}$ decreases linearly with the quantum gravity-correction parameter $c_6$. For instance, $\Delta r_\text{peak} = 0.314$ for $s = \ell = 0$, and $\Delta r_\text{peak} = 0.259$ for $s = \ell = 2$, as $c_6$ varies from $-0.015$ to $0.005$. 
Since the values of the QN modes can be obtained by the behavior of the effective potential in the quantum gravitational corrected near-BH region, then the shape of the
effective potential, and by such means the configuration of the de Laval nozzle, far
from the BH is less significant for the QN modes. Hence, experimental phenomena such 
as the reflection of waves from boundaries and surface friction are expected not to influence the observed apparatus.

In all plots in Fig.~\ref{fig:veff}, the peak of the effective potential reaches a maximum nearer to [farther from] the nozzle throat when compared to the Schwarzschild solution, for $c_6<0$ [$c_6>0$]. 
There is one singular aspect regarding the corrections of the effective potential due to the parameter $ c_6 $ driving quantum gravity corrections. For scalar $s$-wave perturbations, with $s =\ell = 0$ in Fig.~\ref{fig:veffa}, for negative values of $c_6$, the higher the absolute value of $c_6$, the higher the peak is, while for positive values of $c_6$ the higher the absolute value of $c_6$, the lower the peak of the effective potential is. We can see that the ascent of the peak of the effective potential is very sharp, becoming even sharper for higher values of the absolute value of $c_6<0$.
A similar picture is verified for $s =0$, $\ell = 1$ in Fig.~\ref{fig:veffb1}, however, there is an almost imperceptible variance of the peak height as a function of $c_6$. The opposite behavior is verified for all other cases in Figs.~\ref{fig:veffb2}--\ref{fig:veffe2}: for negative values of $c_6$, the higher the absolute value of $c_6$, the lower the peak is, while for positive values of $c_6$ the higher the absolute value of $c_6$, the higher the peak of the effective potential is. Despite this peculiarity, scalar $s$-wave perturbations, with $s =\ell = 0$ in Fig.~\ref{fig:veffa}, have asymptotic values almost identical at the nozzle exit, where the effective potential is also negligible. For $s =0, \ell = 1$  
and  $s =0, \ell = 1$ the asymptotic value of the effective potential starts to differ, for substantially different values of $c_6$, being the effective potential not negligible whatsoever. For all other values of $s$ and $\ell$ here analyzed, a similar behavior is observed. 

The geometric profile of the de Laval nozzle can now be determined, as its effective potential governing perturbations replicates the one for the quantum-corrected Schwarzschild metric~\eqref{ck}. This correspondence enables the derivation of the nozzle critical spatial configuration, as demonstrated numerically in Fig.~\ref{fig:shape}.

The longitudinal profile of the de Laval nozzle, particularly its maximum cross-sectional area $\scalebox{.85}{\textsc{A}}_{\scalebox{.63}{\textsc{max}}}$ (for instance, at $x=20$), is very influenced by the multipole number $\ell$ and spin $s$. As illustrated in Fig.~\ref{fig:shape}, larger values of $\ell$ make the nozzle to expand. In fact,  for $s = 0$ the cross-sectional area  $\scalebox{.85}{\textsc{A}}_{\scalebox{.63}{\textsc{max}}}$ increases from $6.22$ (for $\ell = 0$ in  Fig.~\ref{fig:shapea}) to $72.17$ (for $\ell = 1$ in  Fig.~\ref{fig:shapeb1}), and to $1101.28$ (for $\ell = 2$ in  Fig.~\ref{fig:shapec1}). On the other hand, the value of spin $s$ reduces the cross-sectional area $\scalebox{.85}{\textsc{A}}_{\scalebox{.63}{\textsc{max}}}$, though less prominently. For $\ell = 1$, increasing $s$ from $0$ to $1$ decreases $\scalebox{.85}{\textsc{A}}_{\scalebox{.63}{\textsc{max}}}$ by $40\%$, from $72.17$ to $43.28$, which is depicted in Figs.~\ref{fig:shapeb1} and \ref{fig:shapeb2}. Similarly, for $\ell = 2$, a $27\%$ reduction in the cross-sectional area $\scalebox{.85}{\textsc{A}}_{\scalebox{.63}{\textsc{max}}}$ occurs with higher values of $s$, with less geometric sensitivity at larger multipoles as shown in Figs.~\ref{fig:shapec1}--\ref{fig:shapec3}. This relationship directly impacts the fluid dynamics analog underlying the quantum-corrected BH metric~\eqref{ck}.  Indeed, wider nozzles corresponding to $\ell \gg 0$ enhance the fluid flow velocity and the thrust by reducing backpressure, while higher spin values  ($s > 0$) moderate this effect through increased flow resistance. The $\ell$-dominance aligns with gravitational analogs, where higher multipoles correlate with stronger spacetime curvature perturbations. Computational and numerical results in Fig.~\ref{fig:shape} confirm that $\ell$-driven expansions exceed spin-induced contractions by an order of magnitude, showing the multipole capacity in nozzle shaping. These geometric modulations suggest that optimizing nozzle performance requires careful balancing of $\ell$ and $s$ parameters, with $\ell$ offering greater control over thrust generation and $s$ providing fine-tuning on the fluid flow features. 

One also concludes from the plots in Figs.~\ref{fig:shapee1} and \ref{fig:shapee2}, regarding fermionic perturbations, that quantum gravity effects encoded in the parameter $c_6$ are more perceptible for negative values of $c_6$. Also, the 
higher the absolute value of $c_6$, the more evident the difference between the quantum gravitational corrections to the nozzle geometry and the standard geometry provided by the Schwarzschild solution. In the fermionic perturbation case, positive values of $c_6$ that widen the nozzle shape with respect to the Schwarzschild solution are derisive, and any effect of quantum-corrections onto the nozzle geometry essentially makes the cross-sectional area shrink proportionally to the absolute value of $c_6$. The higher the absolute value of $c_6$, the less steep the nozzle cross-sectional area increases along the longitudinal coordinate.  Figs.~\ref{fig:shapea}--\ref{fig:shapec3} show that bosonic perturbations (scalar, vector, and tensor ones) yield the quantum gravity-corrected nozzle cross-sectional area to decrease [increase] for negative [positive] values of $c_6$. 
 In subsonic flows, sound propagates through the gas flowing along the nozzle axis of symmetry,  from gas inlet to exhaust gas exit. At the nozzle throat, where the cross-sectional area attains the minimum value, the fluid flow velocity becomes locally sonic, with the Mach number equaling unit, which is a situation that characterizes a choked flow. As the de Laval nozzle cross-sectional area increases, the fluid flow expands and increases to a supersonic velocity, where a sound wave does not propagate backward through the gas, as observed with respect to the reference frame of the nozzle. 
The Mach number ${\scalebox{.85}{\textsc{M}}}$ has a dependence on the longitudinal coordinate $x$ along the de Laval nozzle, as illustrated in Fig.~\ref{fig:Mach}. It corroborates to the transition from a subsonic (${\scalebox{.85}{\textsc{M}}} < 1$) to a supersonic (${\scalebox{.85}{\textsc{M}}} > 1$) flow, which numerically occurs precisely \textit{at} the nozzle throat, corresponding to the acoustic horizon. Beyond this point, the diverging nozzle geometry facilitates the fluid flow expansion, driving the Mach number ${\scalebox{.85}{\textsc{M}}}$ to supersonic regimes. For fixed values of $s$ and $\ell$, the ${\scalebox{.85}{\textsc{M}}}$ increment rates increase with larger values of the quantum-correction parameter $c_6$. It is worth emphasizing that only $s=1/2$, as discussed before, breaks the linearity near $c_6\approx 0.002$. The multipole number $\ell$ amplifies the maximum value of the Mach number ${\scalebox{.85}{\textsc{M}}}_{\scalebox{.63}{\textsc{max}}}$. For $s = 0$, ${\scalebox{.85}{\textsc{M}}}_{\scalebox{.63}{\textsc{max}}}$ increases from $5.56$ (for $\ell = 0$ in  Fig.~\ref{fig:Macha}) to $16.04$ (for $\ell = 1$ in  Fig.~\ref{fig:Machb1}), and to $48.20$ (for $\ell = 2$ in  Fig.~\ref{fig:Machc1}). On the other hand, the increase in the spin $s$ reduces the Mach number ${\scalebox{.85}{\textsc{M}}}_{\scalebox{.63}{\textsc{max}}}$, although with less efficacy. For $\ell = 1$, increasing $s$ from $0$ to $1$ makes the Mach number to decrease by $\sim19\%$, from $16.04 \rightarrow 13.00$, as illustrated in  Figs.~\ref{fig:Machb1} and \ref{fig:Machb2}. Similarly, for $\ell = 2$, the Mach number ${\scalebox{.85}{\textsc{M}}}_{\scalebox{.63}{\textsc{max}}}$ is reduced by $22\%$ with higher values of $s$, which consists of a more prominent reduction when compared to $\ell=1$, contrasting with geometric area reductions trends discussed before. An interesting convergence occurs at a point $x > 0$, especially with spin $s=\ell=0$, where the quantum-correction parameter $c_6$ yields almost-identical aerodynamic behavior, with some tiny deviation, implying a regime where the nozzle geometry dominates over quantum corrections. Higher values of $\ell$ correlate with steeper values of the Mach number ${\scalebox{.85}{\textsc{M}}}$ gradients post-throat, reflecting an intensified fluid flow acceleration. One concludes from all the plots in Fig.~\ref{fig:Mach} that quantum gravity effects are more noticeable for positive values of $c_6$.

The pressure, temperature, and density profiles of the fluid, measured relative to reservoir quantities, show distinct dependencies on the parameter $c_6$ within the
quantum-corrected BH metric~\eqref{ck}. While a visual inspection displays small variations in these profiles across different values of $s$ and $\ell$, the comparative analysis presents interesting deviations between extreme values of $c_6$, with lower bound $c_6 = -0.015$ and upper bound $c_6 = 0.005$. For $s = \ell = 0$, relative variations at the longitudinal coordinate $x = 20$ amount to $61.5\%$ for the pressure, $13.9\%$ for the temperature, and $31.1\%$ for the density. The most pronounced variation occurs for fermionic perturbations, $s = \ell = {1}/{2}$, with relative changes of $73.2\%$ in the pressure, $28.1\%$ in the temperature, and $55.8\%$ in the density, marking the largest observed deviation across the cases here scrutinized.  
There is a convergence point $x = \qty(6.697 \pm 0.203)$ for scalar fields ($s = 0$), where the quantum-corrected metric~\eqref{ck}, independently of the values of  $c_6$, aligns with Schwarzschild solutions. This singular point suggests regimes where quantum gravitational corrections~\cite{Calmet:2021lny} yield observables indistinguishable from classical Schwarzschild BHs. This phenomenon is absent in prior analyses of effective potentials or QN modes. For other combinations of integer values for $s$ and $\ell$, maximum variations are on average $40\%$.
All curves converge at $x = 0$ at the nozzle throat, diverging thereafter as $c_6$ modulates expansion rates. Compared to the Schwarzschild solution, all the curves ($p$, $T$, or $\rho$) either grow or shrink up to a certain point and then reverse direction, remaining until the end of the nozzle. For extreme $c_6$ values these points include: for $s = \ell = 0$, $x = \qty(3.046 \pm 0.015)$ (for $c_6 = -0.015$) and $x = \qty(4.561 \pm 0.005)$ (for $c_6 = 0.005$).  When $s = \ell = 1$, we have  $x = \qty(3.571 \pm 0.183)$ (for the lower limit $c_6 = -0.015$) and $x =\qty( 5.451 \pm 0.022)$ (for the upper limit $c_6 = 0.005$).  
In the case where $s = \ell = 2$, we verify that  $x = \qty(5.023 \pm 0.082)$ (for $c_6 = -0.015$) and $x = \qty(6.431 \pm 0.054)$ (for $c_6 = 0.005$).  
Different from what was discussed before, the spin $s$ is more influential than the multipole number $\ell$ on inflection point positions, as evidenced by mixed configurations. In fact, for $s = 0, \ell = 1$ we have $x = \qty(4.660 \pm 0.035)$ for $c_6 = -0.015$ and $x = \qty(5.198 \pm 0.001)$ when the upper limit $c_6 = 0.005$ is attained.  Also, for $s = 1, \ell = 2$ we conclude that  $x = \qty(3.839 \pm 0.039)$  for $c_6 = -0.015$ and $x = \qty(5.613 \pm 0.069)$ for $c_6 = 0.005$.

The relative pressure profiles across all configurations in Fig.~\ref{fig:Pressure} demonstrate asymptotic decay to zero as $x \gg 1$ and approach unity for $x \to -\infty$, confirming that a choked flow at the nozzle throat occurs exclusively when both the stagnation pressure and mass flux meet the threshold required to reach sonic velocities. Below this critical threshold, the system reverts to subsonic Venturi tube dynamics and does not achieve supersonic acceleration. Therefore, operational viability necessitates an entry pressure substantially exceeding ambient conditions, with stagnation pressure dominating the ambient backpressure to sustain a choked flow and supersonic expansion.

In addition, Fig.~\ref{fig:Pressuree1} for the fermionic 
perturbations $s = \ell= {1}/{2}$ is the case that manifests the least quantum gravity effects arising from the $c_6$ that endows the quantum-corrected
metric~\eqref{ck}. In all other cases analyzed and depicted in Fig.~\ref{fig:Pressure}, quantum gravity effects arising from the parameter $c_6$ are perceptible and provide experimental signatures of the quantum-corrected Schwarzschild metric~\eqref{ck}. 

The relative pressure, along the longitudinal $x$-coordinate in a de~Laval nozzle, is also relevant to estimate the exhaust velocity at the nozzle exit. Supersonic flow is well known to be attained only through the diverging portion of the nozzle.  The chamber temperature, which is located at the nozzle
inlet, under isentropic conditions differs little from the stagnation temperature or, for
chemical rocket propulsion, from the combustion temperature. Hence, the exit velocity can be expressed as
\begin{equation}\label{ve}
    u_{\scalebox{.67}{\textsc{e}}} = \sqrt{\frac{RT}{\mu}\frac{2\gamma}{\gamma - 1} } \cdot \sqrt{1 - \qty(\frac{p_0}{p})^{\frac{1-\gamma}{\gamma}}},
\end{equation}
where $\mu$ is the molecular weight of the gas under scrutiny (here we consider $\mu = 28.96$ g/mol for dry air), $T$ is the absolute temperature of the inlet gas, $p_0$ is the total pressure, and $p$ is the relative local pressure to the nozzle throat pressure. Since the ideal gas
constant for any particular gas is inversely proportional to the molecular weight, exhaust velocities strongly depend upon the
ratio of the absolute nozzle entrance temperature, which is close to the combustion temperature, divided by the average molecular mass of the exhaust gas. 
Having the profiles of relative pressure along the $x$-axis, one can determine the exhaust velocity for the quantum-corrected BH metric~\eqref{ck} by applying Eq.~\eqref{ve}.

Now Fig.~\ref{fig:Temperature} shows the relative temperature to the throat as a function of the $x$-axis of the de~Laval nozzle, for the quantum-corrected Schwarzschild
metric~\eqref{ck}. One can realize that again the fermionic perturbations $s = \ell= {1}/{2}$ reveal the most visible alterations driven by the quantum-correction parameter $c_6$, with respect to the Schwarzschild solution. Increasing the absolute value of $c_6$ towards negative values increases the exhaust temperature beyond the nozzle throat,  whose differences compared to the Schwarzschild solution become sharper at the nozzle exit, as depicted in Fig.~\ref{fig:Temperaturee1}. Reinforcing the discussed behavior for the spin $s$ and multipole parameter $\ell$, Fig.~\ref{fig:Temperature} shows that the fluid flow for $s =\ell = 0$, at $x=20$ the de Laval nozzle still has $\qty(15.1\pm1.7)\%$ of the reservoir temperature, whereas the fluid flow for the $s=\ell=1$, in the same point, has about $\qty(3.6\pm0.3)\%$ of the reservoir temperature. For the case of $s=\ell=1/2$, at $x=20$, the fluid has about $\qty(10.4\pm2.0)\%$ of the reservoir temperature. This variation is converted into thrust and can be used for computing the QN modes emitted from the quantum-corrected BH~\eqref{ck}.

Fig.~\ref{fig:Density} depicts the relative density as a function of the longitudinal $x$-axis of the nozzle, for the quantum-corrected BH metric~\eqref{ck}.  

The thermodynamic variables, the nozzle geometry, and the Mach number are the realistic quantities aimed to be measured from a de Laval nozzle in a propulsion laboratory, regarding the quantum-corrected BH metric~\eqref{ck}. 

The total impulse generated by a nozzle is proportional to the total energy released by or into all the propellants utilized by propulsion systems. The power transmitted by the de Laval nozzle is the product of the thrust generated by the nozzle and the fluid flow velocity. 
The thrust generated by the nozzle, in a vacuum, can be described as~\cite{Casadio:2024uwj}
\begin{equation} \label{eq:forcethrust}
    F_{\scalebox{.63}{\textsc{thrust}}} = p_0 {\scalebox{.87}{\textsc{A}}}_{{\scalebox{.65}{$\star$}}}\, \qty[ \sqrt{\frac{2 \gamma^2}{\gamma-1} \qty(\frac{2}{\gamma+1})^{\frac{\gamma+1}{\gamma-1}}}\qty(1-p^\frac{\gamma-1}{\gamma})^{1/2} + \frac{ {\scalebox{.87}{\textsc{A}}}_{\scalebox{.67}{\textsc{e}}}}{{\scalebox{.87}{\textsc{A}}}_{{\scalebox{.65}{$\star$}}}}  \frac{{p_{\scalebox{.67}{\textsc{e}}}}}{p_0}],
\end{equation}
where $p_{\scalebox{.67}{\textsc{e}}}$ denotes the pressure at which the gas exits the de Laval nozzle, ${\scalebox{.87}{\textsc{A}}}_{\scalebox{.67}{\textsc{e}}}$ denotes the area of the nozzle endpoint, and $p$ is the relative local pressure to the throat pressure. 
Eq.~\eqref{eq:forcethrust} shows that the thrust generated by the nozzle is proportional to the throat cross-sectional area, the nozzle inlet pressure, and the pressure ratio across the nozzle. A more detailed numerical analysis indicates that for realistic nozzles at higher external pressures, some flow separation begins to arise within the divergent part of the de Laval nozzle. The cross-sectional area of the exiting supersonic jet will be lower than the cross-sectional area of the nozzle itself, although, to a steady fluid flow the separation typically persists longitudinally.  As external pressure rises, the separation point moves upstream. At the nozzle exit, the separated flow continues to be supersonic in the central region however, it is bounded by an annular subsonic flow. Along their interface, there is a discontinuity, yielding the thrust generated by the nozzle to decrease,  compared to an ideal de Laval nozzle. These discontinuities may generate shock waves~\cite{landau_fluid_1987}.

Another way to quantify the thrust is by the effective exhaust velocity $c$. With this definition, the thrust generated by the nozzle is simplified to 
\begin{equation}
F_{\scalebox{.63}{\textsc{thrust}}}=\dot{m}c,
\end{equation}
where $\dot{m}$ is mass flux.
The effective exhaust velocity is shown in the plots in Fig.~\ref{fig:thrust}.
Following what was discussed, the thrust is highly dependent on the quantum gravity correction parameter $c_6$. In all integer $s$ and $\ell$ cases, the higher the parameter $c_6$ the higher the thrust at the nozzle endpoint is. By knowing the constant value of mass flux, one can determine the thrust generated by the nozzle and measure it in a real model. In all cases, the effective exhaust velocity
increases as a function of the longitudinal nozzle coordinate $x$. 
However, for each fixed value of integer $s$ and $\ell$, for any fixed positive [negative] value of the quantum gravity-correction parameter $c_6$,  the exhaust velocity for the quantum-corrected BH metric~\eqref{ck} increases faster [slower] along $x$ than the Schwarzschild solution. Now, for $s=0, \ell=0$ and for fixed values of $c_6$, the rate of variation of the effective exhaust velocity with respect to the $x$ coordinate is negative, and the rate decrement steeper is for negative values of $c_6$, as plotted in Fig.~\ref{fig:thrusta}. On the other hand, for $s=0, \ell=1$ and for fixed values of $c_6$, the rate of variation of the effective exhaust velocity with respect to the $x$ coordinate is almost constant for negative values of $c_6$ and slightly positive for positive values of $c_6$, as shown in Fig.~\ref{fig:thrustb1}. Similarly to Fig.~\ref{fig:thrusta}, the rate decrement is steeper for negative values of $c_6$. A similar behavior is observed for $s=\ell=1$. For $s=0, \ell=2$ in Fig.~\ref{fig:thrustc1}, 
the rate of variation of the effective exhaust velocity with respect to the $x$ coordinate is always positive, irrespectively the value of $c_6$, and it grows faster for positive values of $c_6$. A drastically different scenario is verified in Fig.~\ref{fig:thruste1} for the case $s=\ell=1/2$, where almost all the values of $c_6$ drive the effective exhaust velocity to vary slower as a function of the $x$ coordinate, the higher the value of $c_6$ is. An analogous result is observed for $s=\ell=1/2$ in Fig.~\ref{fig:thruste2}, where the only difference is a steeper rate of variation as a function of $x$. 

\begin{figure}[H]
    \centering
    \begin{subfigure}{0.49\textwidth}
        \centering
        \includegraphics[width=\linewidth]{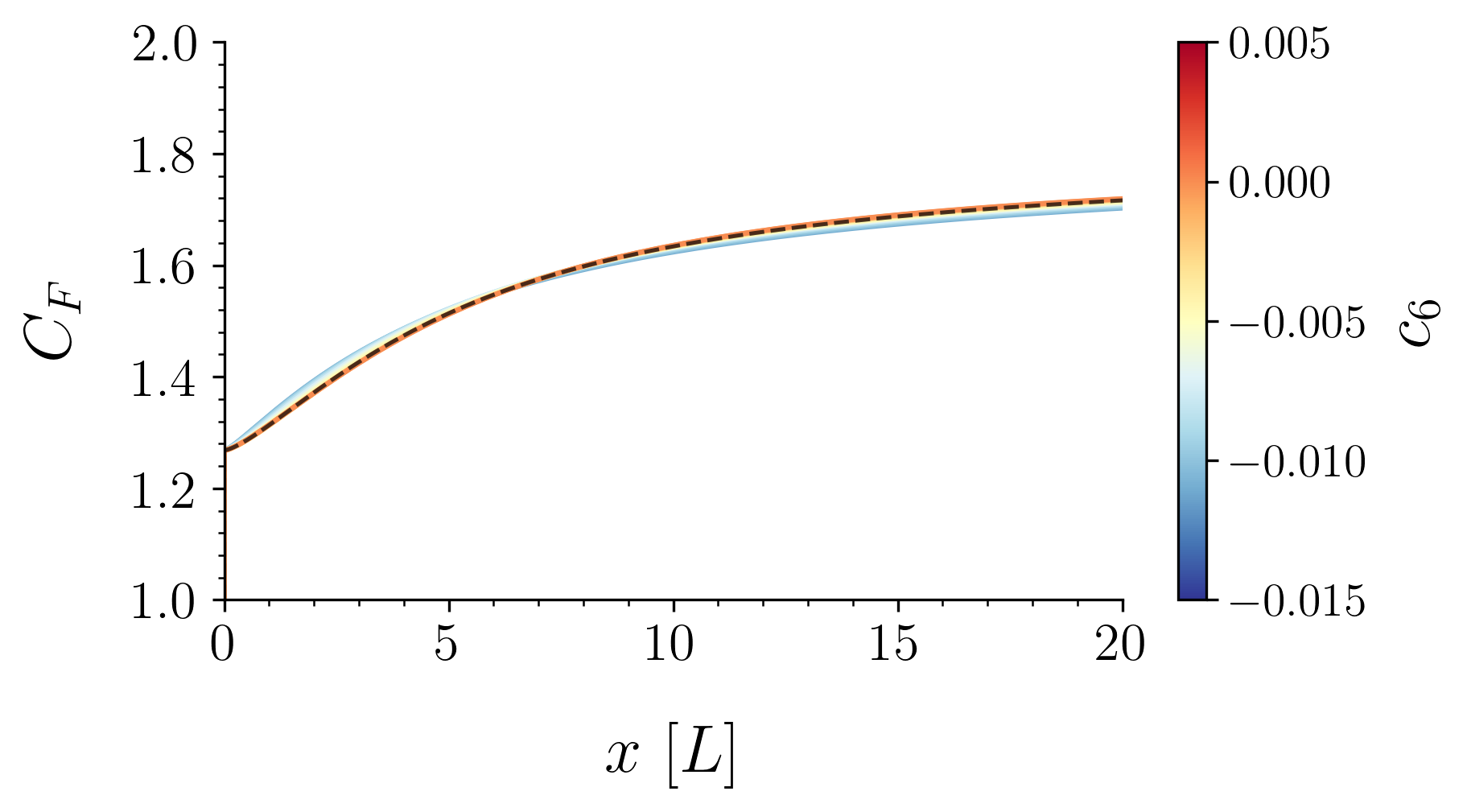}
         \caption{\footnotesize $s = \ell = 0$.}
        \label{fig:C_Fa}
    \end{subfigure}
    \begin{subfigure}{0.49\textwidth}
        \centering
        \includegraphics[width=\linewidth]{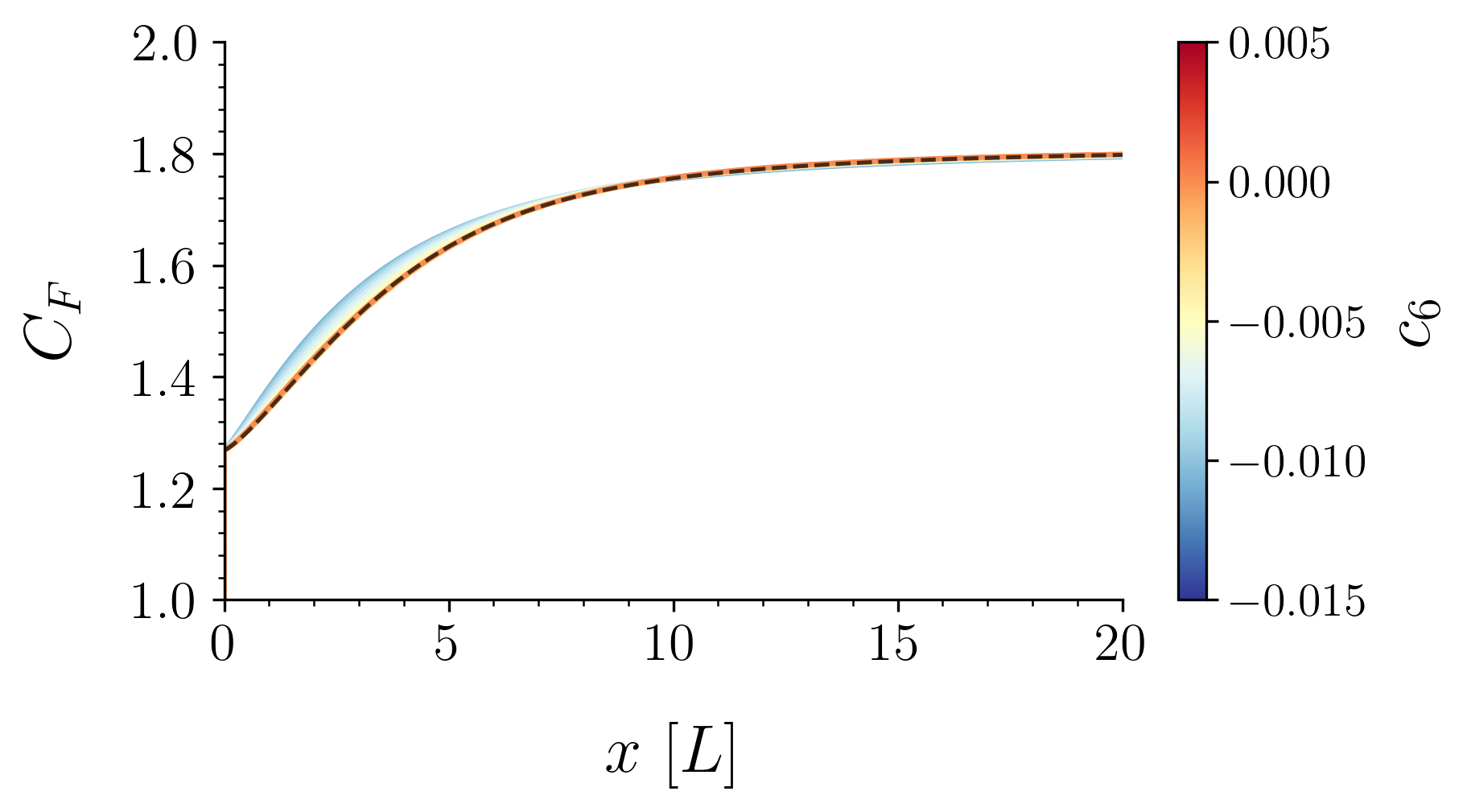}
         \caption{\footnotesize $s = 0$ and $\ell=1$.}
        \label{fig:C_Fb1}  \end{subfigure}
        
    \begin{subfigure}{0.49\textwidth}
        \centering
        \includegraphics[width=\linewidth]{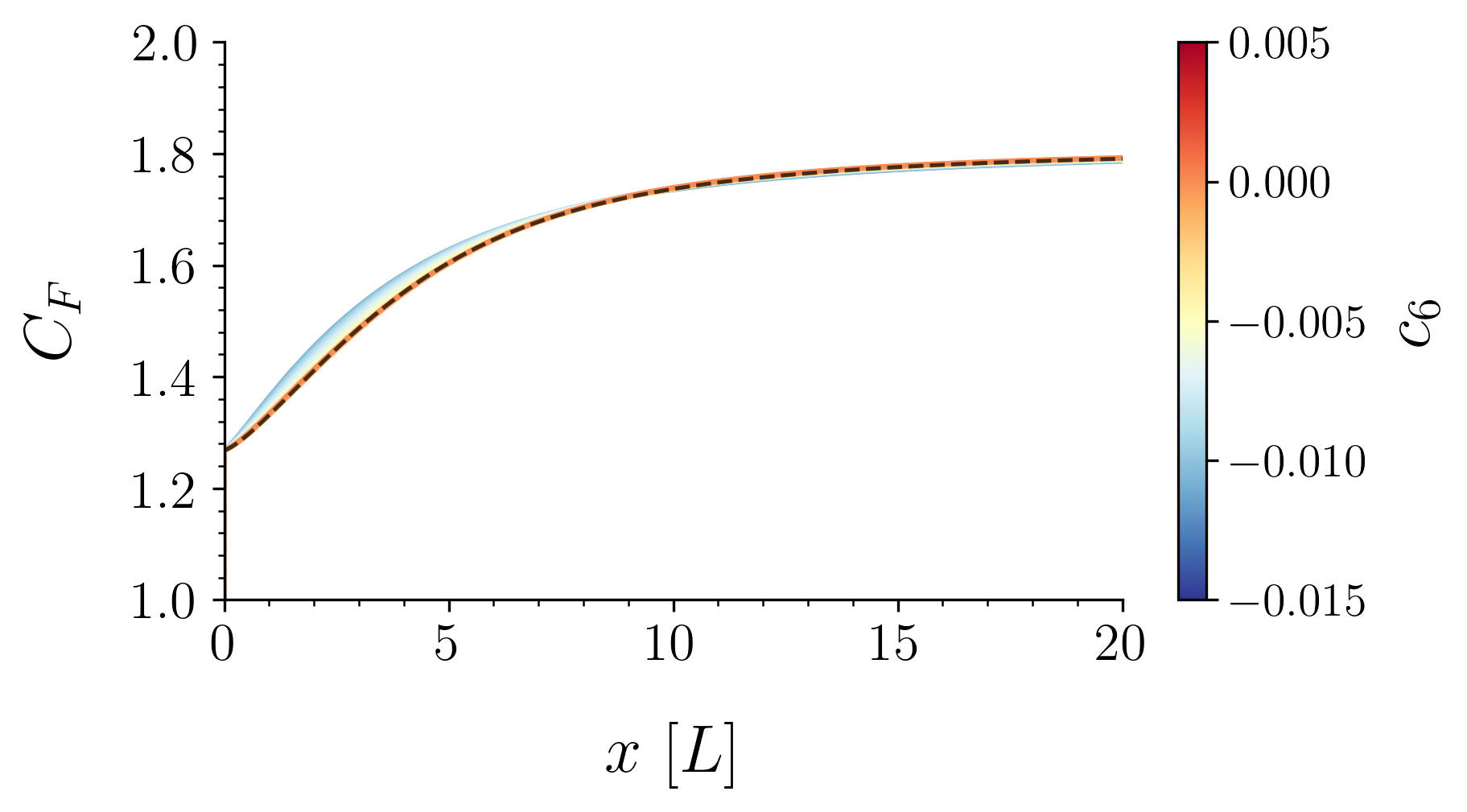}
         \caption{\footnotesize $s = \ell=1$.}
        \label{fig:C_Fb2}
    \end{subfigure}
    \begin{subfigure}{0.49\textwidth}
        \centering
        \includegraphics[width=\linewidth]{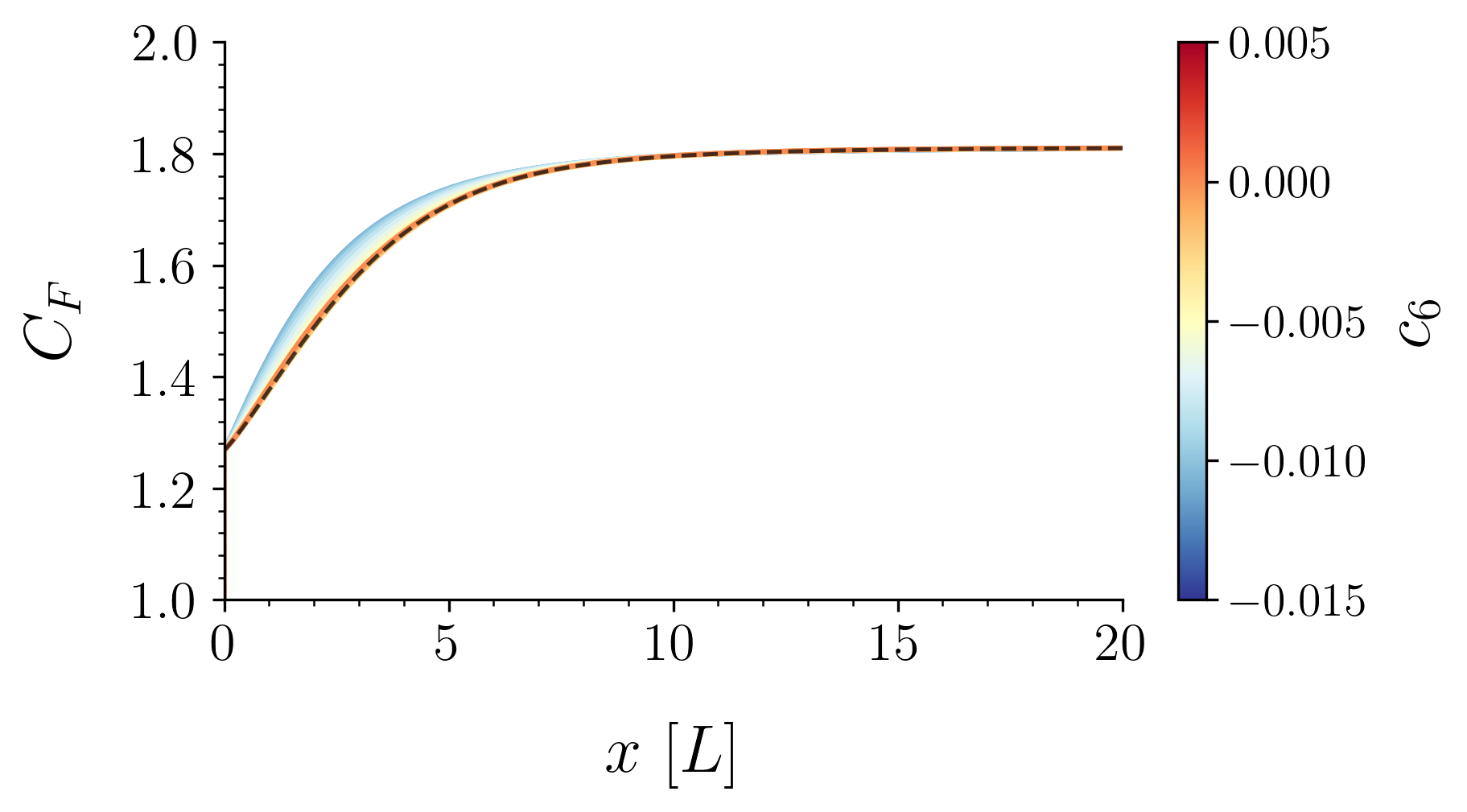}
         \caption{\footnotesize $s = 0$ and $\ell=2$.}
        \label{fig:C_Fc1} 
    \end{subfigure}
    
    \begin{subfigure}{0.49\textwidth}
        \centering\includegraphics[width=\linewidth]{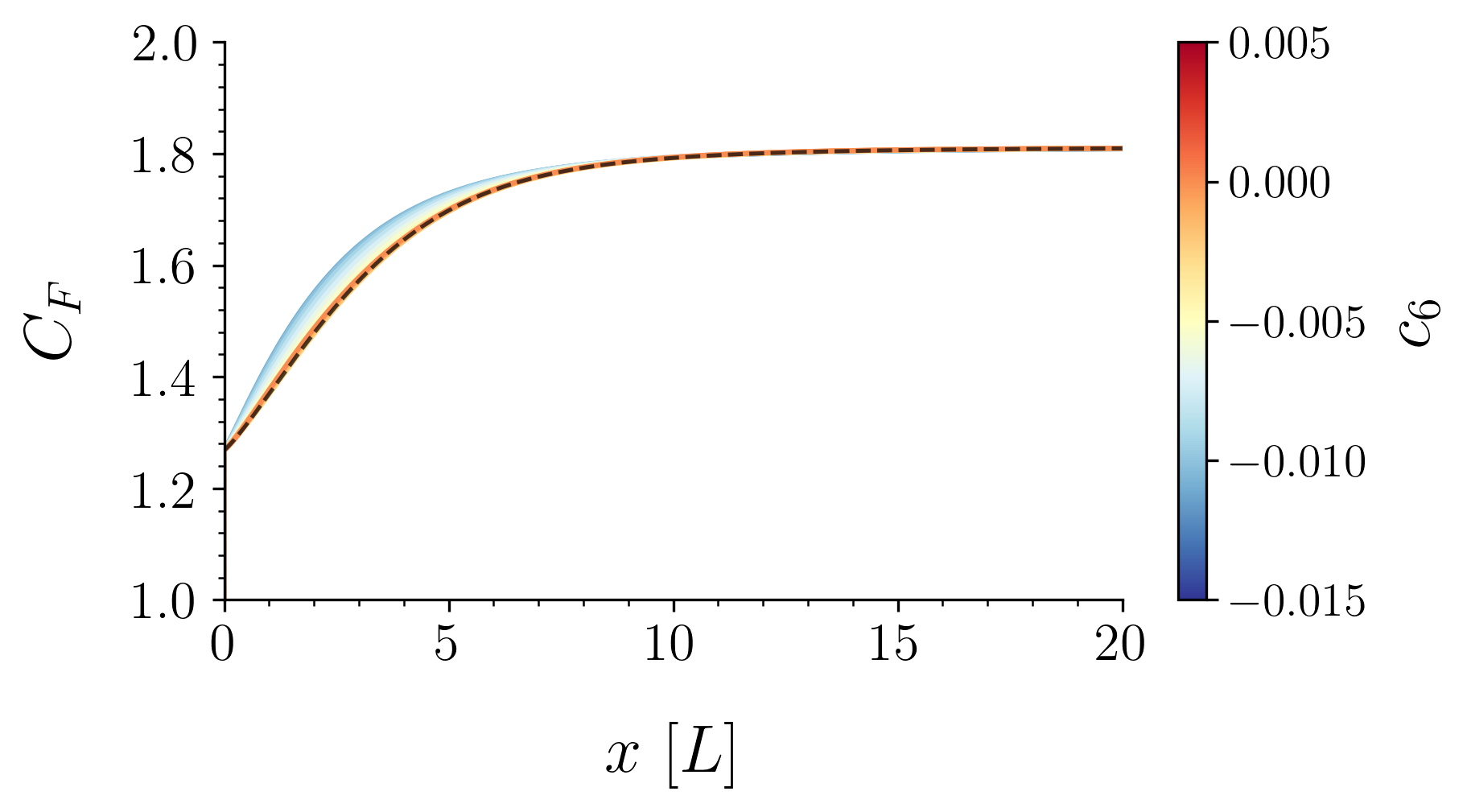}
         \caption{\footnotesize $s = 1$ and $\ell=2$.}
        \label{fig:C_Fc2}
    \end{subfigure}
    \begin{subfigure}{0.49\textwidth}
        \centering\includegraphics[width=\linewidth]{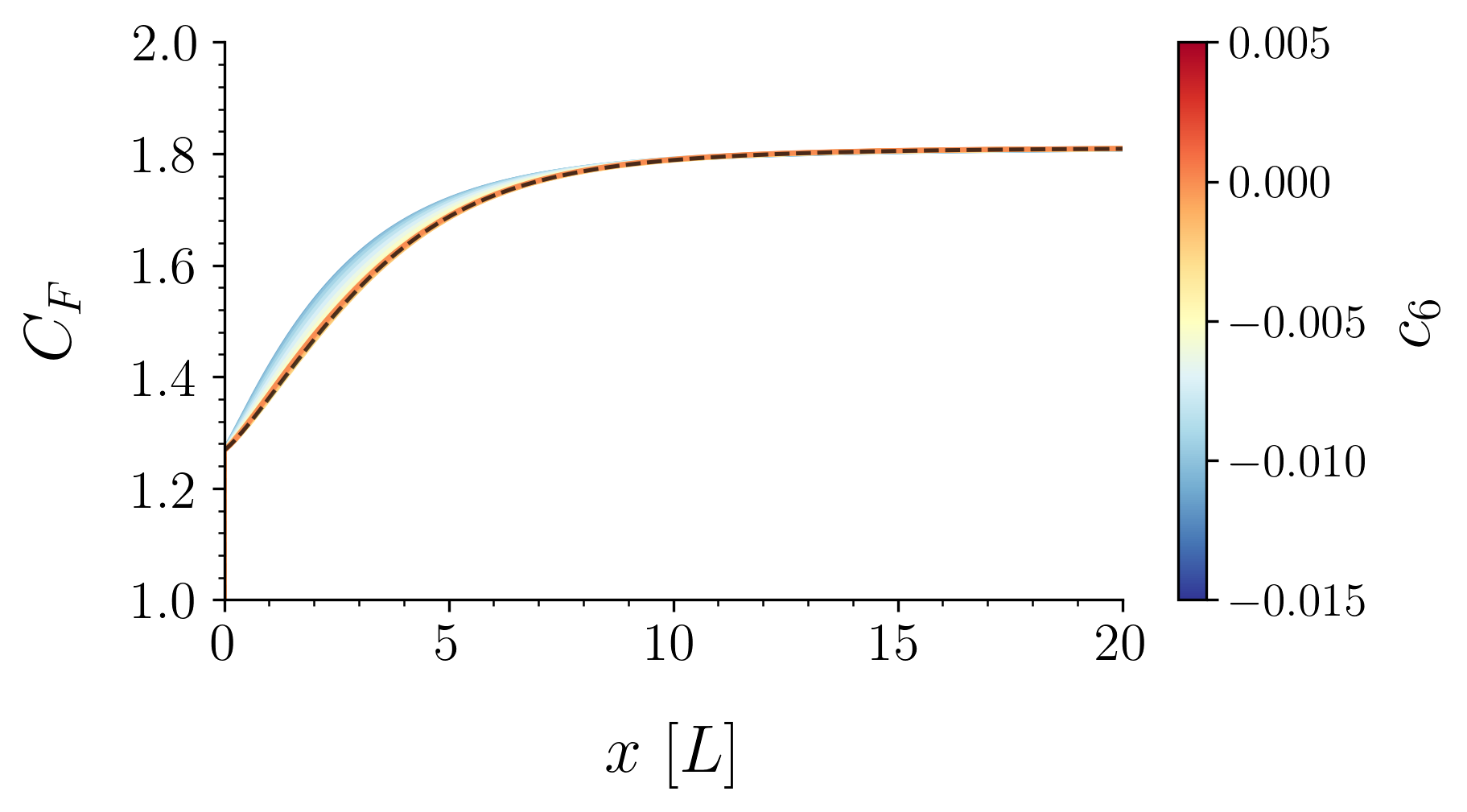}
         \caption{\footnotesize $s= \ell=2$.}
        \label{fig:C_Fc3}
    \end{subfigure} 
    \begin{subfigure}{0.49\textwidth}
        \centering
        \includegraphics[width=\linewidth]{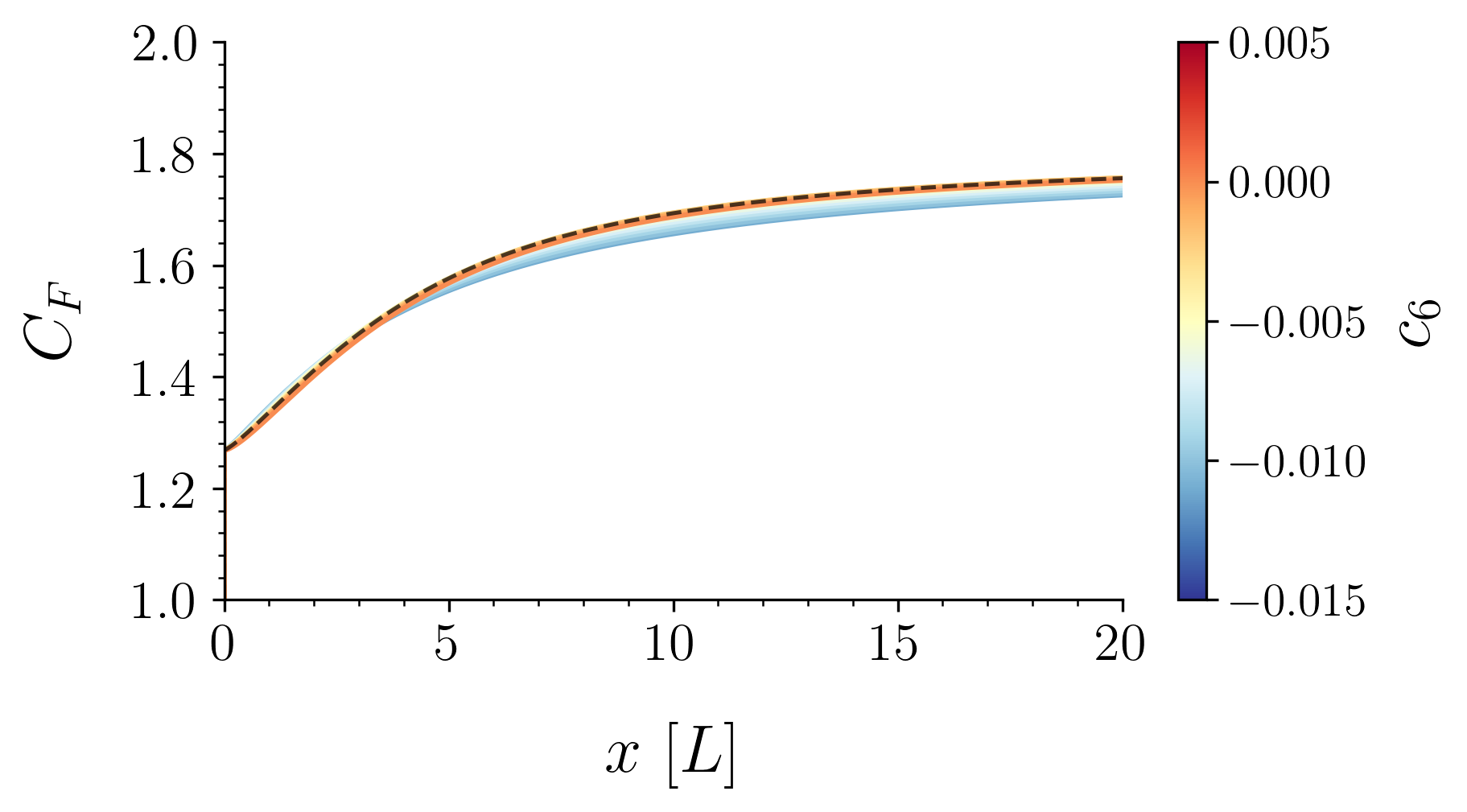}
         \caption{\footnotesize $s = \ell= {1}/{2}$.}
        \label{fig:C_Fe1}
    \end{subfigure}
    \begin{subfigure}{0.49\textwidth}
        \centering
        \includegraphics[width=\linewidth]{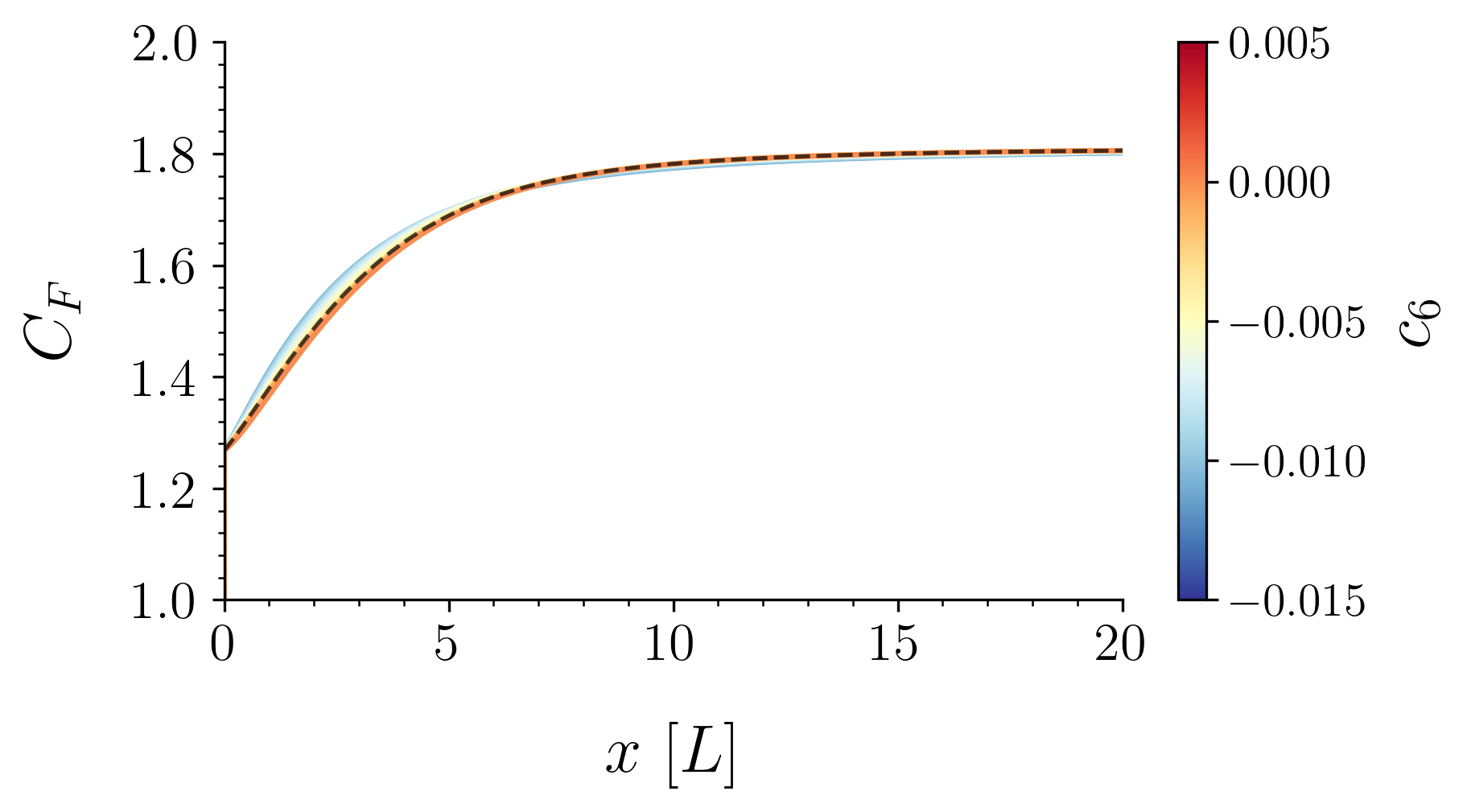}
         \caption{\footnotesize $s = {1}/{2}$ and $\ell= {3}/{2}$.}
        \label{fig:C_Fe2}  \end{subfigure}
     \caption{\footnotesize Thrust coefficient 
    as a function of the longitudinal direction of the de~Laval nozzle, varying $c_6$ for the quantum-corrected Schwarzschild metric~\eqref{ck}. The dashed black line represents the  Schwarzschild solution. The values are calculated using $G_{{\scalebox{.55}{\textsc{N}}}} = M = 1$.}
    \label{fig:C_F}
\end{figure}

The thrust coefficients calculated for all the nozzles are shown in Fig. \ref{fig:C_F}. The efficiency of the nozzle in converting thermal energy into kinetic energy is related to the thrust coefficient, $C_{\scalebox{.63}{{$F$}}}$, defined as
\begin{equation}
C_{\scalebox{.63}{{$\mathit{F}$}}} = \frac{{F_{\scalebox{.63}{\textsc{thrust}}}}}{p_{\scalebox{.67}{\textsc{0}}} {\scalebox{.87}{\textsc{A}}}_{{\scalebox{.65}{$\star$}}}}.
\end{equation} 
The thrust coefficient estimates the thrust that is amplified by the expansion of fluid as it flows through the nozzle, compared to the thrust triggered if the compression chamber were connected only to both the convergent section and the nozzle throat, but not to the divergent section. 
Dividing Eq. \eqref{eq:forcethrust} by the term $p_{\scalebox{.67}{\textsc{0}}} {\scalebox{.87}{\textsc{A}}}_{{\scalebox{.65}{$\star$}}}$, one obtains
\begin{equation}
   C_{\scalebox{.63}{{$\mathit{F}$}}}\qty(x) = \sqrt{\frac{2\gamma^2}{\gamma-1}\qty(\frac{2}{\gamma+1})^{\frac{\gamma+1}{\gamma-1}}}\qty(1-p(x)^\frac{\gamma-1}{\gamma})^{1/2} + {\scalebox{.87}{\textsc{A}}}(x) p(x).
\end{equation}
 Again, we use the cross-sectional area ${\scalebox{.87}{\textsc{A}}}$ and the pressure measured in units of the throat cross-sectional area, ${\scalebox{.87}{\textsc{A}}}_{{\scalebox{.65}{$\star$}}}$, and total pressure, $p_{\scalebox{.67}{\textsc{0}}}$, respectively. The thrust coefficient in de Laval nozzles represents the efficiency of throwing gases out of the nozzle, measuring the de Laval nozzle's capacity to turn internal pressure into velocity at the nozzle exit. A higher value of the thrust coefficient complies with a more effective performance of the de Laval nozzle.

{\color{black}
\subsection{QN modes, eigenfrequencies, and overtones of quantum gravitational corrected black  holes}
}

After scrutinizing the main features of the quantum gravitational corrected analog de Laval nozzle, the QN mode frequencies can now be computed. A complex QN frequency
$\omega_n$ defines a QN mode, appearing in the Schrödinger-like equation~\eqref{quasinormal modes equation}, on the gravitational side, and in Eq.~\eqref{eq:schrodinger acoustic BN}, for its aerodynamic counterpart. Deriving QN modes analytically is often a challenging task, but approximation methods can simplify the process. Among these, the Mashhoon procedure is employed in this work due to its clarity and practicality~\cite{ferrari1984new}. The boundary conditions for QN modes in asymptotically flat BHs, such as the one studied here, are well justified from an astrophysical perspective~\cite{Konoplya:2011qq}. The difficulty in calculating QN modes for many BHs typically arises from the slow decay of the effective potential at radial infinity. It introduces a branch cut and causes GWs to scatter off the effective potential, generating backwards tails. The Mashhoon method bypasses these issues by using a P\"oschl--Teller effective potential
\begin{equation}\label{pote}
V_{{\scalebox{.55}{\textsc{PT}}}}(r_{{\scalebox{.65}{$\star$}}}) = \frac{V_0}{\cosh^2
[\upxi(r_{{\scalebox{.65}{$\star$}}} - r_{\star0})]},
\end{equation}
for  $V_0 = V(r_0)$ being the maximum value of the potential,  
\begin{align}
\upxi = \sqrt{-\frac1{2V_0}\lim_{r\to r_0}\dfrac{\dd^2V}{\dd r_{{\scalebox{.65}{$\star$}}}^2}}
\end{align}
is the inverse of the width of the potential, and the constant $-2V_0\upxi$ indicates the curvature of the potential at its supremum value. 
The  P\"oschl--Teller effective potential~\eqref{pote} decays exponentially in the  $r_{{\scalebox{.65}{$\star$}}} \to \infty$ limit.  
The boundary conditions for the Schrödinger-like equation demand that the wave function vanish at the boundary. As a result, QN modes correspond to bound states for the new $(V_{{\scalebox{.55}{\textsc{PT}}}} \mapsto -V_{{\scalebox{.55}{\textsc{PT}}}})$ effective potential. The QN mode frequencies are given by \cite{ferrari1984new,Konoplya:2011qq,Berti:2009kk}:
\begin{align}
 \omega_n = \pm \sqrt{V_0 - \frac{\upxi^2}{4}} + i\upxi\left(n + \frac{1}{2}\right), n\in\mathbb{N}.\label{ot}
\end{align}
In Eq.~(\ref{ot}), $n \in \mathbb{N}$ denotes the overtone number~\cite{Konoplya:2011qq,Berti:2009kk,Giesler:2019uxc}. 
Including higher overtones yields a precise description of the GW form well before
the fundamental mode dominates. The determination of overtones also extends
the regime over which BH perturbation theory is suitable to a time interval even before the GW peak strain amplitude. In addition, the ringdown of the quantum-corrected BH can be analyzed by the inclusion of higher overtones, providing more accurate estimates of the quantum-corrected BH remnant spin and 
mass~\cite{Giesler:2019uxc,Buonanno:2006ui}. 
The analogy of these overtones is intended to improve the extraction of information from quantum gravity-corrected BH sources from noisy LIGO/Virgo data.

Table~\ref{tabela} presents the QN mode frequencies for the quantum-corrected Schwarzschild metric across different values of $s$, $\ell$, and $n$, as well as various choices of the quantum gravity-correction parameter $c_6$. For fixed $c_6$, increasing $\ell$ leads to larger values of $\Re(\omega_n)$. On the other hand, for fixed values of $c_6$ and $\ell$, higher values of the spin $s$ result in smaller values of the real part $\Re(\omega_n)$ of the QN frequency, which is consistent with earlier observations. For fixed spin $s$ and multipole moment $\ell$, corresponding to fixed $\Re(\omega_n)$ values, higher overtones increase the pure imaginary part of the QN eigenfrequency $\Im(\omega_n)$. Finally, when $s$, $\ell$, and $n$ are fixed integers, increasing $c_6 > 0$ reduces both $\Re(\omega_n)$ and $\Im(\omega_n)$.

\begin{figure}[H]
    \centering
    
    \begin{subfigure}{0.49\textwidth}
        \centering
        \includegraphics[width=\linewidth]{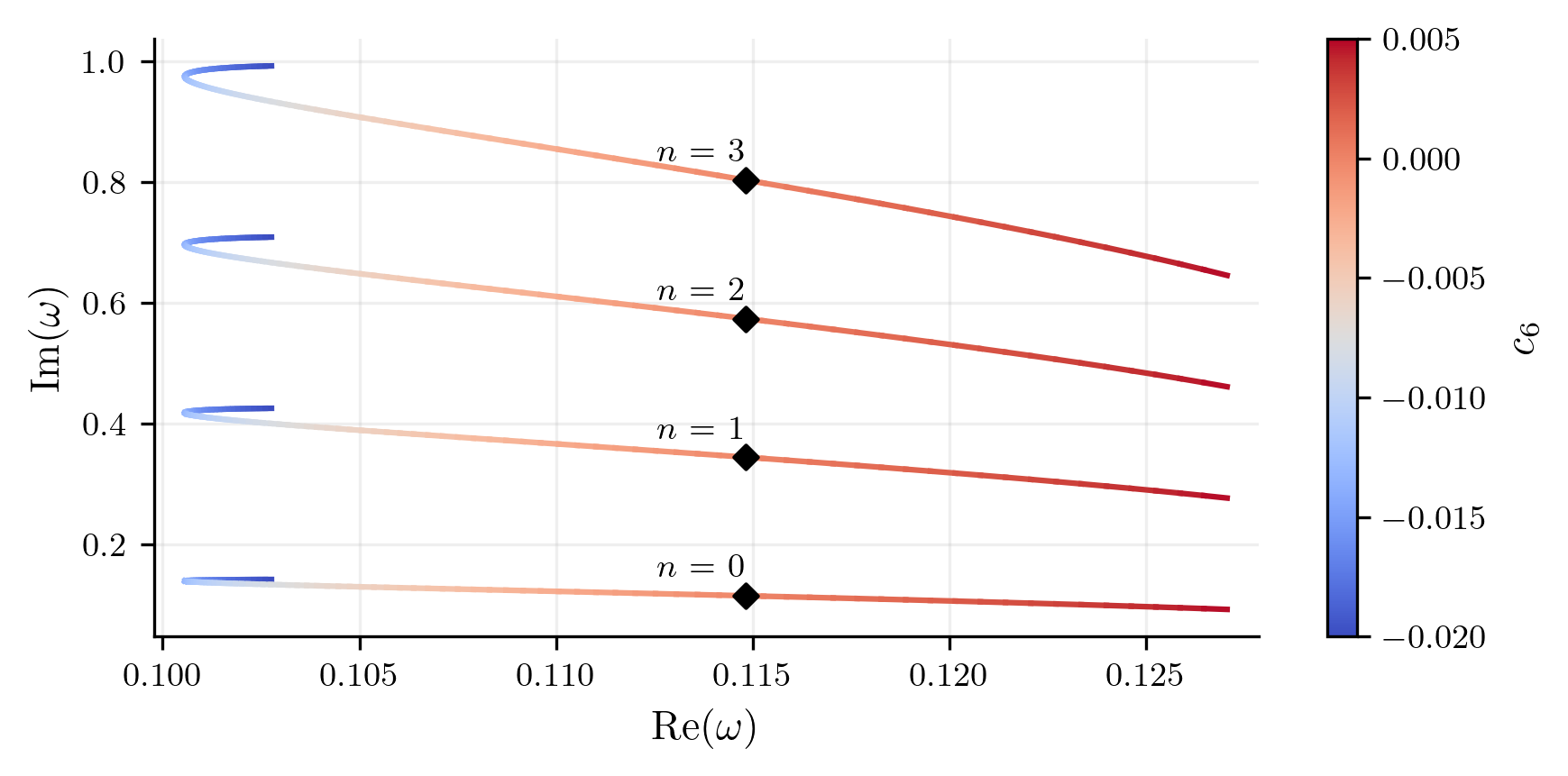}
         \caption{\footnotesize $s = \ell = 0$}
        \label{fig:spin0_l0}  \end{subfigure}
    \begin{subfigure}{0.49\textwidth}
        \centering
        \includegraphics[width=\linewidth]{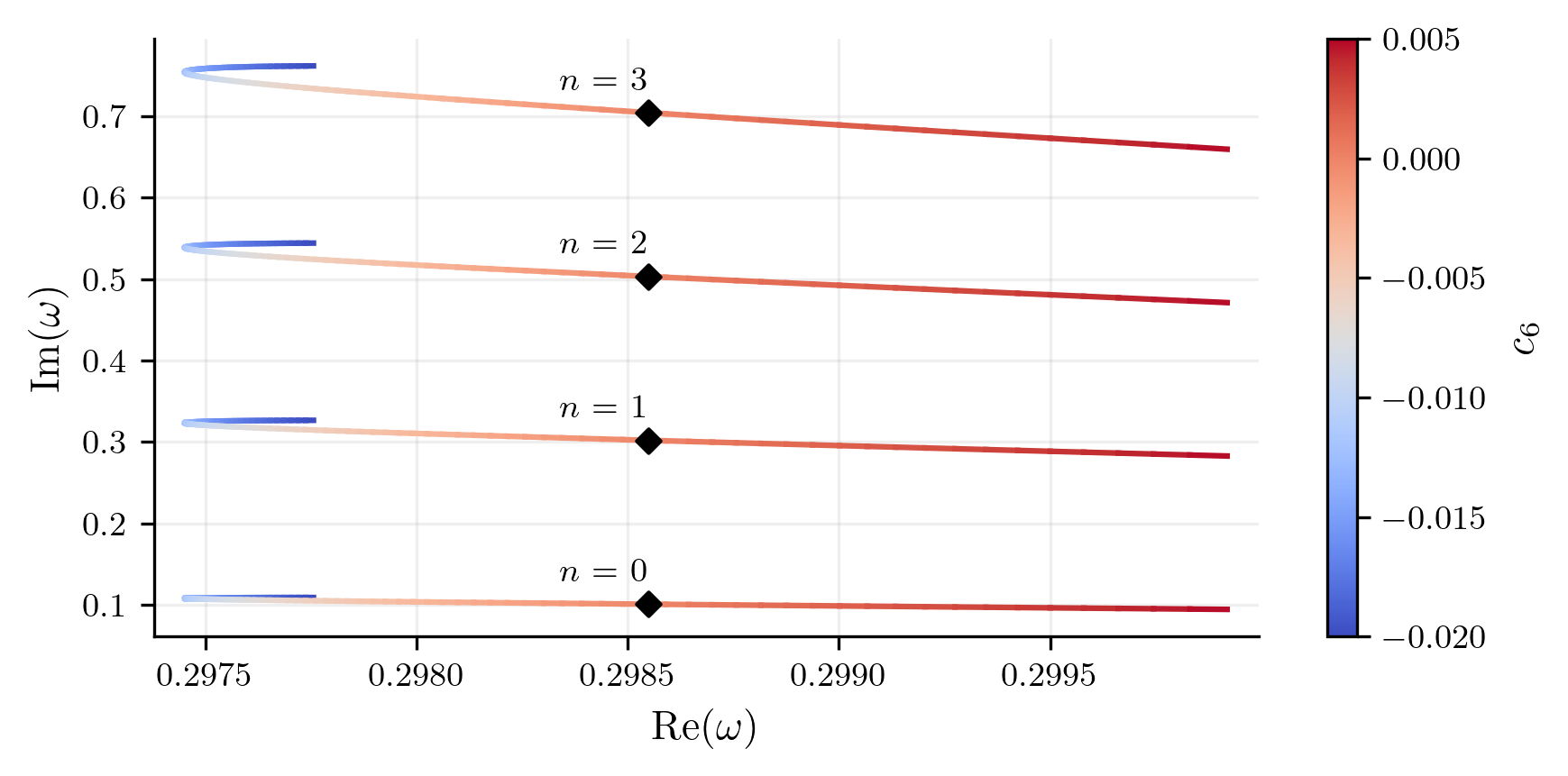}
         \caption{\footnotesize $s = 0$, $\ell = 1$}
        \label{fig:spin0_l1}
    \end{subfigure}


    \begin{subfigure}{0.49\textwidth}
        \centering
        \includegraphics[width=\linewidth]{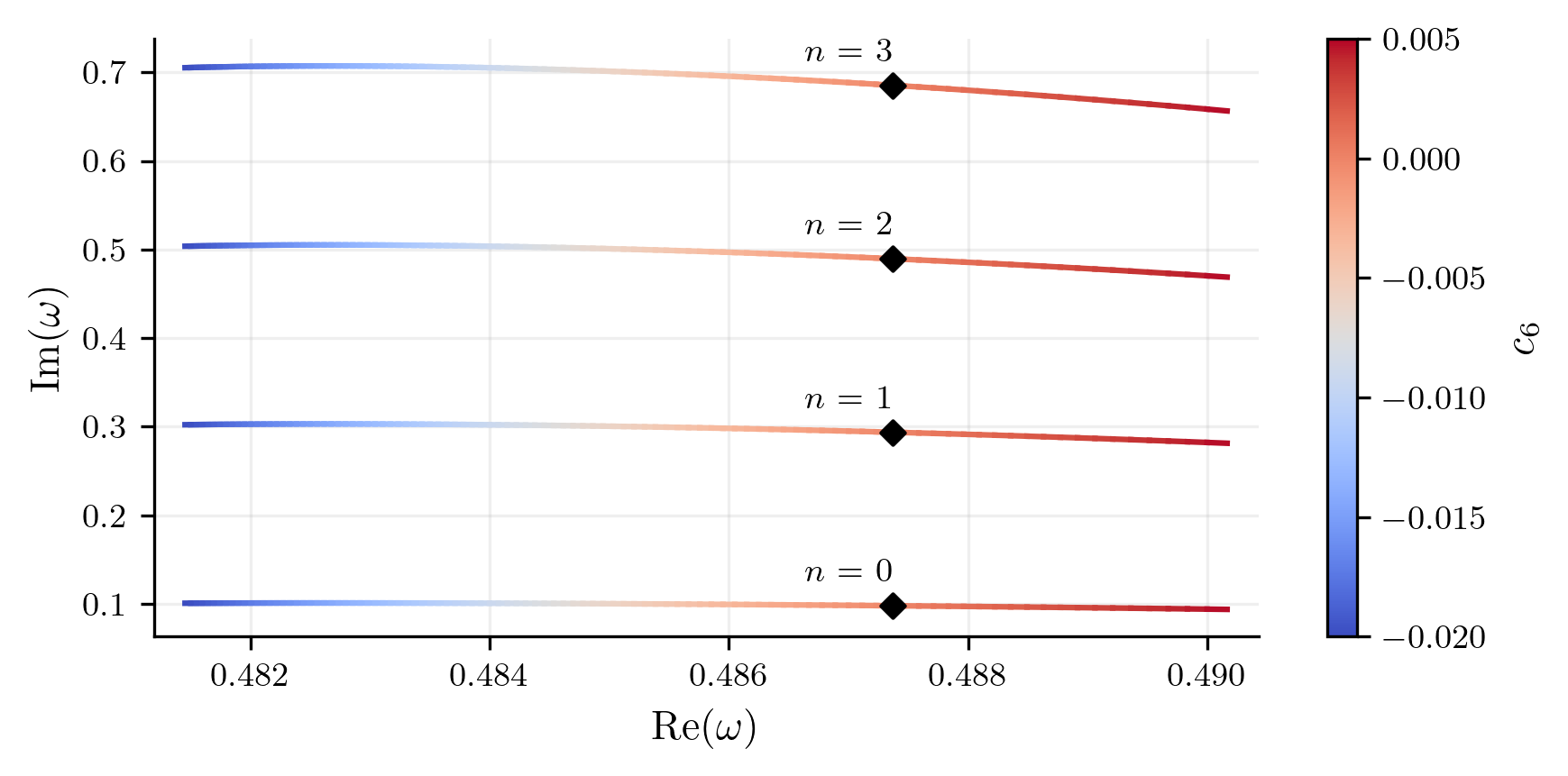}
         \caption{\footnotesize $s = 0$, $\ell = 2$}
        \label{fig:spin0_l2}  \end{subfigure}
    \begin{subfigure}{0.49\textwidth}
        \centering
        \includegraphics[width=\linewidth]{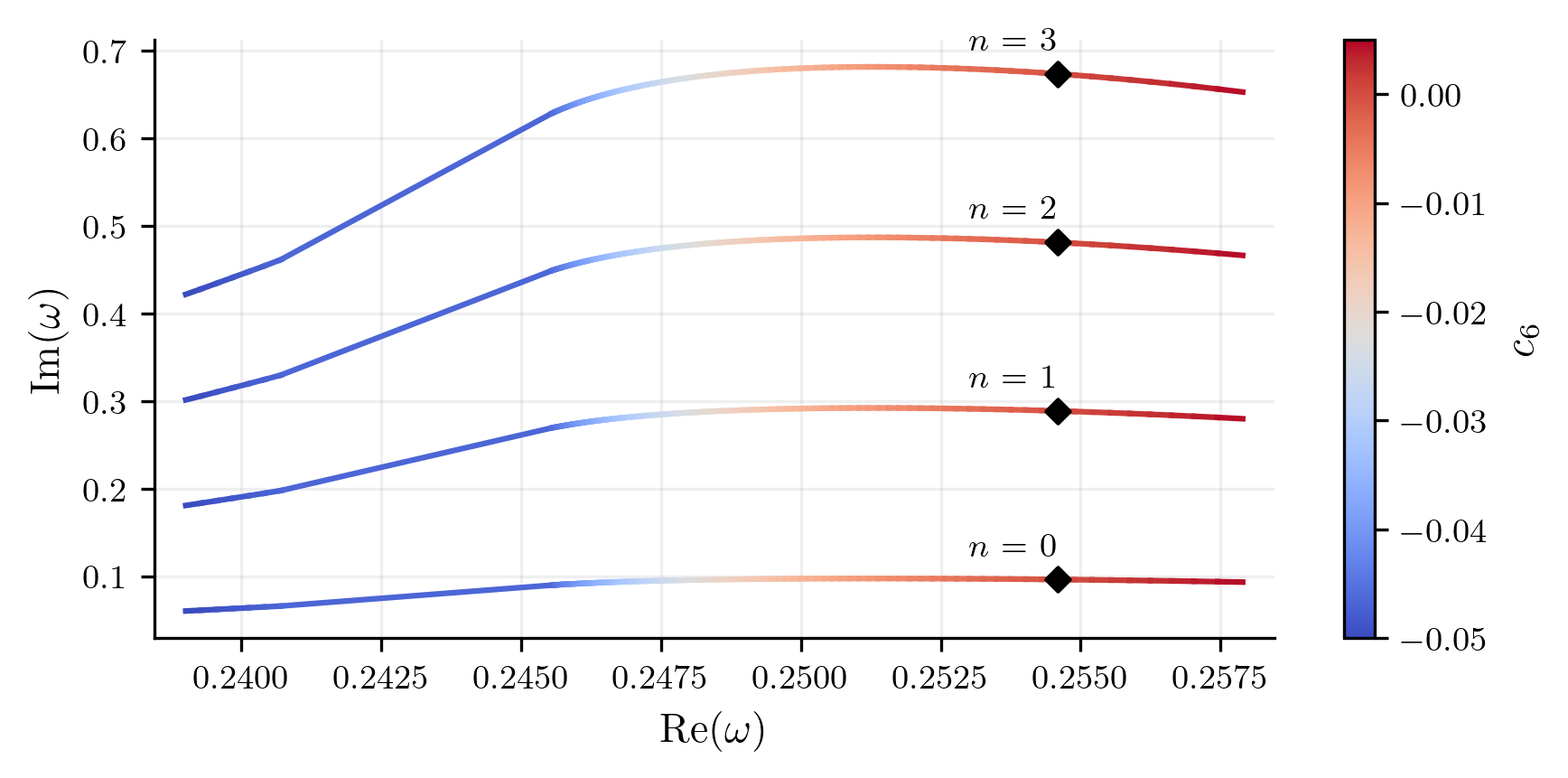}
         \caption{\footnotesize $s =\ell = 1$}
        \label{fig:spin1_l1}
    \end{subfigure}


    \begin{subfigure}{0.49\textwidth}
        \centering
        \includegraphics[width=\linewidth]{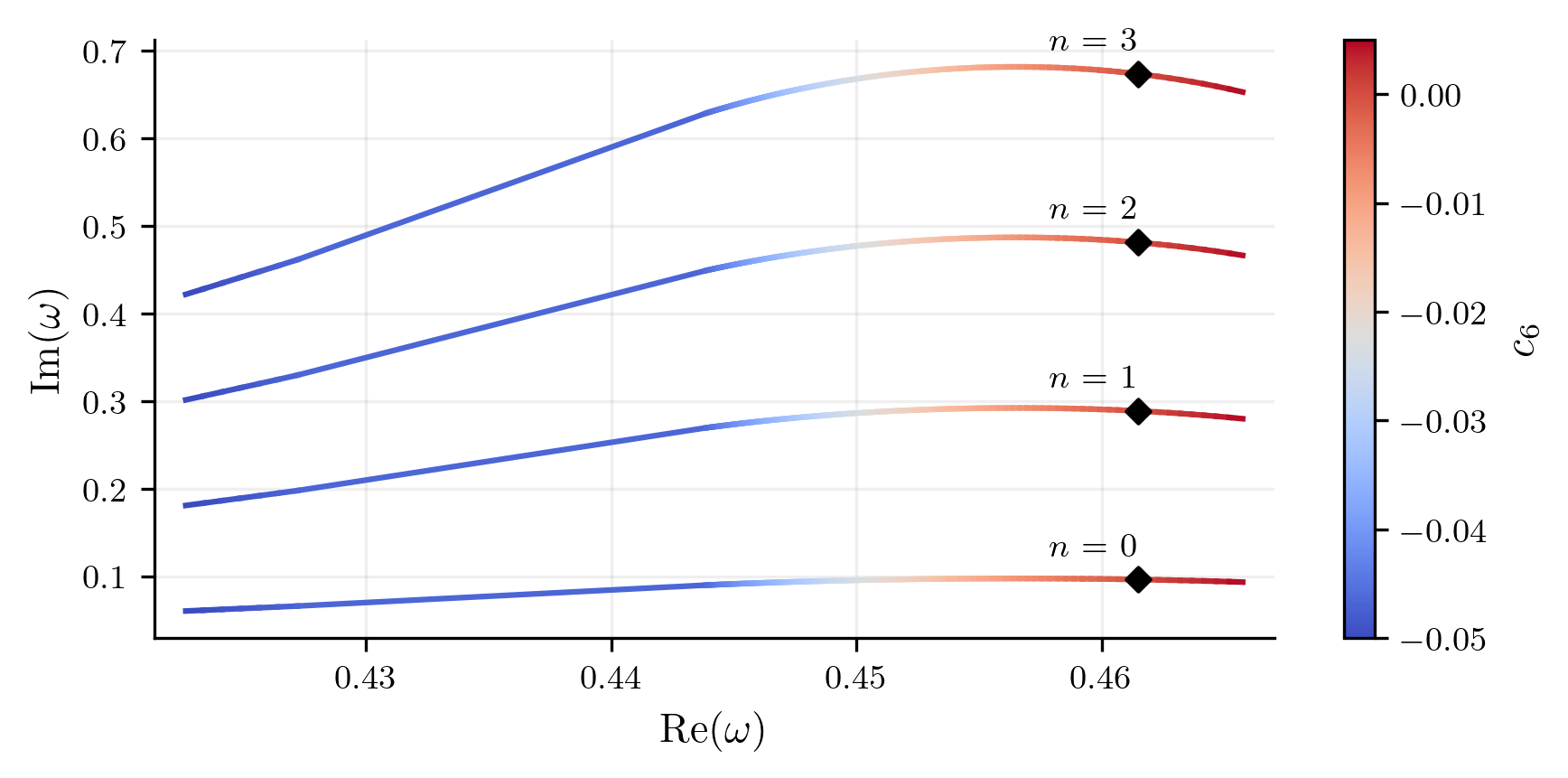}
         \caption{\footnotesize $s = 1$, $\ell = 2$}
        \label{fig:spin1_l2}  \end{subfigure}
    \begin{subfigure}{0.49\textwidth}
        \centering
        \includegraphics[width=\linewidth]{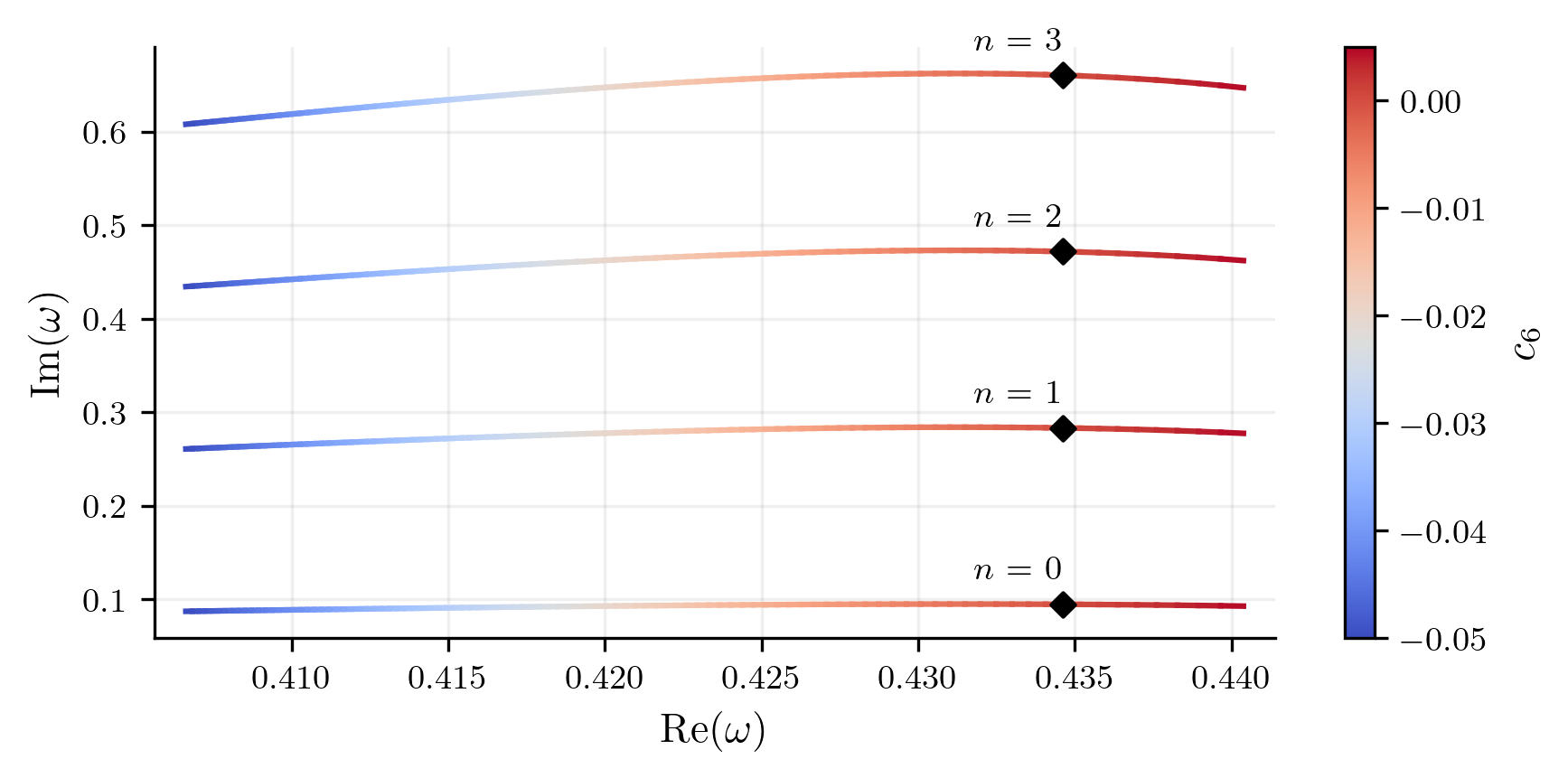}
         \caption{\footnotesize $s =\ell = 2$}
        \label{fig:spin2_l2}
    \end{subfigure}

    
     \caption{\footnotesize \clt{QN modes frequencies $\omega_n$ for spin $s=0,1,2$, with overtones up to $n=3$, and varying $c_6$ for the quantum-corrected Schwarzschild
     metric~\eqref{ck}. The black dot represents the Schwarzschild solution. The values are calculated using $G_{{\scalebox{.55}{\textsc{N}}}} = M = 1$.}}
    \label{fig:qnm012}
\end{figure}

\begin{figure}[H]
    \centering
    
    \begin{subfigure}{0.49\textwidth}
        \centering
        \includegraphics[width=\linewidth]{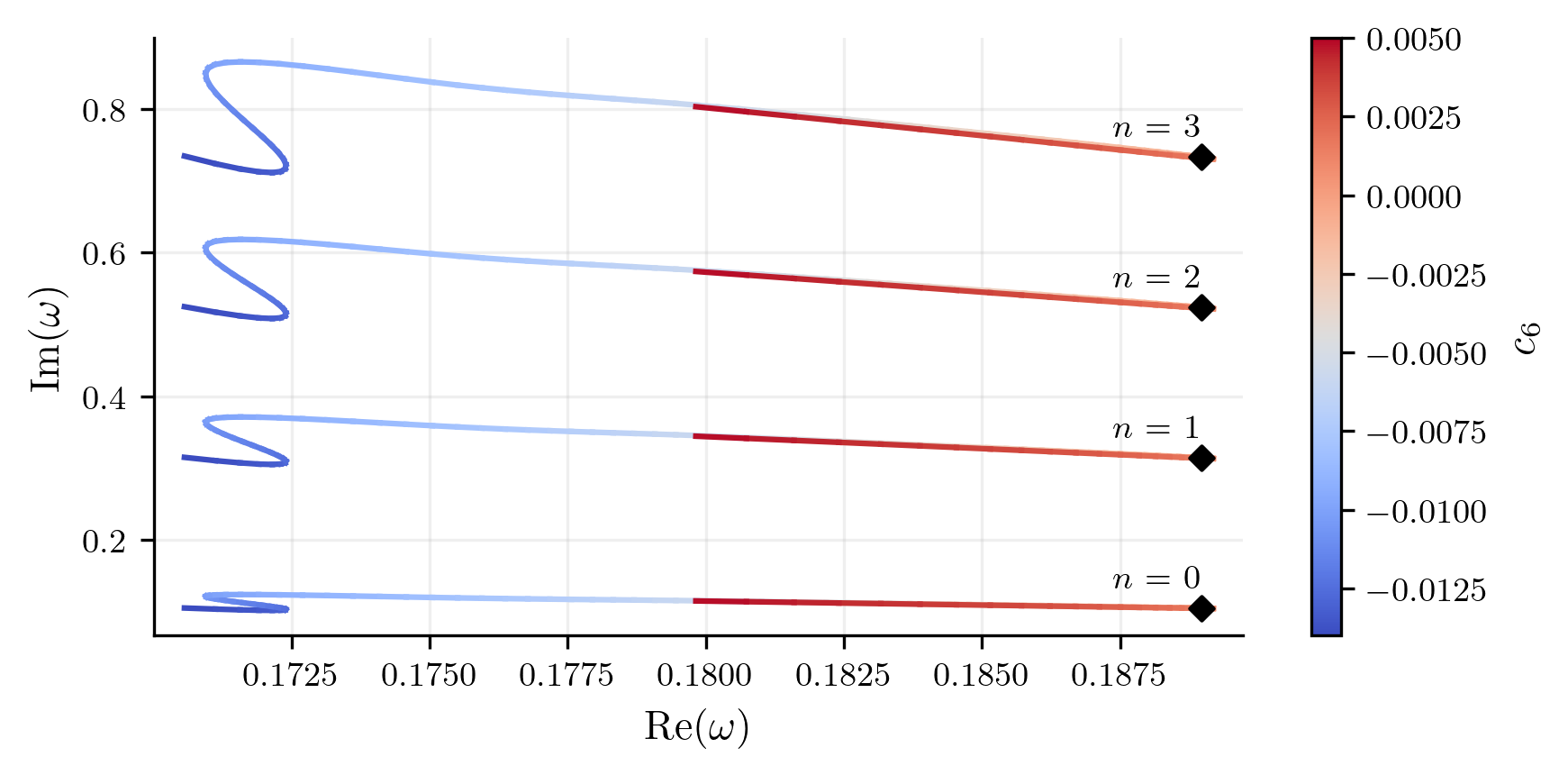}
         \caption{\footnotesize $s = \ell = 1/2$}
        \label{fig:spin0.5_l0.5}  \end{subfigure}
    \begin{subfigure}{0.49\textwidth}
        \centering
        \includegraphics[width=\linewidth]{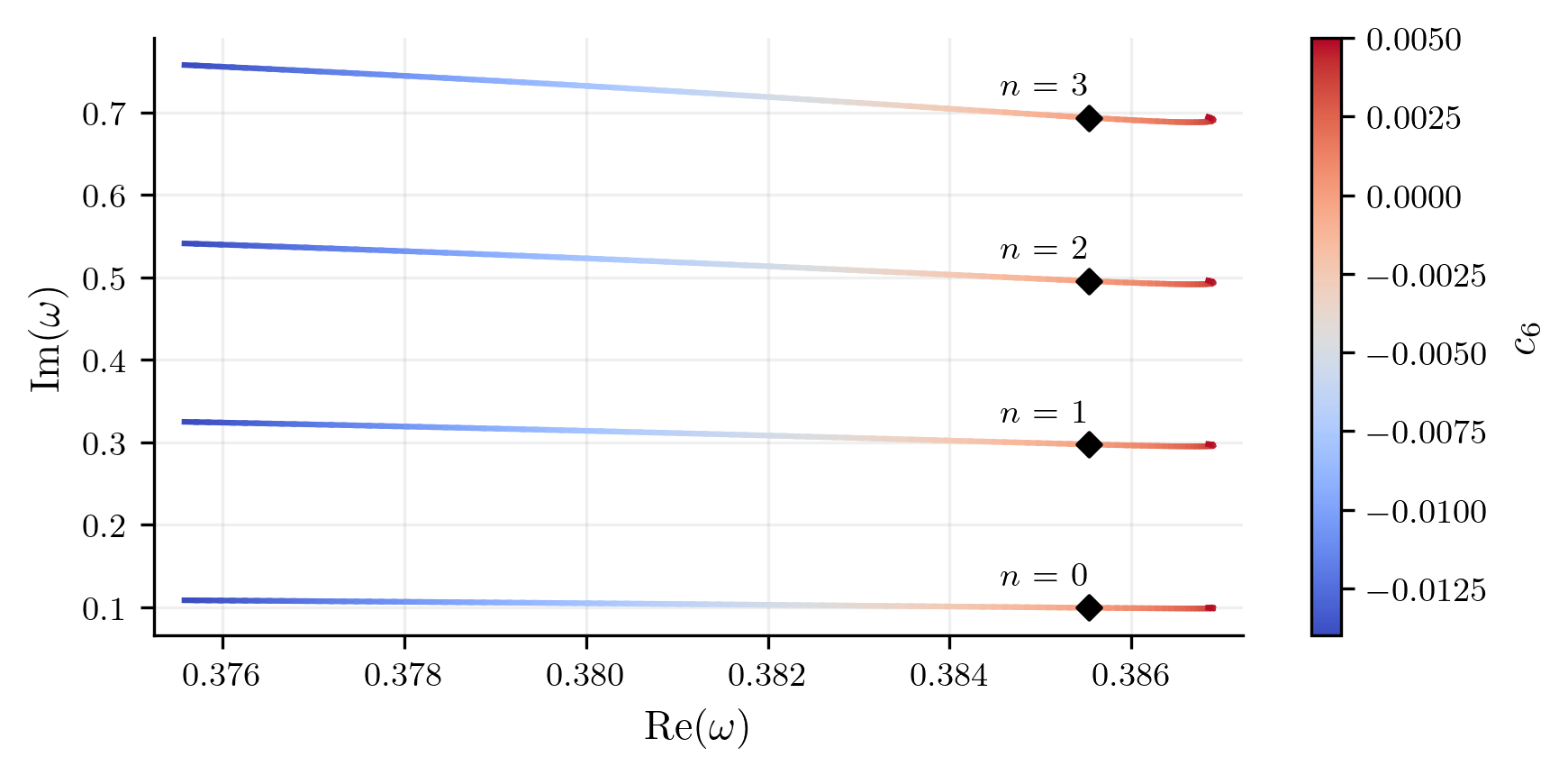}
         \caption{\footnotesize $s = 1/2$, $\ell = 3/2$}
        \label{fig:spin0.5_l1.5}
    \end{subfigure}


    \begin{subfigure}{0.49\textwidth}
        \centering
        \includegraphics[width=\linewidth]{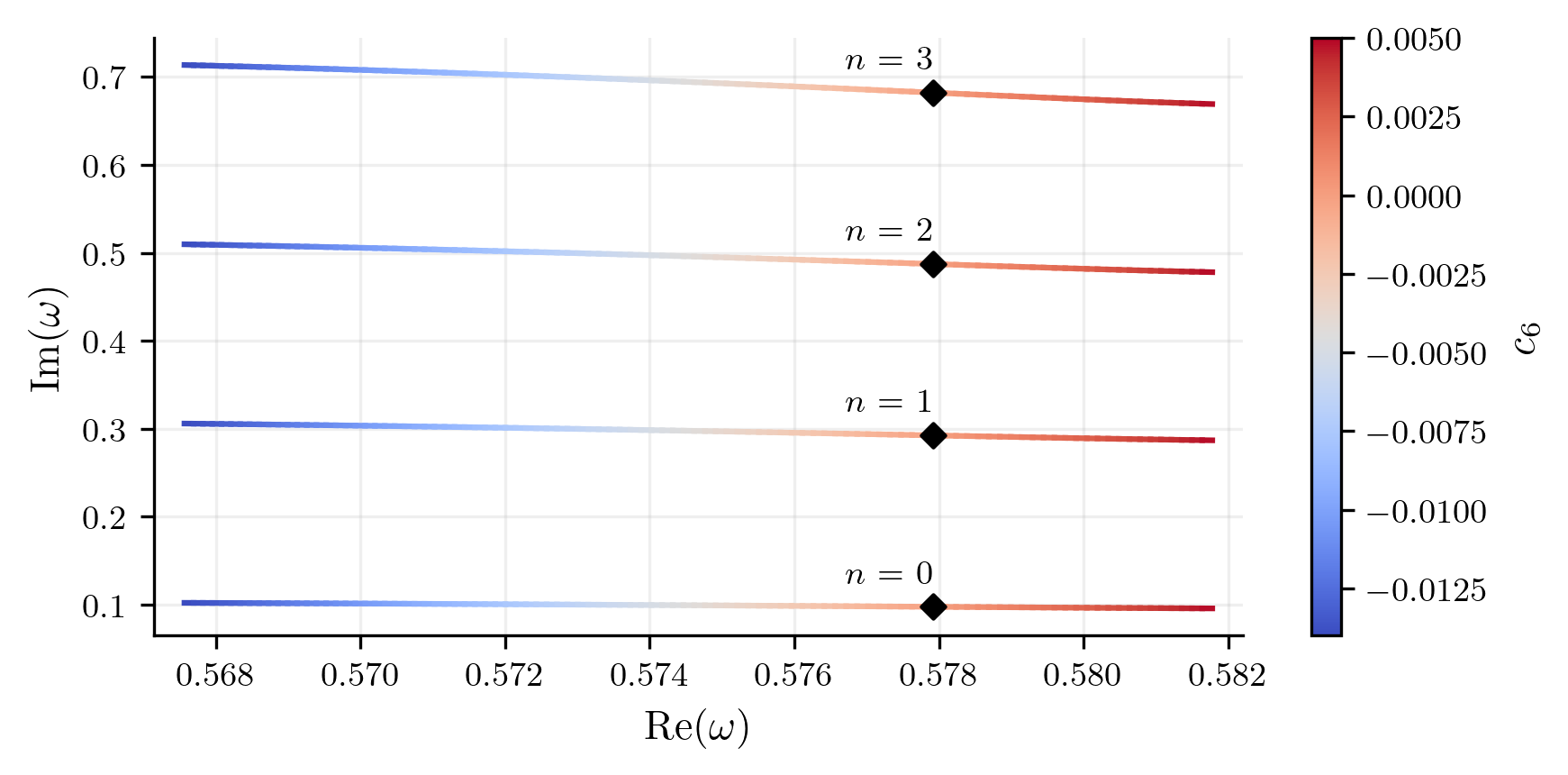}
         \caption{\footnotesize $s = 1/2$, $\ell = 5/2$}
        \label{fig:spin0.5_l2.5}  \end{subfigure}
    
    \begin{subfigure}{0.49\textwidth}
        \centering
        \includegraphics[width=\linewidth]{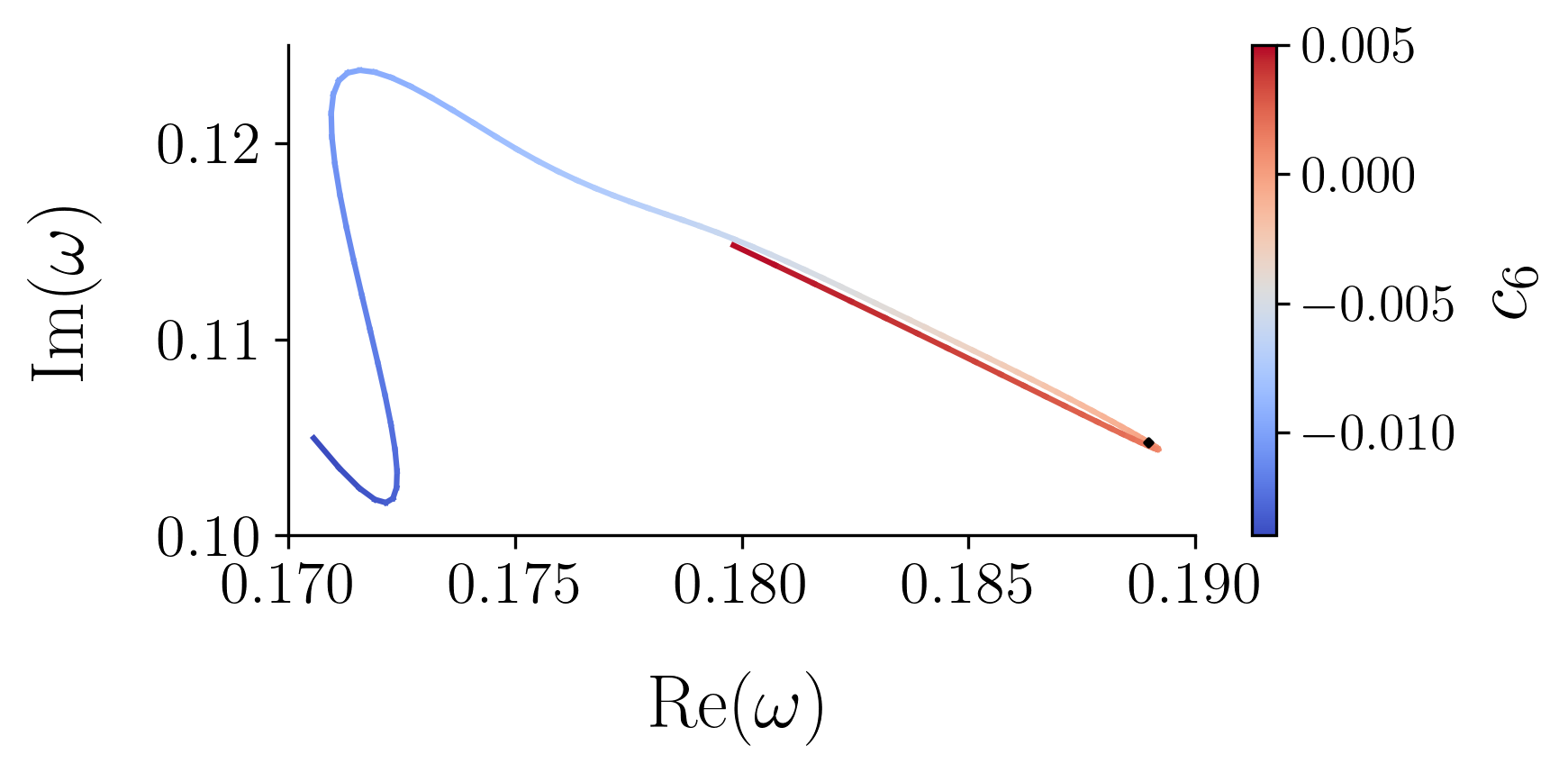}
         \caption{\footnotesize Focus on $s = \ell = {1}/{2}$ and $n=0$.}
        \label{fig:spin0.5focus}  \end{subfigure}
    \begin{subfigure}{0.49\textwidth}
        \centering
        \includegraphics[width=\linewidth]{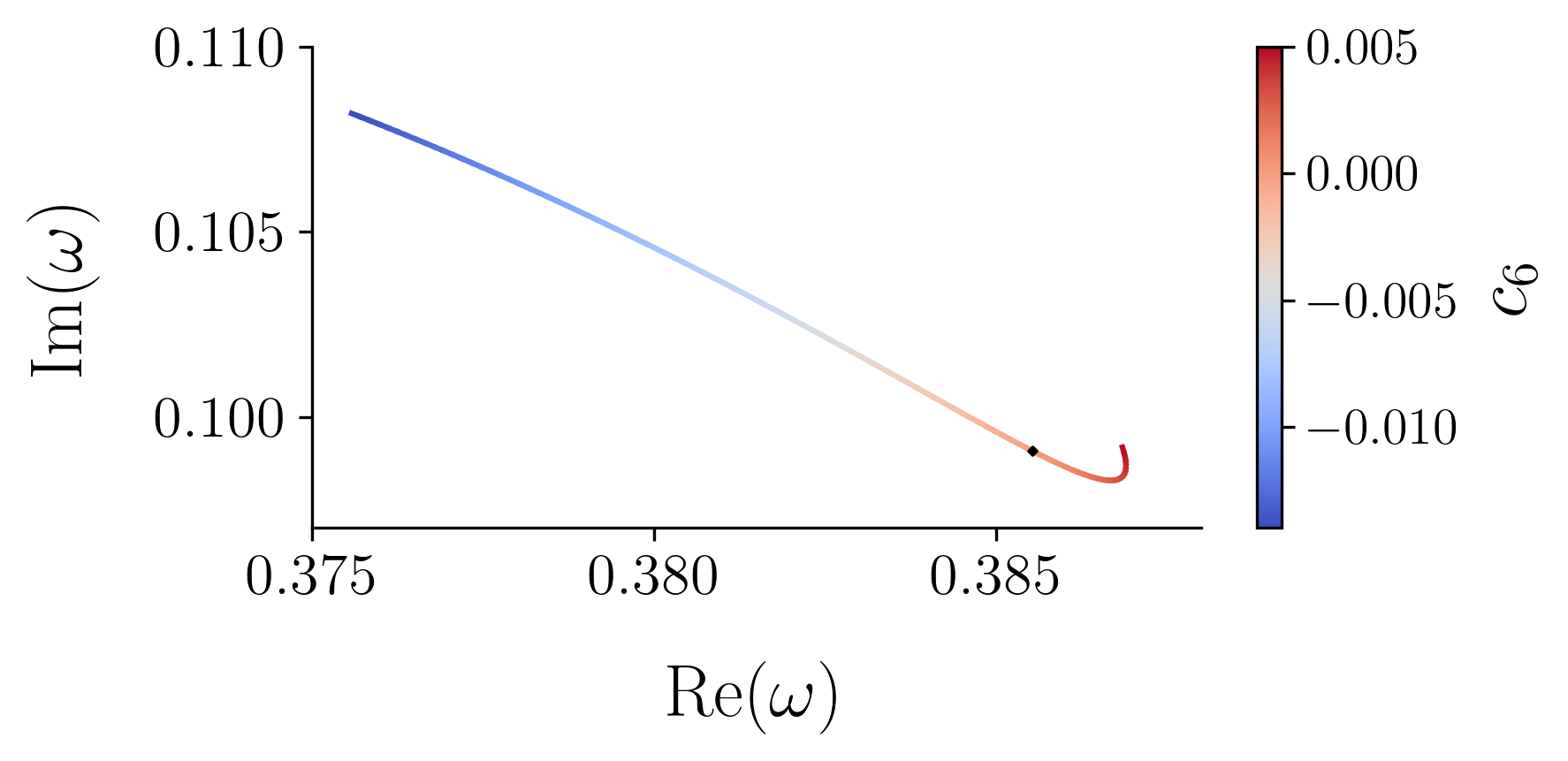}
         \caption{\footnotesize Focus on $s = {1}/{2}$, $\ell = {3}/{2}$ and $n=0$.}
        \label{fig:spin0.5l1.5focus}
    \end{subfigure}
    
     \caption{\footnotesize \clt{QN modes frequencies $\omega_n$ for spin $s=1/2$, with overtones up to $n=3$, and varying $c_6$ for the quantum-corrected Schwarzschild
     metric~\eqref{ck}. The black dot represents the Schwarzschild solution. The values are calculated using $G_{{\scalebox{.55}{\textsc{N}}}} = M = 1$.}}
    \label{fig:qnm0.5}
\end{figure}

As discussed in Ref.~\cite{Okuzumi:2007hf}, the nozzle quality factor $q_n \sim {\rm \Re}(\omega_n)/{\rm \Im}(\omega_n)$ is directly related to the number of oscillation cycles during the damping process. The quality factor can be expressed as a function of the maximum value of the P\"oschl--Teller effective potential as 
\begin{equation}
    q_n \sim \frac{{\rm \Re}(\omega_n)}{{\rm \Im}(\omega_n)} = \frac{1}{ \qty(2n+1)}\sqrt{\frac{4 V_0}{\upxi^2} - 1},
\end{equation}
and higher overtones result in lower values of $q_n$. When the quantum gravity correction parameter vanishes, $c_6 = 0$, the Schwarzschild case is recovered, which aligns with previous results in the literature, such as those presented in Ref.~\cite{Cavalcanti:2022cga}. 
The plots in Fig.~\ref{fig:qnm012} display the QN modes for bosonic perturbations with integer spin $s$, while Figs.~\ref{fig:qnm0.5} show the results for fermionic spin-$1/2$ perturbations, including higher QN mode overtones up to \clt{$n=3$}. The distributions of the QN modes and their overtones in the complex plane illustrate the relationship between the oscillation QN frequency $\Re(\omega)$ and the damping rate $\Im(\omega)$, where the black dots represent the Schwarzschild solution ($c_6 \to 0$) across all bosonic and fermionic perturbation cases here analyzed. 
The QN mode frequencies for the quantum-corrected Schwarzschild BH~\eqref{ck} clearly show deviations, driven by the quantum correction parameter $c_6$,  from their Schwarzschild counterparts, particularly for higher values of overtones $n$. Since QN ringing is typically obscured by noise after only a few damping cycles, it becomes essential to design a de Laval nozzle capable of producing QN modes with higher quality factors to ensure effective detection of QN ringing. In this case, higher quality factors are underdamped systems, combining oscillation at a specific frequency with amplitude decay of the signal. Higher quality factors favor the relative amount of damping to decrease and, in this case,  the quantum gravity-corrected BH can ring with a purer tone for an extended amount of time, which is better from the experimental point of view. 
From Figs.~\ref{fig:qnm012} and \ref{fig:qnm0.5}, it is possible to examine the QN mode spectrum and identify configurations that are more likely to yield quality factors suitable for experimental setups.

\begin{table}[H]
\centering
{\small{\begin{tabular}{||c|c|c||c|c|c|c|c||}
\hline\hline
${s}$ & ${l}$ & $\bf \emph{n}$ & $ {{\scalebox{.9}{${c_6}$}}}=-0.0014$ & $ {{\scalebox{.9}{${c_6}$}}}=-0.006$ & ${{\scalebox{.9}{${c_6}$}}}=0$ & $ {{\scalebox{.9}{${c_6}$}}}=0.003$ & ${{\scalebox{.9}{${c_6}$}}}=0.005$ \\ \hline\hline

\, 0 \,&\, 0 \,&\, 0 \,&\, 0.1006+0.1396$i$ \,&\, 0.1046+0.1304$i$ \,&\, 0.1148+0.1148$i$ \,&\, 0.1221+0.1024$i$ \,&\, 0.1271+0.0923$i$ \,\\ \hline
\, 0 \,&\, 0 \,&\, 1 \,&\, 0.1006+0.4189$i$ \,&\, 0.1046+0.3911$i$ \,&\, 0.1148+0.3445$i$ \,&\, 0.1221+0.3073$i$ \,&\, 0.1271+0.2768$i$ \,\\ \hline
\, 0 \,&\, 0 \,&\, 2 \,&\, 0.1006+0.6981$i$ \,&\, 0.1046+0.6518$i$ \,&\, 0.1148+0.5741$i$ \,&\, 0.1221+0.5122$i$ \,&\, 0.1271+0.4613$i$ \,\\ \hline
\, $\frac{1}{2}$ \,&\, $\frac{1}{2}$ \,&\, 0 \,&\, 0.1706+0.1050$i$ \,&\, 0.1793+0.1156$i$ \,&\, 0.1890+0.1048$i$ \,&\, 0.1869+0.1069$i$ \,&\, 0.1798+0.1148$i$ \,\\ \hline
\, $\frac{1}{2}$ \,&\, $\frac{1}{2}$ \,&\, 1 \,&\, 0.1706+0.3149$i$ \,&\, 0.1793+0.3467$i$ \,&\, 0.1890+0.3143$i$ \,&\, 0.1869+0.3207$i$ \,&\, 0.1798+0.3444$i$ \,\\ \hline
\, $\frac{1}{2}$ \,&\, $\frac{1}{2}$ \,&\, 2 \,&\, 0.1706+0.5248$i$ \,&\, 0.1793+0.5778$i$ \,&\, 0.1890+0.5238$i$ \,&\, 0.1869+0.5344$i$ \,&\, 0.1798+0.5740$i$ \,\\ \hline
\, 0 \,&\, 1 \,&\, 0 \,&\, 0.2975+0.1082$i$ \,&\, 0.2977+0.1052$i$ \,&\, 0.2985+0.1006$i$ \,&\, 0.2993+0.0972$i$ \,&\, 0.2999+0.0942$i$ \,\\ \hline
\, 0 \,&\, 1 \,&\, 1 \,&\, 0.2975+0.3245$i$ \,&\, 0.2977+0.3157$i$ \,&\, 0.2985+0.3019$i$ \,&\, 0.2993+0.2915$i$ \,&\, 0.2999+0.2827$i$ \,\\ \hline
\, 0 \,&\, 1 \,&\, 2 \,&\, 0.2975+0.5408$i$ \,&\, 0.2977+0.5261$i$ \,&\, 0.2985+0.5032$i$ \,&\, 0.2993+0.4859$i$ \,&\, 0.2999+0.4712$i$ \,\\ \hline
\, 1 \,&\, 1 \,&\, 0 \,&\, 0.2497+0.0970$i$ \,&\, 0.2520+0.0973$i$ \,&\, 0.2546+0.0962$i$ \,&\, 0.2564+0.0948$i$ \,&\, 0.2579+0.0933$i$ \,\\ \hline
\, 1 \,&\, 1 \,&\, 1 \,&\, 0.2497+0.2910$i$ \,&\, 0.2520+0.2919$i$ \,&\, 0.2546+0.2887$i$ \,&\, 0.2564+0.2844$i$ \,&\, 0.2579+0.2798$i$ \,\\ \hline
\, 1 \,&\, 1 \,&\, 2 \,&\, 0.2497+0.4850$i$ \,&\, 0.2520+0.4866$i$ \,&\, 0.2546+0.4811$i$ \,&\, 0.2564+0.4740$i$ \,&\, 0.2579+0.4663$i$ \,\\ \hline
\, $\frac{1}{2}$ \,&\, $\frac{3}{2}$ \,&\, 0 \,&\, 0.3756+0.1082$i$ \,&\, 0.3814+0.1032$i$ \,&\, 0.3855+0.0991$i$ \,&\, 0.3868+0.0983$i$ \,&\, 0.3868+0.0992$i$ \,\\ \hline
\, $\frac{1}{2}$ \,&\, $\frac{3}{2}$ \,&\, 1 \,&\, 0.3756+0.3246$i$ \,&\, 0.3814+0.3097$i$ \,&\, 0.3855+0.2972$i$ \,&\, 0.3868+0.2949$i$ \,&\, 0.3868+0.2976$i$ \,\\ \hline
\, $\frac{1}{2}$ \,&\, $\frac{3}{2}$ \,&\, 2 \,&\, 0.3756+0.5410$i$ \,&\, 0.3814+0.5162$i$ \,&\, 0.3855+0.4954$i$ \,&\, 0.3868+0.4915$i$ \,&\, 0.3868+0.4959$i$ \,\\ \hline
\, 0 \,&\, 2 \,&\, 0 \,&\, 0.4827+0.1010$i$ \,&\, 0.4850+0.1002$i$ \,&\, 0.4874+0.0979$i$ \,&\, 0.4889+0.0958$i$ \,&\, 0.4902+0.0938$i$ \,\\ \hline
\, 0 \,&\, 2 \,&\, 1 \,&\, 0.4827+0.3031$i$ \,&\, 0.4850+0.3005$i$ \,&\, 0.4874+0.2937$i$ \,&\, 0.4889+0.2874$i$ \,&\, 0.4902+0.2813$i$ \,\\ \hline
\, 0 \,&\, 2 \,&\, 2 \,&\, 0.4827+0.5052$i$ \,&\, 0.4850+0.5009$i$ \,&\, 0.4874+0.4895$i$ \,&\, 0.4889+0.4789$i$ \,&\, 0.4902+0.4688$i$ \,\\ \hline
\, 1 \,&\, 2 \,&\, 0 \,&\, 0.4537+0.0970$i$ \,&\, 0.4576+0.0973$i$ \,&\, 0.4615+0.0962$i$ \,&\, 0.4639+0.0948$i$ \,&\, 0.4658+0.0933$i$ \,\\ \hline
\, 1 \,&\, 2 \,&\, 1 \,&\, 0.4537+0.2910$i$ \,&\, 0.4576+0.2919$i$ \,&\, 0.4615+0.2887$i$ \,&\, 0.4639+0.2844$i$ \,&\, 0.4658+0.2798$i$ \,\\ \hline
\, 1 \,&\, 2 \,&\, 2 \,&\, 0.4537+0.4850$i$ \,&\, 0.4576+0.4866$i$ \,&\, 0.4615+0.4811$i$ \,&\, 0.4639+0.4740$i$ \,&\, 0.4658+0.4663$i$ \,\\ \hline
\, 2 \,&\, 2 \,&\, 0 \,&\, 0.4236+0.0936$i$ \,&\, 0.4293+0.0945$i$ \,&\, 0.4346+0.0944$i$ \,&\, 0.4379+0.0935$i$ \,&\, 0.4404+0.0924$i$ \,\\ \hline
\, 2 \,&\, 2 \,&\, 1 \,&\, 0.4236+0.2807$i$ \,&\, 0.4293+0.2836$i$ \,&\, 0.4346+0.2831$i$ \,&\, 0.4379+0.2806$i$ \,&\, 0.4404+0.2773$i$ \,\\ \hline
\, 2 \,&\, 2 \,&\, 2 \,&\, 0.4236+0.4678$i$ \,&\, 0.4293+0.4727$i$ \,&\, 0.4346+0.4718$i$ \,&\, 0.4379+0.4676$i$ \,&\, 0.4404+0.4622$i$ \,\\ \hline
\, $\frac{1}{2}$ \,&\, $\frac{5}{2}$ \,&\, 0 \,&\, 0.5676+0.1020$i$ \,&\, 0.5732+0.0998$i$ \,&\, 0.5779+0.0975$i$ \,&\, 0.5803+0.0962$i$ \,&\, 0.5818+0.0956$i$ \,\\ \hline
\, $\frac{1}{2}$ \,&\, $\frac{5}{2}$ \,&\, 1 \,&\, 0.5676+0.3059$i$ \,&\, 0.5732+0.2994$i$ \,&\, 0.5779+0.2924$i$ \,&\, 0.5803+0.2887$i$ \,&\, 0.5818+0.2868$i$ \,\\ \hline
\, $\frac{1}{2}$ \,&\, $\frac{5}{2}$ \,&\, 2 \,&\, 0.5676+0.5098$i$ \,&\, 0.5732+0.4991$i$ \,&\, 0.5779+0.4873$i$ \,&\, 0.5803+0.4812$i$ \,&\, 0.5818+0.4780$i$ \,\\ \hline

\hline\hline
\end{tabular}}}

 \caption{\footnotesize QN modes frequencies $\omega_n$ for the quantum-corrected Schwarzschild metric~\eqref{ck}, for varying value of $c_6$, $s$ and $\ell$, for overtones $n=0,1 \text{ and } 2$. The values are calculated using $G_{{\scalebox{.55}{\textsc{N}}}} = M = 1$.}

\label{tabela}

\end{table}

Fig.~\ref{fig:qnm012} depicts QN modes for integer spins. For scalar perturbations with $s = 0$ in Figs.~\ref{fig:spin0_l0}-\ref{fig:spin0_l2}, the most pronounced effect of the quantum correction parameter $c_6$ is observed in the $\ell = 0$ configuration, where negative $c_6$ values (blue shades) induce a leftward shift in the QN modes distribution. Quantitatively, the most negative $c_6$ value ($\approx -0.020$) reduces the real part $\Re(\omega)$ from approximately 0.115 to 0.101, representing a 12.4\% decrease from the Schwarzschild value. This relevant feature represents a distinct physical signature imprint in the analog GW wave.  This modification significantly affects the quality factor of the fundamental mode ($n = 0$), reducing it by approximately 28\% (from $\approx1$ to 0.725) at the most negative $c_6$ values. The $\ell = 1$ and $\ell = 2$ cases show progressively less sensitivity to $c_6$, with maximum deviations of 8\% and 5\% respectively, suggesting that higher angular momentum configurations are more resilient to quantum corrections in the scalar channel.
Electromagnetic perturbations, corresponding to $s = 1$ in Fig.~\ref{fig:spin1_l1}-\ref{fig:spin1_l2}, demonstrate lesser sensitivity to $c_6$ variations,  compared to the scalar case. For $\ell = 1$, the most negative $c_6$ value ($\approx -0.05$) shifts $\Re(\omega)$ from 0.255 to approximately 0.239, representing a 6.1\% reduction. The quality factor, on the other hand, increases substantially by approximately 50\% for the fundamental mode, from 2.646 to 3.966. The $\ell = 2$ configuration shows more sensitivity to $c_6$, with $\Re(\omega)$ decreasing from 0.461 to approximately 0.423, representing an 8.4\% deviation. 
Gravitational perturbations ($s = 2$, Fig.~\ref{fig:spin2_l2}) QN modes are similar to the $s=1$ case. The most negative $c_6$ ($\approx -0.05$) reduces $\Re(\omega)$ from 0.435 to approximately 0.407, a 6.4\% deviation from Schwarzschild. The quality factor of the fundamental mode increases by approximately 2\%, from 4.606 to 4.681.

Fig.~\ref{fig:qnm0.5} shows the fermionic perturbations ($s = 1/2$, Fig.~\ref{fig:spin0.5_l0.5}-\ref{fig:spin0.5_l2.5}). This case exhibits moderate sensitivity to variations of $c_6$  due to the quantum gravity effects. For $\ell = 1/2$, the most negative $c_6$ ($\approx -0.0125$) produces a reduction in $\Re(\omega)$ from 0.189 to approximately 0.171, a 9.7\% deviation from Schwarzschild. This corresponds to a quality factor reduction of approximately 10\% for the fundamental mode, from 1.804 to 1.623. The effect diminishes for $\ell$ = 3/2 and $\ell$ = 5/2, with maximum deviations of approximately 3\% and 2\% respectively, reinforcing the pattern that higher angular momentum modes display reduced sensitivity to quantum corrections. By taking a close look to the shape of $\Im(\omega) \times \Re(\omega)$ as $c_6$ varies, the non-monotonic shape suggests a maximum value for $\Re(\omega)$ at the threshold $c_6\approx 0.002$ for $\ell=1/2$ (Fig.~\ref{fig:spin0.5focus}) and $c_6\approx 0.004$ for $\ell=3/2$ (Fig.~\ref{fig:spin0.5l1.5focus}), just as discussed with effective potential. This behaviour does not appear in the analysis of higher multipoles $\ell > 3/2$.

Comparing the types of bosonic and fermionic perturbations on the quantum gravitational corrected BH with $c_6 \in \qty[-0.015,0.005]$, in order for all effective potentials to be well behaved, one can determine a rank of sensitivity to quantum corrections. Scalar and fermionic perturbations demonstrate the greatest responsiveness to variations of the quantum correction parameter $c_6$, with a maximum deviation of the quality factor $q_n\sim \Re(\omega)/\Im(\omega)$ about 38\% for scalar and 13\% for fermionic fields, followed by electromagnetic (4.5\%) and gravitational perturbations (4.1\%). 
Additionally, the impact of $c_6$ in $\Re(\omega)$ or $\Im(\omega)$ is prominently smaller, indicating that quantum corrections to Schwarzschild geometry manifest more in the resonant properties of quantum gravitational-corrected BHs rather than in the individual frequency or damping characteristics. These quantitative deviations suggest that scalar and fermionic channels would provide the most robust observational signatures of quantum gravitational effects in BH perturbations, potentially enabling constraints on quantum gravity models through precision measurements of BH ringdown signals.

{\color{black}
\section{Conclusions}
\label{sec5}

Acoustic BHs that carry quantum gravitational corrections at third order in the curvature can be realized through stable analog transonic fluid flows within a de Laval nozzle, with the quantum-corrected Schwarzschild metric~\eqref{ck} employed to define the analog de Laval nozzle properties in laboratory settings. The QN modes and eigenfrequencies of sound waves in this quantum gravitational-corrected analog BH system were computed, demonstrating their potential to experimentally probe quantum gravitational corrections to BH geometry. Crucially, the wave equations governing scalar and fermionic perturbations in the quantum-corrected BH were shown to map directly onto those describing acoustic perturbations in the de Laval nozzle, with equivalent effective potentials underpinning both systems. \blt{We also investigated a vast range of BH masses,  from Planck masses to astrophysical ones, showing that quantum gravitational corrections are more evident in primordial BHs. Despite this, astrophysical and stellar BHs were shown to have their analog de Laval nozzle quantities, like the temperature, pressure, exhaust velocity, Mach number, fluid density, and thrust coefficients to be more prominently modified by negative values of the quantum gravitational correction parameter $c_6$, obeying Eq. (\ref{boundc6}). These quantum gravitational corrections, although somehow tiny, might be macroscopically noticeable in laboratories. Quantum gravitational effects of the nozzle area are almost imperceptible. }
The quantum gravity parameter $c_6$ was found to modulate the nozzle geometric profile, the thermodynamic variables, the nozzle geometry, and aerodynamic features, as the Mach number and the thrust coefficient, as well as the QN mode spectrum with special relevance to their overtones. Variations in the parameter $c_6$ induced shifts in the effective potential peak location (Fig.~\ref{fig:veff}) and altered aerodynamic profiles (Figs.~\ref{fig:shape}--\ref{fig:C_F}). Quality factors $q_n$, derived from QN frequencies, show raised sensitivity to $c_6$ for spin-0 and spin-${1}/{2}$ perturbations, with deviations up to 38\% and 13\%, respectively, compared to electromagnetic and gravitational modes. This emphasizes scalar and fermionic channels as optimal probes for quantum gravity effects in analog experiments. However, spin-2 GW perturbations remain beyond the scope of this hydrodynamic analog. By modelling the ringdown phase of analog acoustic BHs that carry quantum gravitational corrections, with our accurate numerical relativity simulations, we conclude that the fundamental mode alone ($n=0$) does not suffice to recover all the features of the 
quantum gravitational-corrected BH. Higher overtones have also been considered, which modify the QN spectrum and carry signatures of quantum gravity effects. It provides an
unbiased estimate of the quantum-corrected BH remnant. The inclusion of higher overtones permits modelling the quantum-corrected BHs ringdown signal for an arbitrary time interval beyond the peak strain GW amplitude. The higher overtones are shown not to be subdominant for quantum-corrected BHs and play a prominent role in modelling the acoustic BH ringdown. 
\par

A natural extension of this work involves adding rotation to the quantum-corrected BH model. This would need mapping gravitational perturbations of rotating spacetimes onto quasi-one-dimensional transonic flows, potentially enabling laboratory studies of frame-dragging and ergoregion instabilities. Further investigations could also explore other higher-order curvature corrections, implementing quantum gravity effects to refine the nozzle response to Planck-scale physics.

\subsection*{Acknowledgements}
R.C. is partially supported by the INFN grant FLAG and his work has also been carried out
in the framework of activities of the National Group of Mathematical Physics (GNFM, INdAM).
C.N.S.~thanks CAPES (Grant No.~001).
R.d.R~thanks the S\~ao Paulo Research Foundation -- FAPESP (Grants No.~2021/01089-1
and No.~2024/05676-7) and the National Council for Scientific and Technological Development (CNPq) (Grants No. 303390/2019-0 and No. 401567/2023-0).

}

\appendix
\section{Varying the BH mass: quantum gravitational effects for small values of $c_6$}
\label{ap10}
For all figures in this Appendix, for the other values of $s$ and $\ell$ beyond $s$-wave perturbations studied in this paper, the results are quite identical. Therefore, we will focus on the $s$-wave perturbations hereon. 
\clt{All plots in this Section show the relative deviation from Schwarzschild, expressed as a percentage: 
\[\Delta X = 100\% \times \left(\frac{X_{\text{corrected}}}{X_{\text{Schwarzschild}}} - 1\right).\]}

\begin{figure}[H]
    \centering
    \begin{subfigure}{0.49\textwidth}
        \centering
        \includegraphics[width=\linewidth]{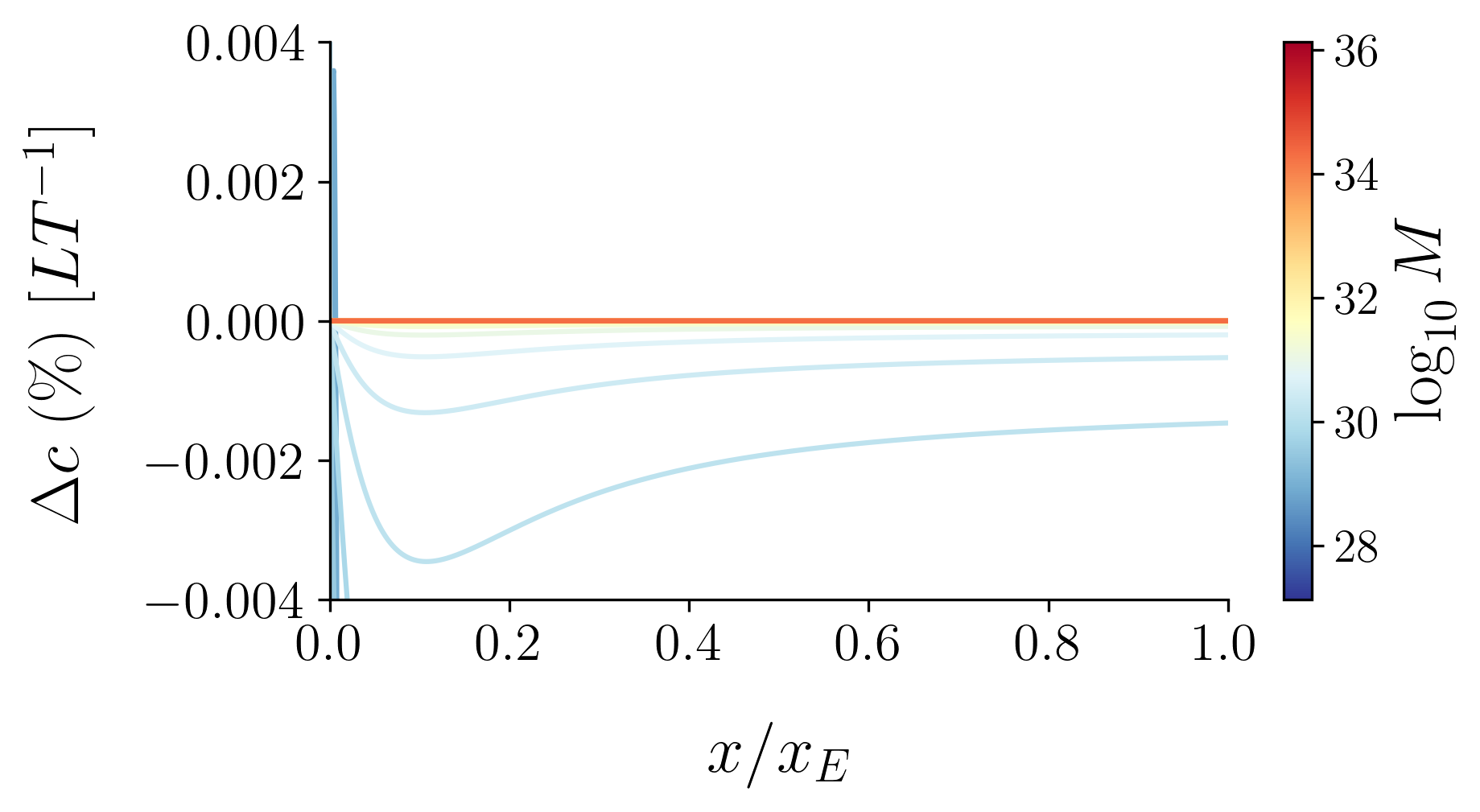}
         \caption{\footnotesize $s = \ell = 0$, $c_6 = 0.005$.}
        \label{fig:veffa101}
    \end{subfigure}
    \begin{subfigure}{0.49\textwidth}
        \centering
        \includegraphics[width=\linewidth]{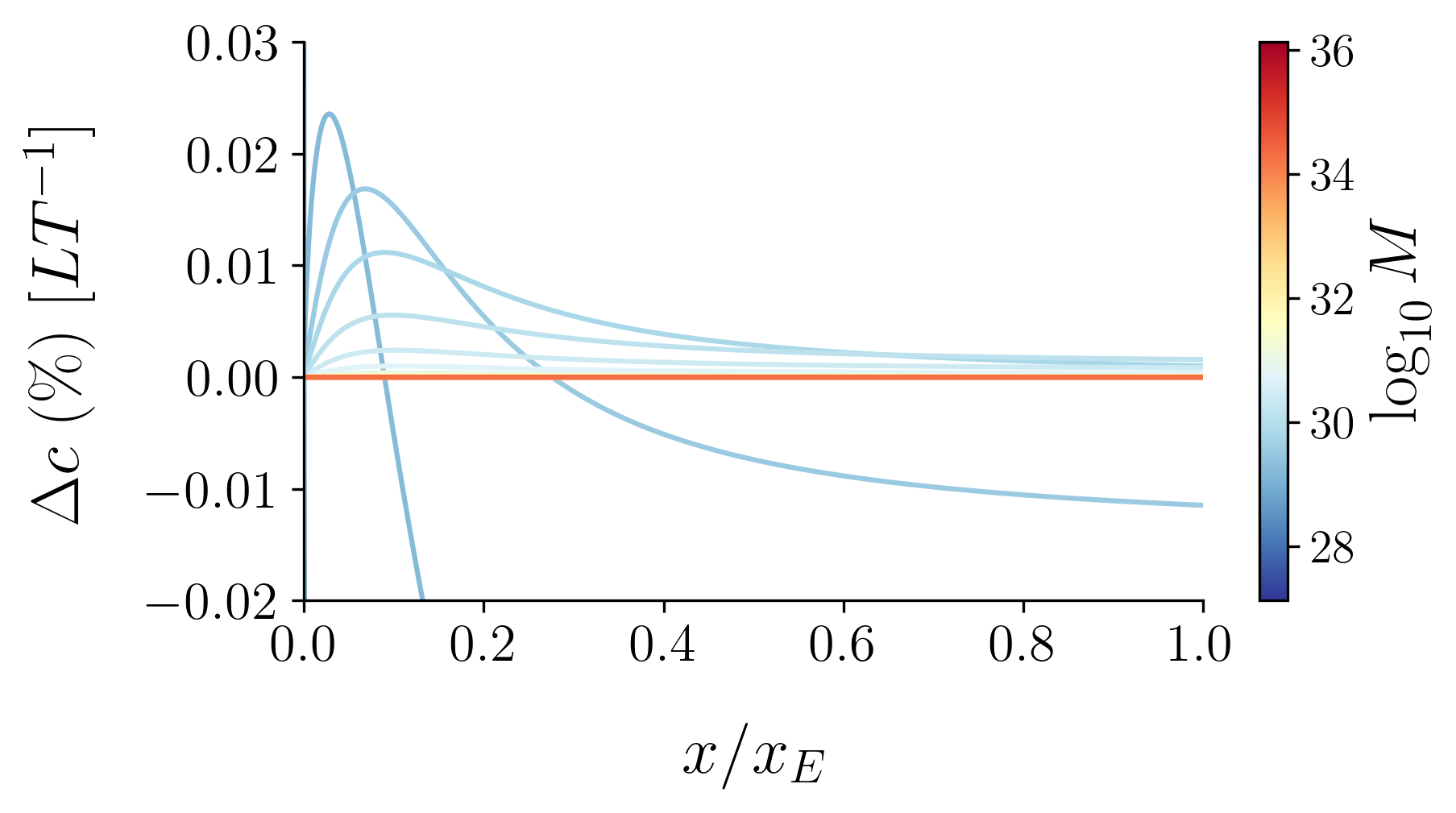}
         \caption{\footnotesize $s = 0$ and $\ell=0$, $c_6 = -0.01$.}
        \label{fig:veffb1101}  \end{subfigure}
     \caption{\footnotesize  Effective exhaust velocity as a function of the longitudinal nozzle coordinate $x$, for  
 the quantum-corrected Schwarzschild metric~\eqref{ck}, with  BH masses in the range $10^{27}$ kg $< M < 10^{36}$ kg. \clt{Relative deviation from Schwarzschild.}}
    \label{fig:C_Fev}
\end{figure}

 Figs. \ref{fig:veffa101}
    and \ref{fig:veffb1101}  illustrate the effective exhaust velocity as a function of the longitudinal nozzle coordinate $x$, for  
$s = \ell = 0$ and $c_6 = 0.005$ (Fig. \ref{fig:veffa101}) [$c_6 = -0.01$ (Fig. \ref{fig:veffb1101})]. 
The exhaust velocity was evaluated for a representative range of BH masses varying systematically in the range \clt{$10^{27}$ kg $< M < 10^{36}$ kg. Fig.~\ref{fig:veffa101} considers the differences of the nozzle exhaust velocity, revealing that the decrease of the BH mass makes the exhaust velocity slightly decrease, mainly near the nozzle throat, with maximal decrease around $M\sim 10^{28}$ kg}. On the other hand, Fig. \ref{fig:veffb1101} regards the negative value $c_6 = -0.01$, illustrating the behaviour of the exhaust velocity. The decrease of the BH mass, \clt{until $M\sim M_\odot$ makes the exhaust velocity slightly increase.} 
Although the asymptotic values of the exhaust velocity are practically independent of the BH mass in  Fig. \ref{fig:veffa101}, Fig.  \ref{fig:veffb1101} depicts a discernible small value ($\ll 1\%$) of the 
exhaust velocity at the nozzle exit.

\begin{figure}[H]
    \centering
    \begin{subfigure}{0.49\textwidth}
        \centering
        \includegraphics[width=\linewidth]{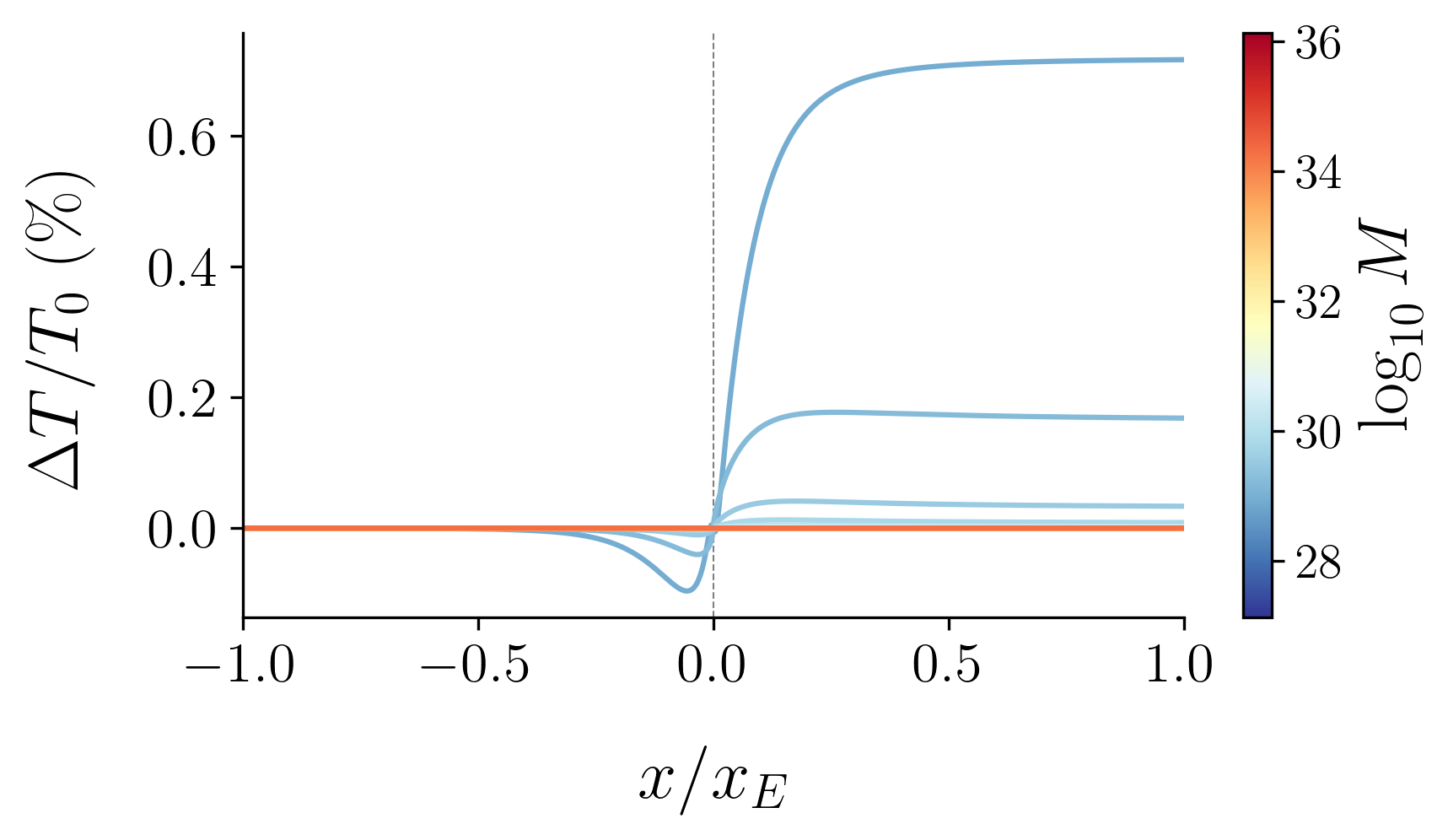}
         \caption{\footnotesize $s = \ell = 0$, $c_6 = 0.005$.}
        \label{fig:veffa101t}
    \end{subfigure}
    \begin{subfigure}{0.49\textwidth}
        \centering
        \includegraphics[width=\linewidth]{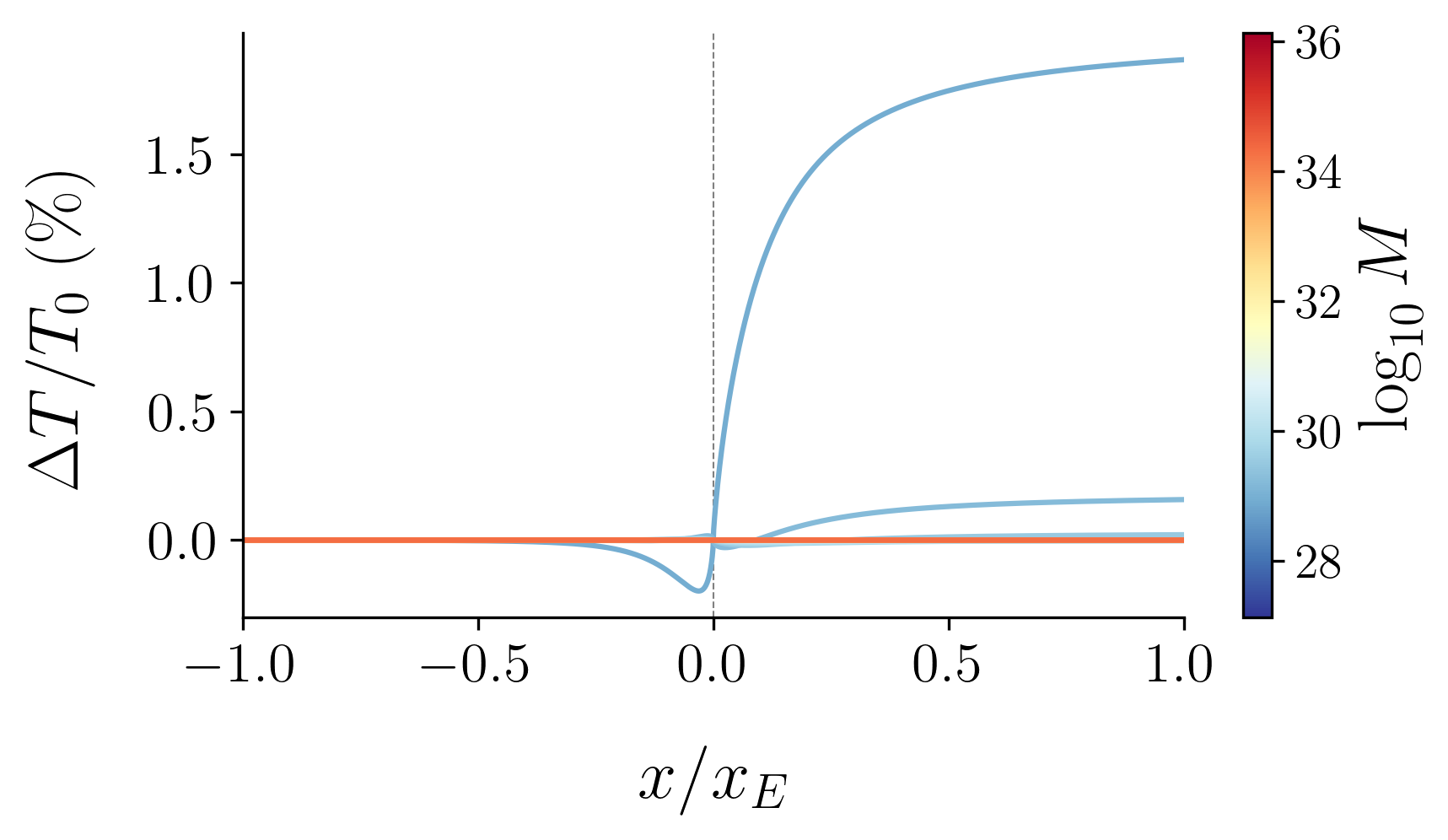}
         \caption{\footnotesize $s = 0$ and $\ell=0$, $c_6 = -0.01$.}
        \label{fig:veffb1101t}  \end{subfigure}
             \caption{\footnotesize  Relative temperature as a function of the longitudinal nozzle coordinate $x$, for  
 the quantum-corrected Schwarzschild metric~\eqref{ck}, with  BH masses in the range $10^{27}$ kg $< M < 10^{36}$ kg.\clt{Relative deviation from Schwarzschild.}}
    \label{fig:C_Ft}
\end{figure}

Now, the relative temperature to the throat can be analysed as a function of the longitudinal nozzle coordinate $x$, for values of $c_6$ obeying Eq. (\ref{boundc6}). 
The value  $c_6 = 0.005$ was picked in Fig. \ref{fig:veffa101t} and $c_6 = -0.01$ is chosen in Fig. \ref{fig:veffb1101t}. The BH mass is evaluated for distinct stellar and astrophysical masses in the range $10^{27}$ kg $< M < 10^{36}$ kg.
\noindent Fig. \ref{fig:veffa101t} shows that for $x>0$, although differences of temperatures are tiny, they are more prominent for stellar BHs, and their temperature present the same profile in the range of mass investigated. One concludes that stellar BHs dictate the analogue nozzles with slightly higher temperatures, regarding the fixed value $c_6=0.005$. Fig. \ref{fig:veffb1101t}, regarding $c_6 = -0.01$, also evinces tiny differences in the fluid flow temperatures in the nozzle \clt{for masses greater than $M\sim M_\odot$. On the other hand, the most prominent changes, for stellar BHs, make the temperature marginally increase along the longitudinal direction in the nozzle. Stellar BHs present the highest variations in the temperature profile, mainly at $x > 0\,x_{\scalebox{.67}{\textsc{e}}}$}.

\begin{figure}[H]
    \centering
    \begin{subfigure}{0.49\textwidth}
        \centering
        \includegraphics[width=\linewidth]{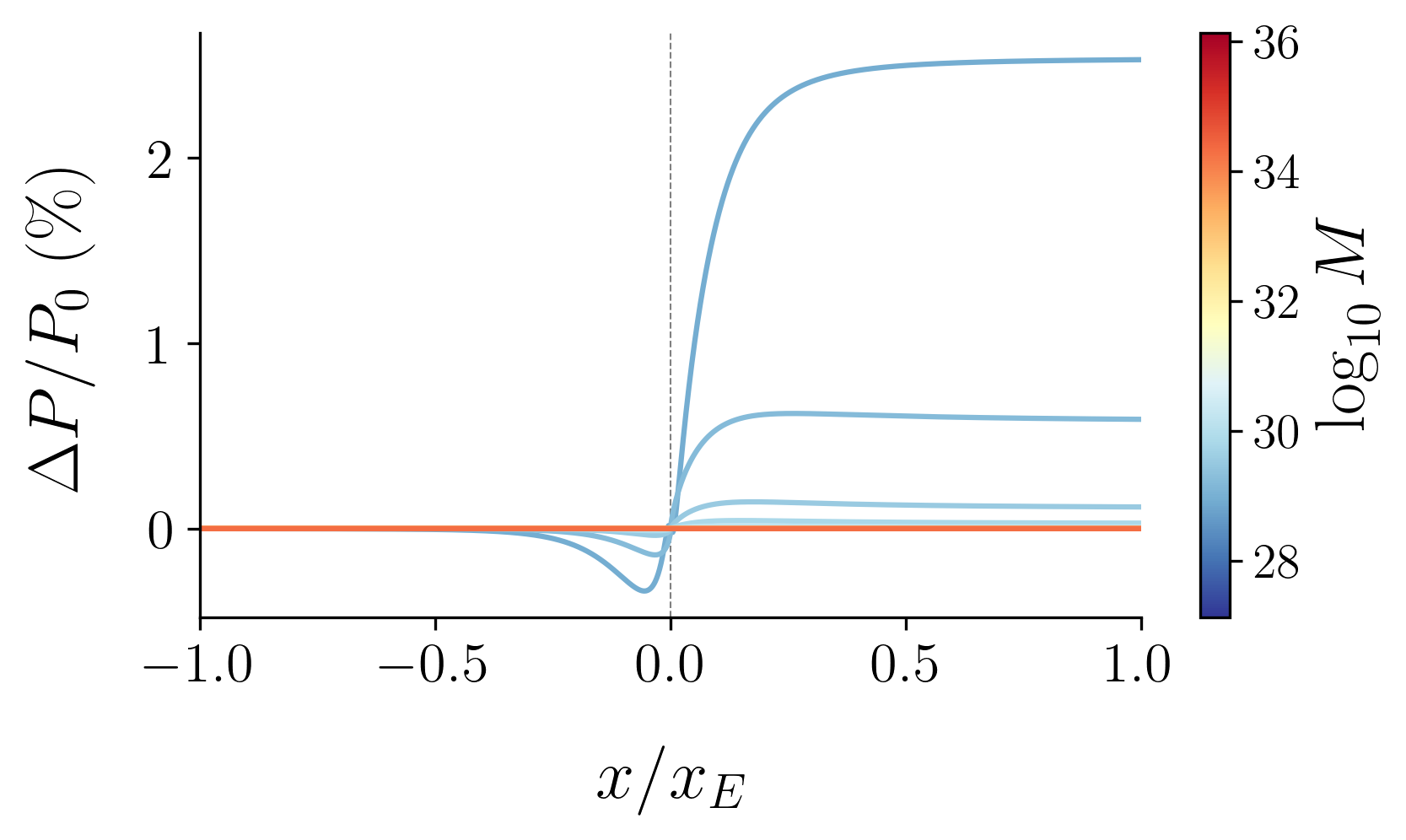}
         \caption{\footnotesize $s = \ell = 0$, $c_6 = 0.005$.}
        \label{fig:veffa1p}
    \end{subfigure}
    \begin{subfigure}{0.49\textwidth}
        \centering
        \includegraphics[width=\linewidth]{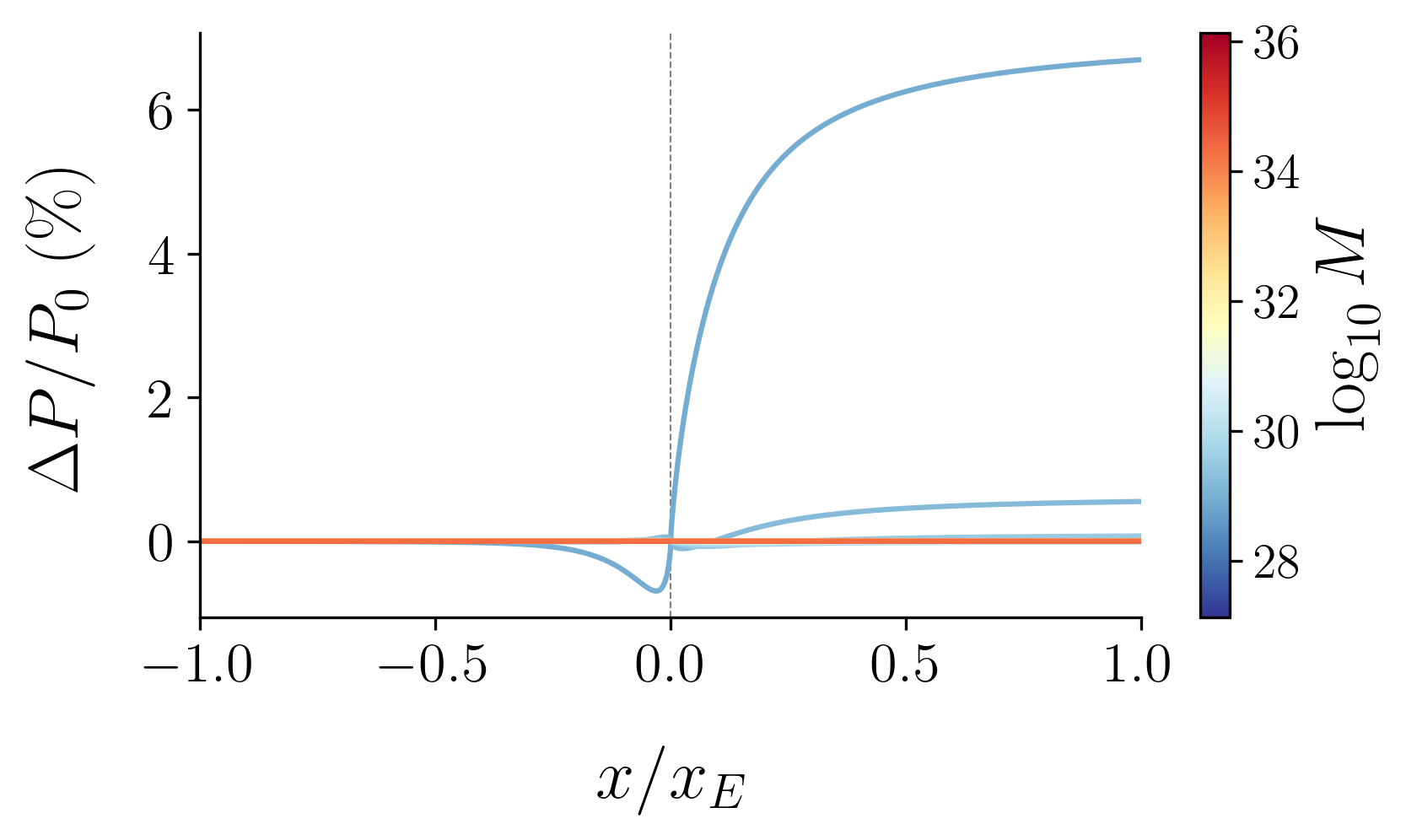}
         \caption{\footnotesize $s = 0$ and $\ell=0$, $c_6 = -0.01$.}
        \label{fig:veffb11p}  \end{subfigure}
\caption{\footnotesize Relative pressure as a function of the longitudinal nozzle coordinate $x$,   
 for the quantum-corrected Schwarzschild metric~\eqref{ck}, with  BH masses in the range $10^{27}$ kg $< M < 10^{36}$ kg.\clt{Relative deviation from Schwarzschild.}}
    \label{fig:pressure_a_1}

\end{figure}

The relative pressure profiles are depicted in Figs. \ref{fig:veffa1p} and \ref{fig:veffb11p} as a function of the longitudinal nozzle coordinate $x$, for distinct stellar and astrophysical masses in the range $10^{27}$ kg $< M < 10^{36}$ kg, respectively for fixed values of $c_6 = 0.005$ and $c_6=-0.01$. As shown in Fig. \ref{fig:veffa1p}, the quantum gravitational corrections are almost unapparent for  $c_6 = 0.005$ and \clt{non-stellar BHs}. Nevertheless, Fig. \ref{fig:veffb11p} represents the quantum gravitational corrections to the relative pressure profile, which are more manifest for stellar BHs and mainly near the throat \clt{(before the horizon)}, at $|x| \lesssim  0.26\,x_{\scalebox{.67}{\textsc{e}}}$.

\begin{figure}[H]
    \centering
    \begin{subfigure}{0.49\textwidth}
        \centering
        \includegraphics[width=\linewidth]{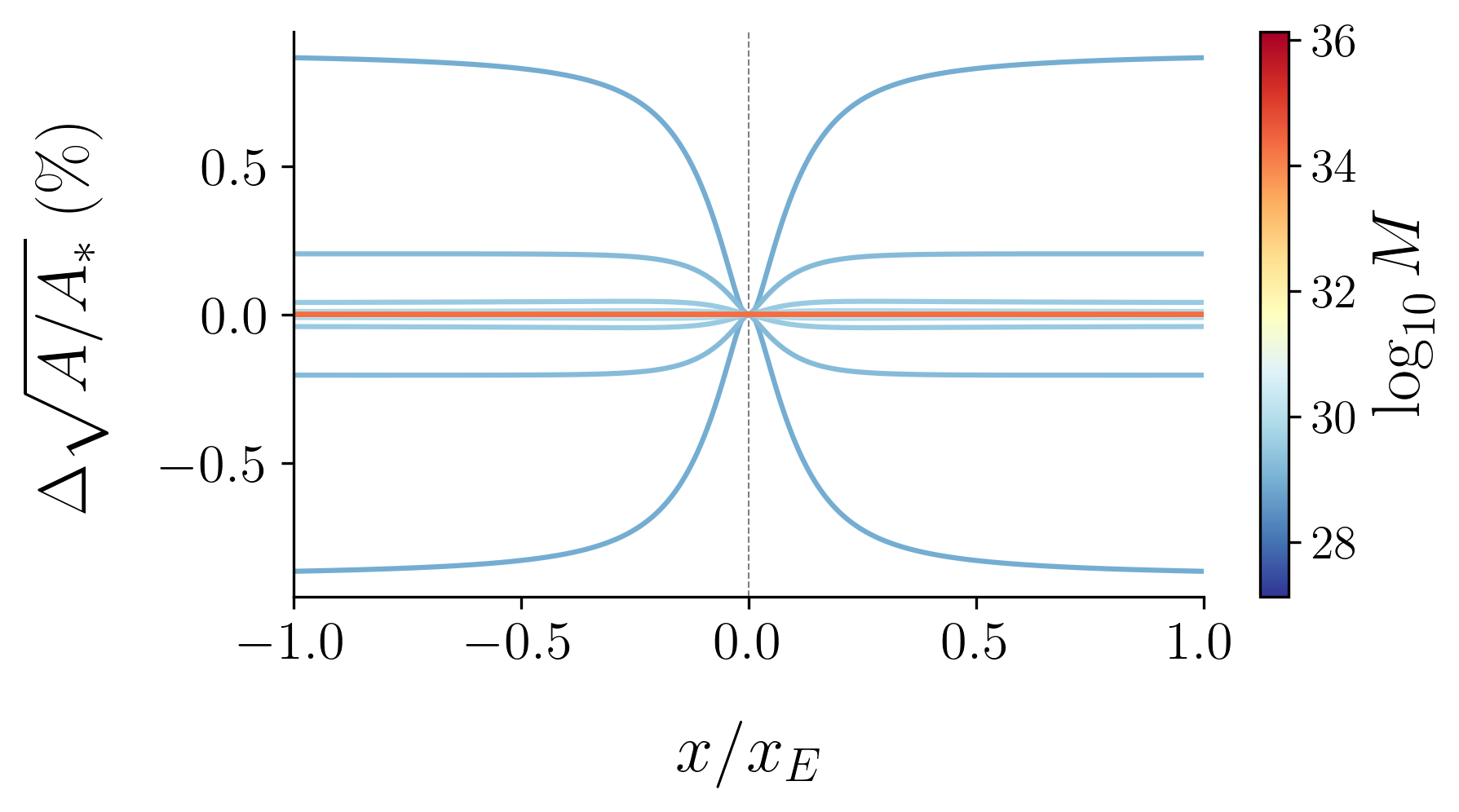}
         \caption{\footnotesize $s = \ell = 0$, $c_6 = 0.005$.}
        \label{fig:veffa1a}
    \end{subfigure}
    \begin{subfigure}{0.49\textwidth}
        \centering
        \includegraphics[width=\linewidth]{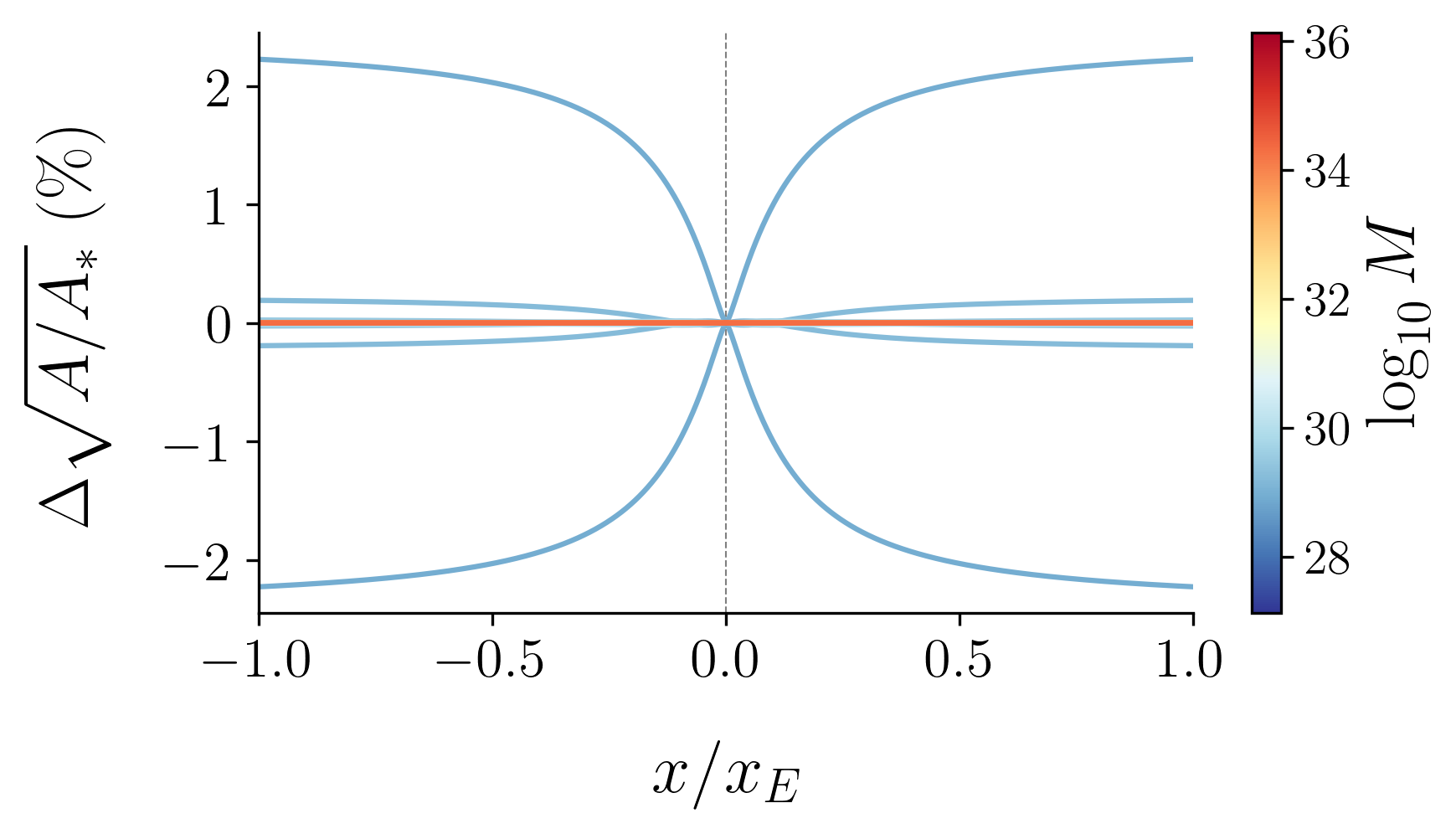}
         \caption{\footnotesize $s = 0$ and $\ell=0$, $c_6 = -0.01$.}
        \label{fig:veffb11a}  \end{subfigure}
\caption{\footnotesize Nozzle shape as a function of the longitudinal nozzle coordinate $x$, for  the quantum-corrected Schwarzschild metric~\eqref{ck}, with  BH masses in the range $10^{27}$ kg $< M < 10^{36}$ kg.\clt{Relative deviation from Schwarzschild.}}
    \label{fig:shape_a_1}

\end{figure}

The quantum gravitational corrections of the nozzle cross-sectional area are illustrated in Figs. \ref{fig:veffa1a} and \ref{fig:veffb11a} as a function of the longitudinal nozzle coordinate $x$, for astrophysical BH masses in the range $10^{27}$ kg $< M < 10^{36}$ kg, respectively for fixed values of $c_6 = 0.005$ and $c_6=-0.01$. From the physical point of view, any quantum gravitational correction in the nozzle area is derisory. 

\begin{figure}[H]
    \centering
    \begin{subfigure}{0.49\textwidth}
        \centering
        \includegraphics[width=\linewidth]{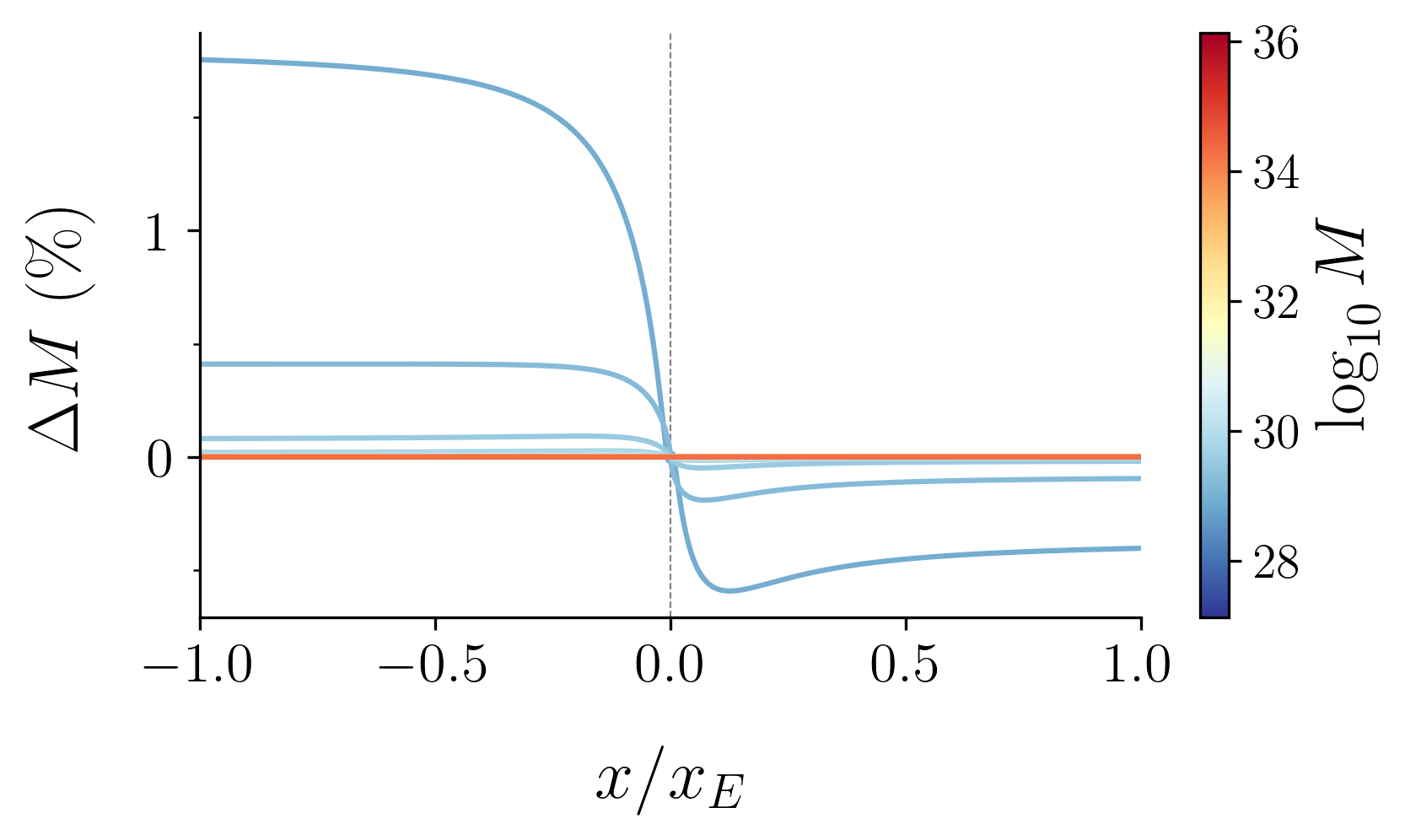}
         \caption{\footnotesize $s = \ell = 0$, $c_6 = 0.005$.}
        \label{fig:veffa1mm}
    \end{subfigure}
    \begin{subfigure}{0.49\textwidth}
        \centering
        \includegraphics[width=\linewidth]{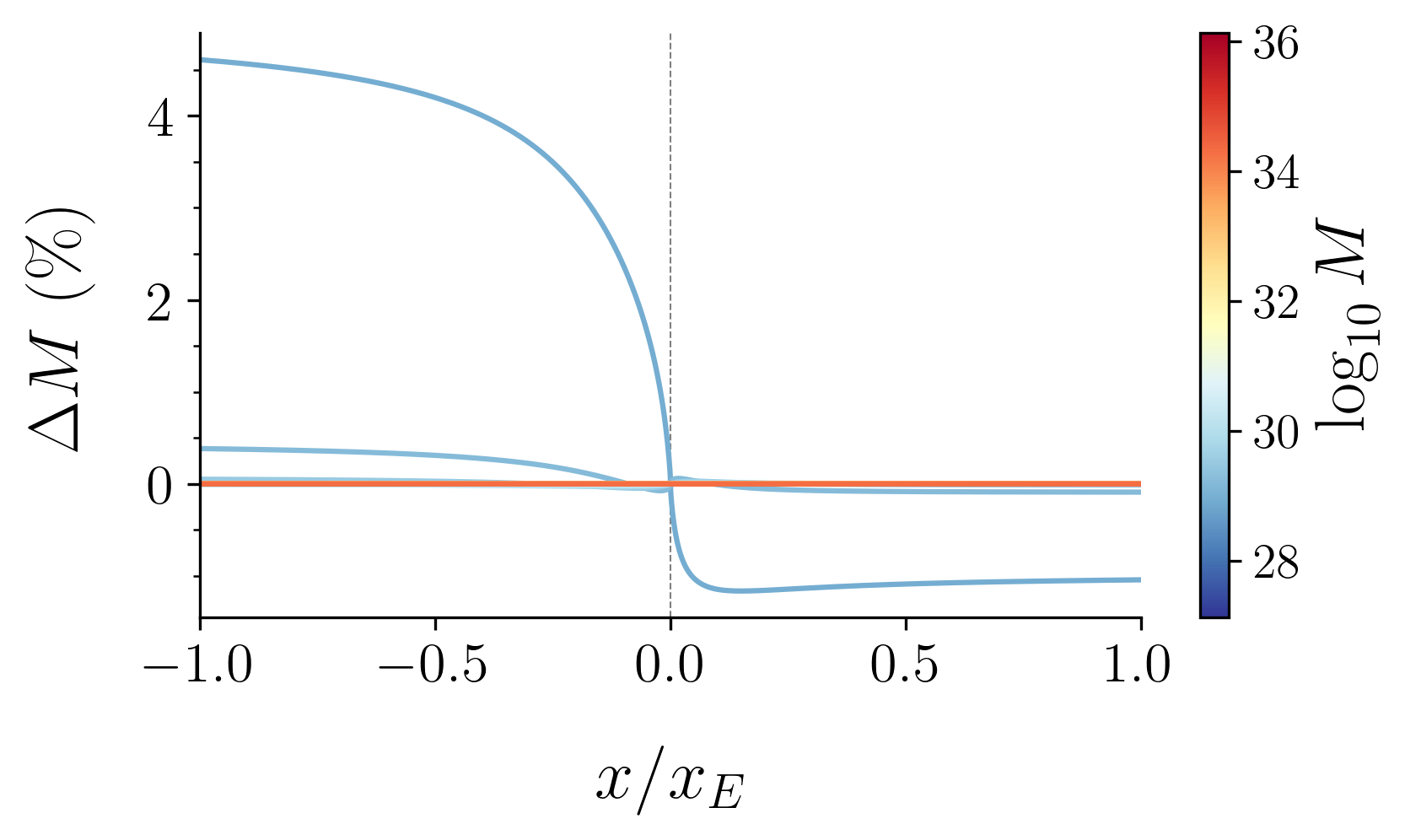}
         \caption{\footnotesize $s = 0$ and $\ell=0$, $c_6 = -0.01$.}
        \label{fig:veffb11mm}  \end{subfigure}
\caption{\footnotesize Mach number as a function of the longitudinal nozzle coordinate $x$, for the quantum-corrected Schwarzschild metric~\eqref{ck}, with  BH masses in the range $10^{27}$ kg $< M < 10^{36}$ kg.\clt{Relative deviation from Schwarzschild.}}
    \label{fig:mach_b}

\end{figure}

The Mach number is shown in Figs. \ref{fig:veffa1mm} and \ref{fig:veffb11mm} as a function of the longitudinal nozzle coordinate $x$, for distinct astrophysical masses in the range $10^{27}$ kg $< M < 10^{36}$ kg, respectively for fixed values of $c_6 = 0.005$ and $c_6=-0.01$. Fig. \ref{fig:veffa1mm} shows that the quantum gravitational corrections are less apparent for  $c_6 = 0.005$, when compared to $c_6 = -0.01$ in Fig. \ref{fig:veffb11mm}, irrespectively of the BH mass. Besides, Figs. \ref{fig:veffa1mm} and \ref{fig:veffb11mm} show that the quantum gravitational corrections are more prominent for stellar BH masses. \clt{Negative values of $c_6$ change slightly more the Mach number of stellar BHs when compared to positive values of $c_6$ (1\%-3\%)}. 

\begin{figure}[H]
    \centering
    \begin{subfigure}{0.49\textwidth}
        \centering
        \includegraphics[width=\linewidth]{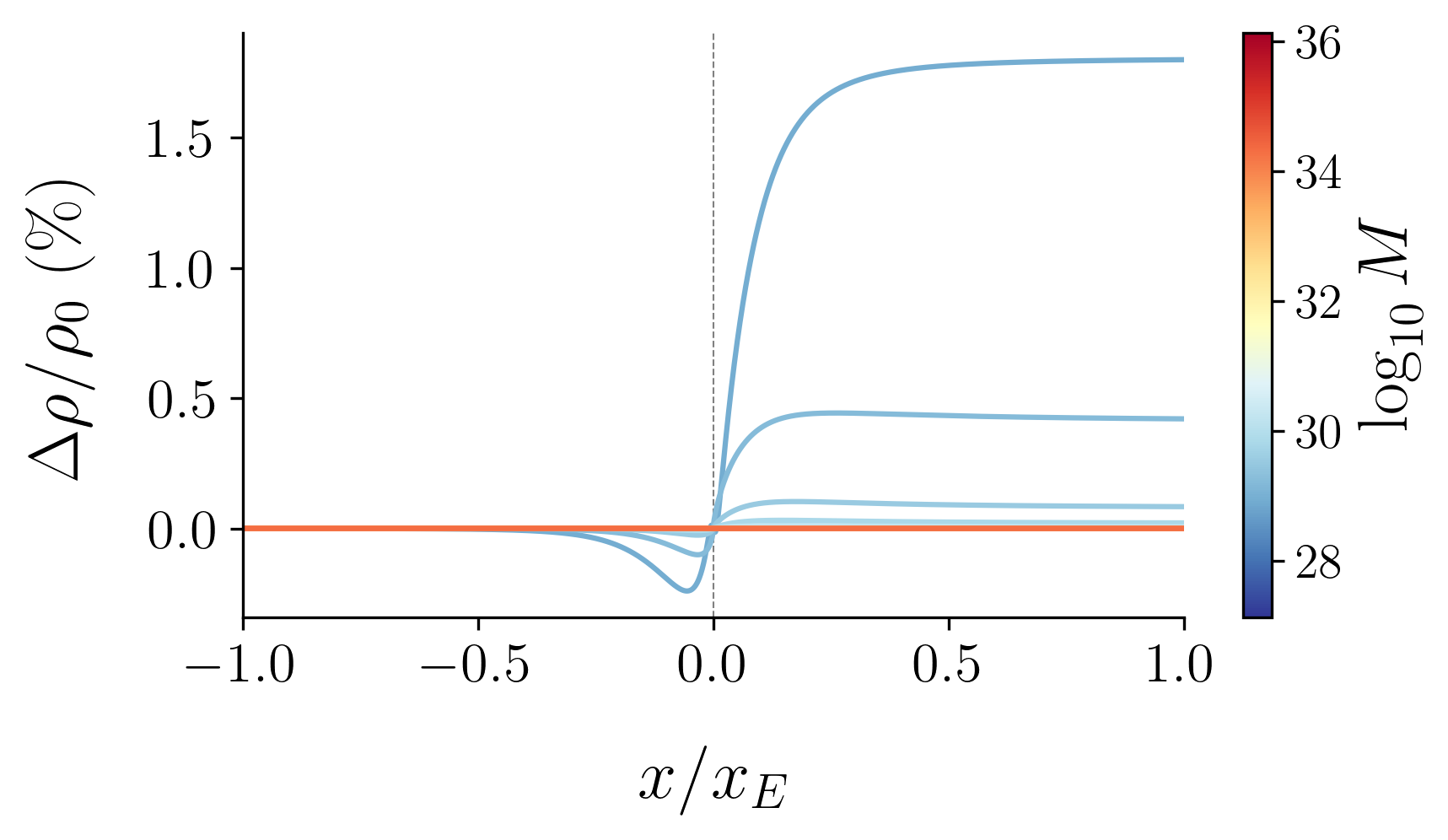}
         \caption{\footnotesize $s = \ell = 0$, $c_6 = 0.005$.}
        \label{fig:veffa1rho}
    \end{subfigure}
    \begin{subfigure}{0.49\textwidth}
        \centering
        \includegraphics[width=\linewidth]{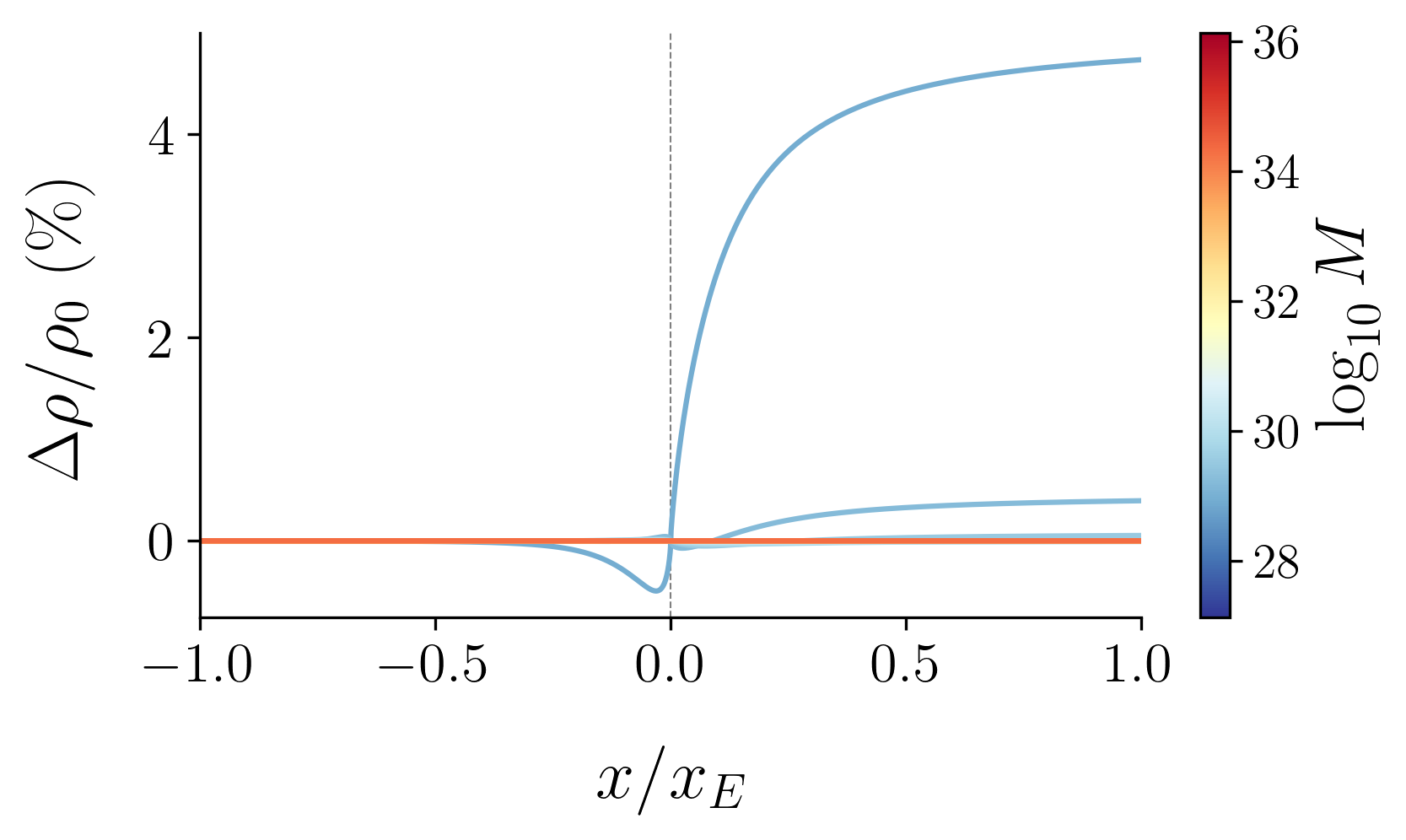}
         \caption{\footnotesize $s = 0$ and $\ell=0$, $c_6 = -0.01$.}
        \label{fig:veffb11rho}  \end{subfigure}
\caption{\footnotesize Relative density  as a function of the longitudinal nozzle coordinate $x$, for the quantum-corrected Schwarzschild metric~\eqref{ck}, with  BH masses in the range $10^{27}$ kg $< M < 10^{36}$ kg.\clt{Relative deviation from Schwarzschild.}}
    \label{fig:density_b}

\end{figure}
\noindent The relative density profiles are depicted in Figs. \ref{fig:veffa1rho} and \ref{fig:veffb11rho} as a function of the longitudinal nozzle coordinate $x$, for distinct astrophysical masses in the range $10^{27}$ kg $< M < 10^{36}$ kg, respectively for fixed values of $c_6 = 0.005$ and $c_6=-0.01$. As depicted in Fig. \ref{fig:veffa1rho}, the quantum gravitational corrections are almost unapparent for  $c_6 = 0.005$, regardless of the BH mass. However,   Fig. \ref{fig:veffb11rho} represents the quantum gravitational corrections to the relative pressure profile, which are more manifest for stellar BHs and mainly near the throat \clt{(before the horizon)}, at $|x| \lesssim  0.29\,x_{\scalebox{.67}{\textsc{e}}}$.

\begin{figure}[H]
    \centering
    \begin{subfigure}{0.49\textwidth}
        \centering
        \includegraphics[width=\linewidth]{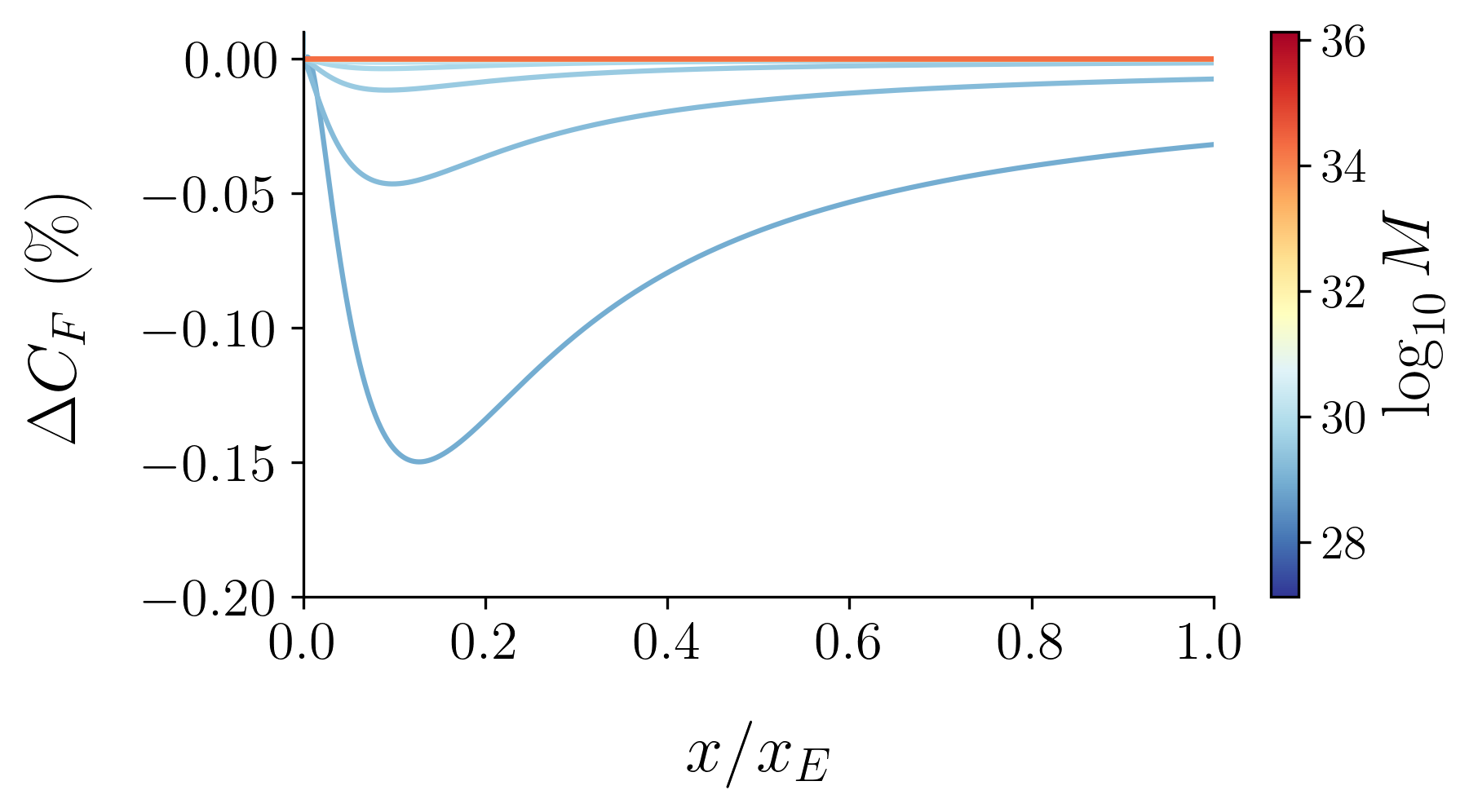}
         \caption{\footnotesize $s = \ell = 0$, $c_6 = 0.005$.}
        \label{fig:veffa1cf}
    \end{subfigure}
    \begin{subfigure}{0.49\textwidth}
        \centering
        \includegraphics[width=\linewidth]{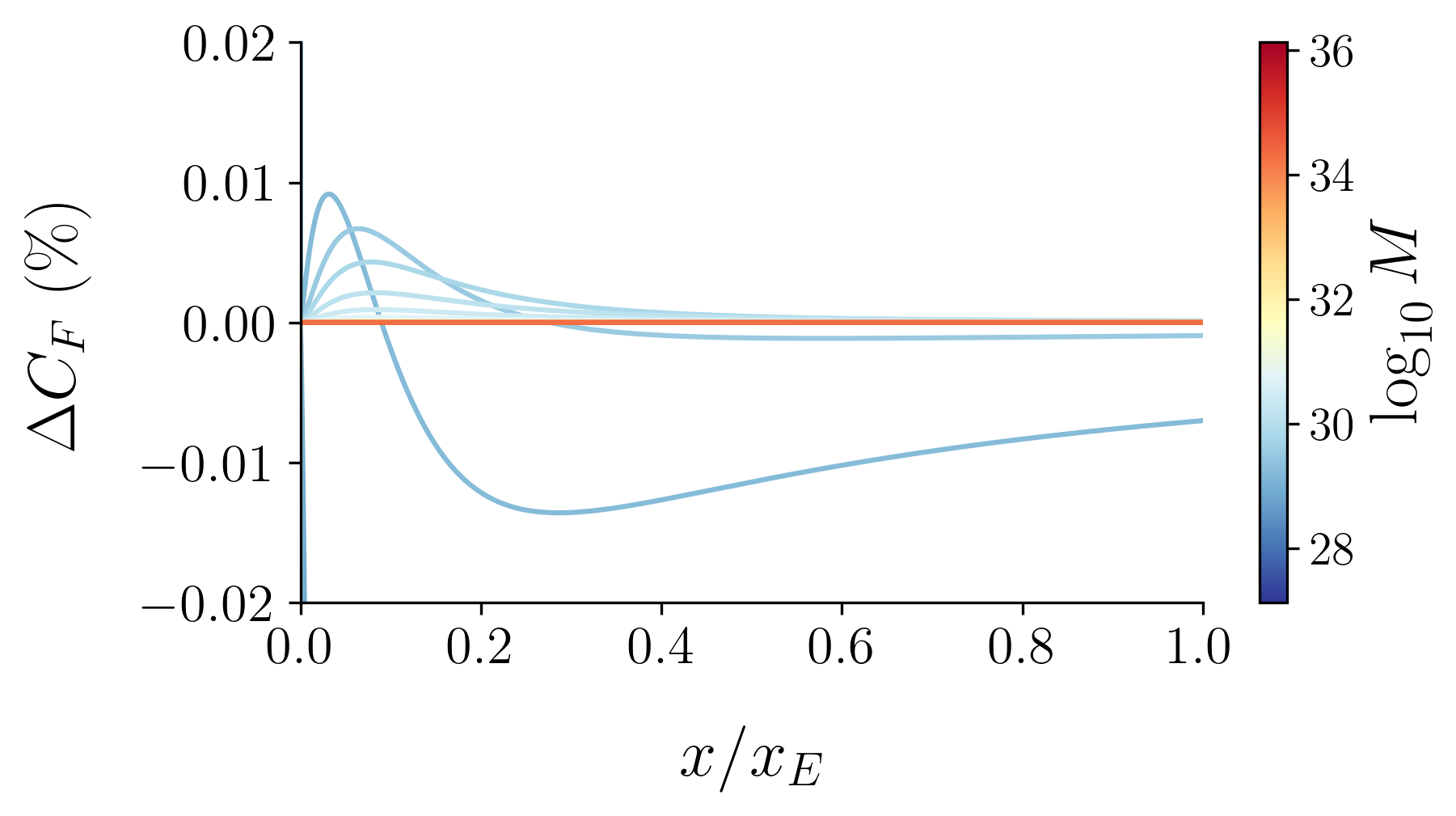}
         \caption{\footnotesize $s = 0$ and $\ell=0$, $c_6 = -0.01$.}
        \label{fig:veffb11cf}  \end{subfigure}
        \caption{\footnotesize Thrust coefficient as a function of the longitudinal nozzle coordinate $x$, for  
 for the quantum-corrected Schwarzschild metric~\eqref{ck}, with  BH masses in the range $10^{27}$ kg $< M < 10^{36}$ kg.\clt{Relative deviation from Schwarzschild.}}
    \label{fig:C_Fp_b}

\end{figure}

   Finally, the thrust coefficient is studied in Figs. \ref{fig:veffa1cf} and \ref{fig:veffb11cf} as a function of the longitudinal nozzle coordinate $x$, for distinct astrophysical masses in the range $10^{27}$ kg $< M < 10^{36}$ kg, respectively for fixed values of $c_6 = 0.005$ and $c_6=-0.01$. 
Fig. \ref{fig:veffa1cf} shows a less steep increase of the thrust coefficient as a function of the longitudinal nozzle coordinate, for stellar BH masses, when compared to astrophysical BHs. The opposite scenario is observed in Fig. \ref{fig:veffb11cf}. 

\bibliography{bib_analog_GD}

\end{document}